\documentclass[a4paper,11pt]{article}
\pdfoutput=1 

\usepackage{jheppub} 

\usepackage[T1]{fontenc} 

 \usepackage{slashed}
 
\usepackage{dsfont}

\usepackage{youngtab}

\usepackage{color}
\definecolor{darkgreen}{rgb}{0,0.5,0}
\definecolor{darkblue}{rgb}{0,0,0.6}
\definecolor{purple}{rgb}{0.4,.2,0.7}
\definecolor{pinko}{rgb}{0.7,0,0.7}

\newcommand{\rf}[1]{$\langle \ref{#1} \rangle$}

\def\CO{{\cal O}}
\def\CA{{\cal A}} 
\def\CG{{\cal G}} 
\def\CH{{\cal H}}  
\def\CZ{{\cal Z}}
\def\IC{{\mathbb C}}

\def\CM{{\cal M}}
\def\tr{\,{\rm tr}\,}
\def\CL{{\cal L}}

\def\IR{{\mathbb R}}

\def\IZ{{\mathbb Z}}

\def\IZ{{\mathbb Z}}
\def\Tr{{\rm Tr}}

\def\so{{\rm so}}
\def\hs{{\rm hs}}

\def\b{{\rm bulk}}

\def\chin{{\hat\chi}}
\def\nn{\nonumber}

\newcommand\pcom[1]{\textcolor{magenta}{#1}}

\newcommand\skp[1]{\pcom{--- xxx ---}}

\def\CD{{\cal D}}
\def\IN{{\mathbb N}}
\def\ms{\overline m_s}
\def\m1{\bar m_1}
\def\CG{{\cal G}}
\def\Vol{vol}
\def\can{{\rm c}}
\def\vg{{\rm \Vol}{(G)}_{\rm PI}}
\def\vc{{\rm \Vol}{(G)}_{\can}}
\def\CP{{\cal P}}
\def\Rac{{\rm Rac}}
\def\Di{{\rm Di}}
 
\def\CN{{\cal N}}

\def\su{{\rm su}}
\def\CN{{\cal N}}
\def\CS{{\cal S}}
\def\CF{{\cal F}}

\def\Ad{A_{d-1}}
\def\ZG{Z_G}
\def\Zchar{Z_{\rm char}}
\def\noir{\times}

\def\pol{P}
\def\inteps{\int}

\def\sp{{s'}}

\def\lint{\mbox{\Large $\int$}}
\def\smint{\mbox{\large $\int$}}

\def\gom{\bar\Omega}
\def\game{\gamma}

\def\ICP{{\mathbb{CP}}}

\def\lieg{{\mathfrak g}}

\def\lcs{l}
\def\bfhalf{{\bf \frac{1}{2}}}

\DeclareFontShape{OT1}{cmr}{mx}{n}{<->cmr10}{}

\title{\boldmath Quantum de Sitter horizon entropy from quasicanonical bulk, edge, sphere and topological string partition functions}

\author[a]{Dionysios Anninos}
\author[b]{Frederik Denef}
\author[b]{Y.T. Albert Law}
\author[b]{Zimo Sun}

\affiliation[a]{Department of Mathematics, King's College London,\\The Strand, London WC2R 2LS, U.K.}
\affiliation[b]{Department of Physics, Columbia University,\\538 West 120th Street, New York, New York 10027, U.S.A.}

\emailAdd{dionysios.anninos@kcl.ac.uk}
\emailAdd{fmd7@columbia.edu}
\emailAdd{yal2109@columbia.edu}
\emailAdd{zs2283@columbia.edu}

\abstract{Motivated by the prospect of  constraining microscopic models, we calculate the exact one-loop corrected de Sitter entropy (the logarithm of the sphere partition function) for every effective field theory of quantum gravity, with particles in arbitrary  spin representations. In doing so, we universally relate the sphere partition function to the quotient of a quasi-canonical bulk and a Euclidean edge partition function, given by integrals of characters encoding the bulk and edge spectrum of the observable universe. Expanding the bulk character splits  the bulk (entanglement) entropy 
into quasinormal mode (quasiqubit) contributions. For 3D higher-spin gravity 
formulated as an sl($n$) Chern-Simons theory, we obtain all-loop exact results. Further to this, we show that the theory has an exponentially large landscape of de Sitter vacua with quantum entropy given by the absolute value squared of a topological string partition function. For generic higher-spin gravity, the formalism succinctly 
relates dS, AdS$^\pm$ and conformal results. Holography is exhibited in 
quasi-exact bulk-edge cancelation. }


\begin{document} 
\maketitle
\flushbottom

\section{Introduction} \label{sec:intro}

\def\SE{S}

\label{sec:motivation-intro}

\begin{figure}
 \begin{center}
   \includegraphics[height=4.8cm]{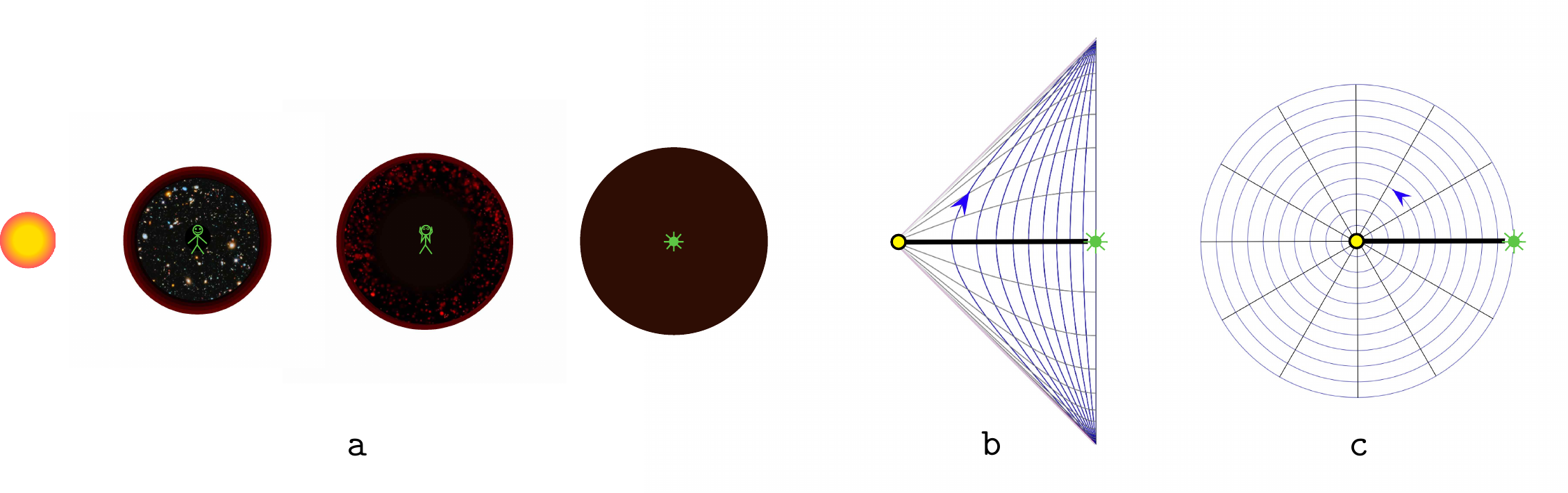} \vskip-5mm
 \caption{ \small \label{fig:dS-intro}  {\tt a:} Cartoon of observable  universe evolving to its maximal-entropy equilibrium state. The horizon consumes everything once seen, growing until it reaches its de Sitter equilibrium area $A$. (The spiky dot is a reference point for ${\tt b,c}$; it will ultimately be gone, too.)          
 {\tt b:} Penrose diagram of dS static patch. 
 {\tt c:} Wick-rotated $(\tt b)$ = sphere. Metric details are given in appendix \ref{app:dSWick} + fig.\ \ref{fig:penrose-app}{\tt c,d}. 
  }
 \end{center} \vskip-5mm
\end{figure}    

As seen by local inhabitants \cite{PhysRevD.15.2738,PhysRevD.15.2752,Spradlin:2001pw,Myers:2002mq,Bousso:2002ju,Loeb:2001dh, Krauss:2007nt} of a cosmology accelerated by a cosmological constant, 
the observable universe is evolving towards a semiclassical equilibrium state asymptotically     indistinguishable from a de Sitter static patch, enclosed by a horizon of area $A=\Omega_{d-1} \ell^{d-1}$, $\ell \propto 1/\sqrt{\Lambda}$, with the de Sitter universe globally in its Euclidean vacuum state. A picture is shown in fig.\ \ref{fig:dS-intro}{\tt b}, and the metric in (\ref{staticpatchmetric})/(\ref{dScoords})S. 
The semiclassical equilibrium state locally maximizes the observable entropy at a value $\CS$ 
semiclassically given by \cite{PhysRevD.15.2752} 
\begin{align}  \label{CSlogCZ}
 \CS = \log \CZ \, ,
\end{align} 
where $\CZ=\int  e^{-S_E[g,\cdots]}$ is the effective field theory Euclidean path integral, 
expanded about the round sphere saddle related by Wick-rotation (\ref{Scoords}) to the de Sitter universe of interest. 
At tree level in Einstein gravity, the familiar area law is recovered: 
\begin{align}\label{treeentropy}
 \CS^{(0)} = \frac{A}{4  G_{\rm N}} \, . 
\end{align}
The interpretation of $\CS$ as a (metastable) equilibrium entropy begs for a microscopic understanding of its origin. By aspirational analogy with the Euclidean AdS partition function for effective field theories with a CFT dual (see \cite{Bhattacharyya:2012ye} for a pertinent discussion), a natural question is: are there effective field theories for which the semiclassical expansion of $\CS$ corresponds to a large-$N$ expansion of a microscopic entropy? Given a proposal, how can it be tested?

In contrast to EAdS, without making any assumptions about the UV completion of the effective field theory, there is no evident extrinsic data constraining the problem.  The sphere has no boundary, all symmetries are gauged, and physically meaningful quantities must be gauge and field-redefinition invariant, leaving little.  
In particular there is no invariant information contained in the tree-level $\CS^{(0)}$ other than its value, which in the low-energy effective field theory merely represents a renormalized coupling constant; an input parameter.  
However, in the spirit of \cite{Banerjee:2010qc, Banerjee:2011jp, Sen:2012kpz, Sen:2014aja, Sen:2012dw,Bhattacharyya:2012ye,Giombi:2013fka, Giombi:2014iua, Gunaydin:2016amv, Giombi:2016pvg, Skvortsov:2017ldz,Gearhart:2010wz}, nonlocal {\it quantum} corrections to $\CS$ 
do offer unambiguous, intrinsic data, directly constraining models. To give a simple example, discussed in more detail under (\ref{CSoneloop3Dsu}), say someone posits that for pure 3D gravity, the sought-after microscopic entropy is $S_{\rm micro} = \log d(N)$, 
where $d(N)$ is the number of partitions of $N$. This is readily ruled out. Both macroscopic and microscopic entropy expansions can uniquely be brought to a form   
\begin{align} \label{comparisonexample}
 \CS = \CS_0 - a \log\CS_0 + b + \mbox{$\sum_{n}$}\, c_n \, \CS_0^{-2n}  \, + \, O(e^{-\CS_0/2}) \, ,
\end{align}   
characterized by absence of odd (=local) powers of $1/\CS^{(0)}$. The microscopic theory predicts $(a,b)=\bigl(2,\log(\pi^2/6\sqrt{3})\bigr)$,  
refuted by the macroscopic one-loop result $(a,b)=\bigl(3,5 \log(2\pi)\bigr)$. Some of the models in \cite{Maldacena:1998ih, Banados:1998tb, Bousso:2001mw, Govindarajan:2002ry, Silverstein:2003jp, Fabinger:2003gp, Parikh:2004wh, Banks:2006rx, Dong:2010pm, Heckman:2011qu, Banks:2013qpa, Neiman:2017zdr, Dong:2018cuv, Arias:2019zug} are sufficiently detailed to be tested along these lines.

In this work, we focus exclusively on collecting macroscopic data, 
more specifically the exact one-loop (in some cases all-loop) corrected $\CS = \log \CZ$.  
The problem is old, and computations for $s\leq 1$ are relatively straightforward, but for higher spin $s\geq2$, sphere-specific complications crop up. Even for pure gravity \cite{Hawking:1976ja,Gibbons:1978ji,Christensen:1979iy,Fradkin:1983mq,Yasuda:1983hk,Allen:1983dg,Polchinski:1988ua,Taylor:1989ua,Vassilevich:1992rk,Volkov:2000ih}, virtually no complete, exact results have been obtained at a level brining tests of the above kind to their full potential.

Building on results and ideas from \cite{Witten:1995gf,Donnelly:2013tia,Donnelly:2014fua,Donnelly:2015hxa,Giombi:2013yva,Tseytlin:2013fca,Joung:2013nma,Joung:2014qya,Sleight:2016dba,Sleight:2016xqq,Basile:2016aen,Basile:2018zoy,Basile:2018acb}, we obtain a universal formula solving this problem in general, for all $d \geq 2$ parity-invariant effective field theories, with matter in arbitrary representations, and general gauge symmetries including higher-spin: 
\begin{equation} \label{mainresult}
  \boxed{\CS^{(1)}   \, = \,  \log \prod_{a=0}^K \frac{\bigl(2 \pi \gamma_a)^{{\rm dim} \, G_a}}{{\rm vol} \, G_a} \, + \, \int_0^\infty \frac{dt}{2t} \biggl( \frac{1+q}{1-q} \,\, \chi^{\rm bos}_{\rm tot}   - \frac{2  \sqrt{q}}{1-q} \,\, 
 \chi^{\rm fer}_{\rm tot}  \biggr)  +  \CS_{\rm ct}}  
\end{equation}   
$q \equiv e^{-t/\ell}$. Below we explain the ingredients in sufficient detail to allow application in practice. A sample of explicit results is listed in (\ref{explexamplestab}). We then summarize the content of the paper by section, with more emphasis on the physics and other results of independent interest.

$G_0$ is the subgroup of (possibly higher-spin) gravitational gauge transformations acting trivially on the $S^{d+1}$ saddle. This includes rotations of the sphere. ${\rm vol} \, G_0$ is the volume for the 
invariant metric normalized such that the standard rotation generators have unit norm, implying in particular ${\rm vol} \, SO(d+2) =  \mbox{(\ref{vcSO})}$.  
The other $G_i$, $i=1,\ldots,K$ are Yang-Mills group factors, with ${\rm vol} \, G_i$ the volume in the metric defined by the trace in the action, as in (\ref{volsuN}). The $\gamma_a$ are proportional to the (algebraically defined) gauge couplings:     
\begin{align} 
 \gamma_0 \equiv \sqrt{\frac{8\pi G_{\rm N}}{A_{d-1}}} = \sqrt{\frac{2 \pi}{\CS^{(0)}}} 
 \, ,  \qquad \qquad \gamma_i \equiv \sqrt{\frac{g_i^2}{2 \pi A_{d-3}}}  \, , 
\end{align} 
with $A_{n} \equiv \Omega_{n} \ell^n$, $\Omega_n = \mbox{(\ref{volSn})}$ for $n \geq 0$, and $A_{-1} \equiv 1/2\pi\ell$ for $\gamma_i$ in $d=2$.  

The functions $\chi_{\rm tot}(t)$ 
are determined by the bosonic/fermionic physical particle spectrum of the theory. They take the form of a ``bulk'' {\it minus} an ``edge'' character: 
\begin{align} \label{chitotchibulkminuschiedge}
 \chi_{\rm tot} = \chi_{\rm bulk} - \chi_{\rm edge} \, .
\end{align}  
The bulk character $\chi_{\rm bulk}(t)$ is defined as follows. Single-particle states on global dS$_{d+1}$ furnish a representation $R$ of the isometry group $SO(1,d+1)$. The content of $R$ is encoded in its Harish-Chandra character $\tilde\chi(g) \equiv {\rm tr}\, R(g)$ (appendix \ref{app:chars}).  
Restricted to $SO(1,1)$ isometries $g=e^{-itH}$ acting as time translations on the static patch, $\tilde\chi(g)$ becomes  $\chi_{\rm bulk}(t) \equiv {\rm tr} \, e^{-i t H}$.
For example for a massive integer spin-$s$ particle it is given by (\ref{chimassivespins}):
\begin{align} \label{massivebulkschar}
 \chi_{\rm bulk,s} = D_s^d \, \frac{q^{\frac{d}{2}+ i \nu} + q^{\frac{d}{2} - i \nu}}{(1-q)^d} \, ,   \qquad \quad q \equiv e^{-|t |/\ell} \, ,
\end{align} 
where $D_s^d$ is the spin degeneracy (\ref{Dsods2}), e.g.\ $D_s^3 = 2s+1$, 
and $\nu$ is related to the mass: 
\begin{equation} \label{numrel}
 s=0: \, \nu^2 = m^2\ell^2-\bigl(\tfrac{d}{2}\bigr)^2  \, , \quad \qquad s \geq 1: \, \nu^2 = m^2\ell^2-\bigl(\tfrac{d}{2}+s-2\bigr)^2  \, .
\end{equation} 
For arbitrary massive matter $\chi_{\rm bulk}$ is given by (\ref{chigenrep}).  
Massless spin-$s$ characters are more intricate, but can be obtained by applying a simple ``flipping'' recipe (\ref{prevsqbrackplusop}) to (\ref{benaive}), or from the general formulae (\ref{excserieschi}) or (\ref{flippedchargenPM}) derived from this. Some low $(d,s)$  examples are 
\begin{equation} \small
\begin{array}{r|ccccccccc}
 (d,s) && (2,1) & (2,2) &&(3,1) & (3,2) && (4,1) & (4,2)  \\
 \hline \\[-5mm]
 \chi_{\rm bulk,s}  
 & 
 & \displaystyle \frac{2 \, q}{(1-q)^2}
 &  0 
 &
 & \displaystyle \frac{6 \, q^2 - 2 \, q^3}{(1-q)^3} 
 & \displaystyle \frac{10 \, q^3 -6 \, q^4}{(1-q)^3}
 &
 & \displaystyle \frac{6 \, q^2}{(1-q)^4}   
 & \displaystyle \frac{10 \, q^2}{(1-q)^4} 
\end{array} \label{bulkcharexmplsspin2}
\end{equation}  
The $q$-expansion of $\chi_{\rm bulk}$ gives the static patch {\it quasinormal} mode degeneracies, its Fourier transform gives the {\it normal} mode spectral density, and the bulk part of (\ref{mainresult}) is the quasicanonical ideal gas partition function at $\beta=2\pi\ell$, as we explain  below (\ref{rhofromchibulk}).  

The edge character $\chi_{\rm edge}(t)$ is inferred from path integral considerations in sections \ref{sec:scalars}-\ref{sec:massless}. It vanishes for spin $s<1$. For integer $s \geq 1$ we get (\ref{chibulkedgeprev}): 
\begin{align} \label{chiedgemassspins}
 \chi_{\rm edge,s} =  N_s \cdot \frac{q^{\frac{d-2}{2}+ i \nu} + q^{\frac{d-2}{2} - i \nu}}{(1-q)^{d-2}} \, , \qquad N_s = D_{s-1}^{d+2} \, ,
\end{align}
e.g.\ $N_1=1$, $N_2=d+2$. Note this is the bulk character of $N_s$ scalars in {\it two lower} dimensions. Thus the edge correction effectively {\it subtracts} the degrees of freedom of $N_s$ scalars living on $S^{d-1}$, 
 the horizon ``edge'' of static time slices (yellow dot in fig.\ \ref{fig:dS-intro}). (\ref{ZPIGENMASS}) yields analogous results for more general matter; e.g.\ ${\tiny \Yvcentermath1 \yng(1,1,1)}$ bulk field  $\to$ ${\tiny \Yvcentermath1 \yng(1,1)}$  edge field, ${\tiny \Yvcentermath1 \yng(2,1)}$ bulk $\to (d+2) \times {\tiny \Yvcentermath1 \yng(1)}$ edge. 
For massless spin-$s$, use (\ref{benaive})-(\ref{prevsqbrackplusop}) or  (\ref{flippedchargenPMedge}). The edge companions of (\ref{bulkcharexmplsspin2}) are 
\begin{equation} \small
\begin{array}{r|ccccccccc}
 (d,s)  && (2,1) & (2,2) && (3,1) & (3,2) && (4,1)  & (4,2)  \\
 \hline \\[-5mm]
 \chi_{\rm edge,s} 
 &  
 & 0
 & 0 
 &
 & \displaystyle \frac{2 \, q}{1-q}
 & \displaystyle \frac{10  \, q^2- 2  \, q^3}{1-q}
 &
 & \displaystyle \frac{2 \, q}{(1-q)^2}  
 & \displaystyle \frac{10 \, q}{(1-q)^2} 
\end{array} \label{edgecharexmplsspin2}
\end{equation}   
The edge correction extends observations of 
\cite{Susskind:1994sm,Kabat:1995eq,Kabat:1995jq,Larsen:1995ax,Dowker:2010bu,Dowker:2010yj,Casini:2011kv,Solodukhin:2011gn,Eling:2013aqa,Donnelly:2014fua,Donnelly:2015hxa,Casini:2015dsg,Buividovich:2008gq, Donnelly:2011hn, Casini:2013rba, Soni:2015yga, Dong:2018seb, Jafferis:2015del,Blommaert:2018rsf}, reviewed in appendix \ref{sec:edgecor}. 

The general closed-form evaluation of the integral in (\ref{mainresult}) is given by (\ref{ZEXACT}) in heat kernel regularization. In even $d$, the finite part is more easily obtained by summing residues.  

Finally, $\CS_{\rm ct}$ in (\ref{mainresult}) is a local counterterm contribution fixed by a 
renormalization condition specified in section \ref{sec:gravETD},
which in practice boils down to $\CS_{\rm ct}(\ell)$ canceling all divergences and finite terms growing polynomially with $\ell$ in $\CS^{(1)}(\ell)$. 

\begin{figure}
 \begin{center}
   \includegraphics[height=3.8cm]{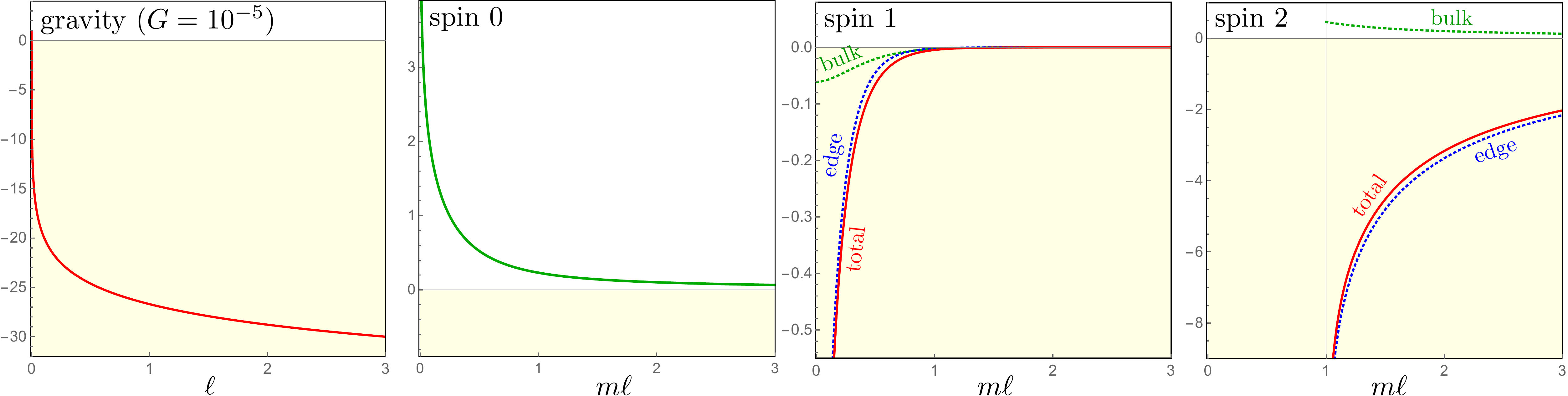} \vskip-3mm
 \caption{ \small \label{fig:Z1plots-intro} Contributions to dS$_3$ one-loop entropy from gravity and massive $s=0,1,2$. 
  }
 \end{center} \vskip-6mm
\end{figure}  

\noindent For concreteness here are some examples readily obtained from (\ref{mainresult}): \vskip-6mm
\def\arraystretch{1.25}
\begin{equation} \small
\begin{array}{r|l}
 \mbox{content} & \CS^{(1)}   \\
 \hline 
  \mbox{3D grav} 
  &  -3 \log \CS^{(0)} + 5 \log(2\pi) \\
  \mbox{3D$\,(s,m)$} 
  & \frac{\pi}{3}\bigl(\nu^3 - (m\ell)^3 + \frac{3 (s-1)^2}{2} m \ell \bigr)
 -2  \sum_{k=0}^{2} \frac{\nu^k}{k!} \, \frac{{\rm Li}_{3-k}(e^{-2 \pi \nu})}{(2\pi)^{2-k}} - s^2 \bigl(\pi (m\ell-\nu) -\log (1-e^{-2 \pi \nu})\bigr) \\
 \mbox{4D grav} 
 & - 5 \log \CS^{(0)} -\frac{571}{45} \log (\ell/L) -\log \frac{8\pi}{3} + \frac{715}{48} -\frac{47}{3} \zeta'(-1) + \frac{2}{3} \zeta'(-3) \\ 
\mbox{5D$\,$su(4)$\,$ym} 
 & -\frac{15}{2} \log (\ell/g^2) - \log \frac{256  \pi^9}{3} \, +
  \frac{75 \, \zeta(3)}{16 \, \pi ^2} + \frac{45 \, \zeta (5)}{16 \, \pi ^4} \\
 \mbox{5D$\,(\, {\tiny \Yvcentermath1 \yng(2,2)}\,,m)$}  
 & -15 \log (2 \pi  m \ell )+\frac{5 \, \zeta (5)}{8 \pi ^4}+\frac{65 \, \zeta (3)}{24 \pi ^2}  \quad (m\ell \to 0) \, , \qquad \frac{5}{12} (m\ell)^4 \, e^{-2 \pi m \ell} \quad (m\ell \to \infty) \\
\mbox{11D grav} 
& -33 \log \CS^{(0)} + \log\bigl( \frac{4! \, 6! \, 8! \, 10!}{2^4} (2\pi)^{63} \bigr) 
+\frac{1998469 \, \zeta (3)}{50400 \,\pi ^2}+\frac{135619 \,\zeta (5)}{60480 \,\pi ^4}-\frac{34463 \,\zeta
   (7)}{3840 \,\pi^6}+\frac{11 \,\zeta (9)}{6 \pi ^8}-\frac{11 \,\zeta (11)}{256 \,\pi ^{10}}\\
\mbox{3D HS$_n$} 
& -(n^2-1) \log \CS^{(0)} + \log\bigl[\frac{1}{n}\bigl(\frac{n(n^2-1)}{6} \bigr)^{n^2-1}  
 \, {\tt G}(n+1)^2 \,
 (2\pi)^{(n-1)(2n+1)}\bigr]
\end{array} \label{explexamplestab} 
\end{equation} \vskip-2mm \noindent
Comparison to previous results for 3D and 4D gravity is discussed under (\ref{ZPIEinst}).\footnote{In the above and in (\ref{mainresult}) we have dropped Polchinski's phase \cite{Polchinski:1988ua} kept in (\ref{ZPIEinst}) and generalized in (\ref{ZPIFINAL}).} 
 
The second line is the contribution of a 3D massive spin-$s$ field, with $\nu$ given by (\ref{numrel}). 
The term $\propto$ $s^2$ is the edge contribution. It is  negative for all $m\ell$ and dominates the bulk contribution (fig.\ \ref{fig:Z1plots-intro}). It diverges  at the unitarity/Higuchi bound $m\ell = s-1$. 
\def\arraystretch{1}

In the 4D gravity example, $L$ is a minimal subtraction scale canceling out of $\CS^{(0)} + \CS^{(1)}$. 
In this case, constant terms in $\CS^{(1)}$ cannot be distinguished from constants in $\CS^{(0)}$ and are as such physically ambiguous.\footnote{Comparing different saddles, unambiguous linear combinations can however be extracted, cf.\ (\ref{lincombinv}).} The term $\alpha_4 \log(\ell/L)$ with $\alpha_4=-\frac{571}{45}$ arises from the log-divergent term $\alpha_4 \log(\ell/\epsilon)$ of the regularized character integral. 

For any $d$, in any theory, the coefficient $\alpha_{d+1}$ of the log-divergent term can simply be read off from the $t \to 0$ expansion of the integrand in (\ref{mainresult}): 
\begin{equation}
 \mbox{integrand} = \cdots + \frac{\alpha_{d+1}}{t} + O(t^0)   
\end{equation}\vskip-1mm
\noindent For a 4D photon, this gives $\alpha_4 = \alpha_{4,\rm bulk} + \alpha_{4,\rm edge} = -\tfrac{16}{45} - \tfrac{1}{3} = -\frac{31}{45}$. 
The bulk-edge split in this case is the same as the split investigated in \cite{Dowker:2010bu,Casini:2015dsg,Soni:2016ogt}. Other illustrations include (partially) massless  spin $s$ around (\ref{gravalphas}), the superstring in (\ref{susystring}), and conformal spin $s$ in (\ref{CHSalphas}).   

3D HS$_n$ = higher-spin gravity with $s=2,3,\ldots,n$ (section \ref{sec:HSCS}). ${\tt G}$ is the Barnes $G$-function. 

\subsection*{Overview}

We summarize the content of sections \ref{sec:thermal}-\ref{sec:CHSgr}, 
highlighting other results of interest, beyond (\ref{mainresult}).  

\subsubsection*{Quasicanonical bulk thermodynamics of the static patch \textnormal{(section \ref{sec:thermal})}}

\begin{figure}
 \begin{center}
   \includegraphics[height=3.5cm]{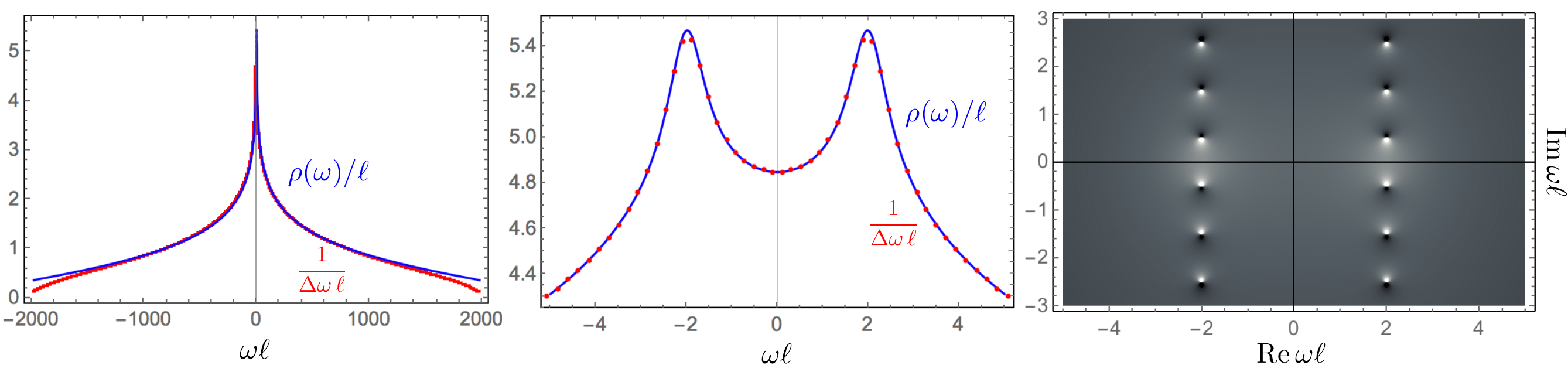} \vskip-3mm
 \caption{\small  Regularized dS$_2$ scalar mode density with $\nu=2$, $\Lambda_{\rm uv} \ell \approx 4000$. Blue line = Fourier transform of $\chi_{\rm bulk}$: $\rho(\omega)/\ell = \frac{2}{\pi} \log(\Lambda_{\rm uv}\ell) -\frac{1}{2\pi} \sum \psi\bigl(\tfrac{1}{2} \pm i \nu \pm i\omega \ell)$. Red dots = inverse eigenvalue spacing of numerically diagonalized $4000 \times 4000$ matrix $H$ in globally truncated model (appendix \ref{sec:DOS}). Rightmost panel = $|\rho(\omega)|$ on complex $\omega$-plane, with quasinormal mode poles at $\omega \ell=\pm i (\frac{1}{2} \pm i\nu + n)$.  
  \label{fig:2dpoles}}
\end{center} \vskip-4mm
\end{figure} 

\begin{figure} 
\begin{center}
   \includegraphics[height=7.9cm]{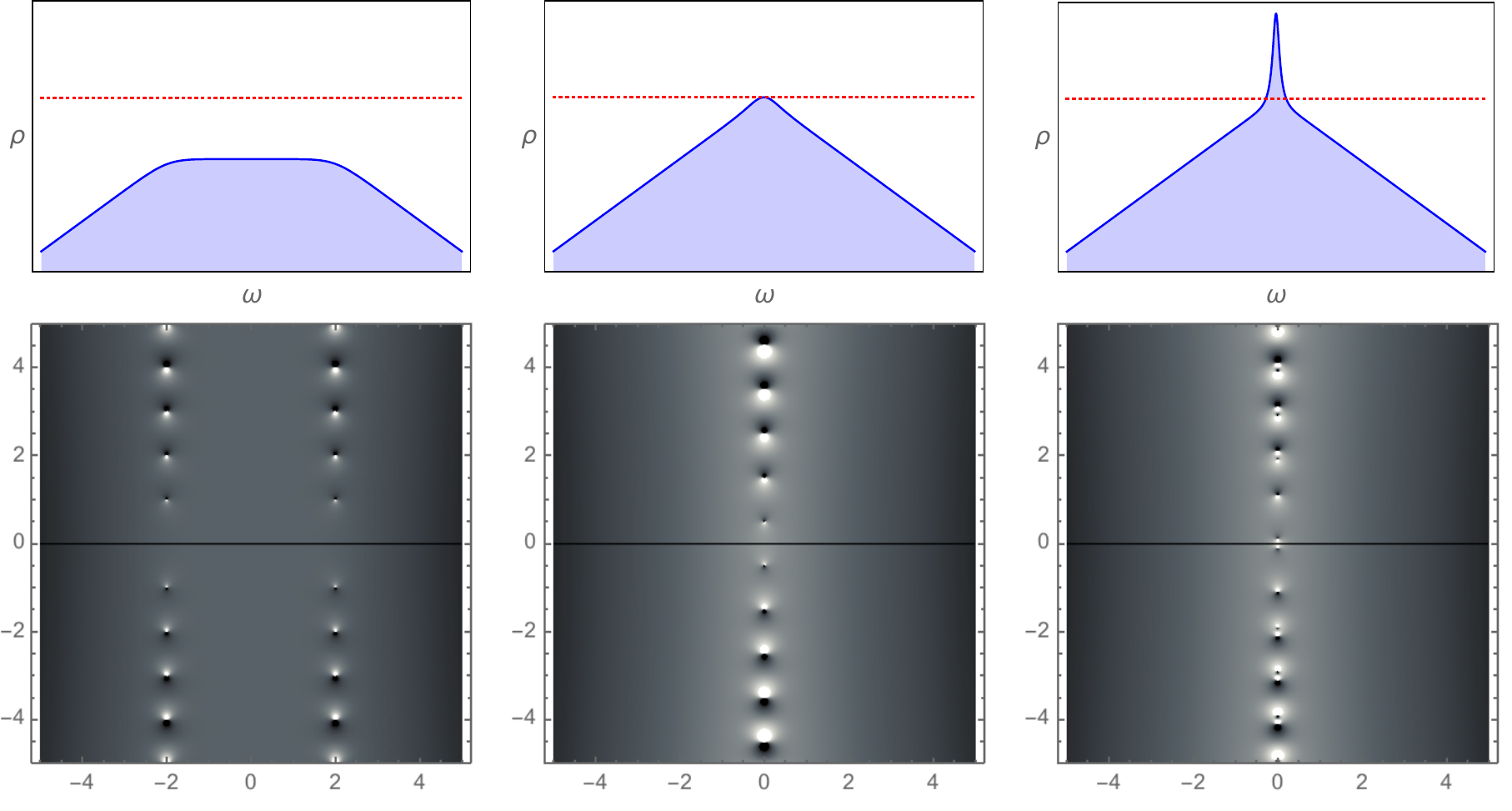}
 \caption{\small Regularized scalar $\rho(\omega)$, $d=2$, $\nu=2,i/2,0.9 \, i$; top: $\omega \in \IR$; bottom: $\omega \in \IC$, showing quasinormal mode poles. See figs.\ \ref{fig:dosdS3}, \ref{fig:poles} for details. 
  \label{fig:summaryDOS}} 
\end{center} \vskip-5mm
\end{figure}

The global dS bulk character $\chi_{\rm bulk}(t) = {\rm tr} \, e^{-i t H}$ locally encodes the quasinormal spectrum and normal mode density of the static patch $ds^2=-(1-r^2/\ell^2) dT^2 + (1-r^2/\ell^2)^{-1} dr^2 + r^2 d\Omega^2$ on which $e^{-itH}$ acts as a time translation $T \to T + t$. 
Its expansion in powers of $q = e^{-|t|/\ell}$, 
\begin{align} \label{qnmexp}
 \chi_{\rm bulk} = \sum_r \, N_r \, q^r \, ,
\end{align} \vskip-2mm \noindent
yields the number $N_r$ of {\it quasinormal} modes decaying as $e^{-r T/\ell}$, in resonance with \cite{Ng:2012xp,Jafferis:2013qia,Sun:2020sgn}. 
The density of {\it normal} modes $\propto e^{-i \omega T}$ is formally given by its Fourier transform 
\begin{align} \label{rhofromchibulk}
 \rho(\omega) \equiv \frac{1}{2\pi} \int_{-\infty}^\infty dt \, \chi_{\rm bulk}(t) \, e^{i \omega t} .  
\end{align} 
Because $\chi_{\rm bulk}$ is singular at $t=0$, this is ill-defined as it stands. However, a standard Pauli-Villars regularization of the QFT  renders it regular (\ref{PVchar}), yielding a manifestly covariantly regularized mode density, 
analytically calculable for arbitrary particle content, including gravitons and higher-spin matter. Some simple examples are shown in figs.\ \ref{fig:2dpoles}, \ref{fig:summaryDOS}. Quasinormal modes appear as resonance poles at $\omega = \pm i r$, seen by substituting (\ref{qnmexp}) into (\ref{rhofromchibulk}). 
  
This effectively solves the problem of making covariant sense of the formally infinite normal mode density universally arising in the presence of a horizon \cite{tHooft:1984kcu}. 
Motivated by the fact that semiclassical information loss can be traced back to this infinity, \cite{tHooft:1984kcu} introduced a rough model getting rid of it by shielding the horizon by a ``brick wall'' (reviewed together with variants in \ref{sec:brickwall}). 
Evidently this alters the physics, introduces boundary artifacts, breaks covariance, and is, unsurprisingly, computationally cumbersome.  The covariantly regularized density (\ref{rhofromchibulk}) suffers none of these problems.

\noindent In particular it makes sense of the {\it a priori} ill-defined canonical ideal gas partition function, 
\begin{align} \label{bulkig}
  \log Z_{\rm can}(\beta)  =  \int_0^\infty d\omega \Bigl( -\rho_{\rm bos}(\omega) \, \log\bigl(e^{\beta \omega/2} - e^{-\beta \omega/2}\bigr) +  \rho_{\rm fer}(\omega) \, \log\bigl(e^{\beta \omega/2} + e^{-\beta \omega/2} \bigr) \Bigr) \, .
\end{align} 
Substituting (\ref{rhofromchibulk}) and integrating out $\omega$, this becomes 
\begin{align} \label{ZthNSchar2prev-intro}
 \boxed{ \log Z_{\rm bulk}(\beta)  =  \int_{0}^\infty \frac{dt}{2t} \biggl( \, \frac{1+e^{-2 \pi t/\beta}}{1-e^{-2 \pi t/\beta}}   \, \, \chi^{\rm bos}_{\rm bulk}(t) \, - \,  \frac{2 \,  e^{-\pi t/\beta}}{1-e^{-2 \pi t/\beta}}  \, \, \chi_{\rm bulk}^{\rm fer}(t) \biggr) } 
\end{align}
At the static patch equilibrium $\beta = 2 \pi \ell$, this is precisely the {\it bulk} contribution to the one-loop Euclidean partition function $\log \CZ^{(1)}$  in (\ref{mainresult}). Although $Z_{\rm bulk}$ is not quite a  standard canonical partition function, calling it a quasicanonical partition function appears apt.  

From (\ref{ZthNSchar2prev-intro}), covariantly regularized quasicanonical bulk thermodynamic quantities  
can be analytically computed for general particle content, as illustrated in section \ref{sec:enentr}. Substituting the expansion (\ref{qnmexp}) 
expresses these quantities as a sum of quasinormal mode contributions, generalizing and refining \cite{Denef:2009kn}. In particular the contribution to the entropy and heat capacity from each physical quasinormal mode is finite and positive (fig.\ \ref{fig:QNMsc}).    

$S_{\rm bulk}$ can alternatively be viewed as a covariantly regularized  entanglement entropy between two hemispheres in the global dS Euclidean vacuum (red and blue lines in figs.\ \ref{fig:penrose}, \ref{fig:penrose-app}). 
In the spirit of \cite{Jafferis:2013qia}, the quasinormal modes can then be viewed as entangled quasiqubits.

\subsubsection*{Sphere partition functions \textnormal{(sections \ref{sec:scalars},\ref{sec:mashsflds},\ref{sec:massless})}}

In sections \ref{sec:scalars}-\ref{sec:massless} we obtain character integral formulae computing exact heat-kernel regularized one-loop sphere partition functions $Z_{\rm PI}^{(1)}$ for general field content, leading to 
(\ref{mainresult}). 

For scalars and spinors (section \ref{sec:scalars}), this is easy. 
For massive spin $s$ (section \ref{sec:mashsflds}), the presence of conformal Killing tensors on the sphere imply naive reduction to a spin-$s$ Laplacian determinant is inconsistent with locality \cite{Tseytlin:2013fca}. The correct answer can in principle be obtained by path integrating the full off-shell action \cite{Zinoviev:2001dt}, but this involves an intricate tower of spin $s'<s$ Stueckelberg fields. Guided by intuition from section \ref{sec:thermal}, we combine locality and unitary constraints with path integral considerations to find the terms in $\log Z$ missed by naive reduction. They turn out to be obtained simply by extending the spin-$s$ Laplacian eigenvalue sum to include its ``subterranean'' levels with formally negative degeneracies, (\ref{ZPItrueformula}). The extra terms capture contributions from unmatched spin $s'<s$ conformal Killing tensor ghost modes in the gauge-fixed Stueckelberg path integral. The resulting sum  yields the bulk$-$edge character integral formula (\ref{formuhs2}). Locality and unitarity uniquely determine the generalization to arbitrary parity-symmetric matter representations, (\ref{ZPIGENMASS}).

In the massless case (section \ref{sec:massless}), new subtleties arise: negative modes requiring contour rotations (which translate into the massless character ``flipping'' recipe 
mentioned  above (\ref{bulkcharexmplsspin2})), and ghost zeromodes which must be omitted and compensated by a carefully normalized group volume division.   
Non-universal factors cancel out, yielding (\ref{ZPIFINAL}) modulo renormalization. 
\subsubsection*{3D de Sitter HS$_n$ quantum gravity and the topological string \textnormal{(section \ref{sec:HSCS})}}

\begin{figure}
 \begin{center}
   \includegraphics[height=3.3cm]{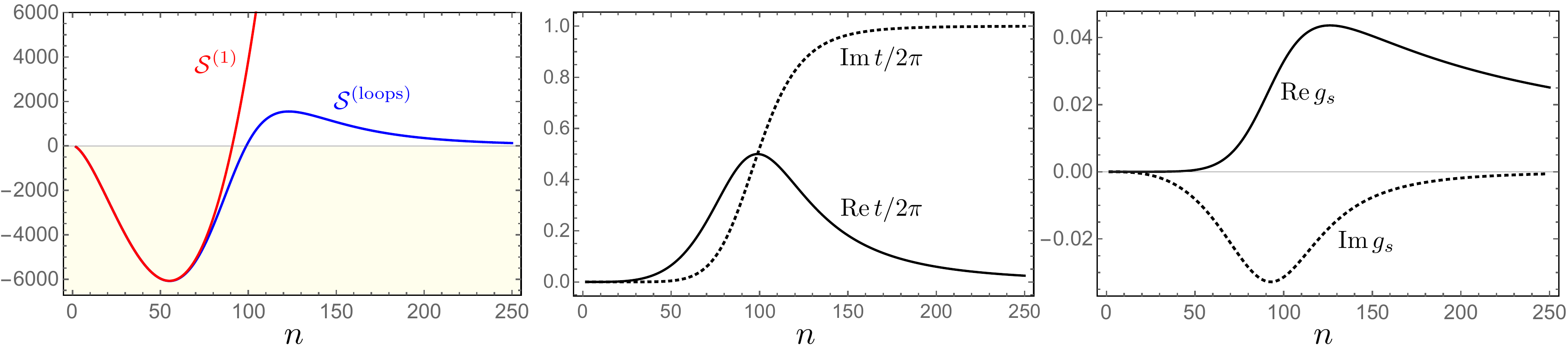} \vskip-2mm
 \caption{ \small \label{fig:CSplots} One- and all-loop entropy corrections, and dual topological string $t$, $g_s$, for 3D  HS$_n$ theory in its maximal-entropy de Sitter vacuum, for different values of $n$ at fixed $\CS^{(0)}=10^8$, $\lcs=0$.   
  }
 \end{center} \vskip-5mm
\end{figure}

The ${\rm sl}(2)$ Chern-Simons formulation of 3D gravity  \cite{Witten:1988hc,Witten:1989ip} can be extended to an ${\rm sl}(n)$ Chern-Simons formulation of $s \leq n$ higher-spin (HS$_n$) gravity \cite{Blencowe:1988gj,Bergshoeff:1989ns,Castro:2011fm,Perlmutter:2012ds}. The action for positive cosmological constant is given by (\ref{actionSL3Dg}). It 
has a real coupling constant $\kappa \propto 1/G_{\rm N}$, and an integer coupling constant $\lcs \in \{0,1,2,\ldots\}$ if a gravitational Chern-Simons term is included. 

This theory has a landscape of dS$_3$ vacua, labeled by partitions $\vec m=\{m_1,m_2,\ldots\}$ of $n$. Different vacua have different values of $\ell/G_{\rm N}$, with tree-level entropy 
\begin{align}
   \CS^{(0)}_{{\vec m}}  = \frac{2\pi \ell}{4G_{\rm N}}\biggr|_{{\vec m}} = 2 \pi \kappa \cdot T_{\vec m} \, , \qquad T_{\vec m} = \mbox{$\frac{1}{6}\sum_a m_a(m_a^2-1)$} \, . 
\end{align} 
The number of vacua grows as $\CN_{\rm vac} \sim e^{2\pi \sqrt{n/6}}$. The maximal entropy vacuum is $\vec m=\{n\}$. 

We obtain the all-loop exact quantum entropy $\CS_{{\vec m}}=\log \CZ_{\vec m}$ 
by analytic continuation $k_\pm \to \lcs \pm i \kappa$  of the  $SU(n)_{k_+} \times SU(n)_{k_-}$ Chern-Simons partition function on $S^3$, (\ref{Zkappaexactsumm}). 
In the weak-coupling limit $\kappa \to \infty$, this reproduces $\CS^{(1)}$ as computed by (\ref{mainresult}) in the metric-like formulation of the theory,   given in (\ref{explexamplestab}) for the maximal-entropy vacuum $\vec m=\{n\}$. 

When $n$ grows large and reaches a value $n \sim \kappa$, the 3D higher-spin  gravity theory becomes strongly coupled. (In the vacuum $\vec m = \{n\}$ this means $n^4 \sim \ell/G_{\rm N}$.) In this regime, Gopakumar-Vafa duality \cite{Gopakumar:1998ki,Marino:2004uf} can be used to express the quantum de Sitter entropy $\CS$ in terms of a weakly-coupled topological string partition function on the resolved conifold, (\ref{ZRtoprel}):
\begin{align} 
 \boxed{\CS_{\vec m} =  \log \left|\tilde Z_{\rm top}(g_s,t) \, e^{-\pi T_{\vec m} \cdot 2 \pi i/g_s} \right|^2 } 
\end{align} 
where $g_s=\frac{2\pi}{n+\lcs+i\kappa}$, and the conifold K\"ahler modulus $t \equiv  \int_{S^2} J + i B = i g_s n = \frac{2 \pi i n}{n+\lcs+i\kappa}$. 


\subsubsection*{Euclidean thermodynamics of the static patch   \textnormal{(section \ref{sec:all-loop})}}

In section \ref{sec:all-loop} we consider the  Euclidean thermodynamics of a QFT on a fixed static patch/sphere background. 
The partition function $Z_{\rm PI}$ is the Euclidean path integral on the sphere of radius $\ell$, the Euclidean energy density is $\rho_{\rm PI} = -\partial_V \log Z_{\rm PI}$, where $V=\Omega_{d+1} \ell^{d+1}$ is the volume of the sphere, and the entropy is   
$S_{\rm PI}=\log Z_{\rm PI} + 2 \pi \ell \, U_{\rm PI} = \log Z_{\rm PI} + V \rho_{\rm PI} = (1-V\partial_V ) \log Z_{\rm PI}$, or 
\begin{align} \label{Eucldefs}
  \boxed{S_{\rm PI} = \bigl(1-\tfrac{1}{d+1} \ell \partial_\ell \bigr) \log Z_{\rm PI}}
\end{align}     
Using the exact one-loop sphere partition functions obtained in sections \ref{sec:scalars}-\ref{sec:massless}, this allows general exact computation of the one-loop Euclidean entropy $S_{\rm PI}^{(1)}$, illustrated in section \ref{fixedexamples}. Euclidean Rindler results are recovered in the limit $m\ell \to \infty$. The sphere computation avoids  introducing the usual conical deficit angle, varying the curvature radius $\ell$ instead. 


For minimally coupled scalars, $S_{\rm PI}^{(1)}=S_{\rm bulk}$, but more generally this is false, due to edge (and other) corrections. Our results thus provide a precise and general version of observations 
made in the work reviewed in appendix \ref{sec:edgecor}. 
Of note, these ``corrections'' actually {\it dominate} the one-loop entropy, rendering it negative, increasingly so as $s$ grows large. 

\subsubsection*{Quantum gravitational thermodynamics \textnormal{(section \ref{sec:gravETD})}}

In section \ref{sec:gravETD} (with details in appendix \ref{app:rendS}), we specialize to theories with dynamical gravity.  Denoting $Z_{\rm PI}$, $\rho_{\rm PI}$ and $S_{\rm PI}$ by $\CZ$, $\varrho$ and $\CS$ in this case,  (\ref{Eucldefs}) trivially implies $\varrho=0$, $\CS = \log \CZ$, reproducing (\ref{CSlogCZ}). All UV-divergences can be absorbed into  renormalized coupling constants, rendering the Euclidean thermodynamics well-defined in an 
effective field theory sense. 

Integrating over the geometry is similar in spirit to integrating over the temperature in statistical mechanics, as one does to extract the microcanonical entropy $S(U)$ from the canonical partition function.\footnote{Along the lines of $S(U) = \log \bigl( \frac{1}{2 \pi i}  \int \frac{d\beta}{\beta} \, {\rm Tr} \, e^{-\beta H + \beta U}   \bigr)$, with contour $\beta = \beta_* + i y$, $y \in \IR$, for any $\beta_*>0$. 
}
The analog of this in the case of interest is    
\begin{align} \label{Srho-intro}
 S(\rho) \equiv \log \int \CD g \, \cdots \, e^{-S_E[g,\ldots]  \, + \, \rho \int \! \sqrt{g} } \, ,
\end{align}  
for some suitable metric path integration contour. In particular $S(0)=\CS$. The analog of the microcanonical $\beta \equiv \partial_U S$ is $V \equiv \partial_\rho S$, 
and the analog of the microcanonical free energy is the Legendre transform $\log Z \equiv  S - V \rho$, satisfying $\rho=-\partial_V \log Z$. 
If we furthermore {\it define} $\ell$ by $\Omega_{d+1} \ell^{d+1} \equiv V$, the relation between $\log Z$, $\rho$ and $S$ is by construction identical to  (\ref{Eucldefs}).

Equivalently, the free energy $\Gamma \equiv - \log Z$ can be thought of as a quantum effective action for the volume. 
At tree level, $\Gamma$ equals the classical action $S_E$ evaluated on the round sphere of radius $\ell$.      
For example for 3D Einstein gravity,  
\begin{align}
 \log Z^{(0)} = \frac{2\pi^2}{8 \pi G} \bigl( -\Lambda \, \ell^3 + 3 \, \ell \bigr) \, , \qquad S^{(0)} = \bigl(1 - \tfrac{1}{3} \ell \partial_\ell \bigr) \log Z^{(0)} = \frac{2 \pi \ell}{4G} \, .
\end{align}  
The tree-level on-shell radius $\ell_0$ maximizes $\log Z^{(0)}$, i.e.\   $\rho^{(0)}(\ell_0) = 0$. 

We define renormalized $\Lambda, G, \ldots$ from the $\ell^{d+1},\ell^{d-1},\ldots$ coefficients in the $\ell \to \infty$  expansion of the quantum $\log Z$, and fix counterterms by equating tree-level and renormalized couplings for the UV-sensitive subset. For 3D Einstein, the renormalized one-loop correction is  
\begin{align}
 \log Z^{(1)} =  - 3 \log \frac{2 \pi \ell}{4G} + 5 \log(2\pi) \,. 
\end{align}    
The {\it quantum} on-shell radius $\bar\ell = \ell_0 + O(G)$ maximizes $\log Z$, i.e.\ $\rho(\bar\ell) = 0$.  
The on-shell entropy 
can be expressed in two equivalent ways to this order: 
\begin{align} \label{Stwoways}
 \boxed{\CS = S^{(0)}(\bar\ell) + S^{(1)} = S^{(0)}(\ell_0) + \log Z^{(1)}} 
\end{align}
This clarifies why the one-loop correction $\CS^{(1)} \equiv \CS - \CS^{(0)}$ to the dS entropy is given by  $\log Z^{(1)}$ rather than $S^{(1)}$: the extra term $-V \rho^{(1)}$ accounts for the change in entropy of the reservoir (= geometry) due to energy transfer to the system (= quantum fluctuations).  

The final result is (\ref{mainresult}).
We work out several examples in detail. We consider higher-order curvature corrections and discuss invariance under local field redefinitions, identifying  the invariants $\CS_M^{(0)}=-S_E[g_M]$ for different saddles $M$ as and their large-$\ell$ expanded quantum counterparts $\CS_M$ as the $\Lambda>0$ analogs  of tree-level and quantum scattering amplitudes, defining invariant couplings and physical observables of the low-energy effective field theory.          

\subsubsection*{dS, AdS$^\pm$, and conformal higher-spin gravity \textnormal(section \ref{sec:CHSgr})}

Massless $\lieg = \hs(\so(d+2))$ higher-spin gravity theories on dS$_{d+1}$ or $S^{d+1}$ \cite{Vasiliev:1990en, Vasiliev:2003ev, Bekaert:2005vh} have infinite spin range and infinite $\dim \lieg$, obviously posing problems for the one-loop  formula (\ref{mainresult}):
\begin{enumerate}
 \item \vskip-2mm Spin sum divergences untempered by the UV cutoff, for example $\dim G = \frac{1}{3} \sum_s s(4s^2-1)$ for $d=3$ and $\chi_{\rm tot} = \sum_s \mbox{{\small $(2s+1)$}} \frac{2 q^2}{(1-q)^4} - \sum_s \frac{s(s+1)(2s+1)}{6} \frac{2 q}{(1-q)^2}$ for $d=4$. 
 \item \vskip-2mm Unclear how to make sense of ${\rm vol} \, G$. 
\end{enumerate} 
\vskip-2mm\noindent We compare the situation to analogous one-loop expressions \cite{Basile:2018zoy,Basile:2018acb,Sun:2020ame} for Euclidean AdS with standard (AdS$^+_{d+1}$) \cite{Giombi:2013fka, Giombi:2014iua, Gunaydin:2016amv, Giombi:2016pvg, Skvortsov:2017ldz} and alternate (AdS$^-_{d+1}$) \cite{Giombi:2013yva} gauge field  boundary conditions, and to the associated conformal higher-spin theory on the boundary $S^d$ (CHS$_d$) \cite{Tseytlin:2013fca,Tseytlin:2013jya}. For AdS$^+$ the above problems are absent, as $\lieg$ is not gauged and $\Delta_s > s$. Like a summed KK tower, the spin-summed bulk character has increased UV dimensionality $d^{\rm bulk}_{\rm eff}=2d-2$. However, the edge character almost completely cancels this, leading to a {\it reduced} $d_{\rm eff} = d-1$ in (\ref{chiAdSstdbe})-(\ref{chi-int}). This realizes a version of a stringy picture painted in \cite{Susskind:1994sm} repainted in fig.\ \ref{fig:worldsheets}. A  HS ``swampland'' is identified: lacking a holographic dual, characterized by $d_{\rm eff}>d-1$.

For AdS$^-$ and CHS, the problems listed for dS all reappear. $\lieg$ is gauged, and the character spin sum divergences are identical to dS, as implied by the relations (\ref{confSdAdSdS}): 
\begin{align} \label{confSdAdSdSintro}
 \boxed{\chi_{s}({\rm CHS}_d) = \chi_s({\rm AdS}_{d+1}^-) - \chi_s({\rm AdS}_{d+1}^+)
 = \chi_s({\rm dS}_{d+1}) - 2 \, \chi_s({\rm AdS}_{d+1}^+) }
\end{align}    
The spin sum divergences are not UV. Their origin lies in low-energy features: an infinite number of quasinormal modes decaying as slowly as $e^{-2T/\ell}$ for $d \geq 4$ (cf.\ discussion below (\ref{QNMcountml})). We see no justification for zeta-regularizing such divergences away. However, in certain supersymmetric extensions, the spin sum divergences cancel in a rather nontrivial way, leaving a finite residual as in (\ref{dis4cancelsusy}). This eliminates problem 1, but leaves problem 2. Problem 2 might be analogous to ${\rm vol} \, G=\infty$ for the bosonic string or ${\rm vol} \, G = 0$ for supergroup Chern-Simons: removed by appropriate insertions. This, and more, is left to future work.

\newpage

\section{Quasicanonical bulk thermodynamics} 

\label{sec:thermal}

\begin{figure}
 \begin{center}
   \includegraphics[height=5.3cm]{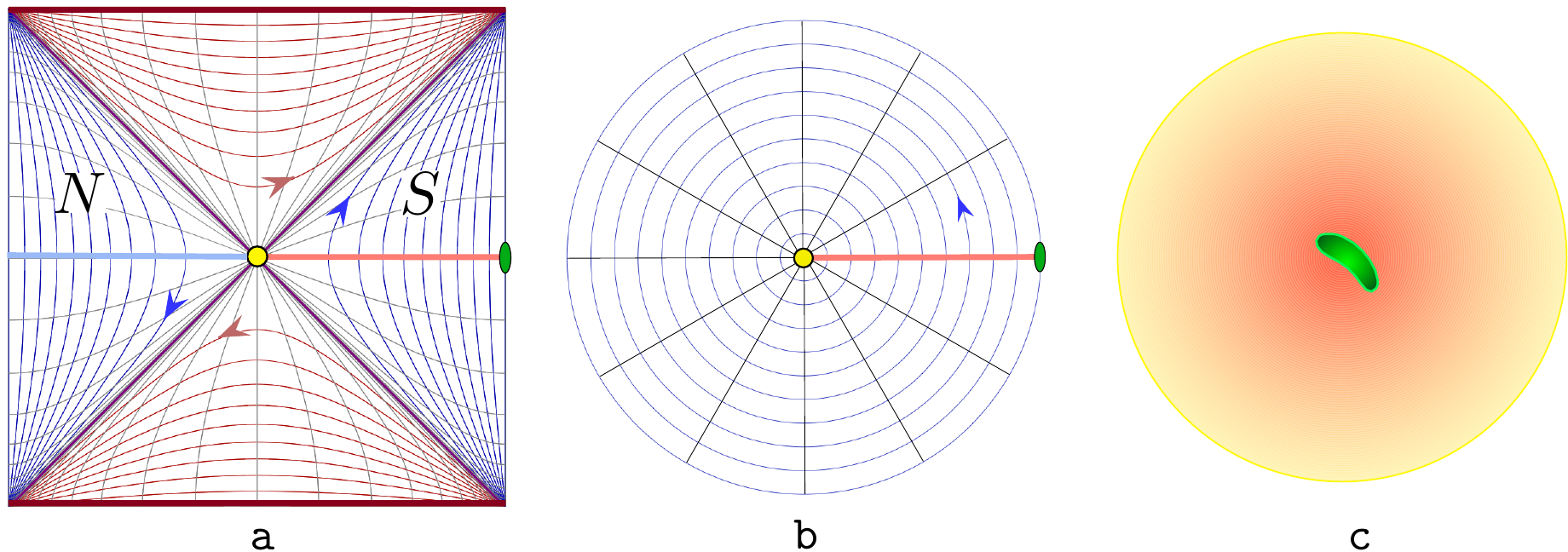}
 \caption{ \small \label{fig:penrose}  {\tt a:} Penrose diagram of global dS, showing  flows of $SO(1,1)$ generator $H=M_{0,d+1}$,  S = southern static patch. {\tt b:} Wick-rotated $S$ = sphere; Euclidean time = angle.  {\tt c:} {\it Pelagibacter ubique} inertial observer in dS with $\ell = 1.2 \, \mu{\rm m}$ finds itself immersed in gas of photons, gravitons and higher-spin particles at a pleasant $30 {}^\circ {\rm C}$. More details are provided in fig.\ \ref{fig:penrose-app} and appendix \ref{app:dSWick}.
  }
 \end{center} \vskip-5mm
\end{figure}

\subsection{Problem and results}

\label{sec:claims}

From the point of view of an inertial observer, such as {\it Pelagibacter~ubique} in fig.\ \ref{fig:penrose}{\tt c}, the global de Sitter vacuum appears thermal \cite{PhysRevD.15.2738,PhysRevD.15.2752,Israel:1976ur}: {\it P.~ubique}  perceives
its universe, the southern static patch (S in fig.\ \ref{fig:penrose}{\tt a}), 
\begin{equation} \label{staticpatchmetric}
 ds^2=-(1-r^2/\ell^2) dT^2 + (1-r^2/\ell^2)^{-1} dr^2 + r^2 d\Omega_{d-1}^2,
\end{equation}
as a static ball of finite volume, whose boundary $r=\ell$ is a horizon at  temperature $T=1/2 \pi \ell$, and whose bulk is populated by field quanta in thermal equilibrium with the horizon.  
{\it P.~ubique} wishes to understand its universe, and figures the easiest thing to understand should be the thermodynamics of its thermal environment in the ideal gas approximation. The partition function of an ideal gas is    
\begin{align} \label{ZIGnogees}
 {\rm Tr} \, e^{-\beta H}  = \exp \int_0^\infty d\omega \Bigl( -
  \rho(\omega)_{\rm bos} \,\log  \bigl( e^{\beta \omega/2} - e^{-\beta \omega/2} \bigr)  + \rho(\omega)_{\rm fer} \, \log \bigl( e^{\beta \omega/2}+e^{-\beta \omega/2} \bigr) \Bigr) ,
\end{align}
where $\rho(\omega) = \rho(\omega)_{\rm bos} + \rho(\omega)_{\rm fer}$ is the density   of bosonic and fermionic single-particle states at energy $\omega$. However to its dismay, it immediately runs into trouble: the dS static patch mode spectrum is continuous and infinitely degenerate, leading to a pathologically divergent density $\rho(\omega) = \delta(0) \sum_{\ell m \cdots}$. It soon realizes the unbounded redshift is to blame, so it imagines a brick wall excising the horizon, or some variant thereof (appendix \ref{sec:brickwall}). Although this allows some progress, it is aware this alters what it is computing and depends on choices. To check to what extent this matters, it tries to work out nontrivial examples. This turns out to be painful. It feels there should be a better way, but its efforts come to an untimely end \cite{pubiqueend}.

Here  we will make sense of the density of states and the static patch bulk thermal partition function in a different way, manifestly preserving the underlying symmetries, allowing general exact results for arbitrary particle content. 
The main ingredient is the Harish-Chandra group character  
(reviewed in appendix \ref{app:chars}) of the $SO(1,d+1)$ representation $R$ furnished by the physical single-particle Hilbert space of the free QFT quantized on {\it global} dS$_{d+1}$. Letting $H$ be the global $SO(1,1)$ generator acting as time translations in the southern static patch and globally as in fig.\ \ref{fig:penrose}{\tt a}, the  character restricted to group elements $e^{-i t H}$ is
\begin{align} \label{HCdef1}
  \chi(t) \equiv {\rm tr}_G \, e^{-i t H} \, . 
\end{align}
Here ${\rm tr}_G$ traces over the {\it global} dS single-particle Hilbert space furnishing $R$. (More generally we denote ${\tr}$ $\equiv$ single-particle trace, ${\rm Tr}$ $\equiv$ multi-particle trace, $G$ $\equiv$ global, $S$ $\equiv$ static patch. Our default units set the dS radius $\boxed{\ell \equiv 1}$.) 

For example for a scalar field of mass $m^2=(\frac{d}{2})^2+\nu^2$, as computed  in  (\ref{chi0delta}),
\begin{align} \label{scachi}
  \chi(t) = \frac{e^{-t \Delta_+}+e^{-t \Delta_-}}{\left|1-e^{-t}\right|^d} \, , \qquad \Delta_\pm = \tfrac{d}{2} \pm i \nu \, .
\end{align} 
For a massive spin-$s$ field this simply gets an additional spin degeneracy factor $D_s^d$, (\ref{chimassivespins}). Massless spin-$s$ characters take a similar but somewhat more intricate form, (\ref{excserieschi})-(\ref{spin1casechis}).

As mentioned in the introduction, (\ref{qnmexp}), the character has a series expansion 
\begin{align} \label{QNMexpchi1}
 \chi(t) = \sum_r N_r \, e^{-r |t|}  
\end{align}
encoding the degeneracy $N_r$ of quasinormal modes $\propto e^{-r T}$ of the dS static patch background. For example expanding the scalar character yields two towers of quasnormal modes with $r_{n \pm}  = \frac{d}{2} \pm i \nu +n$ and degeneracy $N_{n \pm} = {n+d-1 \choose n}$. 

Our main result, shown in \ref{sec:thermalder} below, is the observation that 
\begin{align} \label{ZthNSchar2prev}
 \boxed{\log Z_{\rm bulk}(\beta)  \equiv  \int_{0}^\infty \frac{dt}{2t} \biggl( \, \frac{1+e^{-2 \pi t/\beta}}{1-e^{-2 \pi t/\beta}}   \, \, \chi(t)_{\rm bos} \, - \,  \frac{2 \,  e^{-\pi t/\beta}}{1-e^{-2 \pi t/\beta}}  \, \, \chi(t)_{\rm fer} \biggr)} \, , 
\end{align}
suitably regularized,
provides a physically sensible, manifestly covariant regularization of the static patch bulk thermal partition.  The basic idea is  that $\rho(\omega)$ can be obtained as a well-defined Fourier transform of the covariantly UV-regularized character $\chi(t)$, which upon substitution in the ideal gas formula (\ref{ZIGnogees}) yields the above character integral formula.  
Arbitrary thermodynamic quantities at the horizon equilibrium $\beta=2\pi$ can be extracted from this in the usual way, for example $S_{\rm bulk} = (1-\beta \partial_\beta) \log Z_{\rm bulk}|_{\beta=2\pi}$, 
which can alternatively be interpreted as the ``bulk'' entanglement entropy between the northern and southern $S^d$ hemispheres (red and blue lines fig.\ \ref{fig:penrose}{\tt a}).\footnote{In part because subregion entanglement entropy does not exist in the continuum, an infinity of different notions of it exist in the literature \cite{Lin:2018bud}. Based on  \cite{Soni:2016ogt}, $S_{\rm bulk}$ appears perhaps most akin to the ``extractable''/``distillable'' entropy considered there. Either way, our results are nomenclature-independent.   
}  We work out various  examples of such thermodynamic quantities in section \ref{sec:enentr}. General exact solution are easily obtained. The expansion (\ref{QNMexpchi1}) also allows interpreting the results as a sum over quasinormal modes along the lines of \cite{Denef:2009kn}. 

We conclude this part with some comments on the relation with the Euclidean partition function. As reviewed in appendix \ref{app:thermalZPI}, general physics considerations, or formal considerations based on Wick-rotating the static patch to the sphere  and slicing the sphere path integral along the lines of fig.\ \ref{fig:penrose}{\tt b}, suggests a relation between the one-loop Euclidean path integral $Z_{\rm PI}^{(1)}$ on $S^{d+1}$ and the bulk ideal gas thermal partition function $Z_{\rm bulk}$ at $\beta=2\pi$. More refined considerations suggest 
\begin{align} \label{ZPIexpectZb}
 \log Z^{(1)}_{\rm PI} = \log Z_{\rm bulk} + \mbox{edge corrections} \, ,
\end{align}
where the edge corrections are associated with the $S^{d-1}$ horizon edge of the static patch time slices, i.e.\ the yellow dot in fig.\ \ref{fig:penrose}. The formal slicing argument breaks down here, as does the underlying premise of spatial separability of local field degrees of freedom (for fields of spin $s \geq 1$). Similar considerations apply to other thermodynamic quantities and in other contexts, reviewed in appendix \ref{app:thermalZPI} and more specifically \ref{sec:edgecor}. 

In sections \ref{sec:scalars}-\ref{sec:massless} we will obtain the exact edge corrections by direct computation, logically independent of these considerations, but guided by the physical expectation (\ref{ZPIexpectZb}) and more generally the intuition developed in this section.

\subsection{Derivation}

\label{sec:thermalder}

We first give a formal derivation and then refine this by showing the objects of interest become rigorously well-defined in a manifestly covariant UV regularization of the QFT.  

\subsubsection*{Formal derivation}

 Our starting point is the observation that the thermal partition function ${\rm Tr} \, e^{-\beta  H}$ of a bosonic resp.\ fermionic oscillator of frequency $\omega$ has the integral representation (\ref{integralformula}): 
\begin{equation} \label{integralformula2}
 \begin{aligned}  
  -\log\bigl( e^{\beta \omega/2} - e^{-\beta \omega/2} \bigr)  &=+ \int_0^\infty \frac{dt}{2t} \,  \frac{1+e^{-2 \pi t/\beta}}{1-e^{-2\pi t/\beta}} \, \bigl( e^{-i\omega t} + e^{i \omega t} \bigr) \\
  \log \bigl( e^{\beta \omega/2} + e^{-\beta \omega/2} \bigr) &= -\int_0^\infty \frac{dt}{2t} \, \frac{2 \, e^{-\pi t/\beta}}{1-e^{-2 \pi t/\beta}} \, \bigl( e^{-i\omega t} + e^{i \omega t} \bigr) \, .
 \end{aligned}
\end{equation}
with the pole in the factor $f(t) = c \, t^{-2} + O(t^0)$ 
 multiplying $e^{-i\omega t}+e^{i\omega t}$ resolved by 
\begin{align} \label{iepspres2}
 t^{-2} \to \frac{1}{2} \bigl((t-i\epsilon)^{-2} + (t+i\epsilon)^{-2}\bigr).
\end{align}
Now consider a free QFT on some space of finite volume, viewed as a system $S$ of bosonic and/or fermionic oscillator modes of frequencies $\omega$ with mode (or single-particle) density $\rho_S(\omega)=\rho_S(\omega)_{\rm bos} + \rho_S(\omega)_{\rm fer}$. The system is in thermal equilibrium at inverse temperature $\beta$.  Using the above integral representation, we can write its thermal partition function (\ref{ZIGnogees}) as 
\begin{align} \label{troepie}
 \log {\rm Tr}_{S} \, e^{-\beta H_{S}} = \int_{0}^\infty \frac{dt}{2t} \biggl( \, \frac{1+e^{-2 \pi t/\beta}}{1-e^{-2 \pi t/\beta}}   \, \, \chi_S(t)_{\rm bos} \, - \,  \frac{2 \,  e^{-\pi t/\beta}}{1-e^{-2 \pi t/\beta}}  \, \, \chi_S(t)_{\rm fer} \biggr) \, , 
\end{align}
where we exchanged the order of integration, and we defined
\begin{align} \label{chiSdef}
 \chi_S(t) \equiv \int_0^\infty d\omega \, \rho_S(\omega)  \bigl( e^{-i\omega t} + e^{i \omega t} \bigr)  
\end{align}
We want to apply (\ref{troepie}) to a free QFT on the southern static patch at inverse temperature $\beta$, with the goal of finding a better way to make sense of it than {\it P.~ubique}'s approach. 
To this end, we note that the global dS$_{d+1}$ Harish-Chandra character $\chi(t)$ defined in (\ref{HCdef1}) can formally be written in a similar form by 
using the general property (\ref{chievenarg2}), $\chi(t)=\chi(-t)$: 
\begin{align} \label{chitchit}
\chi(t) = {\rm tr}_G \, e^{-i H t} = \int_{-\infty}^\infty d\omega \, \rho_G(\omega) \, e^{-i \omega t} =  \int_0^\infty d\omega \, \rho_G(\omega) \bigl(e^{-i \omega t} + e^{i \omega t} \bigr) \, ,
\end{align}
This looks like (\ref{chiSdef}), except
$\rho_G(\omega) = {\rm tr}_G \, \delta(\omega-H)$ is the density of single-particle excitations 
of the {\it global} Euclidean vacuum, while $\rho_S(\omega)$ is the density of single-particle excitations 
of the {\it southern} vacuum. The global and southern vacua are very different. Nevertheless, there is a simple kinematic relation between their single-particle creation and annihilation operators: the Bogoliubov transformation (\ref{bogoliubov}) (suitably generalized to $d>0$  \cite{Israel:1976ur}). This provides an explicit one-to-one, inner-product-preserving map between southern and global single-particle states with $H=\omega>0$. Hence, formally,
\begin{align} \label{fofofo}
  \rho_S(\omega) =  \rho_G(\omega) \, \qquad (\omega > 0) \, , \qquad \quad \rho_S(\omega) = 0 \qquad (\omega<0) \, .
\end{align}
While formal in the continuum, this relation becomes precise whenever $\rho$ is rendered effectively finite, e.g.\ by a brick-wall cutoff or by considering finite resolution projections (say if we restrict to states emitted/absorbed by some apparatus built by {\it P.~ubique}).          
   
At first sight this buys us nothing though, as computing $\rho_G(\omega) = {\rm tr}_G \, \delta(\omega-H)$ for say a scalar in dS$_4$ in a basis $|\omega \ell m\rangle_G$ immediately leads to $\rho_G(\omega) = \delta(0) \sum_{\ell m}$, in reassuring but discouraging agreement with {\it P.~ubique}'s result for $\rho_S(\omega)$.  
On second thought however, substituting this into (\ref{chitchit}) leads to a nonsensical $\chi(t) = 2 \pi \delta(t) \delta(0) \sum_{\ell m}$, not remotely resembling the correct expression (\ref{scachi}). 
How could this happen? As explained under (\ref{nonsensechi}), the root cause is the seemingly natural but actually ill-advised idea of computing $\chi(t)={\rm tr}_G \, e^{-i Ht}$ by diagonalizing $H$: despite its lure of seeming simplicity, $|\omega \ell m\rangle_G$ is in fact the {\it worst} possible choice of basis to compute the character trace. Its wave functions on the global future boundary $S^d$ of dS$_{d+1}$ are singular at the north and south pole, exactly the fixed points of $H$ at which the correct computation of $\chi(t)$ in appendix \ref{app:compchar} localizes. 
Although $|\omega \ell m\rangle$ is a perfectly fine basis on the cylinder obtained by a conformal map from sphere, the information needed to compute $\chi$ is irrecoverably lost by this map.

However we can turn things around, and use the properly computed $\chi(t)$ to extract $\rho_G(\omega)$ as its Fourier transform, inverting (\ref{chitchit}). As it stands, this is not really possible, for (\ref{scachi}) implies $\chi(t) \sim |t|^{-d}$ as $t \to 0$, so its Fourier transform does not exist. Happily, this problem is  automatically resolved by standard UV-regularization of the QFT, as we will show explicitly below. For now let us proceed formally, as at this level we have arrived at our desired result: combining (\ref{fofofo}) with (\ref{chitchit}) and (\ref{chiSdef}) implies $\chi_S(t)=\chi(t)$, which by (\ref{troepie}) yields
\begin{align} \label{equafoma}
  {\rm Tr}_S \, e^{-\beta H_S} = Z_{\rm bulk}(\beta) \qquad \mbox{(formal)}
\end{align}
with $Z_{\rm bulk}(\beta)$ as defined in (\ref{ZthNSchar2prev}). The above equation formally gives it its claimed thermal interpretation. In what follows we will make this a bit more precise, and spell out the UV regularization explicitly.

\subsubsection*{Covariant UV regularization of $\rho$ and $Z_{\rm bulk}$}

We begin by showing that $\rho_G(\omega)$ in (\ref{chitchit}) becomes well-defined in a suitable standard UV-regularization of the QFT. 
As in \cite{Demers:1995dq}, it is convenient to consider Pauli-Villars regularization, which is manifestly covariant and has a conceptually transparent   implementation on both the path integral and canonical sides. For e.g.\ a scalar of mass $m^2=(\frac{d}{2})^2+\nu^2$, a possible implementation is adding ${k \choose n}$, $n=1,\ldots,k \geq \frac{d}{2}$ fictitious particles of mass $m^2 = (\frac{d}{2})^2+\nu^2 + n \Lambda^2$ and positive/negative norm for even/odd $n$,\footnote{This is equivalent to inserting a heat kernel regulator $f(\tau \Lambda^2)=\bigl(1-e^{-\tau\Lambda^2} \bigr)^k$ in (\ref{HKSP}), with $k \geq \frac{d}{2}+1$. \label{fn:PV}
} turning the character $\chi_{\nu^2}(t)$ of (\ref{scachi}) into 
\begin{align} \label{PVchar}
 \chi_{\nu^2,\Lambda}(t) = {\rm tr}_{G_\Lambda} \, e^{-i t H} = \sum_{n=0}^k (-1)^n \mbox{\large ${k \choose n}$} \, \chi_{\nu^2+n \Lambda^2}(t). 
\end{align} 
This effectively replaces $\chi(t) \sim |t|^{-d}$ by $\chi_\Lambda(t) \sim |t|^{2k-d}$ with $2k-d \geq 0$, hence, assuming $\chi(t)$ falls off exponentially at large $t$, which is always the case for unitary representations \cite{10.3792/pja/1195522333,10.3792/pja/1195523378,10.3792/pja/1195523460}, $\chi_\Lambda(t)$ has a well-defined Fourier transform, analytic in $\omega$:
\begin{align} \label{PVregdost}
  \rho_{G,\Lambda}(\omega) = \frac{1}{2\pi} \int_{-\infty}^\infty dt \, \chi_\Lambda(t) \, e^{i \omega t}  = \frac{1}{2\pi} \int_{0}^\infty dt \, \chi_\Lambda(t) \, \bigl( e^{i \omega t} + e^{-i \omega t} \bigr) \, .
\end{align}
The above character regularization can immediately be transported to arbitrary massive $SO(1,d+1)$ representations, as their characters $\chi_{s,\nu^2}$  (\ref{chigenrep}) 
only differ from the scalar one by an overall spin degeneracy factor.\footnote{For massless spin-$s$, the PV-regulating characters to add to the physical character (e.g.\ (\ref{charexmpls}),(\ref{chiexcex})) are $\hat\chi_{s,n} = \chi_{s,\nu_\phi^2+n\Lambda^2} - \chi_{s-1,\nu_\xi^2+n \Lambda^2}$ where $\nu_\phi^2 = -(s-2+\frac{d}{2})^2$ and $\nu_\xi^2=-(s-1+\frac{d}{2})^2$, based on (\ref{phixidim}) and (\ref{XphiminXxi}).}

\begin{figure}
\begin{center}
   \includegraphics[height=4.3cm]{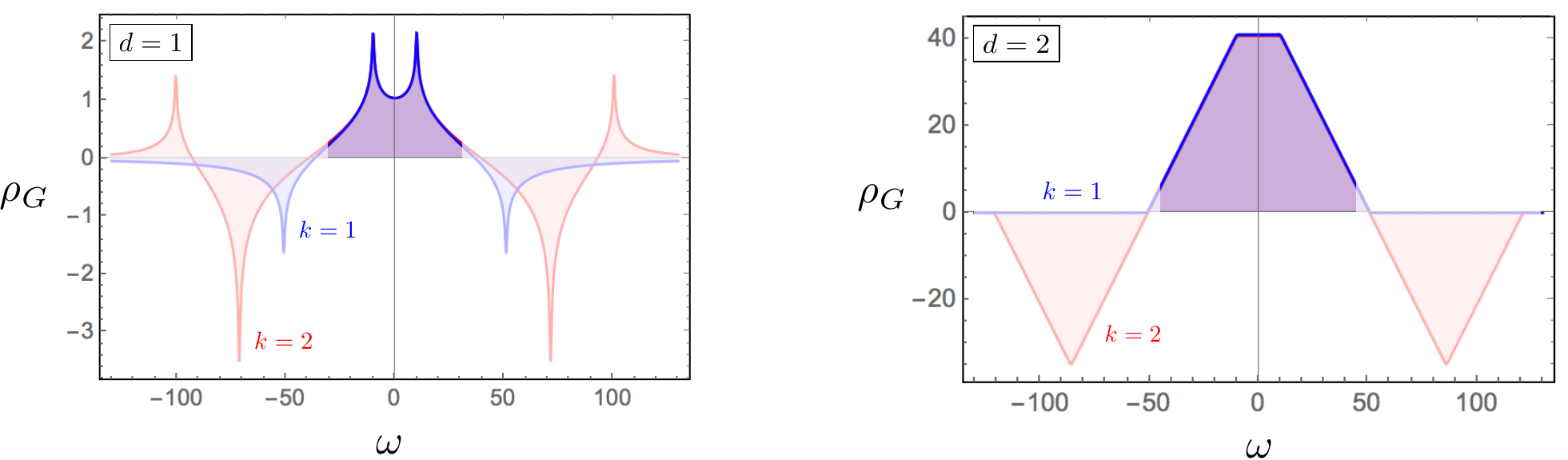}
 \caption{\small $\rho_{G,\Lambda}(\omega)$ for dS$_{d+1}$ scalar of mass $m^2=(\frac{d}{2})^2+\nu^2$, $\nu=10$, for $d=1,2$ in $k=1,2$ Pauli-Villars regularizations (\ref{PVchar}). 
 Faint part is unphysical UV regime $\omega \gtrsim \Lambda$.  
The peaks/kinks appearing at $\omega=\sqrt{\nu^2+n \Lambda^2}$ are related to quasinormal mode resonances \rf{sec:resonances}. 
 \label{fig:dospv}} 
\end{center} \vskip-4mm
\end{figure}

Although we won't need to in practice for computations of thermodynamic quantities (which are most easily extracted directly as character integrals), $\rho_{G,\Lambda}(\omega)$ can be computed explicitly. For the dS$_{d+1}$ scalar, using (\ref{scachi}) regularized with $k=1$, we get for $\omega \ll \Lambda$
\begin{equation} \label{DOSexIR}
\begin{aligned}
  d=1:& \quad \rho_{G,\Lambda}(\omega) = \frac{2}{\pi} \log \Lambda -\frac{1}{2\pi} \sum_{\pm,\pm} \psi\bigl(\tfrac{1}{2} \pm i \nu \pm i\omega)  + O(\Lambda^{-1})  \\
  d=2:& \quad \rho_{G,\Lambda}(\omega) =  \Lambda 
  - \frac{1}{2} \sum_{\pm} (\omega \pm \nu) \coth\bigl(\pi(\omega \pm \nu)\bigr) + O(\Lambda^{-1}) 
\end{aligned}
\end{equation}
where $\psi(x)=\Gamma'(x)/\Gamma(x)$. Denoting the $\Lambda$-independent parts of the above $\omega \ll \Lambda$ expansions by $\tilde\rho_{\nu^2}(\omega)$, the exact $\rho_{G,\Lambda}(\omega)$ for general $\omega$ and $k$ is $\rho_{G,\Lambda}(\omega) = \sum_{n=0}^k (-1)^n {k \choose n} \, \tilde \rho_{\nu^2+n \Lambda^2}(\omega)$, illustrated in fig.\ \ref{fig:dospv} for $k=1,2$. The $\omega \ll \Lambda$ result is independent of $k$ up to rescaling of $\Lambda$. The result for massive higher-spin fields is the same up to an overall degeneracy factor $D_s^d$ from (\ref{chimassivespins}).

To make sense of the southern static patch density $\rho_S(\omega)$ directly in the continuum, we {\it define} its regularized version by mirroring the formal relation (\ref{fofofo}), thus ensuring all of the well-defined features and physics this relation encapsulates are preserved:  
\begin{align} \label{rholambda}
 \rho_{S,\Lambda}(\omega) \equiv \rho_{G,\Lambda}(\omega) = (\ref{PVregdost}) \, \qquad (\omega > 0) \, .
\end{align} 
This definition of the regularized static patch density evidently inherits all of the desirable properties of $\rho_G(\omega)$: manifest general covariance, independence of arbitrary choices such as brick wall boundary conditions, and exact analytic computability.
The physical sensibility of this identification is also supported by the fact that the  quasinormal mode expansion (\ref{QNMexpchi1}) of $\chi(t)$ produces the physically expected static patch quasinormal resonance pole structure  $\rho_S(\omega) = \frac{1}{2\pi} \sum_{r,\pm}  \frac{N_r}{r \pm i \omega}$, cf.\ appendix \ref{sec:resonances}.

Putting things together in the way we obtained the formal relation (\ref{equafoma}), the correspondingly regularized version of the static patch thermal partition function (\ref{troepie}) is then 
\begin{align} \label{ZbulkPVreg}
 \boxed{\log Z_{\rm bulk,\Lambda}(\beta) \equiv \int_{0}^\infty \frac{dt}{2t} \biggl( \, \frac{1+e^{-2 \pi t/\beta}}{1-e^{-2 \pi t/\beta}}   \, \, \chi_\Lambda(t)_{\rm bos} \, - \,  \frac{2 \,  e^{-\pi t/\beta}}{1-e^{-2 \pi t/\beta}}  \, \, \chi_\Lambda(t)_{\rm fer} \biggr) }
\end{align}
Note that if we take $k \geq \frac{d}{2} + 1$, then 
$\chi_\Lambda(t) \sim t^{2k-d}$ with $2k-d \geq 2$ and we can drop the $i\epsilon$ prescription (\ref{iepspres2}). 
$Z_{\rm bulk}$ (or equivalently $\chi$) can be regularized in other ways, including by cutting off the integral at $t=\Lambda^{-1}$, or as in (\ref{chiL}), or by dimensional regularization. For most of the paper we will use yet another variant, defined in section \ref{sec:scalars}, equivalent, like Pauli-Villars, to a manifestly covariant heat-kernel regularization of the path integral.

In view of the above observations, $Z_{\rm bulk,\Lambda}(\beta)$ is naturally interpreted as a well-defined, covariantly regularized and ambiguity-free {\it definition} of the static patch ideal gas thermal partition in the continuum.  
However we refrain from denoting $Z_{\rm bulk}(\beta)$ as ${\rm Tr}_{S,\Lambda} \, e^{-\beta H_S}$, because it is not constructed as an actual sum over states of some definite regularized static patch Hilbert space $\CH_{S,\Lambda}$. This (together with the role of quasinormal modes) is also why we referred to $Z_{\rm bulk}(\beta)$ as a ``quasi''-canonical partition function in the introduction.

\subsection{Example computations} \label{sec:enentr}

In this section we illustrate the use and usefulness of the character formalism by computing some examples of bulk thermodynamic quantities at the equilibrium inverse temperature $\beta=2\pi$ of the static patch. The precise relation of these quantities with their Euclidean counterparts will be determined in \ref{sec:scalars}-\ref{sec:massless} and \ref{sec:all-loop}.


\subsubsection*{Character formulae for bulk thermodynamic quantities at $\beta=2\pi$}
At $\beta=2\pi$, the bulk free energy, energy, entropy and heat capacity are obtained by taking the appropriate derivatives of (\ref{ZthNSchar2prev}) and putting $\beta=2\pi$, using the standard thermodynamic relations $F=-\frac{1}{\beta} \log Z$, $U = -\partial_\beta \log Z$, $S = \log Z + \beta U$,    
$C = - \beta \partial_\beta S$.
Denoting $q \equiv e^{-t}$, 
\begin{align}
 \log Z_{\rm bulk} &= \int_{0}^\infty \frac{dt}{2t} \biggl( \, \frac{1+q}{1-q}   \, \, \chi_{\rm bos} \, - \,  \frac{2 \sqrt{q}}{1-q}  \, \, \chi_{\rm fer} \biggr), \label{chrFbulk} \\
 2 \pi U_{\rm bulk} &= \int_0^\infty \frac{dt}{2}  \biggl( -\frac{2 \sqrt{q}}{1-q} \, \frac{\sqrt{q}}{1-q} \, \chi_{\rm bos} + \frac{1+q}{1-q} \, \frac{\sqrt{q}}{1-q} \, \chi_{\rm fer} \biggr), \label{charintUbulk} 
\end{align}
and similarly for $S_{\rm bulk}$ and $C_{\rm bulk}$.
The characters $\chi$ for general massive representation are given by (\ref{chigenrep}), for massless spin-$s$ representations by (\ref{excserieschi})-(\ref{spin1casechis}), and for partially massless $(s,s')$ representations by (\ref{flippedchargenPM}). Regularization is implicit here. 

\subsubsection*{Leading divergent term}

The leading $t \to 0$ divergence of the scalar character (\ref{scachi}) is $\chi(t) \sim 2/t^d$. For more general representations this becomes $\chi(t) \sim 2n/t^d$ with $n$ the number of on-shell internal (spin) degrees of freedom. The generic leading divergent term of the bulk (free) energy is then given by $F_{\rm bulk}, U_{\rm bulk} \sim -\frac{1}{\pi}(n_{\rm bos} - n_{\rm fer})  \int \!\frac{dt}{t^{d+2}} \sim \pm \Lambda^{d+1} \ell^d$, while for the bulk heat capacity and entropy we get $C_{\rm bulk}, S_{\rm bulk} \sim  (\frac{1}{3}  n_{\rm bos} + \frac{1}{6} n_{\rm fer}) \int \!\frac{dt}{t^{d}} \sim + \Lambda^{d-1} \ell^{d-1}$, where we reinstated the dS radius $\ell$.
In particular $S_{\rm bulk} \sim  + \Lambda^{d-1} \times \mbox{horizon area}$, consistent with an entanglement entropy area law. The energy diverges more strongly because we included the QFT zero point energy term in its definition, which drops out of $S$ and $C$. 

\subsubsection*{Coefficient of log-divergent term}

The coefficient of the logarithmically divergent part of these thermodynamic quantities is universal. 
A pleasant feature of the character formalism is that this coefficient can be read off trivially as the coefficient of the $1/t$ term in the small-$t$ expansion of the integrand, easily computed for any representation. In odd $d+1$, the integrand is  even in $t$, so log-divergences are absent. In even $d+1$, the integrand is odd in $t$, so generically we do get a log-divergence $= a  \log \Lambda$.
For example the $\log Z$ integrand for a dS$_2$ scalar is expanded as 
\begin{align} 
 \frac{1}{2t} \frac{1+e^{-t}}{1-e^{-t}} \, \frac{e^{-t(\frac{1}{2}+i \nu)}+e^{-t(\frac{1}{2}-i \nu)}}{1-e^{-t}} =  \frac{2}{t^3}+\frac{\frac{1}{12}-\nu ^2}{t}+\cdots  \quad \Rightarrow \quad a=\tfrac{1}{12}-\nu^2 \, .
\end{align}
For a $\Delta=\frac{d}{2}+i \nu$ spin-$s$ particle in even $d+1$, the $\log \Lambda$ coefficient for $U_{\rm bulk}$ is similarly read off as $a_{U_{\rm bulk}} =  - D^d_s \frac{1}{\pi (d+1)!}\prod_{n=0}^d (\Delta-n)$. 
For a conformally coupled scalar, $\nu=i/2$, so $a_{U_{\rm bulk}}=0$. Some examples of $a_{S_{\rm bulk}} = a_{\log Z_{\rm bulk}}$ in this case are
\begin{equation} \nn
 \begin{array}{l|ccccccccccc}
  d+1 & 2 & 4 & 6 & 8 & 10 & \cdots & 100 & \cdots & 1000 & \cdots \\
  \hline
  a_{S_{\rm bulk}} & \frac{1}{3} & -\frac{1}{90} & \frac{1}{756} & -\frac{23}{113400} & \frac{263}{7484400} 
  & \cdots &  -8.098 \times 10^{-34} 
   & \cdots  & -3.001 \times 10^{-306} & \cdots
 \end{array}
 \end{equation}

\subsubsection*{Finite part and exact results} 

\noindent $\bullet$ {\it Energy:} For future reference (comparison to previously obtained results in section \ref{sec:all-loop}), we consider dimensional regularization here. The absence of a $1/t$ factor in the integral (\ref{charintUbulk}) for $U_{\rm bulk}$ then allows straightforward evaluation for general $d$. For a 
 scalar of mass $m^2=(\frac{d}{2})^2+\nu^2$, 
\begin{align} \label{Ubulkdimreg}
 U^{\rm fin}_{\rm bulk} = \frac{m^2 \cosh(\pi\nu) \, \Gamma(\frac{d}{2}+i \nu) \, \Gamma(\frac{d}{2}-i \nu)}{2 \pi \, \Gamma(d+2) \cos(\frac{\pi d}{2})}  \qquad \mbox{(dim reg)} \, .
\end{align}
For example for $d=2$, this becomes 
\begin{align} \label{Ufind2}
 U^{\rm fin}_{\rm bulk}= -\frac{1}{12} (\nu ^2+1)  \nu  \coth (\pi  \nu ) \, . 
\end{align}

\vskip2mm \noindent $\bullet$ {\it Free energy:} The UV-finite part of the $\log Z_{\rm bulk}$ integral (\ref{chrFbulk}) for a massive field in even $d$ can be computed simply by extending the integration contour to the real line avoiding the pole, closing the contour and summing residues.
For example for a $d=2$ scalar this gives
\begin{align} \label{Zbubufi2}
 \log Z^{\rm fin}_{\rm bulk} =   \frac{\pi \nu^3}{6} 
 -\sum_{k=0}^2 \frac{\nu^k}{k!} \, \frac{{\rm Li}_{3-k}(e^{-2 \pi \nu})}{(2\pi)^{2-k}} \, ,
\end{align}
where ${\rm Li}_n$ is the polylogarithm, ${\rm Li}_n(x) \equiv \sum_{k=1}^\infty {x^k}/{k^n}$. For future reference, note that
\begin{align} \label{polylogs}
  {\rm Li}_1(e^{-2\pi \nu}) = - \log(1-e^{-2\pi\nu}) \, , \qquad {\rm Li}_0(e^{-2\pi \nu}) = \frac{1}{e^{2 \pi \nu}-1} = \tfrac{1}{2} \coth(\pi \nu) - \tfrac{1}{2} \, .
\end{align} 
For odd $d$, the character does not have an even analytic extension to the real line, so a different method is needed to compute $\log Z_{\rm bulk}$. 
The exact evaluation of arbitrary character integrals, for any $d$ and any $\chi(t)$ is given in (\ref{ZEXACT}) in terms of Hurwitz zeta functions. Simple examples are given in (\ref{dS2exampleZPI})-(\ref{scalarS3}). In (\ref{ZEXACT}) we use the covariant regularization scheme introduced in section \ref{sec:scalars}. Conversion to PV regularization is obtained from the finite part as explained below.  

\vskip2mm \noindent $\bullet$ {\it Entropy}: Combined with our earlier result for the bulk energy $U_{\rm bulk}$, the above also gives the finite part of the bulk entropy $S_{\rm bulk} =\log Z_{\rm bulk} + 2\pi U_{\rm bulk}$. In the Pauli-Villars regularization (\ref{PVchar}), the UV-divergent part is obtained from the finite part by mirrorring (\ref{PVchar}). For example for $k=1$, $S_{\rm bulk,\Lambda} = S_{\rm bulk}^{\rm fin}|_{\nu^2} - S_{\rm bulk}^{\rm fin}|_{\nu^2+\Lambda^2}$. For the $d=2$ example this gives for  $\nu \ll \Lambda$
\begin{align} \label{SLamPV}
 S_{\rm bulk,\Lambda} =  \frac{\pi}{6}(\Lambda-\nu) - \frac{\pi}{3} \, \nu \, {\rm Li}_0(e^{-2 \pi \nu}) - \sum_{k=0}^3 \frac{\nu^k}{k!} \, \frac{{\rm Li}_{3-k}(e^{-2 \pi \nu})}{(2\pi)^{2-k}} \, ,
\end{align}
where we used (\ref{polylogs}).
$S_{\rm bulk}$ decreases monotonically with $m^2=1+\nu^2$. In the massless limit $m \to 0$, it diverges logarithmically: $S_{\rm bulk} = -\log m + \cdots$. For $\nu \gg 1$, $S_{\rm bulk} = \frac{\pi}{6}(\Lambda-\nu)$ up to exponentially small corrections. Thus $S_{\rm bulk} > 0$ within the regime of validity of the low-energy field theory, consistent with its quasi-canonical/entanglement entropy interpretation.  For a conformally coupled scalar $\nu=\frac{i}{2}$, this gives $S^{\rm fin}_{\rm bulk} = \frac{3 \zeta (3)}{16 \pi ^2}-\frac{\log (2)}{8}$.

\subsubsection*{Quasinormal mode expansion} 
Substituting the quasinormal mode expansion (\ref{QNMexpchi1}),
\begin{align}
 \chi(t) = \sum_r N_r e^{-rt}
\end{align} 
in the PV-regularized $\log Z_{\rm bulk}(\beta)$  (\ref{ZbulkPVreg}), rescaling  $t \to \frac{\beta}{2\pi} t$, and using (\ref{QNMcontrib}) gives
\begin{align} \label{ZbulkQNMexp}
 \log Z_{\rm bulk}(\beta) = \sum_r N_r^{\rm bos} \log \frac{\Gamma(b r+1)}{(b\mu)^{br} \sqrt{2 \pi b r}} - N_r^{\rm fer} \log \frac{\Gamma(b r+\frac{1}{2})}{(b\mu)^{br} \sqrt{2 \pi}} \, , \qquad b \equiv \frac{\beta}{2\pi} \, .
\end{align} 
Truncating the integral to the IR part (\ref{QNMcontrib}) is justified because the Pauli-Villars sum (\ref{PVchar}) cancels out the UV part. The dependence on $\mu$ likewise cancels out, as do some other terms, but it is useful to keep the above form. 
At the equilibrium $\beta=2\pi$, $\log Z_{\rm bulk}$ is given by (\ref{ZbulkQNMexp}) with $b=1$. This provides a PV-regularized version of the quasinormal mode expansion of \cite{Denef:2009kn}. Since it is covariantly regularized, it does not require matching to a local heat kernel expansion. Moreover it applies to general particle content, including spin $s \geq 1$.\footnote{The expansion of \cite{Denef:2009kn} pertains to  $Z_{\rm PI}$ for $s \leq \frac{1}{2}$. In the following sections we show $Z_{\rm PI} = Z_{\rm bulk}$ for $s \leq \frac{1}{2}$ but not for $s \geq 1$. Hence in general the QNM expansion of \cite{Denef:2009kn} computes $Z_{\rm bulk}$, not $Z_{\rm PI}$.}

\begin{figure}
\begin{center}
   \includegraphics[height=5.5cm]{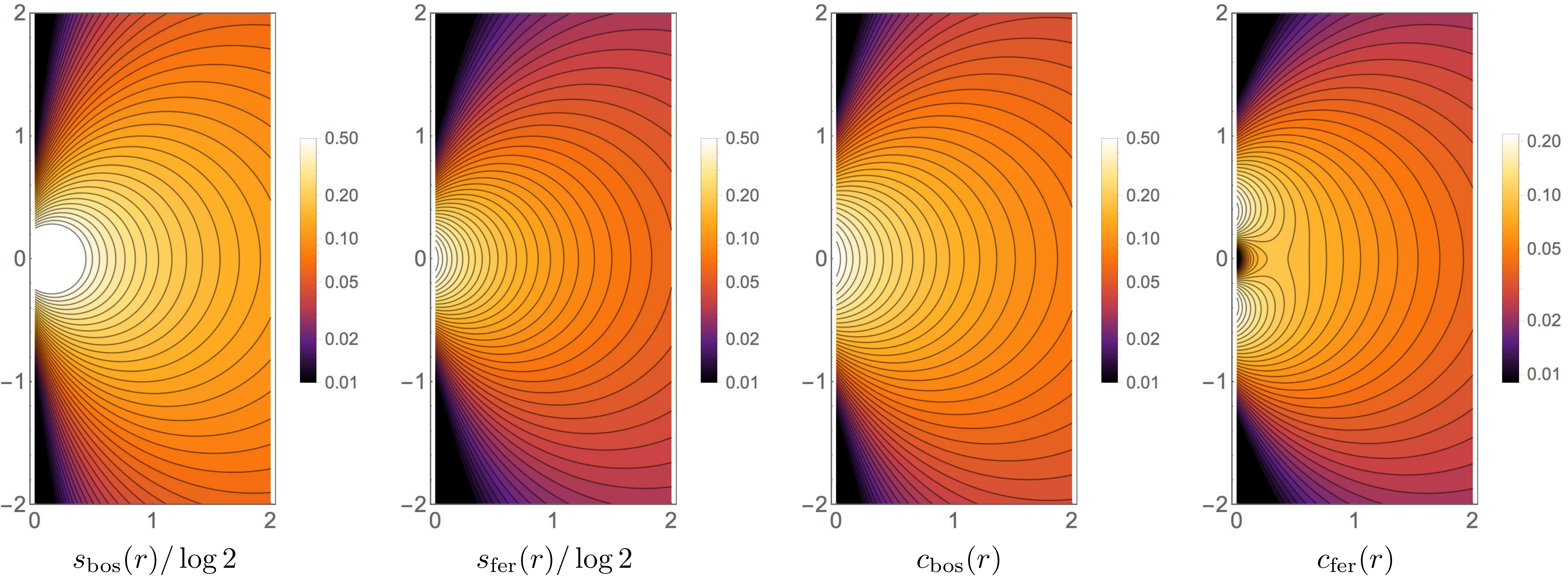}
 \caption{\small Contribution to $\beta=2\pi$ bulk entropy and heat capacity of a quasinormal mode $\propto e^{-r T}$, $r \in \IC$, ${\rm Re} \, r > 0$. Only the real part is shown here because complex $r$ come in conjugate pairs $r_{n,\pm} = \frac{d}{2} + n \pm i \nu$. The harmonic oscillator case corresponds to the imaginary axis. 
 \label{fig:QNMsc}} 
\end{center} \vskip-4mm
\end{figure}

QNM expansions of other bulk thermodynamic quantities are readily derived from (\ref{ZbulkQNMexp}) by taking derivatives $\beta\partial_\beta = b \partial_b = \mu \partial_\mu + r \partial_r$. 
For example $S_{\rm bulk} = (1-\beta\partial_\beta) \log Z_{\rm bulk}|_{\beta=2\pi}$ is 
\begin{align}
 S_{\rm bulk} = \sum_r N^{\rm bos}_r s_{\rm bos}(r) + N^{\rm fer}_r s_{\rm fer}(r)
\end{align}
where the entropy $s(r)$ carried by a single QNM $\propto e^{-r T}$ at $\beta=2\pi$ is given by
\begin{equation}
 \label{sofr}
 s_{\rm bos}(r) 
 = r + (1-r \partial_r) \log \frac{\Gamma(r+1)}{\sqrt{2 \pi r}}  \, , \qquad
 s_{\rm fer}(r) 
 =-r - (1-r \partial_r) \log \frac{\Gamma(r+\frac{1}{2})}{\sqrt{2 \pi}} \, .
\end{equation}
Note  the $\mu$-dependence has dropped out, reflecting the fact that the contribution of each {\it individual} QNM to the entropy is UV-finite, not requiring any regularization. For massive representations, $r$ can be complex, but will always appear in a conjugate pair $r_{n\pm} =\frac{d}{2} + n \pm i \nu$.
Taking this into account, all contributions to the entropy are real and positive for the physical part of the PV-extended spectrum. 
The small and large $r$ asymptotics are
\begin{align}
  r \to 0: \, s_{\rm bos} \to \frac{1}{2} \log \frac{e}{2 \pi r} \, , \quad s_{\rm fer} \to \frac{\log 2}{2} \, , \qquad r \to \infty: \, s_{\rm bos} \to \frac{1}{6 \, r} \, , \quad s_{\rm fer} \to \frac{1}{12 \, r} \, .
\end{align}
The QNM entropies at general $\beta$ are obtained simply by replacing
\begin{align}
  r \to \frac{\beta}{2\pi} \, r \, .
\end{align} 
The entropy of a normal bosonic mode of frequency $\omega$, $\tilde s(\omega) = - \log\bigl(1-e^{-\beta \omega}\bigr) + \frac{\beta \omega}{e^{\beta \omega}-1}$, is recovered for complex conjugate pairs $r_\pm$ in the scaling limit $\beta \to 0$, $\beta \nu = \omega$ fixed, and likewise for fermions. At any finite $\beta$, the $n \to \infty$ UV tail of QNM contributions is markedly different however. Instead of falling off exponentially, if falls off as $s \sim 1/n$. PV or any other regularization effectively cuts off the sum at $n \sim \Lambda \ell$, so since $N_n \sim n^{d-1}$, $S_{\rm bulk} \sim \Lambda^{d-1} \ell^{d-1}$. 
  

The bulk heat capacity $C_{\rm bulk}=-\beta\partial_\beta S_{\rm bulk}$, so the heat capacity of a QNM at $\beta=2\pi$ is 
\begin{align} \label{CbulkQNM} 
 c(r) = -r \partial_r s(r) \, .
\end{align}  
The real part of $s(r)$ and $c(r)$ on the complex $r$-plane are shown in fig.\ \ref{fig:QNMsc}.


\subsubsection*{An application of the quasinormal expansion}

The above QNM expansions are less useful for exact computations of thermodynamic quantities than the direct integral evaluations discussed earlier, but can be very useful in computations of certain UV-finite quantities. A simple example is the following. In thermal equilibrium with a 4D dS static patch horizon, which set of particle species has the largest bulk heat capacity: (A) six conformally coupled scalars + graviton, (B) four photons? The answer is not obvious, as both have an equal number of local degrees of freedom: $6+2 = 4 \times 2 = 8$. One could compute each in full, but the above QNM expansions offers a much easier way to get the answer.  From (\ref{scachi}) and  (\ref{chiexcex}) we read off the scalar and massless spin-$s$ characters:
\begin{align} \label{quizchar}
 \chi_0 = \frac{q+q^2}{(1-q)^3} \, , \qquad \chi_s = \frac{2(2s+1) \, q^{s+1} - 2(2s-1) \, q^{s+2}}{(1-q)^3} \, ,
\end{align} 
where $q = e^{-|t|}$. 
We see $\chi_A - \chi_B = \chi_2 + 6 \, \chi_0 - 4 \, \chi_1 = 6 \, q$, so  $\chi_A$ and $\chi_B$ are {\it almost} exactly equal: A has just 6 more quasinormal modes than B, all with $r=1$. Thus, using (\ref{CbulkQNM}),
\begin{align} 
 C_{\rm bulk}^A - C_{\rm bulk}^B = 6 \cdot c_{\rm bos}(1) = \pi^2 - 9 \, .
\end{align}
Pretty close, but $\pi>3$ \cite{pihist}, so A wins. The difference is $\Delta C \approx 0.87$. Along similar lines, $\Delta S= 6 \,  s_{\rm bos}(1)=3(2 \game + 1 - \log(2 \pi)) \approx 0.95$. 

\subsubsection*{Another UV-finite example: relative entropies of graviton, photon, neutrino}

Less trivial to compute but more real-world in flavor is the following UV-finite linear combination of the 4D graviton, photon, and (assumed massless) neutrino bulk entropies:
\begin{align}  \label{gravneupho}
  S_{\text{graviton}} + \tfrac{60}{7} S_{\text{neutrino}} - \tfrac{37}{7} S_{\text{photon}} = \tfrac{48}{7} \zeta '(-1)-\tfrac{60}{7} \zeta '(-3)+6 \gamma
   +\tfrac{149}{56}-\tfrac{33}{14} \log (2 \pi) \approx 0.61
\end{align} 
Finiteness can be checked from the small-$t$ expansion of the total integrand computing this, and the integral can then be evaluated along the lines of (\ref{ZIRzetanu})-(\ref{zetanuHur}). We omit the details.  
 
\subsubsection*{\bf Vasiliev higher-spin example}
Non-minimal Vasiliev higher-spin gravity on dS$_4$ has a single conformally coupled scalar and a tower of massless spin-$s$ particles of all spins $s=1,2,3,\ldots$. The prospect of having to compute bulk thermodynamics for this theory by brick wall or other approaches mentioned in appendix \ref{sec:brickwall} would be terrifying. Let us compare this to the character approach. The {\it total} character obtained by summing the characters of (\ref{quizchar}) takes a remarkably simple form:   
\begin{align} \label{VasilievThermo}
 \chi_{\rm tot} \, = \,  \chi_0 + \sum_{s=1}^\infty \chi_s \,  = \, 2 \cdot \biggl( \frac{q^{1/2} + q^{3/2}}{(1-q)^2} \biggr)^2 - \frac{q}{(1-q)^2} \, \,  = \, \,  \frac{q+q^3}{(1-q)^4} + 3 \cdot \frac{2 q^2}{(1-q)^4} \, .
\end{align}
The first expression is two times the square of the character of a 3D conformally coupled scalar, plus the character of 3D conformal higher-spin gravity (\ref{charexmplsconfSd}).\footnote{\label{fn:dSdS} For AdS$_4$, the analogous $\chi_{\rm tot}$ equals {\it one} copy of the 3D scalar character squared, reflecting the single-trace spectrum of its holographic dual $U(N)$ model \rf{sec:AdSswamp}. The dS counterpart thus encodes the single-trace spectrum of two copies of this 3D CFT + 3D CHS gravity, reminiscent of \cite{Alishahiha:2004md}. This is generalized by (\ref{confSdAdSdS}).}  
The second expression equals the character of one $\nu=i$ and three $\nu=0$ scalars on dS$_5$. 
Treating the character integral as such, we immediately get, in $k=3$ Pauli-Villars regularization (\ref{PVchar}),
\begin{equation}
\begin{aligned}   
 \log Z_{\rm bulk}^{\rm div} &= a_0  \Lambda ^5+a_2   \Lambda ^3-a_4  \Lambda \, , \qquad & 
 \log Z_{\rm bulk}^{\rm fin} &=\tfrac{\zeta (5)}{4 \pi^4}-\tfrac{\zeta (3)}{24 \pi ^2} \\
S_{\rm bulk}^{\rm div} &= \tfrac{25}{4} a_2 \Lambda^3-\tfrac{103}{20}  a_4 \Lambda   \, , \qquad & S_{\rm bulk}^{\rm fin} &= \tfrac{\zeta (5)}{4\pi ^4} -\tfrac{\zeta(3)}{24 \pi ^2}+\tfrac{1}{20}  \, .
\end{aligned}
\end{equation}
where $a_0=\frac{1-4 \sqrt{2}+3 \sqrt{3}}{10} \, \pi \approx 0.17$, $a_2=-\frac{1-2 \sqrt{2}+\sqrt{3}}{12} \, \pi \approx 0.025$, and $a_4=\frac{3-3 \sqrt{2}+\sqrt{3}}{48} \, \pi \approx 0.032$. The tower of higher-spin particles alters the {\it bulk} UV dimensionality much like a tower of KK modes would. (We will later see edge ``corrections'' rather dramatically alter this.) 




\section{Sphere partition function for scalars and spinors}

\label{sec:scalars}

\subsection{Problem and result}

In this section we consider the one-loop Gaussian Euclidean path integral $Z_{\rm PI}^{(1)}$ of scalar and spinor field fluctuations on the round sphere.
For a free scalar of mass $m^2$ on $S^{d+1}$,
\begin{align}
 Z_{\rm PI} = \int \CD \phi \, e^{-\frac{1}{2} \int \phi (-\nabla^2+m^2) \phi} \, ,
\end{align}
A convenient UV-regularized version is defined using standard heat kernel methods \cite{Vassilevich:2003xt}:
\begin{align} \label{HKSP}
 \log Z_{\rm PI,\epsilon}  =  \int_0^\infty \frac{d\tau}{2\tau} \, e^{-\epsilon^2/4\tau} \, \Tr \, e^{-\tau(-\nabla^2 + m^2)} \, .
\end{align}
The insertion $e^{-\epsilon^2/4\tau}$ implements   a UV cutoff at length scale $\sim \epsilon$. We picked this regulator  for convenience in the derivation below. We could alternatively insert the PV regulator of footnote \ref{fn:PV}, which would reproduce the PV regularization (\ref{PVchar}). However, being uniformly applicable to all dimensions, the above regulator is more useful for the purpose of deriving general evaluation formulae, as in appendix \ref{app:renpr}. 

In view of (\ref{ZPIexpectZb}) we wish to compare $Z_{\rm PI}$ to the corresponding Wick-rotated dS static patch bulk thermal partition function $Z_{\rm bulk}(\beta)$ (\ref{ZthNSchar2prev}), at the equilibrium inverse temperature $\beta=2\pi$. Here and henceforth, $Z_{\rm bulk}$ by default means $Z_{\rm bulk}(2\pi)$:   
\begin{align} \label{Zbulkdef}
 \log Z_{\rm bulk}  \equiv  \int_{0}^\infty \frac{dt}{2t} \biggl( \, \frac{1+e^{-t}}{1-e^{-t}}   \, \, \chi(t)_{\rm bos} \, - \,  \frac{2 \,  e^{-t/2}}{1-e^{-t}}  \, \, \chi(t)_{\rm fer} \biggr)
\end{align}
Below we show that for free scalars and spinors, 
\begin{align} \label{ZPIisZbulkss}
 \boxed{Z_{\rm PI} = Z_{\rm bulk}}
\end{align}
with the specific regularization (\ref{HKSP}) for $Z_{\rm PI}$ mapping to a specific regularization (\ref{ZPIreg}) for $Z_{\rm bulk}$. The relation is exact, for any $\epsilon$. This makes the physical expectation (\ref{ZPIexpectZb}) precise, and shows that for scalars and spinors, there are in fact no edge corrections. 

In appendix \ref{app:renpr} we provide a simple recipe for extracting both the UV and IR parts in the $\epsilon \to 0$ limit in the above regularization, directly from the unregularized form of the character formula (\ref{Zbulkdef}). This yields the general closed-form solution \boxed{(\ref{ZEXACT})} for the regularized $Z_{\rm PI}$ in terms of Hurwitz zeta functions. The heat kernel coefficient invariants are likewise read off from the character using (\ref{HKCFORM}). For simple examples see (\ref{dS2exampleZPI}), (\ref{scalarS3}), (\ref{kzerocase})-(\ref{kzerocaseFer}).  

\subsection{Derivation}\label{scalar PI derivation}

\begin{figure} 
 \begin{center}
   \includegraphics[height=4.5cm]{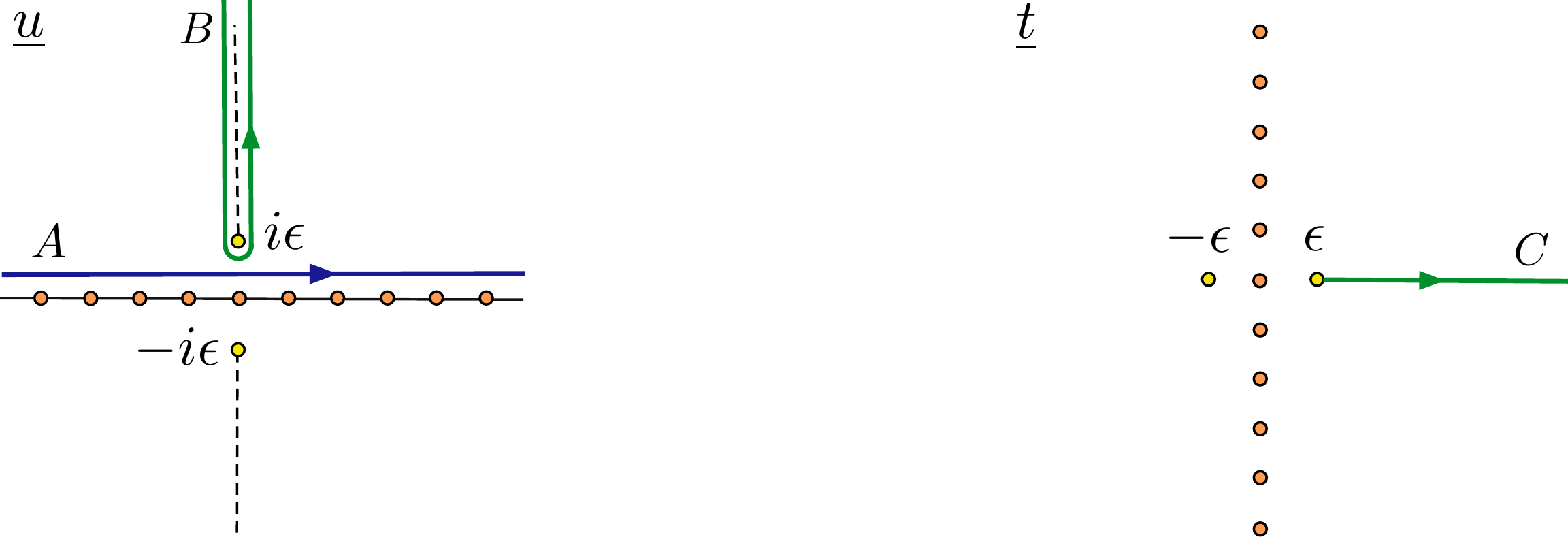}
 \caption{\small Integration contours for $Z_{\rm PI}$. Orange dots are poles, yellow dots branch points.
 \label{fig:contour-PI-deriv}}
 \end{center} \vskip-5mm
\end{figure}

The derivation is straightforward:

\vskip2mm
\noindent{\bf Scalars:}\\ \noindent
The eigenvalues of $-\nabla^2$ on a sphere of radius $\ell \equiv 1$ are $\lambda_n = n(n+d)$, $n\in \IN$, with degeneracies $D^{d+2}_n$ given by (\ref{Dsods2}), that is $D_{n}^{d+2} = {n+d+1 \choose d+1} - {n + d -1 \choose d+1}$. 
Thus (\ref{HKSP}) can be written as 
\begin{align} \label{HKSP2}
 \log Z_{\rm PI} &=  \int_0^\infty \frac{d\tau}{2\tau} \, e^{-\epsilon^2/4\tau} e^{-\tau \nu^2}\sum_{n=0}^\infty D^{d+2}_n  \, e^{-\tau (n+\frac{d}{2})^2} \, , \qquad \nu \equiv \sqrt{m^2-\tfrac{d^2}{4}}\, .
\end{align}
To perform the sum over $n$, we use the Hubbard-Stratonovich trick, i.e.\ we write
\begin{align} 
 \sum_{n=0}^\infty D^{d+2}_n  \, e^{-\tau (n+\frac{d}{2})^2}
  = \int_A du \, \frac{e^{-u^2/4\tau}}{\sqrt{4\pi \tau}} \, f(u) \, ,
  \qquad f(u) \equiv \sum_{n=0}^\infty D^{d+2}_n  \, e^{iu(n+\frac{d}{2})} \, .
\end{align}
with integration contour  $A=\IR + i \delta$, $\delta>0$, as shown in fig.\ \ref{fig:contour-PI-deriv}. The sum evaluates to
\begin{align}
 f(u)  = \frac{1+e^{iu}}{1-e^{iu}} \, \frac{e^{i \frac{d}{2} u}}{(1-e^{iu})^d} \, ,
\end{align}
We first consider the case  $m> \frac{d}{2}$, so $\nu$ is real and positive. Then, keeping ${\rm Im}\, u = \delta  < \epsilon$, we can perform the $\tau$-integral first in (\ref{HKSP2}) to get
\begin{align}
 \log Z_{\rm PI} =  \int_A \frac{du}{2 \sqrt{u^2+\epsilon^2}} \, e^{-\nu \sqrt{u^2+\epsilon^2}} \, f(u)\, .
\end{align} 
Deforming the contour by folding it up along the two sides of the branch cut to contour $B$ in fig.\ \ref{fig:contour-PI-deriv}, changing variables $u=it$ and using that the square root takes opposite signs on both sides of the cut, we transform this to an integral over $C$ in fig.\ \ref{fig:contour-PI-deriv}:
\begin{align} \label{ZPIreg}
 \log Z_{\rm PI} = \int_\epsilon^\infty \! \frac{dt}{2\sqrt{t^2-\epsilon^2}} \, \frac{1+e^{-t}}{1-e^{-t}} \, \frac{e^{-\frac{d}{2} t + i \nu \sqrt{t^2-\epsilon^2}}+ e^{-\frac{d}{2} t - i \nu \sqrt{t^2-\epsilon^2}}}{(1-e^{-t})^d} \, ,
\end{align}
The result for $0<m \leq \frac{d}{2}$, i.e.\ $\nu = i \mu$ with $0 \leq \mu<\frac{d}{2}$ can be obtained from this by analytic continuation. 
Putting $\epsilon = 0$, this formally becomes
\begin{align} \label{zoenk}
 \log Z_{\rm PI} =  \int_0^\infty \frac{dt}{2t} \, \frac{1+e^{-t}}{1-e^{-t}} \, \chi(t) \, , \qquad \chi(t) = \frac{e^{-(\frac{d}{2}-i \nu) t}+ e^{-(\frac{d}{2} + i \nu)t}}{(1-e^{-t})^d}   \, ,
\end{align}
which we recognize as (\ref{Zbulkdef}) with $\chi(t)$ the scalar character  (\ref{scachi}). Thus we conclude that for scalars, $Z_{\rm PI} = Z_{\rm bulk}$, with $Z_{\rm PI}$ regularized as in (\ref{HKSP}) and $Z_{\rm bulk}$ as in (\ref{ZPIreg}).

\vskip3mm
\noindent{\bf Spinors:}\\ \noindent 
For a Dirac spinor field of mass $m$ we have $Z_{\rm PI} = \int \CD \psi \, e^{-\int \bar\psi (\slashed{\nabla} + m) \psi}$. The relevant formulae for spectrum and degeneracies for general $d$ can be found in appendices \ref{app:Weyldim} and \ref{sec:EPIC}. For 
concreteness we just consider the case $d=3$ here, but the conclusions are valid for Dirac spinors in general. The spectrum of $\slashed{\nabla}+m$ on $S^4$ is $\lambda_n = m \pm (n+2)i$,  $n \in \IN$, with degeneracy $D^5_{n+\frac{1}{2},\frac{1}{2}}=4{n+3 \choose 3}$, so $Z_{\rm PI}$ regularized as in (\ref{HKSP}) is given by
\begin{align} \label{HKSPnr}
\log Z_{\rm PI}=-\int_0^\infty\frac{d\tau}{\tau} e^{-\epsilon^2/4\tau} \, \sum_{n=0}^\infty \mbox{$4{n+3 \choose 3}$} \, e^{-\tau((n+2)^2+m^2)}
\end{align}
Following the same steps as for the scalar case, this can be rewritten as
\begin{align}\label{PIspinor}
\log Z_{\text{PI}}=-\int_\epsilon^\infty \! \frac{dt}{2 \sqrt{t^2-\epsilon^2}} \, \frac{2 \, e^{-t/2}}{1-e^{-t}} \cdot 4 \cdot \frac{e^{-\frac{3}{2} t+im \sqrt{t^2-\epsilon^2}}+e^{-\frac{3}{2} t-im \sqrt{t^2-\epsilon^2}}}{(1-e^{-t})^3} \, .
\end{align}
Putting $\epsilon=0$, this formally becomes
\begin{align} \label{zoenkfermi}
 \log Z_{\rm PI} = - \int_0^\infty \frac{dt}{2t} \, \frac{2 \, e^{-t/2}}{1-e^{-t}}  \, \, \chi(t) \, , \qquad \chi(t) =  4 \cdot \frac{e^{-(\frac{3}{2}+im)t}+e^{-(\frac{3}{2}-im)t}}{(1-e^{-t})^3} \, ,
\end{align}
which we recognize as the fermionic (\ref{Zbulkdef}) with $\chi(t)$ the character of the $\Delta=\frac{3}{2} + i m$ unitary $SO(1,4)$ representation carried by the single-particle Hilbert space of a Dirac spinor quantized on dS$_4$, given by twice the  character (\ref{chigenrep}) of the irreducible representation $(\Delta,S)$ with $S=(\frac{1}{2})$.  Thus we conclude $Z_{\rm PI} = Z_{\rm bulk}$. The comment below (\ref{fermionformula}) generalizes this to all $d$.




\section{Massive higher spins}  

\label{sec:mashsflds}

We first formulate the problem, explaining why it is not nearly as simple as one might have hoped, and then state the result, which turns out to be much simpler than one might have feared. The derivation of the result is detailed in appendix \ref{sec:EPIspins}.

\subsection{Problem} \label{sec:probmassive}


Consider a massive spin-$s \geq 1$ field, more specifically a totally symmetric tensor field $\phi_{\mu_1 \cdots \mu_s}$ on dS$_{d+1}$ satisfying the Fierz-Pauli equations of motion:
\begin{align}
\bigl(-\nabla^2+\ms^2 \bigr) \phi_{\mu_1 \cdots \mu_s} =0 \, , \qquad 
\nabla^\nu \phi_{\nu \mu_1 \cdots \mu_{s-1}} = 0 \, , \qquad
{\phi^\nu}_{\nu \mu_1 \cdots \mu_{s-2}} = 0  \, . \label{FPeom}
\end{align}
Upon quantization, the global single-particle Hilbert space furnishes a massive spin-$s$ representation of $SO(1,d+1)$ with $\Delta = \frac{d}{2} + i \nu$, related to the effective mass $\ms$ appearing above (see e.g.\ \cite{Sleight:2017krf}), and to the more commonly used definition of mass $m$ (see e.g.\ \cite{Hinterbichler:2016fgl}) as
\begin{align} \label{convmdef}
 \ms^2  = (\tfrac{d}{2})^2 + \nu^2 + s \, , \qquad m^2=(\tfrac{d}{2}+s-2)^2 + \nu^2 = (\Delta+s-2)(d+s-2-\Delta) \, .
\end{align}
Then $m=0$ for the photon, the graviton and their higher-spin generalizations, and for $s=1$, $m$ is the familiar spin-1 Proca mass.

The massive spin-$s$ {\it bulk} thermal partition function is immediately obtained by  substituting the massive spin-$s$ character (\ref{chimassivespins}) into the character formula (\ref{Zbulkdef}) for $Z_{\rm bulk}$. For $d\geq 3$,\footnote{For $d=2$, the single-particle Hilbert space splits into $(\Delta,\pm s)$ with $D_{\pm s}^2 = 1$, so $D^2_s \to \sum_\pm D^2_{\pm s} = 2$.}  
\begin{align} \label{Zthnus}
 \log Z_{\b} = \int_0^\infty \frac{dt}{2t} \frac{1+q}{1-q} \,\, D^{d}_{s}  \cdot \frac{q^{\frac{d}{2}+i\nu} +q^{\frac{d}{2}-i\nu} }{(1-q)^d} \, , \qquad q=e^{-t} \, ,
\end{align}
with spin degeneracy factor read off from (\ref{Dsods2}) or (\ref{SOKlowdim}).  

The corresponding free massive spin-$s$ Euclidean path integral on $S^{d+1}$ takes the  form
\begin{align} \label{massiveZPIdefin}
 Z_{\rm PI} = \int \CD \Phi \, e^{-S_E[\Phi]}.
\end{align}
where $\Phi$ includes at least $\phi$. However it turns out that in order to write down a local, manifestly covariant action for massive fields of general spin $s$, one also needs to include a tower of auxiliary Stueckelberg fields of all spins $s'<s$ \cite{Zinoviev:2001dt}, generalizing the familiar Stueckelberg action (\ref{SStueck1}) for massive vector fields. These come with gauge symmetries, which in turn require the introduction of a gauge fixing sector, with ghosts of all spins $s'<s$. The explicit form of the action and gauge symmetries is known, but intricate \cite{Zinoviev:2001dt}. 

Classically, variation of the action with respect the Stueckelberg fields merely  enforces the transverse-traceless (TT) constraints in (\ref{FPeom}), after which the gauge symmetries can be used to put the Stueckelberg fields equal to zero. One might therefore hope the intimidating off-shell $Z_{\rm PI}$  (\ref{massiveZPIdefin}) likewise collapses to just the path integral $Z_{\rm TT}$ over the TT modes of $\phi$ with kinetic term given by the equations of motion (\ref{FPeom}). This is easy to evaluate. The TT eigenvalue spectrum on the sphere follows from $SO(d+2)$ representation theory. As detailed in eqs.\ (\ref{ZPITTmassive})-(\ref{ZPIperpformula}), we can then follow the same steps as in section \ref{sec:scalars}, ending up with\footnote{For $d=2$, $D^4_{n, s} \to \sum_\pm D^4_{n,\pm s} = 2 D^4_{n,s}$.}
\begin{align} \label{ZTTformulaaa}
 \log Z_{\rm TT} = \int_0^\infty \frac{dt}{2t} \, \bigl( q^{i\nu} +q^{-i\nu}  \bigr) \sum_{n \geq s} D^{d+2}_{n, s} \, q^{\frac{d}{2}+n}  \, .
\end{align}
Here $D^{d+2}_{n, s}$ is the dimension of the $SO(d+2)$ representation labeled by the two-row Young diagram $(n,s)$, given explicitly by the dimension formulae in appendix \ref{app:Weyldim}. 
 
Unfortunately, $Z_{\rm TT}$ is {\it not} equal to $Z_{\rm PI}$ on the sphere. The easiest way to see this is to consider an example in odd spacetime dimensions, such as (\ref{dis4sis1}), and observe the result has a logarithmic divergence. A manifestly covariant local QFT path integral on an odd-dimensional sphere cannot possibly have logarithmic divergences. Therefore $Z_{\rm PI} \neq Z_{\rm TT}$. 
The appearance of such nonlocal divergences in $Z_{\rm TT}$ can be traced to the existence of (normalizable) zeromodes in tensor decompositions on the sphere \cite{Fradkin:1983mq,Tseytlin:2013fca}. For example the decomposition $\phi_{\mu} = \phi_\mu^T + \nabla_\mu \varphi$ has the constant $\varphi$ mode as a zeromode,  $\phi_{\mu\nu} = \phi_{\mu\nu}^{\rm TT} + \nabla_{(\mu} \varphi_{\nu)} + g_{\mu\nu} \varphi$ has conformal Killing vector zeromodes, and $\phi_{\mu_1 \cdots \mu_s} = \phi_{\mu_1 \cdots \mu_s}^{\rm TT} + \nabla_{(\mu_1} \varphi_{\mu_2 \cdots \mu_s)} + g_{(\mu_1\mu_2} \varphi_{\mu_3 \cdots \mu_s)}$ has rank $s-1$ conformal Killing tensor zeromodes. 
As shown in \cite{Fradkin:1983mq,Tseytlin:2013fca}, this implies $\log Z_{\rm TT}$ contains a nonlocal UV-divergent term $c_s \,  \log \Lambda$, where $c_s$ is the number of rank $s-1$ conformal Killing tensors. This divergence cannot be canceled by a local counterterm. Instead it must be canceled by contributions from the non-TT part. Thus, in principle, the full off-shell path integral must be carefully evaluated to obtain the correct result.
Computing $Z_{\rm PI}$ for general $s$ on the sphere is not as easy as one might have hoped. 

\subsection{Result}

Rather than follow a brute-force approach, we obtain $Z_{\rm PI}$ in appendix \ref{sec:EPIspins} by a series of relatively simple observations. In fact, upon evaluating the sum in (\ref{ZTTformulaaa}), writing it in a way that brings out a term $\log Z_{\rm bulk}$ as in (\ref{Zthnus}), and observing a conspicuous finite sum of terms bears full responsibility for the inconsistency with locality, the answer suggests itself right away: the non-TT part restores locality simply by canceling this finite sum. This turns out to be equivalent to the non-TT part effectively extending the sum $n \geq s$ in (\ref{ZTTformulaaa}) to $n \geq -1$:\footnote{For $d=2$ use (\ref{ZPIGENMASS}): $D^{4}_{n, s} \to \sum_{\pm} D^{4}_{n,\pm s} = 2  D^{4}_{n,s}$ for $n > -1$, and $D^{4}_{-1,s}\to \sum_{\pm} \frac{1}{2} D^{4}_{-1,\pm s} = D^4_{-1,s}=D^4_{s-1}$.} 
\begin{align} \label{ZPItrueformula}
 \log Z_{\rm PI} = \int_0^\infty \frac{dt}{2t} \, \bigl( q^{i\nu} +q^{-i\nu}  \bigr) \sum_{n \geq -1} D^{d+2}_{n, s} \, q^{\frac{d}{2}+n}  \, ,
\end{align}
where $D^{d+2}_{n,s}$ is given by the explicit formulae in appendix \ref{app:Weyldim}, in particular (\ref{coinapp}). For $n<s$, this is no longer the dimension of an $SO(d+2)$ representation, but it can be rewritten as {\it minus} the dimension of such a representation, as $D^{d+2}_{n,s}=- D^{d+2}_{s-1,n+1}$. 
This extension also turns out to be exactly what is needed for consistency with the unitarity bound (\ref{Unbounds}) and more refined unitarity considerations. A limited amount of explicit path integral considerations combined with the observation that the coefficients $D^{d+2}_{s-1,n+1}$ count conformal Killing tensor mode mismatches between ghosts and longitudinal modes then suffice to establish this is indeed the correct answer. We refer to appendix \ref{sec:EPIspins} for details. 

Using the identity (\ref{coin}), we can write this in a rather suggestive form: 
\begin{align} \label{formuhs2}
 \boxed{\log Z_{\rm PI} =  \log Z_{\rm bulk} - \log Z_{\rm edge} = \int_0^\infty \frac{dt}{2t} \, \frac{1+q}{1-q} \bigl( \chi_{\rm bulk}- \chi_{\rm edge} \bigr) }\,  \, ,
\end{align}
where
 $\chi_{\rm bulk}$ and $\chi_{\rm edge}$ are explicitly given by\footnote{For $d=2$, $D_s^2 \to \sum_\pm D_s^2 =2$ in $\chi_{\rm bulk}$  as in (\ref{Zthnus}). $\chi_{\rm edge}$ remains unchanged.} 
\begin{align} \label{chibulkedgeprev}
 \boxed{\chi_{\rm bulk} \equiv  D_s^d \, \frac{q^{\frac{d}{2}+i \nu}+q^{\frac{d}{2} - i \nu}}{(1-q)^d} \, , \qquad \chi_{\rm edge} \equiv  D_{s-1}^{d+2} \, \frac{q^{\frac{d-2}{2}+i \nu}+q^{\frac{d-2}{2} - i \nu}}{(1-q)^{d-2}}}  
\end{align}
The $\log Z_{\rm bulk}$ term is the character integral for the bulk partition function (\ref{Zthnus}). Strikingly, the correction $\log Z_{\rm edge}$ {\it also} takes the form a character  integral, but with an ``edge'' character $\chi_{\rm edge}$ in {\it two lower} dimensions. By our results of section \ref{sec:scalars} for scalars, $Z_{\rm edge}$ effectively equals the Euclidean path integral of $D_{s-1}^{d+2}$ scalars of mass $\tilde m^2 = \bigl(\tfrac{d-2}{2}\bigr)^2 + \nu^2$ on $S^{d-1}$: 
\begin{align}
  Z_{\rm edge} =  \lint \CD \phi \, e^{-\frac{1}{2} \int_{S^{d-1}} \phi^a (-\nabla^2 + \tilde m^2) \phi^a}  \, , \qquad a=1,\ldots, D_{s-1}^{d+2} \, ,
\end{align}
In particular this gives 1 scalar for $s=1$ and $d+2$ scalars for $s=2$. The $S^{d-1}$ is naturally identified as the static patch horizon, the edge of the global dS spatial $S^d$ hemisphere at time zero, the yellow dot in fig.\ \ref{fig:penrose}. Thus (\ref{formuhs2}) realizes in a precise way the somewhat vague physical expectation (\ref{ZPIexpectZb}). 
Notice the relative minus sign here and in (\ref{formuhs2}): the edge corrections effectively {\it subtract} degrees of freedom.  
We do not have a physical interpretation of these putative edge scalars for general $s$ along the lines of the work reviewed in appendix \ref{sec:QFTcons}. Some clues are that their multiplicity equals the number of conformal Killing tensor modes of scalar type appearing in the derivation in appendix \ref{sec:EPIspins} (the {\tiny $\yng(3)$}-modes for $s=4$ in (\ref{CKTbranching})), and that they become massless 
at the unitarity bound $\nu = \pm i(\frac{d}{2}-1)$, eq.\ (\ref{Unbounds}), where a partially massless field emerges with a scalar gauge parameter.   

Independent of any interpretation, we can summarize the result (\ref{formuhs2})-(\ref{chibulkedgeprev}) as
\begin{align} \label{ZPIbulkedgesummary}
 \log Z_{\rm PI}^{d+1}(s) = \log Z_{\rm bulk}^{d+1}(s) - D_{s-1}^{d+2} \, \log Z_{\rm PI}^{d-1}(0) \, . 
\end{align} 

\subsubsection*{Examples} 

For a $d=2$ spin-$s \geq 1$ field of mass $m^2=(s-1)^2+\nu^2$, $\log Z_{\rm PI} = \int \frac{dt}{2t} \frac{1+q}{1-q} ( \chi_{\rm bulk} - \chi_{\rm edge})$ with
\begin{align}\label{mass spin chars}
 \chi_{\rm bulk}=2 \, \frac{q^{1+i \nu}+q^{1-i \nu}}{(1-q)^2}, \qquad \chi_{\rm edge} = s^2 (q^{i\nu}+q^{-i \nu}) \,  .
\end{align}
That is, $Z_{\rm PI}=Z_{\rm bulk}/Z_{\rm edge}$, with the finite part of $\log Z_{\rm bulk}$ explicitly given by twice (\ref{Zbubufi2}), and with $Z_{\rm edge}$ equal to the Euclidean path integral of $D^4_{s-1} = s^2$ harmonic oscillators of frequency $\nu$ on $S^1$, naturally identified with the $S^1$ horizon of the dS$_3$ static patch, with finite part 
\begin{align}
  Z_{\rm edge}^{\rm fin} = \biggl( \frac{e^{-\pi \nu}}{1-e^{-2\pi \nu}} \biggr)^{s^2}  \, .
\end{align}
The heat-kernel regularized $Z_{\rm PI}$ 
is then, restoring $\ell$ and recalling $\nu=\sqrt{m^2 \ell^2 - (s-1)^2}$,
\begin{equation} \label{exmasssd3}
 \log Z_{\rm PI} = 2  \biggl(\!\frac{\pi \nu^3}{6} 
 -\sum_{k=0}^2 \frac{\nu^k}{k!} \, \frac{{\rm Li}_{3-k}(e^{-2 \pi \nu})}{(2\pi)^{2-k}}  - \frac{\pi \nu^2 \ell}{4 \epsilon} + \frac{\pi \ell^3}{2 \epsilon^3} \biggr) - s^2 \biggl(\!-\pi  \nu - \log(1-e^{-2 \pi  \nu }) +\frac{\pi \ell}{\epsilon } \biggr)
\end{equation}
The $d=3$ spin-$s$ case is worked out as another example in (\ref{ZPIexactmssd3}).

\subsubsection*{General massive representations}

(\ref{ZPItrueformula}) has a natural generalization, presented in appendix \ref{sec:EPIC}, to arbitrary parity-invariant massive $SO(1,d+1)$ representations $R=\oplus_a (\Delta_a,S_a)$, $\Delta_a = \frac{d}{2} + i \nu_a$, $S_a=(s_{a1},\ldots,s_{ar})$: 
\begin{align} \label{ZPIGENMASS}
 \boxed{\log Z_{\rm PI} = \int_0^\infty \frac{dt}{2t} \sum_a (-1)^{F_a}  \bigl( q^{i\nu_a} +q^{-i\nu_a}  \bigr)  \sum_{n \in \frac{F_a}{2}  + \IZ} \Theta\bigl(\tfrac{d}{2}+n \bigr) \, D^{d+2}_{n,S_a} \, q^{\frac{d}{2}+n}}  
\end{align}
where $\Theta(x)$ is the Heaviside step step function with $\boxed{\Theta(0) \equiv \tfrac{1}{2}}$, and $F_a=0,1$ for bosons resp.\ fermions. This is the unique TT eigenvalue sum extension consistent with locality and unitarity constraints. 
As in the $S=(s)$ case, this can be rewritten as a bulk-edge decomposition $\log Z_{\rm PI} = \log Z_{\rm bulk} - \log Z_{\rm edge}$.     
For example, using (\ref{hookmagic}) and the notation explained above it, the analog of (\ref{ZPIbulkedgesummary}) for an $S=(s,1^m)$ field becomes  
\begin{align} \label{hookformula}
 \log Z^{d+1}_{\rm PI}(s,1^m) = \log Z^{d+1}_{\rm bulk}(s,1^m) - D_{s-1}^{d+2} \, \log Z^{d-1}_{\rm PI}(1^m) \, ,
\end{align}  
so here $Z_{\rm edge}$ is the path integral of $D_{s-1}^{d+2}$ massive {\it $m$-form  fields} living on the $S^{d-1}$ edge. In particular this implies the recursion relation $\log Z^{d+1}_{\rm PI}(1^p) = \log Z^{d+1}_{\rm bulk}(1^p) - \log Z^{d-1}_{\rm PI}(1^{p-1})$.    
Similarly for a spin $s=k+\frac{1}{2}$ Dirac fermion, in the notation explained under table (\ref{SOKlowdim}) 
\begin{align} \label{fermionformula}
 \log Z_{\rm PI}^{d+1}(s,\mbox{$\bfhalf$}) = \log Z^{d+1}_{\rm bulk}(s,\mbox{$\bfhalf$}) - \tfrac{1}{2} D_{s-1,\bfhalf}^{d+2} \, \log Z^{d-1}_{\rm PI}(\mbox{$\bfhalf$}) \, ,
\end{align}  
where $Z_{\rm bulk}$ now takes the form of the {\it fermionic} part of (\ref{Zbulkdef}), with $\chi_{\rm bulk}$ as in (\ref{chigenrep}) with $D^d_S=2 \, D^d_{s,\bfhalf}$, the factor 2 due to the field being Dirac. The edge fields are Dirac spinors. Note that because $D^{d+2}_{-\frac{1}{2},\bfhalf} = 0$, the above implies in particular $Z_{\rm PI}(\bfhalf)=Z_{\rm bulk}(\bfhalf)$. 

We do not have a systematic group-theoretic or physical way of identifying the edge field content. For evaluation of $Z_{\rm PI}$ using (\ref{ZEXACT}), this identification is not needed however. Actually the original expansions (\ref{ZPItrueformula}), (\ref{ZPIGENMASS}) are more useful for this, as illustrated in (\ref{startcomp})-(\ref{ZPIexactmssd3}). 


\def\res{\rm res}




\section{Massless higher spins} \label{sec:massless}

\subsection{Problems} \label{sec:outlineprob}

\subsubsection*{Bulk thermal partition function $Z_{\rm bulk}$}

Massless spin-$s$ fields on dS$_{d+1}$ are in many ways quite a bit more subtle than their massive spin-$s$ counterparts. This manifests itself already at the level of the characters $\chi_{{\rm bulk},s}$ needed to compute the bulk ideal gas thermodynamics along the lines of section \ref{sec:thermal}. The $SO(1,d+1)$ unitary representations furnished by their single-particle Hilbert space belong to the discrete series for $d=3$ and to the exceptional series for $d \geq 4$ \cite{Basile:2016aen}. The corresponding characters, discussed in appendix \ref{sec:Zbulkmassless}, are  more intricate than their massive (principal and complementary series) counterparts. A brief look at the general formula (\ref{excserieschi}) or the table of examples (\ref{chiexcex}) suffices to make clear they are far from intuitively obvious --- as is, for that matter, the identification of the representation itself.  
Moreover, \cite{Basile:2016aen} reported their computation of the exceptional series characters disagrees with the original results in \cite{10.3792/pja/1195522333,10.3792/pja/1195523378,10.3792/pja/1195523460}.   

As noted in section \ref{sec:thermal} and appendix \ref{sec:resonances}, the expansion $\chi_{{\rm bulk}}(q) = \sum_q N_k q^k$ can be interpreted as counting the number $N_k$ of static patch quasinormal modes decaying as $e^{-k T/\ell}$. This gives some useful physics intuition for the peculiar form of these characters, explained in appendix \ref{sec:Zbulkmassless}. The characters $\chi_{{\rm bulk},s}(q)$ can  in principle be computed by explicitly constructing and counting {\it physical} quasinormal modes of a massless spin-$s$ field. This is a rather nontrivial problem, however. 

Thus we see that for massless fields, complications appear already in the computation of $Z_{\rm bulk}$. Computing $Z_{\rm PI}$ adds even more complications, due to the presence of negative and zero modes in the path integral. Happily, as we will see, the complications of the latter turn out to be the key to resolving the complications of the former. Our final result for $Z_{\rm PI}$ confirms the identification of the representation made in  \cite{Basile:2016aen} and the original results for the corresponding characters in \cite{10.3792/pja/1195522333,10.3792/pja/1195523378,10.3792/pja/1195523460}. This is explicitly verified by counting quasinormal modes in \cite{Sun:2020sgn}.

\subsubsection*{Euclidean path integral $Z_{\rm PI}$}

We consider massless spin-$s$ fields in the metric-like formalism, that is to say   
totally symmetric double-traceless fields $\phi_{\mu_1 \cdots \mu_s}$, with linearized gauge transformation 
\begin{align} \label{linGT}
 \delta^{(0)}_\xi \phi_{\mu_1 \cdots \mu_s} = \alpha_s \nabla_{(\mu_1} \xi_{\mu_2 \cdots \mu_s)} \, ,
\end{align}
with $\xi$ is traceless symmetric in its $s' = s-1$ indices, and $\alpha_s$ picked by convention.\footnote{As explained in appendix \ref{sec:Gvol}, for compatibility with certain other conventions we adopt, we will pick $\alpha_s \equiv \sqrt{s}$ with symmetrization conventions such that $\phi_{(\mu_1 \cdots \mu_s)}=\phi_{\mu_1 \cdots \mu_s}$. \label{fn:alphas}} 

\noindent We use the notation $s' \equiv s-1$ as it makes certain formulae more transparent and readily generalizable to the partially massless ($0\leq s' < s$) case.  The dimensions of $\phi_s$ and $\xi_{s'}$ are
\begin{align} \label{phixidim}
 \mbox{\large $\frac{d}{2}$} + i \nu_\phi = \Delta_\phi =s'+d-1 \, , \qquad \mbox{\large $\frac{d}{2}$} + i \nu_\xi = \Delta_\xi  = s+d-1  \, .
\end{align}
Note that this value of $\nu_\phi$ assign a mass $m=0$ to $\phi$ according to (\ref{convmdef}). 
The Euclidean path integral of a collection of (interacting) gauge fields $\phi$ on $S^{d+1}$ is formally given by
\begin{align}  \label{formalZPI0}
 Z_{\rm PI} = \frac{\int \CD \phi \, e^{-S[\phi]}}{{\rm \Vol}(\CG)}
\end{align}
where $\CG$ is the group of local gauge transformations. At the one-loop (Gaussian) level $S[\phi]$ is the quadratic Fronsdal action \cite{Fronsdal:1978rb}. Several complications arise compared to the massive case:
\begin{enumerate}
 \item For $s \geq 2$, the Euclidean path integral has negative (``wrong sign'' Gaussian) modes, generalizing the well-known issue arising for the conformal factor in Einstein gravity \cite{Gibbons:1978ac}. These can be dealt with by rotating field integration contours. A complication on the sphere is that rotations at the local field level ensuring positivity of short-wavelength modes causes a finite subset of low-lying modes to go negative, requiring these modes to be rotated back \cite{Polchinski:1988ua}.            
 \item The linearized gauge transformations (\ref{linGT}) have zeromodes: symmetric traceless tensors $\bar\xi_{\mu_1 \ldots \mu_{s-1}}$ satisfying $\nabla_{(\mu_1} \bar\xi_{\mu_2 \cdots \mu_s)}=0$, the Killing tensors of $S^{d+1}$. This requires omitting associated modes from the BRST gauge fixing sector of the Gaussian path integral. As a result, locality is lost, and with it the flexibility to freely absorb various normalization constants into local counterterms without having to keep track of nonlocal residuals.
 \item At the {\it nonlinear} level, the Killing tensors generate a subalgebra of the gauge algebra. The structure constants of this algebra are determined by the TT cubic couplings of the interacting theory \cite{Joung:2013nma}. At least when it is finite-dimensional, as is the case for Yang-Mills, Einstein gravity and the 3D higher-spin gravity theories of section \ref{sec:HSCS}, the Killing tensor algebra exponentiates to a group $G$. For example for Einstein gravity, $G=SO(d+2)$. To compensate for the zeromode omissions in the path integral, one has to divide by the volume of $G$. The appropriate measure determining this volume is inherited from the path integral measure, and depends on the UV cutoff and the coupling constants of the theory. Precisely relating the path integral volume $\vg$ to the ``canonical'' $\vc$ defined by a theory-independent invariant metric on $G$ requires considerable care in defining and keeping track of normalization factors.  
\end{enumerate}
Note that these complications do {\it not} arise for massless spin-$s$ fields on AdS with standard boundary conditions. In particular the algebra generated by the (non-normalizable) Killing tensors in this case is a global symmetry algebra, acting nontrivially on the Hilbert space. 

These problems are not insuperable, but they do require some effort. A brute-force path integral computation correctly dealing with all of them for general higher-spin theories is comparable to pulling a molar with a plastic fork: not impossible, but necessitating the sort of stamina some might see as savage and few would wish to witness. The character formalism simplifies the task, and the transparency of the result will make generalization obvious.  

\subsection{Ingredients and outline of derivation} \label{sec:ingredientsoutlineder}

We derive an exact formula for $Z_{\rm PI}$ in appendix \ref{sec:masslessZPI}-\ref{sec:Gvol}. 
In what follows we merely give a rough outline, just to give an idea what the origin is of various ingredients appearing in the final result. To avoid the $d=2$ footnotes of section \ref{sec:mashsflds} we assume $\boxed{d \geq 3}$ in what follows.

\subsubsection*{Naive characters } 

Naively applying the reasoning of section \ref{sec:mashsflds} to the massless case, one gets a character formula of the form (\ref{formuhs2}), with ``naive'' bulk and edge characters $\hat \chi$ given by
\begin{align} \label{XphiminXxi}
 \hat\chi \equiv \chi_{\phi} - \chi_{\xi} \, ,
\end{align}
where $\chi_\phi$, $\chi_\xi$ are the massive bulk/edge characters for the spin-$s$, $\Delta=s'+d-1$ field $\phi$ and the spin-$s'$, $\Delta=s+d-1$ gauge parameter (or ghost) field $\xi$, recalling $s' \equiv s-1$. The subtraction $-\chi_\xi$ arises from the BRST ghost path integral. 
More explicitly, from (\ref{chibulkedgeprev}),
\begin{equation} \label{benaive}
\begin{aligned} 
 \chin_{{\rm bulk},s}  &=  D_s^d \, \frac{q^{\sp+d-1}+q^{1-\sp}}{(1-q)^d} - D_\sp^d \, \frac{q^{s+d-1}+q^{1-s}}{(1-q)^d} \\
 \chin_{{\rm edge},s} &=  D_{s-1}^{d+2} \, \frac{q^{\sp+d-2}+q^{-\sp}}{(1-q)^{d-2}} - D_{\sp-1}^{d+2} \, \frac{q^{s+d-2}+q^{-s}}{(1-q)^{d-2}} \, . 
\end{aligned}
\end{equation}
For example for $s=2$ in $d=3$,
\begin{align} \label{spin2exa}
 \hat \chi_{{\rm bulk},2} = \frac{5 \, ( q^3+1 ) -3 \, ( q^4 +  q^{-1} )}{(1-q)^3} \, , \qquad
 \hat \chi_{{\rm edge},2} = \frac{5 \, (   q^2 + q^{-1} ) - (q^3 + q^{-2})}{1-q} \, .
\end{align}
Because of the presence of non-positive powers of $q$, $\hat\chi_{\rm bulk}$ is manifestly {\it not} the character of any unitary representation of $SO(1,d+1)$. 
 Indeed, the character integral (\ref{formuhs2}) using these naive $\hat \chi$ is badly IR-divergent, due to the presence of non-positive powers of $q$.  

\subsubsection*{Flipped characters} 

In fact this pathology is nothing but the character integral incarnation of the negative and zeromode mode issues of the path integral mentioned under (\ref{formalZPI0}). The zeromodes must be omitted, and the negative modes are dealt with by contour rotations. These prescriptions turn out to translate to a certain ``flipping'' operation at the level of the characters.  
More specifically the flipped character $[\hat \chi]_+$ is obtained from  
$\hat \chi=\sum_k c_k q^k$ by flipping $c_k q^k \to -c_k q^{-k}$ for $k<0$ and dropping the $k=0$ terms:
\begin{align} \label{prevsqbrackplusop}
 \bigl[\hat\chi \bigr]_+ = \Bigl[ \sum_k c_k \, q^k \Bigr]_+ \equiv \sum_{k<0} (-c_k) \, q^{-k} + \sum_{k>0} c_k \, q^k \, = \hat \chi - c_0 - \sum_{k<0} c_k \bigl(q^{k} + q^{-k} \bigr) \, .
\end{align}
For example for $s=2$ in $d=3$, starting from (\ref{spin2exa}) and observing $\hat\chi_{\rm bulk} = -3 \, q^{-1} - 4 + \cdots$ and $\hat\chi_{\rm edge} = - q^{-2}+4 \, q^{-1} + 4 + \cdots$, we get
\begin{align}
  \bigl[\chin_{{\rm bulk},2} \bigr]_+ &= \chin_{{\rm bulk},2}  + 3(q^{-1}+q) + 4 =  \frac{10 \, q^{3} - 6 \, q^4}{(1-q)^3} \, \nn \\
  \bigl[\chin_{{\rm edge},2}\bigr]_+  &= \chin_{{\rm edge},2}  + (q^{-2}+q^2) -4 (q^{-1}+q)  - 4 = \frac{10 \, q^2 - 2 \, q^3}{1-q} \, .
\end{align}
Explicit expressions for general $d$ and $s$ are given by  $[\hat \chi_{\rm bulk}]_+ = \rm (\ref{flippedchargenPM})$ and $[\hat \chi_{\rm edge}]_+ = \rm (\ref{flippedchargenPMedge})$.   
Some simple examples are
\begin{equation} \small
\begin{array}{l|l|l|l}
 d & s  & \bigl[\hat \chi_{{\rm bulk},s}\bigr]_+ \cdot (1-q)^d & \bigl[\hat \chi_{{\rm edge},s}\bigr]_+ \cdot (1-q)^{d-2} \\
 \hline
 2 & \geq 2 & 0 & 0 \\
 3 & \geq 1 & 2(2s+1) \, q^{s+1} -2  (2s-1) \, q^{s+2} & \frac{1}{3} s (s+1) (2 s+1)  \, q^s-\frac{1}{3} (s-1) s (2 s-1) \, q^{s+1}   \\
 4 & \geq 1 & 2(2s+1) \, q^2 & \frac{1}{3}s(s+1)(2s+1) \, q \\
 \geq 3 & 1 & d \, (q^{d-1}+q)  - q^d+1 + (1-q)^d  & q^{d-2} + 1 - (1-q)^{d-2} 
\end{array} \label{charexmpls}
\end{equation}

\subsubsection*{Contributions to $Z_{\rm PI}$} 

To be more precise, after implementing the appropriate contour rotations and zeromode subtractions, we get the following expression for the path integral:
\begin{align} \label{ZPIingr}
  Z_{\rm PI} =  \frac{1}{\vg} \prod_s \left(\CA_s \, i^{-\pol_s} \, Z_{{\rm char},s} \right)^{n_s} \, .
\end{align}
where $n_s$ is the number of massless spin-$s$ fields in the theory, 
and the different factors appearing here are defined as follows:
\begin{enumerate}
\item $Z_{{\rm char},s}$ is defined by the character integral
\begin{align} \label{Zchcha}
  \log Z_{{\rm char},s} \equiv \int_0^\noir \frac{dt}{2t} \, \frac{1+q}{1-q} \, \Bigl(  \bigl[\hat \chi_{{\rm bulk},s}\bigr]_+ - \bigl[\hat \chi_{{\rm edge},s}\bigr]_+ - 2 \, D^{d+2}_{s-1,s-1} \Bigr) \, ,
\end{align}
where $\int_0^\noir$  means $\int_0^\infty$ with the IR divergence due to the constant term removed: 
\begin{align} \label{IRsubtrdef}
  \int_0^\noir \frac{dt}{t} \, f(t) \,\, \equiv \,\, \lim_{L \to \infty} \,\,  \int_0^\infty \frac{dt}{t} \, f(t) \, e^{-t/L} \,\, - \,\, f(\infty) \, \log L  \, .
\end{align}
The flipped $\bigl[\hat \chi_{{\rm bulk},s}\bigr]_+$ turns out to be precisely the massless spin-$s$ exceptional series character $\chi_{{\rm bulk},s}$: (\ref{flippedchargenPM}) = (\ref{excserieschi}). Thus the $\chi_{\rm bulk}$ contribution = ideal gas partition function $Z_{\rm bulk}$, pleasingly consistent with the physics picture. The second term is an edge correction as in the massive case. The third term has no massive counterpart, tied to the presence of gauge zeromodes: 
$D_{s-1,s-1}^{d+2}$ counts rank $s-1$ Killing tensors on $S^{d+1}$.

\item  $\CA_s$ is due to the zeromode omissions. Denoting $M=2e^{-\game}/\epsilon$ as in (\ref{simplereg}),
\begin{align} \label{CASMAS}
  \log \CA_s \equiv  D^{d+2}_{s-1,s-1} \int_0^{\noir} \frac{dt}{2t}  \bigl(2 + q^{2s+d-4} + q^{2s+d-2} \bigr)  = \frac{1}{2} D^{d+2}_{s-1,s-1} \log \mbox{\large$\frac{M^4}{(2s+d-4)(2s+d-2)}$}
\end{align}  
This term looks ugly. Happily, it will drop out of the final result.

\item $i^{-\pol_s}$ is the spin-$s$ generalization of Polchinski's phase of the one-loop path integral of Einstein gravity on the sphere \cite{Polchinski:1988ua}. It arises because every negative mode contour rotation adds a phase factor $-i$ to the path integral. Explicitly, 
\begin{align} \label{prepolphase}
 \pol_s  
 = \sum_{n=0}^{s-2} D_{s-1,n}^{d+2} + \sum_{n=0}^{s-2} D_{s-2,n}^{d+2} 
 =  D_{s-1,s-1}^{d+3} - D_{s-1,s-1}^{d+2} + D^{d+3}_{s-2,s-2} \, .
\end{align}  
In particular $\pol_1=0$, $\pol_2=D^{d+2}_{1} + D^{d+2}_0 = d+3$ in agreement with \cite{Polchinski:1988ua}. For $d+1=4$, $P_s=\frac{1}{3} s \left(s^2-1\right)^2$ and $i^{-P_s}=1,-1,1,1,1,-1,1,1,\ldots$. For $d+1=2 \text{ mod } 4$, $i^{-P_s}=1$.

\item $\vg$ is discussed below.      

\end{enumerate}


\subsubsection*{Volume of $G$} 

As mentioned under (\ref{formalZPI0}), $G$ is the subgroup of gauge transformations generated by the Killing tensors $\bar\xi_{s-1}$ in the parent {\it interacting} theory on the sphere. Equivalently it is the subgroup of gauge transformations leaving the background invariant. For Einstein gravity, we have a single massless $s=2$ field $\phi_2$. The Killing vectors $\bar \xi_1$ generate diffeomorphisms rigidly rotating the sphere, hence $G=SO(d+2)$. For $SU(N)$ Yang-Mills, we have $N^2-1$ massless $s=1$ fields $\phi^a_1$. The $N^2-1$ Killing scalars $\bar \xi_0^a$ generate constant $SU(N)$ gauge transformations, hence $G=SU(N)$.\footnote{or a quotient thereof, such as $SU(N)/\IZ_N$, depending on other data such as additional matter content. Here and in other instances, we will not try to be precise about the global structure of $G$.} For the 3D higher-spin gravity theories introduced in section \ref{sec:HSCS}, we have massless fields $\phi_s$ of spin $s=2,\ldots,n$. The Killing tensors $\bar\xi_{s-1}$ turn out to generate $G=SU(n)_+ \times SU(n)_-$. 

$\vg$ is the volume of $G$ according to the QFT  path integral measure. We wish to relate it to a  ``canonical'' $\vc$. We use the word ``canonical'' in the sense of defined in a theory-independent way. 
We  determine $\vg/\vc$ given our normalization conventions in appendix \ref{sec:Gvol}. Below we summarize the most pertinent definitions and results.

For Einstein gravity, the Killing vector Lie algebra is $\lieg = \so(d+2)$. Picking a standard basis $M_{IJ}$ satisfying $[M_{IJ},M_{KL}] = \delta_{IK} M_{JL} + \delta_{JL} M_{IK} - \delta_{IL} M_{JK} - \delta_{JK} M_{IL}$, we define the ``canonical'' bilinear form $\langle \cdot|\cdot \rangle_\can$ on $\lieg$ to be the unique invariant bilinear normalized such that
\begin{align} \label{canonicalnorm12}
 \langle M_{IJ} | M_{IJ} \rangle_\can \equiv 1 \, \qquad (I \neq J, \text{no sum}) \, .
\end{align}
This invariant bilinear on $\lieg=\so(d+2)$ defines an invariant metric $ds^2_\can$ on $G=SO(d+2)$. Closed orbits generated by $M_{IJ}$ then have length $\oint ds_c = 2\pi$, and $\vc$ is given by (\ref{vcSO}). 

For higher-spin gravity, the Killing tensor Lie algebra $\lieg$ contains $\so(d+2)$ as a subalgebra with generators $M_{IJ}$.  We define $\langle \cdot|\cdot \rangle_c$ on $\lieg$ to be the unique $\lieg$-invariant bilinear form \cite{Joung:2013nma,Joung:2014qya} normalized by  (\ref{canonicalnorm12}). $\vc$ is defined using the corresponding metric $ds^2_c$ on $G$.

The Killing tensor commutators  are determined by the local gauge algebra $[\delta_\xi,\delta_{\xi'}] = \delta_{[\xi,\xi']}$ as in \cite{Joung:2013nma}. For Einstein or HS gravity, in our conventions (canonical $\phi$ + footnote \ref{fn:alphas}), this gives for the $\so(d+2)$ Killing vector (sub)algebra of $\lieg$
\begin{align} \label{GNGNdef}
 [\bar \xi_1,\bar\xi_1' ] = \sqrt{16 \pi G_{\rm N}} \, [\bar \xi_1',\bar\xi_1]_{\text{Lie}} \, ,
\end{align} 
where $[\cdot,\cdot]_{\text{Lie}}$ is the standard vector field Lie bracket. In Einstein gravity, $G_{\rm N}$ is the Newton constant. In Einstein + higher-order curvature corrections (section \ref{sec:gravETD}) or in higher-spin gravity we take it to {\it define} the Newton constant. It is related to a ``central charge'' $C$ in (\ref{CGNewt}).

Building on \cite{Joung:2013nma,Sleight:2016xqq}, we find the bilinear $\langle \cdot|\cdot \rangle_\can$ determining $\vc$ can then be written as 
\begin{align} \label{predsPIdscHSsum}
 \bigl\langle \bar\xi | \bar\xi \bigr\rangle_{\can} = \frac{4G_{\rm N}}{\Ad} \sum_s \sum_{\alpha=1}^{n_s} (2s+d-4)(2s+d-2) \, \int_{S^{d+1}} \bar\xi^{(\alpha)}_{s-1} \cdot \bar\xi^{(\alpha)}_{s-1} \, ,
\end{align}
where $A_{d-1} = {\rm vol}(S^{d-1})$ is the dS horizon area. On the other hand, the path integral measure computing $\vg$ is derived from the bilinear $\langle \xi|\bar\xi\rangle_{\rm PI} = \frac{M^4}{2\pi} \int \bar\xi \cdot \bar\xi$. From this we can read off the ratio $\vc/\vg$: an awkward product of factors determined by the HS algebra. This turns out to cancel the awkward eigenvalue product of (\ref{CASMAS}), up to a universal factor: 
\begin{align} \label{collapse}
 \frac{\vc}{\vg} \prod_s \CA^{n_s}_s = 
 \biggl(\frac{8 \pi G_{\rm N}}{\Ad}\biggr)^{\frac{1}{2} \dim G} \, ,  
\end{align} 
for all theories covered by \cite{Joung:2013nma}, i.e.\ all parity-invariant HS theories consistent at cubic level. 

For Yang-Mills, $\vc$ is computed using the metric $ds^2_c$ on $\lieg$ defined by the canonically normalized YM action $S=:\frac{1}{4} \int \langle F|F \rangle_\can$. For example for $SU(N)$ YM with $S=-\frac{1}{4} \int {\rm Tr}_N \, F^2$, this gives $\vc = {\rm vol}(SU(N))_{{\rm Tr}_N} = (\ref{volsuN})$.  A similar but simpler computation gives the analog of (\ref{collapse}). See appendix \ref{sec:Gvol} for details on all of the above. 

\def\itm{\noindent $\bullet$ }

\subsection{Result and examples}
\label{sec:masslessfinalres}

Thus we arrive at the following universal formula for the one-loop Euclidean path integral for parity-symmetric (higher-spin) gravity and Yang-Mills gauge theories on $S^{d+1}$, $d \geq 3$: 
\begin{align} \label{ZPIFINAL}
 \boxed{Z^{(1)}_{\rm PI} =  i^{-\pol} \, \prod_{a=0}^K \frac{\gamma_a^{\dim G_a}}{{\rm vol} \, G_a} \cdot \exp \int_0^\noir \frac{dt}{2t} \frac{1+q}{1-q} \, \bigl( \chi_{\rm bulk} - \chi_{\rm edge} - 2 \dim G \bigr) }
\end{align}
\itm $G=G_0 \times G_1 \times \cdots G_K$ is the subgroup of (higher-spin) gravitational and Yang-Mills gauge transformations acting trivially on the background,   
\begin{align} \label{gammadefs}
 \gamma_0 \equiv \sqrt{\frac{8 \pi G_{\rm N}}{\Ad}} \, , \qquad \gamma_1 \equiv \sqrt{\frac{g_1^2}{2 \pi A_{d-3}}}  \, , \qquad \cdots
\end{align}
where $A_n \equiv \Omega_n \ell^n$, $\Omega_n = (\ref{volSn})$, 
 the gravitational and YM coupling constants $G_{\rm N}$ and $g_1,\ldots,g_K$ are defined by the canonically normalized $\so(d+2)$ and YM gauge algebras as explained around (\ref{GNGNdef}), and ${\rm vol} \, G_a$ is the canonically normalized volume of $G_a$, defined in the same part. 

\itm For a theory with $n_s$ massless spin-$s$ fields
\begin{align}
 \chi = \sum_s n_s \chi_s  \, , \qquad \dim G = \sum_s n_s D_{s-1,s-1}^{d+2} \, , \qquad P = \sum_s n_s P_s \, ,
\end{align}   
where $\chi_s = [\hat \chi_s]_+$ are the flipped versions (\ref{prevsqbrackplusop}) of the naive characters (\ref{benaive}), with examples in (\ref{charexmpls}) and general formulae in (\ref{flippedchargenPM}) and (\ref{flippedchargenPMedge}), and 
$P_s = (\ref{prepolphase})$ is the spin-$s$ generalization of the $s=2$ phase $P_2=d+3$ found in \cite{Polchinski:1988ua}.


\itm The heat-kernel regularized integral can be evaluated using (\ref{ZEXACT}), as spelled out in appendix \ref{app:masslesssol}. For odd $d+1$, the finite part can alternatively be obtained by summing residues.\\ 
$\int_0^\noir$ means integration with the IR log-divergence from the constant $-2 \dim G$ term removed as in (\ref{IRsubtrdef}). 
The constant term contribution is then  $\dim G \cdot \bigl( c\, \epsilon^{-1} + \log(2\pi) \bigr)$, so when keeping track of linearly divergent terms is not needed, one can replace (\ref{ZPIFINAL}) by 
\begin{align} \label{ZPIFINAL2}
Z^{(1)}_{\rm PI} =  i^{-\pol} \, \prod_a \frac{(2\pi \gamma_a)^{\dim G_a}}{{\rm vol} \, G_a} \cdot \exp \int_0^\infty \frac{dt}{2t} \frac{1+q}{1-q} \, 
\bigl( \chi_{\rm bulk} - \chi_{\rm edge} \bigr)  \qquad \mbox{(mod $\epsilon^{-1}$)}
\end{align}

\itm The case $d=2$ requires some minor amendments,  discussed in appendix \ref{app:dS3charform}: for $s \geq 2$, nothing changes except $P_s$, and $\chi=0$, resulting in (\ref{HSdS3reg}). Yang-Mills gives (\ref{ZPIYMd3}), or mod $\epsilon^{-1}$ (\ref{YM3D}), equivalent to putting $A_{-1} \equiv 1/2\pi\ell$ in (\ref{gammadefs}), and Chern-Simons (\ref{Z1LCS}). 

\itm The above can be extended to more general theories. For examples $(s,s')$ partially massless gauge fields have characters given by (\ref{flippedchargenPM}) and (\ref{flippedchargenPMedge}), and contribute $D_{s-1,s'}^{d+2}$ to $\dim G$. Fermionic counterparts can be derived following the same steps, with $\hat\chi_{\rm edge}$ given by (\ref{fermionformula}). Fermionic $(s,s')$ PM fields give negative contributions $-D_{s-1,s',\frac{1}{2},\ldots,\frac{1}{2}}^{d+2}$ to $\dim G$. 



\subsubsection*{Example: coefficient $\alpha_{d+1}$ of log-divergent term}

The heat kernel coefficient $\alpha_{d+1}$, i.e.\ the coefficient of the log-divergent term of $\log Z$, can be read off simply as the coefficient of the $1/t$ term in the small-$t$ expansion of the integrand. As explained in \ref{app:masslesssol}, we can just use the original, naive integrand $\hat F(t)=\frac{1}{2t} (\hat \chi_{\rm bulk}-\hat \chi_{\rm edge})$ for this purpose, obtained from (\ref{benaive}). For e.g.\ a massless spin-$s$ field on $S^4$ this immediately  gives $\alpha_4^{(s)} = -\frac{1}{90} \left(75 \, s^4-15 \, s^2+2\right)$, in agreement with eq.\ (2.32) of \cite{Tseytlin:2013fca}. For $s=1,2$,
\begin{equation} \label{gravalphas}
\begin{array}{c|ccccccc}
 d & 3 & 5 & 7 & 9 & 11 & 13 & 15 \\
 \hline
  \alpha^{(1)}_{d+1} & -\frac{31}{45} & -\frac{1271}{1890} & -\frac{4021}{6300} & -\frac{456569}{748440} &
   -\frac{1199869961}{2043241200} & -\frac{893517041}{1571724000} &
   -\frac{17279945447657}{31261590360000} \\
 \alpha^{(2)}_{d+1} & -\frac{571}{45} & -\frac{3181}{140} & -\frac{198851}{5670} & -\frac{74203873}{1496880}
   & -\frac{75059846731}{1135134000} & -\frac{114040703221}{1347192000} &
   -\frac{821333912103503}{7815397590000} \\
\end{array}
\end{equation}
Another case of general interest  
is a partially massless field with $(s,s')=(42,26)$ on $S^{42}$:
\begin{align}
 \alpha_{42}^{(42,26)} &= -\tfrac{5925700837995152105818399547396345088821635783305199815444602762021561970991151
   947221547}{5348867203248512743202760066455665920000000000} \sim - 10^{42} \nn
\end{align}

\subsubsection*{Example: $SU(4)$ Yang-Mills on $S^5$}

As a simple illustration and test of (\ref{ZPIFINAL}), consider $SU(4)$ YM theory on $S^5$ of radius $\ell$ with action $S=\frac{1}{4g^2}\! \int {\rm Tr}_4 \, F^2$, so $G=SU(4)$, $n_1=\dim G = 15$, $\vc=\frac{(2\pi)^9}{6}$ as given by (\ref{volsuN}), $\gamma=\sqrt{\frac{g^2}{(2 \pi)^2 \ell}}$, 
 and $P=0$.  Bulk and edge characters are read off from table (\ref{charexmpls}).  
Thus
\begin{align}
 \log Z^{(1)}_{\rm PI} = \log \frac{\bigl(g/\sqrt{\ell}\bigr)^{15}}{(2\pi)^{15}  \cdot \frac{(2\pi)^9}{6}} + 15 \cdot \int_0^\noir \frac{dt}{2t} \frac{1+q}{1-q} \biggl( \frac{6 \, q^2}{(q-1)^4} - \frac{2 \, q}{(q-1)^2} - 2 \biggr) \, .
\end{align}
The finite part can be evaluated by simply summing residues, similar to (\ref{Zbubufi2}):
\begin{align} \label{MaxwellS5}
 \log Z_{\rm PI}^{\rm fin} = \log \frac{\bigl(g/\sqrt{\ell}\bigr)^{15}}{{\frac{1}{6}(2\pi)^9}} + 15 \cdot \Bigl( \, \frac{5 \, \zeta (3)}{16 \, \pi ^2} +\frac{3 \, \zeta (5)}{16 \, \pi ^4} \, \Bigr) \, .
\end{align}
The $U(1)$ version of this agrees with \cite{Giombi:2015haa} eq.\ (2.27). We could alternatively use (\ref{ZEXACT}) as in \ref{app:masslesssol}, which includes the UV divergent part: $\log Z^{(1)}_{\rm PI} = \log Z_{\rm PI}^{\rm fin} + 15\bigl( \frac{9 \pi }{8} \epsilon^{-5} \ell^5-\frac{5 \pi }{8}\epsilon^{-3} \ell^3-\frac{7 \pi }{16} \epsilon^{-1} \ell \bigr)$.  


  


\def\LR{L}

\subsubsection*{Example: Einstein gravity on $S^3$, $S^4$ and $S^5$}

The exact one-loop Euclidean path integral for Einstein gravity on the sphere can be worked out similarly.  The $S^3$ case is obtained in (\ref{ZPIS3Einst}). The $S^4$ and $S^5$ cases are detailed in \ref{app:masslesssol}, with results including UV-divergent terms given in (\ref{HSdS3reg}), (\ref{fullZchar}), (\ref{dS5Einst}). The finite parts are:
\begin{align} \label{Zfinde}
 Z_{\rm PI}^{\rm fin} = i^{-\pol} \cdot \frac{1}{\vc}  \, 
 \biggl( \frac{8 \pi G_{\rm N}}{A_{d-1}}\biggr)^{\frac{1}{2}\dim G} \cdot \Zchar^{\rm fin} \, ,
\end{align}
\begin{equation} 
\begin{array}{l|l|l|l|l|l}
S^{d+1}&i^{-\pol}&\vc&A_{d-1}&\!\dim \!G\!& \log \Zchar^{\rm fin} \\
 \hline 
S^3 & - i & (2\pi)^4 &2 \pi \ell & 6 & 6 \, \log(2\pi) \\ 
S^4 & -1 & \frac{2}{3} (2\pi)^6 & 4 \pi \ell^2 & 10 & - \frac{571}{45}  \log(\ell/L) 
  +\frac{715}{48} -\log 2 - \frac{47}{3}  \zeta'(-1) +\frac{2}{3}  \zeta'(-3)  \\
S^5 & i & \frac{1}{12} (2\pi)^9 & 2\pi^2 \ell^3 & 15 & 15 \log (2 \pi ) + \frac{65 \, \zeta (3)}{48 \, \pi ^2}+\frac{5 \, \zeta (5)}{16 \, \pi ^4}
\end{array} \label{ZPIEinst}
\end{equation}


\vskip3mm \noindent {\it Checks:} We rederive the $S^3$ result in the Chern-Simons formulation of 3D gravity 
\cite{Witten:1988hc} in appendix \ref{app:EinstCS}, and find precise agreement, with the phase matching for odd framing of the Chern-Simons partition function (it vanishes for even framing). 
The coefficient $-\frac{571}{45}$ of the log-divergent term of the $S^4$ result agrees with \cite{Christensen:1979iy}. The phases agree with \cite{Polchinski:1988ua}. The powers of $G_{\rm N}$ agree with zeromode counting arguments of \cite{Witten:1995gf,Giombi:2013yva}. {The full one-loop partition function on $S^4$ was calculated using zeta-function regularization in \cite{Volkov:2000ih}. Upon correcting an error in the second number of their equation (A.36) we find agreement.} As far as we know, the zeta-function regularized $Z_{\rm PI}^{(1)}$ has not been explicitly computed before for $S^{d+1}$, $d \geq 4$.

\subsubsection*{Higher-spin theories}

Generic Vasiliev higher-spin gravity theories have infinite spin range and $\dim G=\infty$, evidently posing problems for (\ref{ZPIFINAL}). We postpone discussion of this case to section \ref{sec:CHSgr}. Below we consider a 3D higher-spin gravity theory with finite spin range $s=2,\ldots,n$.


\def\hsn{{\rm hs}_n^3}
\section{3D HS$_n$ gravity and the topological string}

\label{sec:HSCS}

As reviewed in appendix \ref{app:EinstCS}, 3D Einstein gravity with positive cosmological constant in Lorentzian or Euclidean signature can be formulated as an $SL(2,\IC)$ resp.\ $SU(2) \times SU(2)$ Chern-Simons theory \cite{Witten:1988hc}.\footnote{Or more precisely an $SO(1,3) = SL(2,\IC)/\IZ_2$ or $SO(4)=\bigl(SU(2) \times SU(2)\bigr)/\IZ_2$ CS theory. For the higher-spin extensions, we could similarly consider quotients. We will use the unquotiented groups here.} 
This has a natural extension to an $SL(n,\IC)$ resp.\ $SU(n) \times SU(n)$ Chern-Simons theory, discussed in appendix \ref{app:HSCS}, which can be viewed as an $s \leq n$ dS$_3$ higher-spin gravity theory, analogous to the AdS$_3$ theories studied e.g.\ in \cite{Blencowe:1988gj,Bergshoeff:1989ns,Castro:2011fm,Perlmutter:2012ds,Campoleoni:2010zq,Ammon:2011nk}. The  Lorentzian/Euclidean actions $S_L/S_E$ are
\begin{align} \label{actionSL3Dg}
 S_L = i S_E = (\lcs + i \kappa) \, S_{\rm CS}[\CA_+] + (\lcs - i \kappa) \, S_{\rm CS}[\CA_-] \, , \qquad \lcs \in \IN, \quad \kappa \in \IR^+ \, ,
\end{align}
where $S_{\rm CS}[\CA] =  \frac{1}{4 \pi} \int {\rm Tr}_n\bigl( \CA \wedge d \CA  + \tfrac{2}{3} \CA \wedge \CA \wedge \CA \bigr)$ and $\CA_\pm$ are ${\rm sl}(n)$-valued connections with reality condition $\CA_\pm^* = \CA_\mp$ for the Lorentzian theory and $\CA_{\pm}^\dagger = \CA_{\pm}$ for the Euclidean theory. 

The Chern-Simons formulation allows all-loop exact results, providing a useful check of our result (\ref{ZPIFINAL}) for $Z_{\rm PI}^{(1)}$ obtained in the metric-like formulation. Besides this, we observed a number of other interesting features, collected in appendix \ref{app:HSCS}, and summarized below. 

\subsubsection*{Landscape of vacua \textnormal{(\ref{sec:landscape})}}
The theory has a set of dS$_3$ vacua (or round $S^3$ solutions in the Euclidean theory), corresponding to different embeddings of ${\rm sl}(2)$ into ${\rm sl}(n)$, labeled by $n$-dimensional representations 
\begin{align}
 R = \oplus_a {\bf m_a} \, , \qquad n = \sum_a m_a \, .
\end{align} of $\su(2)$, i.e.\ by partitions of $n=\sum_a m_a$. 
The radius in Planck units $\ell/G_{\rm N}$ and $Z^{(0)} = e^{-S_E}$ depend on the vacuum $R$ as
\begin{align} \label{SESESE}
  \log Z^{(0)} = \frac{2 \pi \ell}{4 G_N} = 2 \pi \kappa \, T_R \, , \qquad  T_R = \frac{1}{6} \sum_a  m_a (m_a^2-1) \, .
\end{align} 
Note that $\CS^{(0)} = \log Z^{(0)}$ takes the standard Einstein gravity horizon entropy form. The entropy is maximized for the principal embedding, i.e.\ $R={\bf n}$, for which $T_{\bf n} = \tfrac{1}{6} n(n^2-1)$.
The number of vacua equals the number of partitions of $n$: 
\begin{align}
 \CN_{\rm vac} \sim e^{2 \pi \sqrt{n/6}} \, . 
\end{align}
For, say,  $n \sim 2 \times 10^5$, we get $\CN_{\rm vac} \sim 10^{500}$, with maximal entropy $\CS^{(0)}|_{R={\bf n}} \sim 10^{15} \kappa$.  
 
\subsubsection*{Higher-spin algebra and metric-like field content \textnormal{(\ref{sec:hsa})} } 
 
As worked out in detail for the AdS analog in \cite{Campoleoni:2010zq}, the fluctuations of the Chern-Simons connection for the principal embedding vacuum $R={\bf n}$ correspond in a metric-like description to a set of massless spin-$s$ fields with $s=2,3,\ldots,n$. The Euclidean higher-spin algebra is $\su(n)_+ \oplus \su(n)_-$, which exponentiates to $G=SU(n)_+ \times SU(n)_-$. The higher-spin field content of the $R={\bf n}$ vacuum can also be inferred from the decomposition of $\su(n)$ into irreducible representations of $\su(2)$, with $S \in \su(2)$ acting on $L \in \su(n)$ as $\delta L = \epsilon [R(S),L]$, to wit,
\begin{align} \label{princdecomp}
 ({\bf n^2-1})_{{\rm su}(n)} = \sum_{r=1}^{n-1} \, ({\bf 2r+1})_{{\rm su}(2)} \, .
\end{align} 
The $({\bf 2r+1},{\bf 1})$ and $({\bf 1},{\bf 2r+1})$ of $\so(4)=\su(2)_+ \oplus \su(2)_-$ correspond to rank-$r$ self-dual and anti-self-dual Killing tensors on $S^3$, the zeromodes of (\ref{linGT}) for a massless spin-$(r+1)$ field, confirming  $R={\bf n}$ has $n_s=1$ massless spin-$s$ field for $s=2,\ldots,n$. For different vacua $R$, one gets  decompositions different from (\ref{princdecomp}), associated with different field content. For example for $n=12$ and $R={\bf 6} \oplus {\bf 4} \oplus {\bf 2}$, we get $n_1=2$, $n_2=7$, $n_3=8$, $n_4=6$, $n_5=3$, $n_6=1$.  

\subsubsection*{One-loop and all-loop partition function \textnormal{(\ref{sec:zpim}-\ref{sec:epics})} } 

In view of the above higher-spin interpretation, we can compute the one-loop Euclidean path integral on $S^3$ for $\lcs=0$ from our general formula (\ref{ZPIFINAL}) for higher-spin gravity theories in the metric-like formalism. The dS$_3$ version of (\ref{ZPIFINAL}) is worked out in (\ref{pols3D})-(\ref{HSdS3}), and applied to the case of interest in (\ref{ZPIkappa}), using (\ref{SESESE}) to convert from $\ell/G_{\rm N}$ to $\kappa$. The result is 
\begin{align} \label{ZPIkappasumm}
 Z_{\rm PI}^{(1)} =   i^{n^2-1} \cdot \frac{\bigl(2 \pi/\sqrt{\kappa}\bigr)^{\dim G}}{{\rm vol}(G)_{{\rm Tr}_n}} \, , 
\end{align}
where ${\rm vol}(G)_{{\rm Tr}_n} = \bigl( \sqrt{n} \prod_{s=2}^n (2\pi)^s/\Gamma(s)\bigr)^2$  as in (\ref{volsuN}). 

This can be compared to the weak-coupling limit of the all-loop expression (\ref{ZR0cs2})-(\ref{ZRframed}), obtained from the known exact partition function of $SU(n)_{k_+} \times SU(n)_{k_-}$ Chern-Simons theory on $S^3$ by analytic continuation $k_{\pm} \to \lcs \pm i \kappa$,   
\begin{align} \label{Zkappaexactsumm}
 Z(R)_r = e^{i r \phi} \cdot \left| \frac{1}{\sqrt{n}}  \frac{1}{(n+\lcs+i\kappa)^{\frac{n-1}{2}}} \prod_{p=1}^{n-1} \Bigl( 2  \sin \frac{\pi p}{n + \lcs + i \kappa} \Bigr)^{(n-p)}   \right|^2 \cdot e^{2 \pi \kappa T_R} \, .
\end{align}
Here $\phi = \frac{\pi}{4} \sum_{\pm} c(\lcs \pm i\kappa)$ with $c(k) \equiv (n^2-1)\bigl(1-\frac{n}{n+k} \bigr)$, and $r \in \IZ$ labels the choice of framing needed to define the Chern-Simons theory as a QFT, discussed in more detail below (\ref{ZCSframed}). Canonical framing corresponds to $r=0$. $Z(R)$ is interpreted as the all-loop quantum-corrected Euclidean partition function of the dS$_3$ static patch in the vacuum $R$. 

The weak-coupling limit $\kappa \to \infty$ of 
 (\ref{Zkappaexactsumm}) precisely reproduces (\ref{ZPIkappasumm}), with the phase matching for odd framing $r$. Alternatively this can be seen more directly by a slight variation of the computation leading to (\ref{Z1LCS}). This provides a  check of (\ref{ZPIFINAL}), in particular its normalization in the metric-like formalism, and of the interpretation of (\ref{actionSL3Dg}) as a higher-spin gravity theory.


\subsubsection*{Large-$n$ limit and topological string dual \textnormal{(\ref{sec:top})} } 

Vasliev-type ${\rm hs}({\rm so}(d+2))$ higher-spin theories (section \ref{sec:CHSgr}) have infinite spin range but finite $\ell^{d-1}/G_{\rm N}$. To mimic this case, consider the $n \to \infty$ limit of the theory at $\lcs=0$. The semiclassical expansion is reliable only if $n \ll \kappa$. Using $\ell/G_{\rm N} \sim \kappa T_R$, this translates to $n \, T_R \ll \ell/G_{\rm N}$, which becomes $n^4 \ll \ell/G_{\rm N}$ for the principal vacuum $R={\bf n}$, and $n \ll \ell/G_{\rm N}$ at the other extreme for $R = {\bf 2} \oplus {\bf 1} \oplus \cdots \oplus {\bf 1}$. Either way, the Vasiliev-like limit $n \to \infty$ at fixed $\CS^{(0)} = 2\pi \ell/4G_{\rm N}$ is strongly coupled.    

However (\ref{Zkappaexactsumm}) continues to make sense in any regime, and in particular {\it does} have a weak coupling expansion in the $n \to \infty$ 't Hooft limit. 
Using the large-$n$ duality between $U(n)_k$ Chern-Simons on $S^3$ and closed topological string theory on the resolved conifold \cite{Gopakumar:1998ki,Marino:2004uf}, the partition function (\ref{Zkappaexactsumm}) of de Sitter higher-spin quantum gravity in the vacuum $R$ can be expressed in terms of the weakly-coupled topological string partition function $\tilde Z_{\rm top}$, 
(\ref{ZRisZtopsquared}):
\begin{align} \label{ZRtoprel}
 \boxed{Z(R)_0 = \left|\tilde Z_{\rm top}(g_s,t) \, e^{-\pi T_R \cdot 2\pi i/g_s}  \right|^2}
\end{align}
where (in the notation of \cite{Marino:2004uf}) the string coupling constant $g_s$ and the resolved conifold K\"ahler modulus 
$t \equiv \int_{S^2} J + i B$  are given by  
\begin{align}\label{topstring3d}
 g_{s}  
  = \frac{2\pi}{n + \lcs + i\kappa} \, , \qquad 
 t = i g_s n = \frac{2 \pi i n}{n + \lcs + i\kappa} 
  \, .
\end{align}
Note that $\bigl| e^{-\pi T_R \cdot 2\pi i/g_s} \bigr|^2 = e^{2 \pi \kappa T_R} = e^{\CS^{(0)}}$, and that $\kappa>0$ implies $\int_{S^2} J > 0$ and ${\rm Im} \, g_s \neq 0$. The dependence on $n$ at fixed $\CS^{(0)}$ is illustrated in fig.\ \ref{fig:CSplots}. 
We leave further exploration of the dS quantum gravity - topological string duality suggested by these observations to future work.

\def\lb{\bar{\ell}}
\def\rhopi{\rho_{\rm PI}}

\section{Euclidean thermodynamics} \label{sec:all-loop}
In section \ref{sec:enentr} we defined and computed the bulk partition function, energy and entropy of the static patch ideal gas. In this section we define and compute their Euclidean counterparts, building on the results of the previous sections.

\subsection{Generalities} \label{sec:ETDfix}


Consider a QFT on a dS$_{d+1}$ background with curvature radius $\ell$. Wick-rotated to the round sphere metric $g_{\mu\nu}$ of radius $\ell$ 
\rf{app:dSWick}, we get the Euclidean  partition function:
\begin{align} \label{ZPIbare}
 Z_{\rm PI}(\ell) \equiv \mbox{\Large $\int$} \CD \Phi \, e^{-S_E[\Phi]} 
\end{align}
where $\Phi$ collectively denotes all fields. 
The quantum field theory is to be thought of here as a (weakly) interacting low-energy effective field theory with a UV cutoff $\epsilon$.  

Recalling the path integral definition (\ref{ZPIint1}) of the Euclidean vacuum $|O\rangle$ paired with its dual $\langle O|$ as $Z_{\rm PI}=\langle O|O\rangle$, the Euclidean expectation value of the stress tensor is  
\begin{align} \label{Tmunuexpval}
 \langle T_{\mu\nu} \rangle \equiv  \frac{\langle O|T_{\mu\nu}|O\rangle}{\langle O|O\rangle} = - \frac{2}{\sqrt{g}} \frac{\delta}{\delta g^{\mu\nu}} \log Z_{\rm PI} = - \rhopi \, g_{\mu\nu} \, ,
\end{align}
The last equality, in which $\rhopi$ is a constant, follows from $SO(d+2)$ invariance of the round sphere background. Denoting the volume of the sphere by $V = {\rm vol}(S^{d+1}_\ell) = \Omega_{d+1} \ell^{d+1}$, 
\begin{align} \label{rhodefgen}
 \boxed{-\rhopi V = \tfrac{1}{d+1} \mbox{$\int$}\! \sqrt{g} \, \bigl\langle  T^\mu_\mu \bigr\rangle  = \tfrac{1}{d+1} \ell \partial_\ell \log Z_{\rm PI}  = V \partial_V \log Z_{\rm PI}} 
\end{align} 
Reinstating the radius $\ell$,  the sphere metric  in the $S$ coordinates of (\ref{Scoords}) takes the form 
\begin{align}
 ds^2 = (1-r^2/\ell^2) d\tau^2 + (1-r^2/\ell^2)^{-1} dr^2 + r^2 d\Omega^2 \, ,
\end{align}
where $\tau \simeq \tau+2\pi \ell$. Wick rotating $\tau \to i T$ yields the static patch metric. Its horizon at $r=\ell$ has inverse temperature $\beta=2\pi\ell$. On a constant-$T$ slice, the vacuum expectation value of the Killing energy density corresponding to translations of $T$ equals $\rhopi$ at the location $r=0$ of the inertial observer. Away from $r=0$, it is redshifted by a factor $\sqrt{1-r^2/\ell^2}$. The Euclidean vacuum expectation  value $U_{\rm PI}$ of the total static patch energy then equals $\rhopi \sqrt{1-r^2/\ell^2}$ integrated over a constant-$T$ slice: 
\begin{align} \label{UPIcomputation}
 U_{\rm PI} = \rhopi \, \Omega_{d-1} \int_0^\ell dr \, r^{d-1} = \rhopi \, v \, , \qquad v = \frac{\Omega_{d-1} \ell^d}{d} =\frac{V}{2 \pi \ell} \, . 
\end{align}
Note that $v$ is the volume of a $d$-dimensional ball of radius $\ell$ in flat space, so effectively we can think of $U_{\rm PI}$ as the energy of an ordinary ball of volume $v$ with energy density $\rhopi$. 

Combining (\ref{rhodefgen}) and (\ref{UPIcomputation}), the Euclidean energy on this background is obtained as 
\begin{align} \label{UPIdef}
  \boxed{2 \pi \ell \, U_{\rm PI} =   V  \rhopi = - \tfrac{1}{d+1} \, \ell \partial_\ell \log Z_{\rm PI}} 
\end{align}
and the corresponding Euclidean entropy $S_{\rm PI} \equiv \log Z_{\rm PI} + \beta \, U_{\rm PI}$ is 
\begin{align} \label{SPIdef}
 \boxed{S_{\rm PI} = \bigl(1 - \tfrac{1}{d+1} \ell \partial_\ell \bigr) \log Z_{\rm PI} = \bigl(1-V \partial_V \bigr) \log Z_{\rm PI}} 
\end{align}
$S_{\rm PI}$ can thus be viewed as the Legendre transform of $\log Z_{\rm PI}$ trading $V$ for $\rhopi$:  
\begin{align} \label{LegendrelZtoS}
  d\log Z_{\rm PI} = - \rhopi \, dV, \qquad S_{\rm PI} = \log Z_{\rm PI} +  V \rhopi \, , \qquad  dS_{\rm PI} = V d\rhopi \,.
\end{align}
The above differential relations express the first law of (Euclidean) thermodynamics for the system under consideration: using $V=\beta v$ and $\rhopi=U_{\rm PI}/v$, they can be rewritten as
\begin{align} \label{firstlawagain}
 d \log Z_{\rm PI} = - U_{\rm PI} \, d\beta - \beta \rhopi \, dv \, , \qquad  dS_{\rm PI} =  \beta \, d U_{\rm PI} - \beta \rhopi \, dv\, . 
\end{align}
Viewing $v$ as the effective thermodynamic volume as under (\ref{UPIcomputation}), these take the familiar form of the first law, with pressure $p=-\rho$, the familiar cosmological vacuum equation of state. 

The expression (\ref{SPIdef}) for the Euclidean entropy and (\ref{LegendrelZtoS}) naturally generalize to Euclidean partition functions $Z_{\rm PI}(\ell)$ for {\it arbitrary} background geometries $g_{\mu\nu}(\ell) \equiv \ell^2 \tilde g_{\mu\nu}$ with volume $V(\ell) = \ell^{d+1} \tilde V$. In contrast,   the expression (\ref{UPIdef}) for the Euclidean energy is specific to the sphere. A generic geometry has no isometries, so there is no notion of Killing energy to begin with. On the other hand, the density $\rhopi$ appearing in (\ref{LegendrelZtoS}) does generalize to arbitrary backgrounds. The last equality in (\ref{Tmunuexpval}) and the physical interpretation of $\rhopi$ as a Killing energy density no longer apply, but (\ref{rhodefgen}) remains valid.

\subsection{Examples} \label{fixedexamples}
 
\subsubsection*{Free $d=0$ scalar}

To connect to the familiar and to demystify the ubiquitous ${\rm Li}_n(e^{-2\pi\nu})=\sum_k e^{-2\pi k\nu}/k^n$ terms encountered later, consider a scalar of mass $m$ on an $S^1$ of radius $\ell$, a.k.a.\ a harmonic oscillator of frequency $m$ at $\beta=2\pi \ell=V$. 
Using (\ref{kzerocase}) and applying (\ref{UPIdef})-(\ref{SPIdef}) with $\nu(\ell) \equiv m \ell$, 
\begin{equation} \label{harmoscEucl}
\begin{aligned}
   \log Z_{\rm PI} &= 
   \frac{\pi\ell}{\epsilon} - \pi\nu + {\rm Li}_1(e^{-2\pi\nu}) \\
   2 \pi \ell \, U_{\rm PI} = V \rhopi &= -\frac{\pi \ell}{\epsilon} + \pi \nu \coth(\pi\nu)  
   \\
 S_{\rm PI} &=   
 {\rm Li}_1(e^{-2 \pi \nu}) + 2 \pi \nu \, {\rm Li}_0(e^{-2 \pi \nu}) \, , 
\end{aligned}  
\end{equation}
Mod $\Delta E_0 \propto -\epsilon^{-1}$, these are the textbook canonical formulae turned into polylogs by (\ref{polylogs}).
  

\def\mo{m}
\def\mef{m_{\rm eff}}

\subsubsection*{Free scalar in general $d$}  
The Euclidean action of a free scalar on $S^{d+1}$ is 
\begin{align} \label{scalaractionfull}
 S_E[\phi] = \frac{1}{2} \int \! \! \sqrt{g} \, \phi \bigl(-\nabla^2 + \mo^2 + \xi R\bigr) \phi \, ,
\end{align}
with $R=d(d+1)/\ell^2$ the $S^{d+1}$ Ricci scalar. The total effective  mass $\mef^2 = \bigl((\frac{d}{2})^2 + \nu^2\bigr)/\ell^2$ is
\begin{align} \label{teffmass}
 \mef^2 = \mo^2 + \xi R 
 \qquad \Rightarrow \qquad \nu = \sqrt{(\mo \ell)^2 - \eta } \, , \quad \eta \equiv \mbox{\large $\bigl(\frac{d}{2}\bigr)^2$} - d(d+1) \, \xi \, .
\end{align}
Neither $Z_{\rm PI}$ nor the bulk thermodynamic quantities of section \ref{sec:thermal}  distinguish between the $\mo^2$ and $\xi R$ contributions to $\mef^2$, but $U_{\rm PI}$ and $S_{\rm PI}$ do, due to the $\partial_\ell$ derivatives in (\ref{UPIdef})-(\ref{SPIdef}). This results in an additional explicit dependence on $\xi$, as
\begin{align}
 \ell \partial_\ell \log Z_{\rm PI} = \bigl(-\epsilon \partial_\epsilon +  J \cdot \nu \partial_\nu \bigr) \log Z_{\rm PI} \, , \qquad J = \frac{\ell \partial_\ell \nu}{\nu} = \frac{(\mo \ell)^2}{\nu^2} = \frac{\nu^2+\eta}{\nu^2} \, .
\end{align} 
For the minimally coupled case $\xi=0$, the Euclidean and bulk thermodynamic quantities agree, but in general not if $\xi \neq 0$.  
To illustrate this we consider the $d=2$ example. Using (\ref{Zbubufi2}) and (\ref{scalarS3}), restoring $\ell$, and putting $\nu \equiv \sqrt{(\mo \ell)^2-\eta}$ with $\eta=1- 6 \xi$,
\begin{align} \label{Zbubufi3}
 \log Z_{\rm PI} = \frac{\pi \ell^3}{2 \epsilon^3} - \frac{\pi \nu^2 \ell}{4 \epsilon} \,\, + \frac{\pi \nu^3}{6} 
 -\sum_{k=0}^{2} \frac{\nu^k}{k!} \, \frac{{\rm Li}_{3-k}(e^{-2 \pi \nu})}{(2\pi)^{2-k}}    \, .
\end{align}
The corresponding Euclidean energy $U_{\rm PI} = \rhopi \, \pi \ell^2$  (\ref{UPIdef}) is given by  
\begin{align} \label{UPIscalar}
 2 \pi \ell \, U_{\rm PI} = V \rhopi =  
 -\frac{\pi \ell^3}{2 \epsilon ^3}+\frac{\pi (\nu^2+\tfrac{2}{3}\eta ) \ell}{4 \epsilon}-\frac{\pi}{6}   (\nu^2+\eta) \nu \coth (\pi  \nu )
\end{align}
where $V={\rm vol}(S^3_\ell) = 2 \pi^2 \ell^3$. For minimal coupling $\xi=0$ (i.e.\ $\eta=1$), $U^{\rm fin}_{\rm PI}$ equals $U^{\rm fin}_{\rm bulk}$ (\ref{Ufind2}), but not for $\xi \neq 0$. For general $d,\xi$, $U_{\rm PI}^{\rm fin}$ is given by (\ref{Ubulkdimreg}) with the overall factor $m^2$ the mass $\mo^2$ appearing in the action rather than $\mef^2$, 
in agreement with \cite{Dowker:1975tf,Candelas:1975du} or (6.178)-(6.180) of \cite{birrell_davies_1982}.    
The entropy $S_{\rm PI} = \log Z_{\rm PI} + 2\pi \ell \, U_{\rm PI}$ (\ref{SPIdef}) is 
\begin{align} \label{SPIscalar}
 S_{\rm PI} = \frac{\pi\eta}{6} \Bigl(\frac{\ell}{\epsilon} -\nu \coth(\pi \nu) \Bigr)  
  - \sum_{k=0}^{3} \frac{\nu^k}{k!} \, \frac{{\rm Li}_{3-k}(e^{-2 \pi \nu})}{(2\pi)^{2-k}} \,  \, ,
\end{align}
where we used $\coth(\pi\nu) = 1 + 2 \, {\rm Li}_0(e^{-2 \pi \nu})$ (\ref{polylogs}). 
Since $Z_{\rm PI} = Z_{\rm bulk}$ in general for scalars and $U_{\rm PI}=U_{\rm bulk}$ for minimally coupled scalars, $S_{\rm PI} = S_{\rm bulk}$ for minimally coupled scalars. Indeed, after conversion to Pauli-Villars regularization, (\ref{SPIscalar}) equals (\ref{SLamPV}) if $\eta=1$. 
As a check on the results, the first law $dS_{\rm PI} = V d\rhopi$ (\ref{LegendrelZtoS}) can be verified explicitly.

In the $\mo \ell \to \infty$ limit, $S_{\rm PI} \to \frac{\pi}{6} \eta (\epsilon^{-1}-\mo) \ell$, reproducing the well-known scalar one-loop Rindler entropy correction computed by a Euclidean path integral on a conical geometry \cite{Susskind:1994sm,Callan:1994py,Kabat:1995eq,Kabat:1995jq,Larsen:1995ax,Solodukhin:2011gn}. Note that $S_{\rm PI} < 0$ when $\eta<0$. Indeed as reviewed in the Rindler context in appendix \ref{sec:edgecor}, $S_{\rm PI}$ does {\it not} have a statistical mechanical interpretation on its own. Instead it must be interpreted as a correction to the large positive classical gravitational horizon entropy. We discuss this in the de Sitter context in section \ref{sec:gravETD}.

A pleasant feature of the sphere computation is that it avoids replicated or conical geometries: instead of varying a deficit angle, we vary the sphere radius $\ell$, preserving manifest $SO(d+2)$ symmetry, and allowing straightforward exact computation of the Euclidean entropy directly from $Z_{\rm PI}(\ell)$, for arbitrary field content.


\def\ls{l_{s}}  

\subsubsection*{Free 3D massive spin $s$} 
Recall from (\ref{exmasssd3}) that for a $d=2$ massive spin-$s \geq 1$ field of mass $m$, the bulk part of $\log Z_{\rm PI}$ is twice that of a $d=2$ scalar (\ref{Zbubufi3}) with $\nu=\sqrt{(m \ell)^2-\eta}$, $\eta = (s-1)^2$, while the edge part is $-s^2$ times that of a $d=0$ scalar, as in (\ref{harmoscEucl}), with the important difference however that $\nu = \sqrt{(m\ell)^2-\eta}$ instead of $\nu=m\ell$. Another important difference with (\ref{harmoscEucl}) is that in the case at hand, (\ref{UPIdef}) stipulates $V  \rho_{\rm PI} = 2 \pi \ell \, U_{\rm PI}=-\frac{1}{d+1} \ell \partial_\ell \log Z_{\rm PI}$ with $d=2$ instead of $d=0$. As a result, for the bulk contribution, we can just copy the scalar formulae (\ref{UPIscalar}) and (\ref{SPIscalar}) for $U_{\rm PI}$ and $S_{\rm PI}$ setting $\eta=(s-1)^2$, while for the edge contribution we get something rather different from the harmonic oscillator energy and entropy (\ref{harmoscEucl}):  
\begin{align}\label{3dmassives}
  V \rho_{\rm PI} &= 2 \times (\ref{UPIscalar}) 
- s^2 \bigl( -\tfrac{\pi}{3} \tfrac{1}{\epsilon} \ell +\tfrac{\pi}{3} \bigl( \nu^2 + \eta \bigr) \nu^{-1} \coth(\pi\nu)  \bigr) \\
 S_{\rm PI} &= 2 \times (\ref{SPIscalar})  
 - s^2 \bigl(  \tfrac{2\pi}{3} (\tfrac{1}{\epsilon} \ell - \nu) + \tfrac{\pi}{3} \eta  \nu^{-1} \coth(\pi \nu)   + {\rm Li_1}(e^{-2 \pi \nu}) +\tfrac{2 \pi}{3} \nu \, {\rm Li}_0(e^{-2 \pi \nu})  \bigr) 
 \end{align}
The edge contribution renders $S_{\rm PI}$ {\it negative} for all $\ell$. 
In particular, in the  $m \ell \to \infty$ limit, $S_{\rm PI} \to \tfrac{\pi}{3} \bigl((s-1)^2 - 2 s^2 \bigr) \,   \bigl(\epsilon^{-1}  -m  \bigr) \ell \to -\infty$: although the bulk part gives a large positive contribution for $s \geq 2$, the edge part gives an even larger negative contribution. Going in the opposite direction, to smaller $m\ell$, we hit the $d=2$, $s \geq 1$ unitarity bound at $\nu=0$, i.e.\ at $m \ell = \sqrt{\eta} = s-1$. Approaching this bound, the bulk contribution remains finite, while the edge part diverges, again negatively. For $s=1$, $S_{\rm PI} \to \log (m\ell)$, due to the ${\rm Li}_1(e^{-2\pi\nu})$ term, while for $s\geq 2$, more dramatically, we get a pole $S_{\rm PI} \to -\frac{s^2(s-1)}{6}\bigl(m\ell-(s-1) \bigr)^{-1}$, due to the $\eta \nu^{-1} \coth(\pi \nu)$ term. Below the unitarity bound, i.e.\ when $\ell < (s-1)/m$, $S_{\rm PI}$ becomes complex. To be consistent as a perturbative low-energy effective field theory valid down to some length scale $\ls$, massive spin-$s \geq 2$ particles on dS$_3$ must satisfy $m^2 > (s-1)^2/\ls^2$.   

\def\GN{G_{\rm N}}

\subsubsection*{Massless spin 2} 
From the results and examples in section \ref{sec:masslessfinalres}, $\log Z_{\rm PI}^{(1)} = \log Z_{\rm PI,div}^{(1)} + \log Z_{\rm PI,fin}^{(1)} - \frac{(d+3)\pi}{2} \, i$, 
\begin{align} \label{ZPIgraviton}
 \log Z_{\rm PI,fin}^{(1)}(\ell) = -\frac{D_d}{2}  \log \frac{A(\ell)}{4\GN} + \alpha^{(2)}_{d+1} \log \frac{\ell}{\LR} + K_{d+1}  
\end{align}
$D_d = \dim \so(d+2) = \frac{(d+2)(d+1)}{2}$, $A(\ell)= \Omega_{d-1}  \ell^{{d-1}}$,  $\alpha^{(2)}_{d+1}=0$ for even $d$ and given by (\ref{gravalphas}) for odd $d$.  $L$ is an arbitrary length scale canceling out of the sum of finite and divergent parts, and $K_{d+1}$ an exactly computable numerical constant. Explicitly for $d=2,3,4$, from (\ref{ZPIEinst}):  
\begin{equation} \label{Cdcoeff}
\begin{array}{l|l|l}
d & \log Z_{\rm PI,div}^{(1)}  & \log Z_{\rm PI,fin}^{(1)}  \\
 \hline 
2 & 0  - \frac{9 \pi}{2} \frac{1}{\epsilon}    \ell  & -3 \log (\frac{\pi}{2\GN} \ell) + 5 \log(2\pi) \\ 
3 &\frac{8}{3} \frac{1}{\epsilon^4}   \ell^4 -\frac{32}{3} \frac{1}{\epsilon^2}   \ell^2  -\frac{571}{45} \log(\frac{2 e^{-\game}}{\epsilon} L)   & - 5 \log(\frac{\pi}{\GN} \ell^2) -\frac{571}{45} \log(\frac{1}{L} \ell) 
  -\log(\frac{8 \pi}{3})  +\frac{715}{48}  - \frac{47 \, \zeta'(-1)}{3}   +\frac{2 \, \zeta'(-3)}{3}    \\
4 & \frac{15  \pi }{8} \frac{1}{\epsilon^5} \ell^5 -\frac{65  \pi}{24} \frac{1}{\epsilon^3} \ell^3 -\frac{105  \pi }{16} \frac{1}{\epsilon}   \ell  & -\frac{15}{2} \log(\frac{\pi^2}{2\GN} \ell^3) + \log(12) + \frac{27}{2} \log (2 \pi ) + \frac{65 \, \zeta (3)}{48 \, \pi^2}+\frac{5 \, \zeta (5)}{16 \, \pi^4}
\end{array} 
\end{equation}
The one-loop energy and entropy (\ref{UPIdef})-(\ref{SPIdef}) are split accordingly. The finite parts are  
\begin{align} \label{SPIfingen}
 S_{\rm PI,fin}^{(1)} = \log Z_{\rm PI,fin}^{(1)}  + V \rho_{\rm fin}^{(1)} \, , \qquad V \rho_{\rm fin}^{(1)} = \tfrac{1}{2} \tfrac{d-1}{d+1} \, D_d  - \tfrac{1}{d+1} \alpha_{d+1}^{(2)} \, ,
\end{align} 
where as always $2 \pi \ell \, U = V \rho$ with $V=\Omega_{d+1} \ell^{d+1}$. For $d=2,3,4$:
\begin{equation} \label{SPIdivex}
\begin{array}{l|l|l|l}
d & V \rho_{\rm div}^{(1)} &  V \rho_{\rm fin}^{(1)} & S_{\rm PI,div}^{(1)} \\
 \hline 
2 
& 0 + \frac{3 \pi}{2} \frac{1}{\epsilon} \ell
& 1
& -3 \pi \frac{1}{\epsilon}  \ell
 \\ 
3 
& -\frac{8}{3 }\frac{1}{\epsilon^4} \ell^4+\frac{16}{3}\frac{1}{\epsilon^2} \ell^2
&  \frac{5}{2}  + \frac{571}{180}
& -\frac{16}{3} \frac{1}{\epsilon^2}  \ell^2 -\frac{571}{45} \log (\frac{2 e^{-\game}}{\epsilon} L)
 \\
4 
& -\frac{15 \pi}{8} \frac{1}{\epsilon^5} \ell^5+\frac{13 \pi}{8}\frac{1}{\epsilon^3} \ell^3+\frac{21 \pi}{16} \frac{1}{\epsilon} \ell
& \frac{9}{2} 
&-\frac{13  \pi}{12} \frac{1}{\epsilon^3} \ell^3 - \frac{21  \pi}{4} \frac{1}{\epsilon}  \ell  
\end{array} 
\end{equation}


\vskip3mm 

\noindent Like their quasicanonical bulk counterparts, the Euclidean  quantities obtained here are UV-divergent, and therefore ill-defined 
from a low-energy effective field theory point of view. 
However if the metric itself, i.e.\ gravity, is dynamical, these the UV-sensitive terms can be absorbed into standard renormalizations of the gravitational coupling constants, rendering the Euclidean thermodynamics finite and physically meaningful. 
We turn to this next.

\section{Quantum gravitational thermodynamics} \label{sec:gravETD}

In section \ref{sec:all-loop} we considered the Euclidean thermodynamics of  effective field theories on a fixed background geometry. In general the Euclidean partition function and entropy depend on the choice of background metric; more specifically on the background sphere radius $\ell$.    
Here we specialize to  field theories which include the metric itself as a dynamical field, i.e.\ we consider gravitational effective field theories. 
We denote $Z_{\rm PI}$, $\rho_{\rm PI}$ and $S_{\rm PI}$ by $\CZ$, $\varrho$ and $\CS$ in this case: 
\begin{align}
 \CZ = \lint \CD g \, \cdots \, e^{-S_E[g,\ldots]} \, , \qquad S_E[g,\ldots] = \frac{1}{8 \pi G} \int \!\! \sqrt{g} \, \bigl(\Lambda - \tfrac{1}{2} R + \cdots \bigr)  \, .
\end{align} 
The geometry itself being dynamical, we have $\partial_\ell \CZ = 0$, so 
(\ref{UPIdef})-(\ref{SPIdef}) reproduce (\ref{CSlogCZ}):
\begin{align} \label{UPISPIgrav}
 \varrho = 0  \, , \quad \CS = \log \CZ \, ,  
\end{align}
We will assume $d \geq 2$, but it is instructive to first consider $d=0$, i.e.\ 1D quantum gravity coupled to quantum mechanics on a circle. 
Then $\CZ =  \int \frac{d\beta}{2\beta} \, {\rm Tr} \, e^{-\beta H}$, where $\beta$ is the circle size and $H$ is the Hamiltonian of the quantum mechanical system shifted by the 1D cosmological constant.    
 To implement the conformal factor contour rotation of \cite{Gibbons:1978ac} implicit in (\ref{UPISPIgrav}), we pick an integration contour $\beta=2 \pi \ell + i y$ with $y \in \IR$ and $\ell > 0$ the background circle radius. Then $\CZ = \pi i \, \CN(0)$ where $\CN(E)$ is the number of states with $H<E$. 
This being $\ell$-independent implies $\varrho=0$. A general definition of microcanonical entropy is $S_{\rm mic}(E) = \log \CN(E)$. Thus, modulo the content-independent $\pi i$ factor in $\CZ$, $\CS=\log\CZ$ is the microcanonical entropy at zero energy in this case. 

Of course $d=0$ is very different from the general-$d$ case, as there is no classical saddle of the gravitational action, and no horizon. For $d \geq 2$ and $\Lambda \to 0$, the path integral has a semiclassical expansion about a round sphere saddle or radius $\ell_0 \propto 1/\sqrt{\Lambda}$, and $\CS$ is dominated by the leading tree-level horizon entropy (\ref{treeentropy}). 
As in the AdS-Schwarzschild case reviewed in \ref{app:adscft}, the microscopic degrees of freedom accounting for the horizon entropy, assuming they exist, are invisible in the effective field theory. A natural analog of the dual large-$N$ CFT partition function on $S^1 \times S^{d-1}$ microscopically computing the AdS-Schwarzschild free energy may be some dual large-$N$ quantum mechanics coupled to 1D gravity on $S^1$ microscopically computing the dS static patch entropy.
These considerations suggest interpreting $\CS = \log \CZ$ as a macroscopic  approximation to a  microscopic microcanonical  entropy, 
with the semiclassical/low-energy 
expansion mapping to some large-$N$ expansion.     

The one-loop corrected $\CZ$ is obtained by expanding the action to quadratic order about its sphere saddle. The Gaussian $Z_{\rm PI}^{(1)}$  was computed in previous sections. Locality and dimensional analysis imply that one-loop divergences are  $\propto \int R^n$ with $2n \leq d+1$. Picking counterterms canceling all (divergent and finite) local contributions of this type in the limit $\ell_0 \propto 1/\sqrt{\Lambda} \to \infty$, we get a well-renormalized $\CS=\log \CZ$ to this order. 
Proceeding along these lines would be the most straightforward path to the  computational objectives of this section. 
However, when pondering comparisons to microscopic models, one is naturally led to wondering what the actual physics content is of what has been computed. This in turn leads to small puzzles and bigger questions, such as:
\begin{enumerate}
  \item \vskip-1mm 
  \label{ci1} A natural guess would have been that the one-loop correction to the entropy $\CS$ is given by a renormalized version of the Euclidean entropy $S_{\rm PI}^{(1)}$ (\ref{SPIdef}).  However (\ref{UPISPIgrav}) says it is given by a renormalized version of the free energy $\log Z_{\rm PI}^{(1)}$. In the examples given earlier, these two look rather different. Can these considerations be reconciled?
 \item \vskip-2.4mm \label{ci2} Besides local UV contributions absorbed into renormalized coupling constants determining the tree-level radius $\ell_0$, there will be nonlocal IR vacuum energy contributions (pictorially Hawking radiation in equilibrium with the horizon), shifting the radius from $\ell_0$ to $\bar\ell$ by gravitational backreaction.      The effect would be small, $\bar\ell = \ell_0 + O(G)$, but since the leading-order horizon entropy is $S(\ell) \propto \ell^{d-1}/G$, we have $S(\bar \ell) = S(\ell_0) + O(1)$, a shift at the one-loop order of interest. The horizon entropy term in (\ref{UPISPIgrav}) is $\CS^{(0)} = S(\ell_0)$, apparently not taking this shift into account. Can these considerations be reconciled?    
\item \vskip-2.4mm \label{ci3}   
At any order in the large-$\ell_0$ perturbative expansion, UV-divergences can be absorbed into a renormalization of a finite number of renormalized coupling constants, but for the result to be physically meaningful, these 
must be defined in terms of low-energy physical ``observables'', invariant under diffeomorphisms and local field redefinitions. 
In asymptotically flat space, one can use scattering amplitudes for this purpose. 
These are unavailable in the case at hand. What replaces them?    
\end{enumerate} 
To address these and other questions, we follow a slghtly less direct path, summarized below, and explained in more detail including examples in appendix \ref{app:rendS}. 
 
\vskip3mm \noindent {\bf Free energy/quantum effective action for volume} \vskip1mm
\noindent 
We define an off-shell free energy/quantum effective action $\Gamma(V)=-\log Z(V)$ for the volume, the Legendre transform of the off-shell entropy/moment-generating function $S(\rho)$:\footnote{Non-metric fields in the path integral are left implicit. Note ``off-shell'' = on-shell for c.c.\ $\Lambda'=\Lambda-8\pi G \, \rho$.
} 
\begin{align} \label{SrhotoZLef}
  S(\rho) \equiv \log \lint \CD g \, e^{-S_E[g] + \rho \int\! \sqrt{g}} 
  \, , \qquad \log Z (V)\equiv S - V \rho \, , \qquad V=\partial_\rho S =\bigl\langle \, \mbox{$\int\! \sqrt{g}$} \, \bigr\rangle_\rho \, .
\end{align}
At large $V$, the geometry semiclassically fluctuates about a round sphere. 
Parametrizing the mean volume $V$ by a corresponding mean radius $\ell$ as $V(\ell) \equiv \Omega_{d+1} \ell^{d+1}$, 
we have 
\begin{align} \label{Zexplicit}
  Z(\ell) =  \lint_{\!\!\!\text{tree}} \, d\rho \, 
 \lint \CD g \, e^{-S_E[g] + \rho(\int \! \sqrt{g} - V(\ell))} \,  ,
\end{align} 
where $\int_{\text{tree}} d\rho$  means saddle point evaluation, i.e.\ extremization.
The Legendre transform (\ref{SrhotoZLef}) is the same as (\ref{LegendrelZtoS}), 
so we get thermodynamic relations of the same form as (\ref{UPIdef})-(\ref{LegendrelZtoS}): 
\begin{align} \label{SfromlogZsr}
 dS=V d\rho \, , \quad d \log Z = -\rho \, dV \, , \qquad
 \rho 
 = -\tfrac{1}{d+1} \ell\partial_\ell \log Z \, / \, V \, , \quad S 
  = \bigl(1-\tfrac{1}{d+1} \ell \partial_\ell \bigr) \log Z \, . 
\end{align}        
On-shell quantities are obtained at $\rho=0$, i.e.\ at the minimum $\bar\ell$ of the free energy $-\log Z(\ell)$: 
\begin{equation} \label{quantumonshell}
  \varrho = \rho(\bar\ell) 
  = 0 \, , \qquad \CS = S(\bar\ell) = \log Z(\bar\ell) \, , \qquad \bigl\langle \, \mbox{$\int \! \sqrt{g}$} \, \bigr\rangle = \Omega_{d+1} \bar\ell^{d+1} \, .
\end{equation}   
\vskip1mm \noindent{\bf Tree level} \vskip1mm
\noindent At tree level (\ref{Zexplicit}) evaluates to    
\begin{align}
 \log Z^{(0)}(\ell)=-S_E[g_\ell] \, , \qquad \mbox{$g_\ell$ = round $S^{d+1}$ metric of radius $\ell$} \, ,
\end{align}
readily evaluated for any action using $R_{\mu\nu\rho\sigma}= (g_{\mu\rho} g_{\nu\sigma} - g_{\mu\sigma}g_{\nu\rho})/\ell^2$, taking the general form 
\begin{align} \label{Z05Dell}
 \log Z^{(0)} =  \frac{\Omega_{d+1} \ell^{d+1}}{8 \pi G} \bigl(-\Lambda  + \tfrac{d(d+1)}{2} \, \ell^{-2} + z_1  \, \ls^2 \, \ell^{-4} + z_2  \, \ls^4 \, \ell^{-6} + \cdots \bigr) \, . 
\end{align}  
The $z_{n}$ are $R^{n+1}$ coupling constants and $\ls \ll \ell$ is the length scale of UV-completing physics. 
The off-shell entropy and energy density are obtained from $\log Z^{(0)}$ as in (\ref{SfromlogZsr}).
\begin{equation} \label{S0expansion5D}
 S^{(0)}
 = \frac{\Omega_{d-1} \ell^{d-1}}{4G}  \bigl( 1 + s_1  \, \ls^2 \, \ell^{-2} +  \cdots \bigr)  ,  \qquad \rho^{(0)} = \frac{1}{8\pi G} \bigl(\Lambda - \tfrac{d(d-1)}{2} \, \ell^{-2} + \rho_1 \, \ls^2 \, \ell^{-4}  + \cdots\bigr)  
\end{equation}
where $s_n,\rho_n \propto z_n$ and we used $\Omega_{d+1}=\frac{2\pi}{d} \Omega_{d-1}$. 
The on-shell entropy and radius are given by 
\begin{align} \label{onshelltreelevel}
  \CS^{(0)} = S^{(0)}(\ell_0) \, , \qquad \rho^{(0)}(\ell_0)=0 \, ,
\end{align}  
either solved perturbatively for $\ell_0(\Lambda)$ or, more conveniently, viewed as parametrizing $\Lambda(\ell_0)$. 

\vskip3mm \noindent{\bf One loop 
} \vskip1mm
\noindent The one-loop order, (\ref{Zexplicit}) is a by construction  tadpole-free Gaussian path integral, (\ref{ZVdefnosigma}):  
\begin{align} \label{logZzeroplusone}
  \log Z = \log Z^{(0)} + \log Z^{(1)}  \, , \qquad \log Z^{(1)} = \log Z^{(1)}_{\rm PI} + \log Z_{\rm ct} \, ,
\end{align} 
with $Z_{\rm PI}^{(1)}$ as computed in sections \ref{sec:mashsflds}-\ref{sec:massless} and $\log Z_{\rm ct}(\ell)=-S_{E,\rm ct}[g_\ell]$ 
a polynomial counterterm. We define renormalized coupling constants as the coefficients of the $\ell^{d+1-2n}$ terms in the $\ell \to \infty$ expansion of $\log Z$, and fix $\log Z_{\rm ct}$ by equating tree-level and renormalized coefficients of the polynomial part, 
which amounts 
to  the renormalization condition 
\begin{align} \label{rrrprrr}
  \lim_{\ell \to \infty} \partial_\ell \log Z^{(1)} = 0 \, ,
\end{align}
 in even $d+1$ supplemented by $\log Z_{\rm ct}(0) \equiv -\alpha_{d+1} \log(2 e^{-\gamma} L/\epsilon)$, 
 implying $L \partial_L \log Z^{(0)} = \alpha_{d+1}$.

\vskip1mm \noindent {\it Example:} 3D Einstein gravity + minimally coupled scalar (\ref{sec:3dscipip}), putting $\nu \equiv \sqrt{m^2\ell^2-1}$, 
\begin{align} \label{logZ1example}
 \log Z^{(1)} = 
-3 \log \frac{2\pi \ell}{4G}  + 5 \log(2\pi) \,\,
 - \,\, \sum_{k=0}^{2} \frac{\nu^k}{k!} \, \frac{{\rm Li}_{3-k}(e^{-2 \pi \nu})}{(2\pi)^{2-k}} + \frac{\pi\nu^3}{6}  
 \,\, - \,\, \frac{\pi m^3 \ell^3}{6}  + \frac{\pi m \ell}{4}  \, . 
\end{align}
The last two terms are counterterms. The first two are nonlocal graviton terms. The scalar part is $O(1/m\ell)$ for $m\ell \gg 1$ but goes nonlocal at $m\ell \sim 1$, approaching $- \log(m\ell)$ for $m\ell \ll 1$. 

\vskip2mm \noindent Defining $\rho^{(1)}$ and $S^{(1)}$ from $\log Z^{(1)}$ as in (\ref{SfromlogZsr}), and the quantum on-shell  $\bar\ell = \ell_0 + O(G)$  as in (\ref{quantumonshell}), the quantum entropy can be expressed in two  equivalent ways, (\ref{CS1ex})-(\ref{CSEinstf1}): 
\begin{align} \label{ABcases}
 { A}: \,\, \CS  
 = S^{(0)}(\bar\ell) + S^{(1)}(\bar\ell) + \cdots \, , \qquad
 { B}: \,\, \CS  
 = S^{(0)}(\ell_0) + \log Z^{(1)}(\ell_0) + \cdots
\end{align}  
where the dots denote terms neglected in the one-loop approximation. 
This 
simultaneously answers  questions \ref{ci1} and \ref{ci2} on our list, reconciling intuitive  $({ A})$ and (\ref{UPISPIgrav})-based (${ B}$) expectations. 
To make this physically obvious, consider the quantum static patch as two subsystems, geometry (horizon) + quantum fluctuations (radiation), with total energy $\propto \rho=\rho^{(0)} + \rho^{(1)}=0$. If $\rho^{(0)}=0$, the horizon entropy is $S^{(0)}(\ell_0)$. But here we have $\rho=0$, 
so the horizon entropy is actually $S^{(0)}(\bar\ell)=S^{(0)}(\ell_0) + \delta S^{(0)}$, where by the first law (\ref{SfromlogZsr}), $\delta S^{(0)}=V \delta\rho^{(0)}= - V \rho^{(1)}$. 
Adding the radiation entropy $S^{(1)}$ and recalling $\log Z^{(1)}=S^{(1)}-V \rho^{(1)}$ yields $\CS = { A} = { B}$.  
Thus ${ A} = { B}$ is just the usual small+large = system+reservoir approximation, the horizon being the reservoir, and the Boltzmann factor $e^{-V \rho^{(1)}} = e^{-\beta U^{(1)}}$ in $Z^{(1)}$ accounting for the reservoir's entropy change due to energy transfer to the system.    


Viewing the quantum contributions as (Hawking) radiation has its  picturesque merits and correctly conveys their nonlocal/thermal character, e.g.\ ${\rm Li}(e^{-2\pi\nu}) \sim e^{-\beta m}$ for $m\ell \gg 1$ in (\ref{logZ1example}), but might incorrectly convey a presumption of positivity of $\rho^{(1)}$ and $S^{(1)}$. Though positive for minimally coupled scalars (fig.\ \ref{fig:Z1plot}), they are in fact {\it negative} for higher spins (figs.\ \ref{fig:Z1plotHS}, \ref{fig:Z1plotHS4D}), due to edge and group volume contributions. 
Moreover, although the negative-energy backreaction causes the horizon to grow,  partially compensating the negative $S^{(1)}$ by a positive $\delta S^{(0)}=-V \rho^{(1)}$, the former still wins: $\CS^{(1)} \equiv \CS-\CS^{(0)} = S^{(1)} - V \rho^{(1)} = \log Z^{(1)} < 0$.



\vskip5mm \noindent {\bf Computational recipe and examples} \vskip1mm
\noindent
For practical purposes, (B) is the more useful expression in (\ref{ABcases}). Together with (\ref{onshelltreelevel}) computing $\CS^{(0)}$, the exact results for $Z^{(1)}_{\rm PI}$ obtained in previous sections (with $\gamma_0 = \sqrt{2\pi/\CS^{(0)}}$,  see (\ref{gamma0S0}) below), and the renormalization prescription outlined above, it immediately gives 
\begin{align}
  \CS=\CS^{(0)} + \CS^{(1)} + \cdots \, , \qquad \CS^{(0)}=S^{(0)}(\ell_0) \, , \qquad  \CS^{(1)} = \log Z^{(1)}(\ell_0)
\end{align}  
in terms of the renormalized coupling constants, for general effective field theories of gravity coupled to arbitrary matter and gauge fields.        

For 3D gravity, this gives $\CS=\CS^{(0)}-3 \log \CS^{(0)} + 5 \log(2\pi) + O(1/\CS^{(0)})$. We work out and plot several other concrete examples in appendix \ref{sec:exampilezz}: 3D Einstein gravity + scalar (\ref{sec:3dscipip}, fig.\ \ref{fig:Z1plot}), 3D massive spin $s$ (\ref{sec:3dmsps}, fig.\ \ref{fig:Z1plotHS}), 2D scalar (\ref{sec:2dsc}), 4D massive spin $s$ (\ref{sec:4dms}, fig.\ \ref{fig:Z1plotHS4D}), and 3D,4D,5D gravity (including higher-order curvature corrections) (\ref{sec:hdgee}).  Table \ref{explexamplestab} 
in the introduction lists a few more sample results.

\vskip3mm \noindent {\bf Local field redefinitions, invariant coupling constants and physical observables} \vskip2mm
\noindent     



\noindent Although the higher-order curvature corrections to the tree-level dS entropy $\CS^{(0)}=S^{(0)}(\ell_0)$ (\ref{S0expansion5D}) seem superficially similar to curvature corrections to the entropy of black holes in asymptotically flat space \cite{Wald:1993nt,Iyer:1994ys}, there are no charges or other asymptotic observables available here to endow them with physical meaning. Indeed, they have no intrinsic low-energy physical meaning at all, 
as they can be removed order by order in the $\ls/\ell$ expansion 
by a metric field redefinition, 
bringing the entropy to pure Einstein form (\ref{treeentropy}). 
In $Z^{(0)}(\ell)$ (\ref{Z05Dell}), this amounts to setting all $z_n \equiv 0$ 
by a redefinition $\ell \to \ell \sum_n c_{n} \ell^{-2n}$ (\ref{ellredef}). 
The value of $\CS^{(0)}=\max_{\ell \gg \ls} \log Z^{(0)}(\ell)$  remains of course unchanged, providing the unique field-redefinition invariant combination of the coupling constants $G, \Lambda \mbox{(or $\ell_0$)}, z_1,z_2,\ldots$.  




Related to this, as discussed in \ref{sec:hdgee}, caution must be exercised when porting the one-loop graviton contribution in (\ref{ZPIFINAL}) or (\ref{ZPIgraviton}): $\GN$ appearing in $\gamma_0=\sqrt{8\pi \GN/A}$ is the  algebraically defined Newton constant (\ref{GNGNdef}), as opposed to $G$ defined by the Ricci scalar coefficient $\frac{1}{8 \pi G}$ in the low-energy effective action. The former is field-redefinition invariant; the latter is not. In Einstein frame ($z_n=0$) the two definitions coincide, 
hence in a general frame \vskip-6mm  
\begin{align} \label{gamma0S0}
  \gamma_0 = \sqrt{2\pi/\CS^{(0)}} \, .
\end{align} 
Since $\log \CS^{(0)} = \log \frac{A}{4G} + \log(1+ O(\ls^2/\ell_0^2))$, this distinction matters only at $O(\ls^2/\ell_0^{2})$, however. 

In $d=2$, $\CS^{(0)}$ is in fact the only invariant gravitational coupling: 
because the Weyl tensor vanishes identically, any 3D parity-invariant effective gravitational action can be brought to Einstein form by a field redefinition.   In the Chern-Simons formulation  
of \ref{app:EinstCS}, 
$\CS^{(0)}=2\pi\kappa$.   
In $d \geq 3$, the Weyl tensor vanishes on the sphere, but not identically. As a result, there are coupling constants not picked up by the sphere's $\CS^{(0)}=-S_E[g_{\ell_0}]$. 
Analogous  $\CS^{(0)}_M \equiv -S_E[g_M]$ for different saddle geometries $g_M$, approaching Einstein metrics in the limit $\Lambda \propto \ell_0^{-2} \to 0$, can be used instead to probe them, and analogous $\CS_M \equiv  \log \CZ_M$ expanded about $g_M$ provide quantum observables. Section \ref{sec:other} provides a few more details, and illustrates extraction of unambiguous linear combinations of the 4D one-loop correction for 3 different $M$. 
 

This provides the general picture we have in mind as the answer, in principle, to question \ref{ci3} on our list below (\ref{UPISPIgrav}): the tree-level $\CS^{(0)}_M$ are the analog of tree-level scattering amplitudes, and the analog of quantum scattering amplitudes are the quantum $\CS_M$. 


\vskip3mm \noindent {\bf Constraints on microscopic models} \vskip2mm
\noindent
%
%
For pure 3D gravity $\CS^{(0)}=\frac{2\pi}{4G}\bigl(\ell_0+s_1 \ell_0^{-1}+s_2 \, \ell_0^{-3}+\cdots\bigr)$, and to one-loop order we have (\ref{CSres}):
\begin{align} \label{CSoneloop3Dsu}
 \CS = \CS^{(0)} -3 \log \CS^{(0)} + 5 \log(2\pi) + \cdots \, .
\end{align}
Granting\footnote{This does not affect the 1-loop based conclusions below, but  does affect the $c_n$. One could leave $l$ general.}  (\ref{Z0canfram}) with $l=0$ gives the all-loop expansion of pure 3D gravity, taking into account $G \equiv SO(4)$ here while $G \equiv SU(2) \times SU(2)$ there, to all-loop order,    
\begin{equation} \label{alllll}
 \CS = \CS_0 + \log \left| \sqrt{\tfrac{4}{2+i \, \CS_0/{2\pi}}} \, \sin \bigl(\tfrac{\pi}{2 + i\, {\CS_0}/{2\pi}}\bigr)  \right|^2  = \CS_0 - 3 \log \CS_0 + 5 \log(2\pi) + \mbox{$\sum_n$}\, c_n \, \CS_0^{-2n} 
\end{equation} 
where $\CS_0 \equiv \CS^{(0)}$ to declutter notation. 
Note all quantum corrections are strictly nonlocal, i.e.\ no odd powers of $\ell_0$ appear, reflected in the absence of odd powers of $1/\CS_0$. 

Though outside the scope of this paper, let us illustrate how such results 
may be used to constrain microscopic models identifying large-$\ell_0$ and large-$N$ expansions in some way. Say a  
modeler posits a model consisting of $2N$ spins $\sigma_i=\pm 1$ with $H\equiv\sum_i \sigma_i = 0$. The microscopic entropy is $S_{\rm mic} = \log {2N \choose N} = 2\log 2 \cdot N - \frac{1}{2} \log(\pi N) + \sum_{n} c'_n  N^{1-2n}$. There is a unique identification of $\CS_0$ bringing this in a form with the same analytic/locality structure as (\ref{alllll}), to wit, 
$\CS_0=  \log 4 \cdot N   + \sum_{n} c'_n   N^{1-2n}$, resulting in   
\begin{equation}
 S_{\rm mic} \, = \, \CS_0 - \tfrac{1}{2} \log \CS_0 + \log(\tfrac{\pi}{2\log 2}) +\mbox{$\sum_{n}$}\, c_n'' \, \CS_0^{-2n} \, , 
\end{equation}
where $c_1''=-\frac{1}{8} \log 2, c_2''=\frac{3}{64} (\log 2)^2+\frac{1}{48} (\log 2)^3, \ldots$, fully failing to match (\ref{alllll}), starting at one loop. The model is ruled out. 

A slightly more sophisticated modeler might posit $S_{\rm mic} = \log d(N)$, where $d(N)$ is the $N$-th level degeneracy of a chiral boson on $S^1$. 
To leading order $S_{\rm mic} \approx 2\pi \sqrt{N/6} \equiv K$. Beyond, 
 $S_{\rm mic} = K - a' \log K + b' + \sum_n c_n' K^{-n} + O(e^{-K/2})$, where $a'=2$, $b'=\log(\pi^2/6\sqrt{3})$ and $c_n'$ given by \cite{10.1112/plms/s2-17.1.75}. Identifying  $\CS_0=K+\sum_{n} c'_{2n-1} K^{-(2n-1)}$ brings this to the form (\ref{alllll}), yielding $S_{\rm mic} = \CS_0 - a' \log\CS_0 + b' + \sum_n c_n'' \CS_0^{-2n} + O(e^{-\CS_0/2})$, with $c_1''= -\frac{5}{2}$, $c_2''=\frac{37}{12}$, $\ldots$ --- ruled out. 


We actually did not need the higher-loop corrections at all to rule out the above models. In higher dimensions, or coupled to more fields, one-loop constraints moreover become increasingly nontrivial, evident in (\ref{explexamplestab}). For pure 5D gravity (\ref{CSres}),
\begin{equation}
 \CS= \CS^{(0)} -\frac{15}{2} \log \CS^{(0)} + \log(12) + \frac{27}{2} \log (2 \pi ) + \frac{65 \, \zeta (3)}{48 \, \pi^2}+\frac{5 \, \zeta (5)}{16 \, \pi^4} \, .
\end{equation}     
It would be quite a miracle if a microscopic model managed to match this.

\newcommand{\ph}[1]{#1^\text{phys}}

\section{dS, AdS$\pm$, and conformal higher-spin gravity} \label{sec:CHSgr}

Vasiliev higher-spin gravity theories  \cite{Vasiliev:1990en, Vasiliev:2003ev, Bekaert:2005vh} have infinite spin range and an infinite-dimensional higher-spin algebra, $\lieg = \hs(\so(d+2))$, leading to divergences in the one-loop sphere partition function formula (\ref{ZPIFINAL}) untempered by the UV cutoff. In this section we take a closer look at these divergences. We contrast the situation to AdS with standard boundary conditions (AdS$+$), where the issue is entirely absent,  
and we point out that, on the other hand, for AdS with alternate HS boundary conditions (AdS$-$) as well as conformal higher-spin (CHS) theories, similar issues arise. We end with a discussion of their significance. 

\subsection{dS higher-spin gravity} \label{sec:dSHSgr}




Nonminimal type A Vasiliev gravity on dS$_{d+1}$ has a tower of massless spin-$s$ fields for all $s \geq 1$ and a $\Delta=d-2$ scalar. 
We first consider $d=3$. The total bulk and edge characters are obtained by summing  (\ref{charexmpls}) and adding the scalar, as we did for the bulk part  in (\ref{VasilievThermo}): 
\begin{align} \label{VasilievThermo2}
 \chi_{\rm bulk} \, =  \, 2 \cdot \biggl( \frac{q^{1/2} + q^{3/2}}{(1-q)^2} \biggr)^2 - \frac{q}{(1-q)^2} \, , \qquad \chi_{\rm edge} = 2 \cdot \biggl( \frac{q^{1/2} + q^{3/2}}{(1-q)^2} \biggr)^2 \, .
\end{align}
Quite remarkably, the bulk and edge contributions almost exactly cancel:
\begin{align} \label{chicancel3d}
 \chi_{\rm bulk} - \chi_{\rm edge} = - \frac{q}{(1-q)^2} \, .
\end{align}
For $d = 4$ however, we see from (\ref{charexmpls}) that due to the absence of overall $q^s$ suppression factors,  
 the total bulk and edge characters each diverge separately by an overall multiplicative factor: 
\begin{align} \label{chid4}
 \chi_{\rm bulk} = \sum_{s} (2s+1) \cdot \frac{ 2\, q^2}{(1-q)^4} \, , \qquad  \chi_{\rm edge} = \sum_{s} \tfrac{1}{6}s(s+1)(2s+1) \cdot \frac{2\,q}{(1-q)^2} \, .
\end{align} 
This pattern persists for all $d \geq 4$, as can be seen from the explicit form of bulk and edge characters in (\ref{chiexcex}), (\ref{flippedchargenPM}), (\ref{flippedchargenPMedge}). 
For any $d$, there is moreover an infinite-dimensional group volume factor in (\ref{ZPIFINAL}) to make sense of, involving a divergent factor $(\ell^{d-1}/G_{\rm N})^{\dim G/2}$ and the volume of an object of unclear mathematical existence \cite{Monnier:2014tfa}.

Before we continue the discussion of what, if anything, to make of this, we consider AdS$\pm$ and CHS theories within the same formalism. Besides independent interest, this will make clear the issue is neither intrinsic to the formalism, nor to de Sitter.

\subsection{AdS$\pm$ higher-spin gravity} \label{sec:AdSswamp}

\subsubsection*{AdS characters for standard and alternate HS boundary conditions}

Standard boundary conditions on massless higher spin fields $\varphi$ in AdS$_{d+1}$ lead to quantization such that spin-$s$ single-particle states transform in a UIR of ${\rm so}(2,d)$ with primary dimension $\Delta_\varphi=\Delta_+ = s+d-2$. Higher-spin Euclidean AdS one-loop partition functions with these boundary conditions were computed in \cite{Giombi:2013fka, Giombi:2014iua, Gunaydin:2016amv, Giombi:2016pvg, Skvortsov:2017ldz}. In \cite{Giombi:2013yva}, the Euclidean one-loop partition function for alternate boundary conditions ($\Delta_\varphi=\Delta_-=2-s$) was considered. In the EAdS$+$ case, the complications listed under (\ref{formalZPI0}) are absent, but for EAdS$-$ close analogs do appear.

EAdS path integrals can be expressed as character integrals \cite{Sun:2020ame,Basile:2018zoy,Basile:2018acb}, in a form exactly paralleling the formulae and bulk/edge picture of the present work \cite{Sun:2020ame}.\footnote{In this picture, EAdS is viewed as the Wick-rotated AdS-Rindler wedge, with dS$_d$ static patch boundary metric, as in \cite{Keeler:2014hba, Keeler:2016wko}. The bulk character is $\chi \equiv {\rm tr}_G \, q^{i H}$, with $H$ the {\it Rindler} Hamiltonian, {\it not} the global AdS  Hamiltonian. Its $q$-expansion counts {\it quasinormal modes} of the Rindler wedge. The one-loop results are interpreted as corrections to the gravitational thermodynamics of the AdS-Rindler horizon \cite{Sun:2020ame,Keeler:2014hba, Keeler:2016wko}.} The AdS analog of the dS  bulk and edge characters (\ref{chibulkedgeprev}) for a {\it massive} spin-$s$ field $\varphi$ with $\Delta_{\varphi} = \Delta_\pm$ is \cite{Sun:2020ame}
\begin{align} \label{XdefAdS}
 \chi^{\rm AdS \pm}_{\rm bulk,\varphi} \equiv  D_s^d \, \frac{q^{\Delta_{\pm}}}{(1-q)^d} \, , \qquad \chi^{\rm AdS \pm}_{\rm edge,\varphi} \equiv  D_{s-1}^{d+2} \, \frac{q^{\Delta_{\pm}-1}}{(1-q)^{d-2}} \, ,
\end{align} 
where $\Delta_- = d-\Delta_+$. Thus, as functions of $q$,  
\begin{align} \label{dSAdSrelation}
 \chi_\varphi^{\rm dS} = \chi^{\rm AdS+}_\varphi + \chi^{\rm AdS-}_\varphi \, . 
\end{align}
The AdS analog of (\ref{XphiminXxi}) for a massless spin-$s$ field $\phi_s$ with gauge parameter field $\xi_{s'}$ is 
\begin{align} \label{XphiminXxiAdS}
  \hat\chi_s^{{\rm AdS}\pm} \equiv \chi^{\rm AdS\pm}_{\phi} - \chi^{\rm AdS\pm}_{\xi} \, ,
\end{align}
where $\Delta_{\phi,+} = s'+d-1$, $\Delta_{\xi,+} = s+d-1$, $s' \equiv s-1$. 
More explicitly, analogous to (\ref{benaive}),
\begin{align}
\hat\chi^{\rm AdS+}_{{\rm bulk},s} &= \frac{D_{s}^d \, q^{s'+d-1} - D_{s'}^d \, q^{s+d-1}}{(1-q)^d} \, , &
\hat\chi^{\rm AdS+}_{{\rm edge},s} &= \frac{D^{d+2}_{s-1} \, q^{s'+d-2} - D^{d+2}_{s'-1} \, q^{s+d-2}}{(1-q)^{d-2}} \label{AdSpluschar} \\
\hat\chi^{\rm AdS-}_{{\rm bulk},s}  &=   \frac{D_s^d \, q^{1-s'} - D_{s'}^d \, q^{1-s}}{(1-q)^d} \, , &
 \hat\chi^{\rm AdS-}_{{\rm edge},s} &=   \frac{D_{s-1}^{d+2} \, q^{-s'} - D_{s'-1}^{d+2} \, q^{-s}}{(1-q)^{d-2}} \, . \label{AdSminchar}
\end{align} 
The presence of non-positive powers of $q$ in $\chi^{\rm AdS-}$ has a similar path integral interpretation as in the dS case summarized in section \ref{sec:ingredientsoutlineder}. The necessary negative mode contour rotation and zeromode subtractions are again implemented at the character level by flipping characters. In particular
the proper $\chi_s$ to be used in the character formulae for EAdS$\pm$ are 
\begin{align}
 \chi_s^{\rm AdS-} = \bigl[\hat\chi_s^{\rm AdS-} \bigr]_+ \, , \qquad \chi_s^{\rm AdS+} = \bigl[\hat\chi_s^{\rm AdS+} \bigr]_+ = \hat\chi_s^{\rm AdS+}  \, ,
\end{align}
with $[\hat \chi]_+$ defined as in (\ref{prevsqbrackplusop}). The omission of Killing tensor zeromodes for alternate boundary conditions must be compensated by a a division by the volume of the residual {\it gauge} group $G$ generated by the Killing tensors. Standard boundary conditions on the other hand kill these Killing tensor zeromodes: they are not part of the dynamical, fluctuating degrees of freedom. The group $G$ they generate acts nontrivially on the Hilbert space as a global symmetry group. 

\subsubsection*{AdS$+$}

For standard boundary conditions, the character formalism reproduces the original results of \cite{Giombi:2013fka, Giombi:2014iua, Gunaydin:2016amv, Giombi:2016pvg, Skvortsov:2017ldz} by two-line computations \cite{Sun:2020ame}. We consider some examples: 

For nonmimimal type A Vasiliev with $\Delta_0=d-2$ scalar boundary conditions, dual to the free $U(N)$ model, using (\ref{AdSpluschar}) and  the scalar $\chi_0=q^{d-2}/(1-q)^d$, the following total bulk and edge characters are readily obtained: 
\begin{align} \label{chiAdSstdbe}
 \chi_{\rm bulk}^{{\rm AdS}+} =  \sum_{s=0}^\infty \chi_{{\rm bulk},s}^{{\rm AdS}+} = 
 \biggl( \frac{q^{\frac{d}{2}-1} + q^{\frac{d}{2}}}{(1-q)^{d-1}} \biggr)^2 \, , \quad 
  \chi_{\rm edge}^{{\rm AdS}+} = \sum_{s=0}^\infty \chi_{{\rm edge},s}^{{\rm AdS}+} = 
 \biggl( \frac{q^{\frac{d}{2}-1} + q^{\frac{d}{2}}}{(1-q)^{d-1}} \biggr)^2 \, .
\end{align}
The total bulk character takes the singleton-squared form expected from the Flato-Fronsdal theorem \cite{Flato:1978qz}. More interestingly, the edge characters sum up to exactly the same. Thus the generally negative nature of edge ``corrections'' takes on a rather dramatic form here: 
\begin{align}\label{chiAdSstdbe2}
  \chi_{\rm tot}^{{\rm AdS}+} = \chi_{\rm bulk}^{{\rm AdS}+}-\chi_{\rm edge}^{{\rm AdS}+} = 0  \qquad \Rightarrow \qquad \log Z_{\rm PI}^{{\rm AdS}+} = 0 \, .
\end{align}
As $Z^{{\rm AdS}+}_{\rm bulk}$ has an Rindler bulk ideal gas interpretation analogous to the static patch ideal gas of section \ref{sec:thermal} \cite{Sun:2020ame}, the exact bulk-edge cancelation on display here is reminiscent of analogous one-loop bulk-edge cancelations expected in string theory according to the qualitative picture reviewed in appendix \ref{app:rindler}. 

For minimal type A, dual to the free $O(N)$ model, the sum yields an expression which after rescaling of integration variables $t \to t/2$ is effectively equivalent to the $\so(2,d)$ singleton character, which is also the $\so(1,d)$ character of a conformally coupled ($\nu=i/2$) scalar on $S^d$. Using (\ref{zoenk}), this means $Z_{\rm PI}^{{\rm AdS}+}$ equals the sphere partition function on $S^d$, immediately implying the $N \to N-1$ interpretation of \cite{Giombi:2013fka, Giombi:2014iua, Gunaydin:2016amv, Giombi:2016pvg, Skvortsov:2017ldz}.   

For nonminimal type A with $\Delta_0=2$ scalar boundary conditions, dual to an interacting U(N) CFT, the cancelation is {\it almost} exact but not quite: 
\begin{align} \label{chi-int} \chi_{\rm tot}^{{\rm AdS}+} = \frac{\sum_{k=2}^{d-3} \, q^k}{(1-q)^{d-1}} \, . \end{align} 

\subsubsection*{AdS$+$ higher-spin swampland}

In the above examples it is apparent that although the spin-summed $\chi_{\rm bulk}$ has increased effective UV-dimensionality $d^{\rm bulk}_{\rm eff} = 2d-2$, as if we summed KK modes of a compactification manifold of dimension $d-2$, the edge subtraction collapses this back down to a net $d_{\rm eff}=d-1$, {\it decreasing} the original $d$.
Correspondingly, the UV-divergences of $Z^{(1)}_{\rm PI}$ are not those of a $d+1$ dimensional bulk-local theory, but rather of a $d$-dimensional {\it boundary}-local theory. In fact this peculiar property appears necessary for quantum consistency, in view of the non-existence of a nontrivially interacting local {\it bulk} action \cite{Sleight:2017pcz}. 
It appears to be true for all AdS$+$ higher spin theories with a known holographic dual \cite{Sun:2020ame}, but not for all classically consistent higher-spin theories. Thus it appears to be some kind of AdS higher-spin ``swampland'' criterion:
\begin{align} \label{swamplandcrit}
  \mbox{AdS$_{d+1}$ HS theory has holographic dual} \qquad \Rightarrow \qquad d_{\rm eff} = d-1  \, .
\end{align}
Higher-spin theories violating this criterion do exist. Theories with a tower of massless spins $s\geq 2$ and an a priori undetermined number $n$ of real scalars can be constructed in AdS$_3$ \cite{Vasiliev:1999ba,Gaberdiel:2010pz}. Assuming all integer spins $s \geq 2$ are present, the total character sums up to 
\begin{align}
 \chi_{\rm tot} = \frac{2 \, q^2}{(1-q)^2} - \frac{4 \, q}{(1-q)^2} + \sum_{i=1}^n \frac{q^{\Delta_i}}{(1-q)^3} \, . 
\end{align}  
For $t \to 0$ diverges as $\chi_{\rm HS} \sim (n-2)/t^2 + O(1/t)$. To satisfy (\ref{swamplandcrit}), the number of scalars must be $n=2$. This is inconsistent with the $n=4$ AdS$_3$ theory originally conjectured in \cite{Gaberdiel:2010pz} to be dual to a minimal model CFT$_2$, but consistent with the amended conjecture of \cite{Chang:2011mz,Gaberdiel:2012ku,Gaberdiel:2012uj}.

\subsubsection*{AdS$-$}

For alternate boundary conditions, one ends up with a massless higher-spin character formula similar to (\ref{ZPIFINAL}). The factor $\gamma^{\dim G}$ in (\ref{ZPIFINAL}) is consistent  with $\log Z^{{\rm AdS}-}_{\rm PI} \propto (G_{\rm N})^{\frac{1}{2} \sum_s N^{\rm KT}_{s-1}}$ found in \cite{Giombi:2013yva}. 
(\ref{dSAdSrelation}) implies the massless AdS$\pm$ and dS bulk and edge characters are related as 
\begin{align} \label{dSAdSpmrel}
 \boxed{\chi_s^{{\rm AdS}-}  = \chi_s^{\rm dS} - \chi_s^{{\rm AdS}+}}
\end{align}
hence we can read off the appropriate flipped $\chi_s^{{\rm AdS}-}=[\hat\chi_s^{{\rm AdS}-}]_+$ characters from our earlier explicit results (\ref{flippedchargenPM}) and (\ref{flippedchargenPMedge}) for $\chi_s^{\rm dS}$. 
Just like in the dS case, the final result involves divergent spin sums when the spin range is infinite. 

\subsection{Conformal higher-spin gravity}

\subsubsection*{Conformal HS characters}

Conformal (higher-spin) gravity theories \cite{FRADKIN1985233} have (higher-spin extensions of) diffeomorphisms and local Weyl rescalings as gauge symmetries.  If one does not insist on a local action, a general way to construct such theories is to view them as induced theories, obtained by integrating out the degrees of freedom of a conformal field theory coupled to a general background metric and other background fields. In particular one can consider a free $U(N)$ CFT$_d$ in a general metric and higher-spin source background. For even $d$, this results in a local action, which at least at the free level can be rewritten as a theory of towers of partially massless fields with standard kinetic terms \cite{Tseytlin:2013jya,Tseytlin:2013fca}. 
Starting from this formulation of CHS theory on $S^d$ (or equivalently dS$_d$), using our general explicit formulae for partially massless higher-spin field characters (\ref{flippedchargenPM}) and (\ref{flippedchargenPMedge}), and summing up the results, we find  
\begin{align} \label{confSdAdSdS}
 \boxed{\chi_s^{\rm CdS_d} = \chi_s^{{\rm AdS_{d+1}}-} - \chi_s^{{\rm AdS_{d+1}}+}
 = \chi_s^{{\rm dS_{d+1}}} - 2 \, \chi_s^{{\rm AdS_{d+1}}+}} 
\end{align}
where $\chi_s^{\rm C dS_d}$ are the CHS bulk and edge characters and the second equality uses (\ref{dSAdSpmrel}). 
Since we already know the explicit dS and AdS HS bulk and edge characters, this relation also provides the explicit CHS bulk and edge characters. For example  
\begin{equation} \label{charexmplsconfSd} 
\begin{array}{l|l|l|l}
 d & s  & \chi^{\rm CdS_d}_{{\rm bulk},s} \cdot (1-q)^d & \chi^{\rm CdS_d}_{{\rm edge},s} \cdot (1-q)^{d-2} \\
 \hline
 2 & \geq 2 & -4 q^s(1-q) & -2\bigl(s^2 q^{s-1} - (s-1)^2 q^s \bigr) \\
 3 & \geq 1 & 0 & 0   \\
 3 & 0 & -q(1-q) & 0 \\
 4 & \geq 0 & 2 (2 s\!+\!1) \, q^2 + 2 s^2 q^{s+3}-2 (s\!+\!1)^2 q^{s+2}  & 
 \frac{s (s+1) (2 s+1)}{3}  \, q + \frac{(s-1) s^2 (s+1)}{6} 
   \, q^{s+2}-\frac{s (s+1)^2 (s+2)}{6}  \, q^{s+1} \\
 5 & \geq 0 & \frac{(s+1) (2 s+1) (2 s+3)}{3} \, q^2(1-  q) &  \frac{s (s+1) (s+2) (2 s+1) (2 s+3)}{30} \, q (1-q)
\end{array} 
\end{equation}
The bulk $SO(1,d)$ $q$-characters $\chi^{\rm CdS_d}_{{\rm bulk},s}$ computed from (\ref{confSdAdSdS}) agree with the $\so(2,d)$ $q$-characters obtained in \cite{Beccaria:2014jxa}. Edge characters were not derived in \cite{Beccaria:2014jxa}, as they have no role in the  thermal $S^1 \times S^{d-1}$ CHS partition functions studied there.\footnote{A priori the interpretation of the bulk characters in (\ref{charexmplsconfSd}) and those in \cite{Beccaria:2014jxa} is different. Their mathematical equality is a consequence of the enhanced $\so(2,d)$ symmetry allowing to map $S^d \to \IR \times S^{d-1}$.} 

The one-loop Euclidean path integral of the CHS theory on $S^d$ is given by  (\ref{ZPIFINAL}) using the bulk and edge CHS characters $\chi_s^{\rm CdS_d}$ and with $G$ the CHS symmetry group generated by the conformal Killing tensors on $S^d$ (counted by $D^{d+3}_{s-1,s-1}$). 
 The coefficient of the log-divergent term, the Weyl anomaly of the CHS theory, is extracted as usual, by reading off the coefficient of the $1/t$ term in the small-$t$ expansion of the integrand in (\ref{ZPIFINAL}), or more directly from the  ``naive'' integrand $\frac{1}{2t} \frac{1+q}{1-q} \, \hat \chi$. 
 For example for conformal $s=2$ gravity on $S^2$ coupled to $D$ massless scalars, also known as bosonic string theory in $D$ spacetime dimensions, we have $\dim G = \sum_\pm D^{4}_{1,\pm 1} = 6$, generating $G=SO(1,3)$, and from the above table (\ref{charexmplsconfSd}), 
\begin{align} \label{bosstringchar}
 \chi_{\rm tot}    =D \cdot \frac{1+q}{1-q} - \frac{4 q^2}{1-q} + 2 (4 q - q^2) \, .
\end{align}
The small-$t$ expansion of the integrand in (\ref{ZPIFINAL}) for this case is
\begin{align} \label{bosstringan}
 \frac{1}{2t} \frac{1+q}{1-q} \bigl( \chi_{\rm tot} - 12 \bigr)  \to \frac{2(D-2)}{t^3} + \frac{D-26}{3 \, t} + \cdots \, ,
\end{align}
reassuringly informing us the critical dimension for the bosonic string is $D=26$. 
Adding a massless $s=\frac{3}{2}$ field, we get 2D conformal supergravity. For half-integer conformal spin $s$, $\chi_{\rm bulk}=-4 q^s/(1-q)$ and $\chi_{\rm edge} = -2\bigl( (s-\frac{1}{2})(s+\frac{1}{2}) q^{s-1} - (s-\frac{3}{2})(s-\frac{1}{2}) q^{s} \bigr)$.  
Furthermore adding $D'$ massless Dirac spinors, the total fermionic character is  
\begin{align}
 \chi_{\rm tot}^{\rm fer} = D' \cdot \frac{2 \, q^{1/2}}{1-q} - \frac{4 \, q^{3/2}}{1-q} + 4 \, q^{1/2} \, .
\end{align}
The symmetry algebra has $\sum_\pm D^4_{\frac{1}{2},\pm \frac{1}{2}}=4$ fermionic generators, contributing negatively to $\dim G$ in (\ref{ZPIFINAL}). 
Putting everything together,
\begin{align} \label{susystring}
 \frac{1}{2t} \, \frac{1+q}{1-q} \bigl( \chi^{\rm bos}_{\rm tot} - 2(6-4) \bigr) - \frac{1}{2t} \, \frac{\sqrt{q}}{1-q} \, \chi^{\rm fer}_{\rm tot}    \to 
 \frac{2 (D-D')}{t^3}+\frac{2 D+D'-30}{6 \, t}+ \cdots \, ,
\end{align}   
from which we read off supersymmetry + conformal symmetry requires $D'=D=10$. 

More systematically, the Weyl anomaly $\alpha_{d,s}$ can be read off by expanding  $\frac{1}{2t} \frac{1+q}{1-q} \, \hat \chi^{\rm CS^d}$ with $\hat \chi^{\rm CS^d} = \hat\chi^{{\rm AdS_{d+1}}-} - \hat\chi^{{\rm AdS_{d+1}}+}$ given  by  (\ref{AdSpluschar})-(\ref{AdSminchar}) for integer $s$. For example,
\begin{equation} \label{CHSalphas}
\begin{array}{l|l}
 d & -\alpha_{d,s} \\
 \hline
 2 & \frac{2(6 s^2-6 s+1)}{3} \\
 4 & \frac{s^2 (s+1)^2 (14 s^2+14 s+3)}{180}  \\
 6 & \frac{(s+1)^2 (s+2)^2 (22 s^6+198 s^5+671 s^4+1056 s^3+733 s^2+120
   s-50)}{151200} \\
 8 & \frac{(s+1) (s+2)^2 (s+3)^2 (s+4) (150 s^8+3000 s^7+24615 s^6+106725 s^5+261123
   s^4+351855 s^3+225042 s^2+31710 s-14560)}{2286144000} 
\end{array}
\end{equation}
This reproduces the $d=2,4,6$ results of \cite{Tseytlin:2013fca,Tseytlin:2013jya} and generalizes them to any $d$. 

\subsubsection*{Physics pictures}

Cartoonishly speaking, the character relation (\ref{confSdAdSdS}) translates to one-loop partition function relations of the form $Z^{{\rm CS}^d} \sim Z^{{\rm EAdS_{d+1}}-} / Z^{{\rm EAdS_{d+1}}+}$ and $Z^{S^{d+1}} \sim Z^{{\rm CS}^d} \bigl(Z^{{\rm EAdS_{d+1}}+}\bigr)^2$. The first relation can then be understood as a  consequence of the holographic duality between AdS$_{d+1}$ higher-spin theories and free CFT$_d$ vector models \cite{Tseytlin:2013fca,Giombi:2013yva,Tseytlin:2013jya}, while the second relation can be understood as an expression at the Gaussian/one-loop level of $Z^{S^{d+1}} \sim \int \CD\sigma \, \bigl|\psi_{\rm HH}(\sigma)\bigr|^2$, where $\psi_{\rm HH}(\sigma)=\psi_{\rm HH}(0) \, e^{-\frac{1}{2} \sigma K \sigma + \cdots}$ is the late-time dS Hartle-Hawking wave function, related by analytic continuation to the EAdS partition function with boundary conditions $\sigma$ \cite{Maldacena:2002vr}. The factor $\bigl(Z^{{\rm EAdS}_{d+1}+} \bigr)^2$ can then be identified with the bulk one-loop contribution to $|\psi_{\rm HH}(0)|^2$, and $Z^{{\rm cnf}\, S^{d}}$ with $\int \CD\sigma \, e^{- \sigma K \sigma}$, along the lines of \cite{Giombi:2013yva}. 
Along the lines of footnote \ref{fn:dSdS}, perhaps another interpretation of the spin-summed relation (\ref{confSdAdSdS}) exists within the picture of  \cite{Alishahiha:2004md}.

\subsection{Comments on infinite spin range divergences}\label{sec:Vasiliev}



Let us return now to the discussion of section \ref{sec:dSHSgr}. Above we have seen that for EAdS$+$, summing spin characters leads to clean and effortless computation of the one-loop partition function. The group volume factor is absent because the global higher-spin symmetry algebra $\lieg$ generated by the Killing tensors is not gauged. The character spin sum converges, and no additional regularization is required beyond the UV cutoff at $t \sim \epsilon$ we already had in place. The underlying reason for this is that in AdS$+$, the minimal energy of a particle is bounded below by its spin, hence a UV cutoff is effectively also a spin cutoff. In contrast, for dS, AdS$-$ and CHS theories alike, $\lieg$ is gauged, leading to the group volume division factor, and moreover, for $d \geq 4$, the quasinormal mode levels (or energy levels for CHS on $\IR \times S^{d-1}$) are infinitely degenerate, not bounded below by spin, leading to character spin sum divergences untempered by the UV cutoff. The geometric origin  of quasinormal modes decaying as slowly as $e^{-2T/\ell}$ for every spin $s$ in $d \geq 4$ was explained below (\ref{QNMcountml}).  

One might be tempted to use some form of zeta function regularization to deal with divergent sums $\sum_s \chi_s$ such as (\ref{chid4}), which amounts to inserting a convergence factor $\propto e^{-\delta s}$ and discarding the divergent terms in the limit $\delta \to 0$. This might be justified if the discarded divergences were UV, absorbable into local counterterms, but that is not the case here. The divergence is due to {\it low}-energy features, the infinite multiplicity of slow-decaying quasinormal modes, analogous to the divergent thermodynamics of an ideal gas in a box with an infinite number of different massless particle species. Zeta function regularization would give a finite result, but the result would be meaningless. 


As discussed at the end of section \ref{sec:HSCS}, the Vasiliev-like\footnote{``Vasiliev-like'' is meant only in a superficial sense here. The higher-spin algebras are rather different  \cite{Joung:2014qya}.} limit of the 3D HS$_n$ higher-spin gravity theory, $n \to \infty$ with $\lcs=0$ and $\CS^{(0)}$ fixed, is strongly coupled as a 3D QFT. Unsurprisingly, the one-loop entropy ``correction'' $\CS^{(1)} = \log Z^{(1)}$ diverges in this limit: writing the explicit expression for the maximal-entropy vacuum $R = {\bf n}$ in (\ref{explexamplestab}) as a function of $\dim G=2(n^2-1)$, one gets $\CS^{(1)} = \dim G \cdot \log \bigl(\dim G /\sqrt{\CS^{(0)}} \bigr) + \cdots \to \infty$. The higher-spin decomposition (\ref{dimdecomp}) might inspire an ill-advised zeta function regularization along the lines of 
 $\dim G = 2\sum_{r=1}^\infty 2r+1 = 4 \, \zeta(-1) + 2 \, \zeta(0) = -\frac{4}{3}$. This gives $\CS^{(1)} = \frac{2}{3} \log \CS^{(0)} + c$  with $c$ a computable constant --- a finite but meaningless answer. 
In fact, using  (\ref{Zkappaexactsumm}), the all-loop quantum correction to the entropy can be seen to {\it vanish} in the limit under consideration, as illustrated in fig.\ \ref{fig:CSplots}. 
As discussed around (\ref{ZRtoprel}), there are more interesting $n \to \infty$ limits one can consider, taking $\CS^{(0)} \to \infty$ together with $n$. 
In these cases, the weakly-coupled description is not a 3D QFT, but a topological string theory. 



Although these and other considerations suggest massless higher-spin theories with infinite spin range cannot be viewed as weakly-coupled field theories on the sphere, one might wonder whether certain quantities might nonetheless be computable in certain (twisted) supersymmetric versions. We did observe some  hints in that direction. 
One example, with details omitted, is the following. First consider the supersymmetric AdS$_5$ higher-spin theory dual to the 4D $\CN=2$ supersymmetric free $U(N)$ model, i.e.\ the $U(N)$ singlet sector of $N$ massless hypermultiplets, each consisting of two  complex scalars and a Dirac spinor. The AdS$_5$ bulk field content is obtained from this following \cite{Basile:2018dzi}. In their notation, the hypermultiplet corresponds to the $\so(2,4)$ representation $\Di + 2 \, \Rac$.  Decomposing $(\Di + 2 \, \Rac) \otimes (\Di + 2 \, \Rac)$ into irreducible $\so(2,4)$ representations gives the  AdS$_5$ free field content: four $\Delta=2$ and two $\Delta=3$ scalars, one $\Delta=3$, $S=(1,\pm 1)$ 2-form field, six towers of massless spin-$s$ fields for all $s \geq 1$, one tower of massless $S=(s,\pm 1)$ fields for all $s \geq 2$, one  $\Delta=\frac{5}{2}$ Dirac spinor, and four towers of massless spin $s=k+\frac{1}{2}$ fermionic gauge fields 
for all $k \geq 1$. 
Consider now the same field content on $S^5$. The bulk and edge characters are obtained paralleling the steps summarized in section \ref{sec:ingredientsoutlineder}, generalized to the present field content using (\ref{hookformula}) and (\ref{fermionformula}). Each individual spin tower gives rise to a badly divergent spin sum similar to (\ref{chid4}). However, a remarkable conspiracy of cancelations between various bosonic and fermionic bulk and edge contributions in the end leads to a finite, unambiguous net integrand:\footnote{The spin sums are performed by inserting a convergence factor such as $e^{-\delta s}$, but the end result is finite and unambiguous when taking $\delta \to 0$, along the lines of $\lim_{\delta \to 0} \sum_{s \in \frac{1}{2} \IN} \, (-1)^{2s} (2s+1) \, e^{-\delta s} = \frac{1}{4}$.} 
\begin{align} \label{dis4cancelsusy}
 \int \frac{dt}{2t} \biggl( \frac{1+q}{1-q} \, \chi_{\rm tot}^{\rm bos} -
 \frac{2\sqrt{q}}{1-q} \, \chi_{\rm tot}^{\rm fer} \biggr) = 
 -\frac{3}{4} \int \frac{dt}{2t} \, \frac{1+q}{1-q} \, \frac{q}{(1-q)^2}  \, .
\end{align}    
Note that the effective UV dimensionality is reduced by {\it two} in this case. 

An analogous construction for $S^4$ starting from the 3D $\CN=2$ $U(N)$ model,   
gives two $\Delta_\pm =1,2$ scalars, a $\Delta=\frac{3}{2}$ Dirac spinor and two massless spin-$1,\frac{3}{2},2,\frac{5}{2},\ldots$ towers, as in \cite{Sezgin:2012ag,Hertog:2017ymy}.   
The fermionic bulk and edge characters cancel and the bosonic part is twice (\ref{chicancel3d}). In this case we moreover get a finite and unambiguous $\dim G = \lim_{\delta \to 0} \sum_{s \in \frac{1}{2} \IN}^\infty \, (-1)^{2s} \, 2 \, D^5_{s-1,s-1} \, e^{-\delta s} = \frac{1}{4}$. 


The above observations are tantalizing, but leave several problems unresolved, including what to make of the supergroup volume ${\rm vol} \, G$. Actually supergroups present an issue of this kind already with a finite number of generators, as their volume is generically zero. In the context of supergroup Chern-Simons theory this leads to indeterminate $0 / 0$ Wilson loop expectation values \cite{Mikhaylov:2014aoa}. In this case the indeterminacy is resolved by a construction replacing the Wilson loop by an auxiliary worldline quantum mechanics \cite{Mikhaylov:2014aoa}. Perhaps in this spirit, getting a meaningful path integral on the sphere in the present context may require inserting an auxiliary ``observer'' worldline quantum mechanics, with a natural action of the higher-spin algebra on its phase space, allowing to soak up the residual gauge symmetries. 
        
One could consider other options, such as breaking the background isometries,  models with a finite-dimensional higher-spin algebra \cite{Boulanger:2011qt, Joung:2015jza, Manvelyan:2013oua, Brust:2016gjy}, models with an $\alpha'$-like parameter breaking the higher-spin symmetries, or models of a different nature, perhaps along the lines of \cite{Anninos:2017eib}, or bootstrapped bottom-up. We leave this, and more, 
to future work.


\acknowledgments

We are grateful to Simone Giombi, Sean Hartnoll, Chris Herzog, Austin Joyce, Igor Klebanov, Marcos Mari\~no, Ruben Monten, Beatrix M\"uhlmann, Bob Penna, Rachel Rosen and Charlotte Sleight for discussions, explanations and past collaborations inspiring this work. FD and DA are particularly grateful to Eva Depoorter and Carolina Martes for near-infinite patience. FD, AL and ZS were supported in part by the U.S.\ Department of Energy grant de-sc0011941. DA is funded by the Royal Society under the grant The Atoms of a deSitter Universe.



\appendix

\section{Harish-Chandra characters} \label{app:chars}

\subsection{Definition of $\chi$} \label{app:chardef}

A central ingredient in this work is the Harish-Chandra group character  for unitary representations $R$ of Lie groups $G$, 
\begin{align}
 \tilde\chi_R(g) \equiv {\rm tr} \, R(g) \, , \qquad g  \in G \, .
\end{align}
More rigorously this should be viewed as a distribution to be integrated against smooth test functions $f(g)$ on $G$. The smeared operators $\int [dg] \, f(g) R(g)$ are trace-class operators, and $\tilde\chi_R(g)$ is always a locally integrable function on $G$, analytic away from its poles \cite{bams/1183520006,bams/1183525024}. 

The group of interest to us is $SO(1,d+1)$, the isometry group of global dS$_{d+1}$, generated by $M_{IJ}$ as defined under (\ref{embeddingspacedS}). The representations of $SO(1,d+1)$ were classified and their characters explicitly computed in \cite{10.3792/pja/1195522333,10.3792/pja/1195523378,10.3792/pja/1195523460}. For a recent review and an extensive dictionary between fields and representations, see \cite{Basile:2016aen}.


For our purposes in this work we only need to consider characters restricted to group elements of the form $g=e^{-i t H}$, where $H= M_{0,d+1}$ generates global $SO(1,1)$ transformations  acting as time translations $T \to T + t$ on the southern static patch (fig.\ \ref{fig:penrose}):   
\begin{align} \label{chitdef}
 \chi(t) \equiv {\rm tr} \, e^{-i t H} \, .
\end{align}
For example for a spin-0 UIR corresponding to a scalar field of mass $m^2=\Delta(d-\Delta)$, as we will explicitly compute below, this takes the form 
\begin{align} \label{formulaforchi2}
 \chi(t) = {\rm tr} \, e^{-itH} = \frac{e^{-t \Delta} + e^{-t \bar\Delta}}{|1-e^{-t}|^d}  \, , \qquad \bar\Delta \equiv d - \Delta \, .
\end{align} 
Putting $\Delta=\frac{d}{2}+i \nu$, we get $m^2 = (\frac{d}{2})^2+\nu^2$, 
 so $m^2>0$ if $\nu \in \IR$ (principal series) or $\nu=i \mu$ with $|\mu| < \frac{d}{2}$ (complementary series). Since $\bar\Delta=d-\Delta=\frac{d}{2} - i \nu$, this implies $\chi(t) = \chi(t)^*$, as follows more generally from $H^\dagger = H$. The absolute value signs moreover ensure $\chi(t)=\chi(-t)$ for all $d$. The latter property holds for any $SO(1,d+1)$ representation:
\begin{align} \label{chievenarg2}
 \chi(-t) = \chi(t) \, .
\end{align}
This follows from the fact that the $SO(1,1)$ boost generator $H=M_{0,d+1}$ can be conjugated to a boost $-H$ in the opposite direction by a 180-degree rotation: $-H = u H u^{-1}$ for e.g.\ $u=e^{i \pi M_{d,d+1}}$, implying $\chi(-t) = {\rm tr} \, e^{i H t} = {\rm tr} \, u \, e^{-i H t} u^{-1} = {\rm tr} \, e^{-i H t} = \chi(t)$. 


\subsection{Computation of $\chi$} \label{app:compchar}

Here we show how characters $\chi(t)={\rm tr} \, e^{-i t H}$ 
can be computed by elementary means. The full characters $\chi(t,\phi) = {\rm tr} \, e^{-itH + i \phi \cdot J}$ can be computed similarly, but we will focus on the former.

\subsubsection*{Simplest example: $d=1$, $s=0$}

We first consider a $d=1$, spin-0 principal series representation with $\Delta = \frac{1}{2} + i \nu$, $\nu \in \IR$. This corresponds to a massive scalar field on dS$_2$ with mass $m^2 = \frac{1}{4} + \nu^2$. This unitary irreducible representation of $SO(1,2)$ can be realized on the Hilbert space of square-integrable wave functions $\psi(\varphi)$ on $S^1$, with standard inner product. 
The circle can be thought of as the future conformal boundary of global dS$_2$ in global coordinates (cf.\ (\ref{dScoords})), which for dS$_2$ becomes $ds^2 = (\cos \vartheta)^{-2} (-d\vartheta^2+d\varphi^2)$.  Kets $|\varphi\rangle$ can be thought of as states produced by a boundary conformal field\footnote{$\CO(\varphi)$ arises from the bulk scalar $\phi(\vartheta,\varphi)$ as $\phi(\frac{\pi}{2}-\epsilon,\varphi) \sim \CO(\varphi) \, \epsilon^\Delta + \bar \CO(\varphi) \, \epsilon^{\bar\Delta}$ in the infinite future $\epsilon \to 0$} $\CO(\varphi)$ of dimension $\Delta=\frac{1}{2}+i \nu$  acting on an $SO(1,2)$-invariant global vacuum state $|{\rm vac}\rangle$ such as the global Euclidean vacuum:
\begin{align} \label{innerprodkets}
 |\varphi\rangle \equiv \CO(\varphi)  |{\rm vac}\rangle \, , \qquad \langle \varphi|\varphi'\rangle = \delta(\varphi-\varphi') \, .
\end{align}
This pairing is $SO(1,2)$ invariant. Normalizable states $|\psi\rangle$ are then superpositions
\begin{align} 
 |\psi \rangle = \int_{-\pi}^\pi d\varphi \, \psi(\varphi) \, |\varphi\rangle \, , \qquad \langle \psi|\psi \rangle = \int_{-\pi}^{\pi} d\varphi \, |\psi(\varphi)|^2 < \infty \, . 
\end{align}
In conventions in which $H$, $P$ and $K$ are hermitian, the Lie algebra of ${\rm so(1,2)}$ is
\begin{align} \label{SO12cofalg}
 [H,P]=i P \, , \qquad [H,K] = - i K \, , \qquad [K,P] = 2 i H \, ,
\end{align}
the action of these generators on kets $|\varphi\rangle$ in the above representation is 
\begin{align} 
 H |\varphi\rangle &= i\bigl(\sin\varphi \, \partial_{\varphi} + \Delta \cos\varphi \bigr) |\varphi\rangle \label{globalHepxr} \\
 P |\varphi\rangle &= i\bigl((1+\cos\varphi) \partial_{\varphi} - \Delta \sin\varphi \bigr) |\varphi\rangle \nn \\
 K |\varphi\rangle &= i\bigl((1-\cos\varphi) \partial_{\varphi} + \Delta \sin\varphi \bigr) |\varphi\rangle \nn \, .
\end{align}
Note that this implies that the action of for example $H$ on wave functions $\psi(\varphi)$  is given by $H|\psi\rangle=\int d\varphi \, \CH \psi(\varphi) \, |\varphi\rangle$ where $\CH \psi(\varphi) =  -i\bigl(\sin\varphi \, \partial_{\varphi} + \bar\Delta \cos\varphi \bigr)\psi(\varphi)$, with $\bar\Delta=1-\Delta=\frac{1}{2}-i\nu$.
One gets simpler expressions after conformally mapping this to planar boundary coordinates $x=\tan \frac{\varphi}{2}$, that is to say changing basis from $|\varphi\rangle_{S^1}$ to $|x\rangle_\IR$, $x \in \IR$, where
\begin{align} \label{planarbasis}
 |x\rangle_{\IR} \equiv \bigl(\tfrac{\partial \varphi}{\partial x} \bigr)^{\Delta} \bigl|\varphi(x) \bigr\rangle_{S^1} = \bigl( \tfrac{2}{1+x^2} \bigr)^\Delta \, \bigl|2\arctan x\bigr\rangle_{S^1} \, , \qquad \langle x|x'\rangle = \delta(x-x') \, .
\end{align}
Then $H,P,K$ take the familiar planar dilatation, translation and special conformal form:  
\begin{align}
 H|x\rangle = i(x \partial_x + \Delta) |x\rangle \,  \qquad P|x\rangle = i \partial_x |x\rangle \, , \qquad K|x\rangle = i(x^2 \partial_x + 2 \Delta x )|x\rangle\, .
\end{align}
In particular this makes exponentiation of $H$  easy:
\begin{align}
 e^{-i t H}|x\rangle = e^{t \Delta}|e^t x \rangle \, .
\end{align}
However one has to keep in mind that planar coordinates miss a point of the global boundary, here the point $\varphi=\pi$. This will actually turn out to be important in the computation of the character. Let us first ignore this though and compute
\begin{align}
 \chi(t)|_{\rm planar} = \int dx\, \langle x|e^{-i t H}|x\rangle = e^{t\Delta} \int dx\, \delta\bigl(x-e^t x\bigr) = e^{t\Delta} \int dx \, \frac{\delta(x)}{|1-e^t|} = \frac{e^{t \Delta}}{|1-e^t|} \, . \nn
\end{align}
We see that the computation localizes at the point $x=0$, singled out because it is a fixed point of $H$. Actually there is another fixed point, which we missed here because it is exactly the point at infinity in planar coordinates. This is clear from  the global version (\ref{globalHepxr}): one fixed point of $H$ is at $\varphi=0$, which maps to $x=0$ and was picked up in the above computation, while the other fixed point is at $\varphi=\pi$, which maps to $x=\infty$ and so was missed. 

This is easily remedied though. The most straightforward way is to repeat the computation in the global boundary basis $|\varphi\rangle$, which is sure not to miss any fixed points. It suffices to consider an infinitesimally small neighborhood of the fixed points. For $\varphi = y \to 0$, we get $H \approx i(y \partial_y + \Delta)$, which coincides with the planar expression, while for $\varphi = \pi + y$ with $y \to 0$, we get $H \approx -i (y \partial_{y} + \Delta)$, which is the same except with the opposite sign. Thus we obtain
\begin{align} \label{chitcompd1}
 \chi(t) = \int d\varphi \, \langle \varphi|e^{-i t H}|\varphi\rangle = \frac{e^{t \Delta}}{|1-e^t|} + \frac{e^{-t \Delta}}{|1-e^{-t}|} \, = \frac{e^{-t \Delta}+e^{-t\bar\Delta}}{|1-e^{-t}|} \, ,
\end{align}
where $\bar\Delta=1-\Delta=\frac{1}{2}-i\nu$, reproducing (\ref{formulaforchi2}) for $d=1$.

For the complementary series $0<\Delta<1$, we have $\Delta^* = \Delta$ instead of $\Delta^* = \bar\Delta \equiv 1-\Delta$, so the conjugation properties of $H$, $D$, $K$ are different. As a result they are no longer hermitian with respect to the inner product (\ref{innerprodkets}), but rather with respect to $\langle \varphi|\varphi'\rangle \propto \bigl(1-\cos(\varphi-\varphi')\bigr)^{-\Delta}$. However we can now define a ``shadow'' bra $(\varphi| \propto \int d\varphi' \bigl(1-\cos(\varphi-\varphi') \bigr)^{-\bar\Delta} \langle \varphi'|$ satisfying $(\varphi|\varphi'\rangle = \delta(\varphi-\varphi')$ and compute the trace as $\chi(t) = \int d\varphi \, ( \varphi|e^{-i t H}|\varphi\rangle$. The computation then proceeds in exactly the same way, with the same result (\ref{chitcompd1}).

\subsubsection*{General dimension and spin} \label{app:chargenrep}

The generalization to $d>1$ is straightforward. Again the trace only picks up contributions from fixed points of $H$. The fixed point at the origin in planar coordinates contributes
$
 e^{t \Delta } \int d^d x \, \delta^d(x-e^t x) = \frac{e^{t\Delta}}{|1-e^t|^d} \, ,
$
while the fixed point at the other pole of the global boundary sphere gives a contribution of the same form but with $t \to -t$. Together we get
\begin{align} \label{chi0delta}
 \chi_{0,\Delta}(t) = \frac{e^{-t\Delta} + e^{-t \bar\Delta}}{|1-e^{-t}|^d} \, ,
\end{align}
where $\bar\Delta=d-\Delta$. 

For massive spin-$s$ representations, the basis merely gets some additional $SO(d)$ spin labels, and the trace picks up a corresponding degeneracy factor, so\footnote{\label{fn:spinsunitarity} Here $\Delta = \frac{d}{2} + i \nu$ with either $\nu \in \IR$ (principal series) or $\nu = i \mu$ with $|\mu|<\frac{d}{2}$ for $s=0$ and $|\mu|<\frac{d}{2}-1$ for $s \geq 1$ (complementary series). For $s=0$ the mass is $m^2 = (\frac{d}{2})^2+ \nu^2 = \Delta(d-\Delta)$ while for $s\geq 1$ it is given by (\ref{convmdef}): $m^2=(\tfrac{d}{2}+s-2)^2 + \nu^2 = (\Delta+s-2)(d-\Delta+s-2)$.}
\begin{align} \label{chimassivespins}
 \chi_{s,\Delta}(t) = D_s^{d} \, \frac{e^{-t\Delta} + e^{-t \bar\Delta}}{|1-e^{-t}|^d} \, , 
\end{align}
where $\bar\Delta=d-\Delta$ as before, and $D_s^d$ is the spin degeneracy, for example $D_s^3 = 2s+1$. More generally for $d>2$ it is the number of totally symmetric traceless tensors of rank $s$:
\begin{align} \label{Dsods2}
D^{d}_{s}=\mbox{\large $\binom{s+d-1}{d-1}-\binom{s+d-3}{d-1}$} 
\end{align}  
(For $d=2$ we get spin $\pm s$ irreps of $SO(2)$ with $D^2_{\pm s}=1$. However both of these appear when quantizing a Fierz-Pauli spin-$s$ field.) Explicit low-$d$ spin-$s$ degeneracies are listed in (\ref{SOKlowdim}).   

The most general massive unitary representation of $SO(1,d+1)$ is labeled by an irrep $S=(s_1,\ldots,s_r)$ of $SO(d)$ (cf.\ appendix \ref{app:Weyldim}) and $\Delta =\frac{d}{2}+i \nu$, $\nu \in \IR$ (principal series) or $\Delta = \frac{d}{2} + \mu$, $|\mu| < \mu_{\rm max}(S) \leq \frac{d}{2}$ (complementary series) \cite{10.3792/pja/1195522333,10.3792/pja/1195523378,10.3792/pja/1195523460,Basile:2016aen}. The spin-$s$ case discussed above corresponds to $S=(s_1,0,\ldots,0)$. The character for general $S$ is 
\begin{align} \label{chigenrep}
 \chi_{S,\Delta}(t) = D_S^{d} \, \frac{e^{-t\Delta} + e^{-t \bar\Delta}}{|1-e^{-t}|^d} \, ,
\end{align}
where the generalized spin degeneracy factor $D_S^d$ is the dimension of the $SO(d)$ irrep $S$, explicitly given for general $S$ in appendix \ref{app:Weyldim}. 

\subsubsection*{Massless and partially massless representations}

(Partially) massless representations correspond to higher-spin gauge fields and are in the exceptional or discrete series. These representations and their characters $\chi(t)$ are considerably more intricate. We give the general expression and examples in appendix \ref{sec:Zbulkmassless} for the massless case. Guided by our path integral results of section \ref{sec:massless}, we are led to a simple recipe for constructing these characters from their much simpler ``naive'' counterparts, spelled out in (\ref{prevsqbrackplusop}). This generalizes straightforwardly to the partially massless case, leading to the explicit general-$d$ formula (\ref{flippedchargenPM}).




\subsection{Importance of picking a globally regular basis} \label{sec:traces}


Naive evaluation of the character trace $\chi(t) = {\rm tr} \, e^{-i t H}$ by diagonalization of $H$ results in nonsense. In this section we explain why:   emphasizing the importance of using a basis on which finite $SO(1,d+1)$ transformations act in a globally regular way. 

\subsubsection*{Failure of computation by diagonalization of $H$}

Naively, one might have thought the easiest way to compute $\chi(t)={\rm tr} \, e^{-i t H}$ would be to  diagonalize $H$ and sum over its eigenstates. The latter are given by $|\omega\sigma\rangle$, where $H=\omega \in \IR$, $\sigma$ labels $SO(d)$ angular momentum quantum numbers, and $\langle \omega \sigma|\omega' \sigma' \rangle = \delta(\omega-\omega') \, \delta_{\sigma \sigma'}$. However this produces a  nonsensical result,
\begin{align} \label{nonsensechi}
 \chi(t) = {\rm tr} \, e^{-it H} \,\, \stackrel{\text{naive}}{=} \,\, \int d\omega \sum_\sigma \langle \omega \sigma | e^{-i t H} |\omega \sigma \rangle = 2 \pi \, \sum_\sigma  \delta(0) \, \delta(t) \, \qquad \mbox{(naive)} \, ,
\end{align}
not even remotely resembling the correct $\chi(t)$ as computed earlier in \ref{app:compchar}. 

Our method of computation there also illuminates why this naive computation fails. To make this concrete, let us go back to the $d=1$ scalar example with $\Delta=\frac{1}{2} + i \nu$, $\nu \in \IR$. Recalling the action of $H$ on wave functions $\psi(\varphi)$ mentioned below (\ref{globalHepxr}), it is straightforward to find the wave functions $\psi_{\omega\sigma}(\varphi)$ of the eigenkets $|\omega\sigma\rangle=\int_{-\pi}^\pi d\varphi \, \psi_{\omega\sigma}(\varphi) \, |\varphi\rangle$ of $H$: 
\begin{align} \label{diagHwave}
 \psi_{\omega\sigma}(\varphi)  = \frac{\Theta(\sigma \sin\varphi)}{\sqrt{2\pi}} \, |\sin\varphi|^{-\bar\Delta}  \left|\tan \frac{\varphi}{2} \right|^{i \omega}  \, , \qquad \omega \in \IR \, , \quad \sigma = \pm 1 \, . 
\end{align}
where $\Theta$ is the step function.
Alternatively we can first conformally map $S^1$ to the ``cylinder'' $\IR \times S^{d-1} = \IR \times S^0$ parametrized by $(T,\Omega)$, $T \in \IR$, $\Omega \in \{-1,+1\} = S^0$, that is to say change basis $|\varphi\rangle_{S^1} \to |T\Omega\rangle_{\IR \times S^0}$.\footnote{
Explicitly 
$T=\log |\tan \frac{\varphi}{2}|$, $\Omega={\rm sign} \, \varphi$, which analogously to the global $\to$ planar map (\ref{planarbasis}) yields $\bigl|T\pm\bigr\rangle_{\IR \times S^0} = (\cosh T)^{-\Delta} \bigl|\pm 2 \arctan e^T \bigr\rangle_{S^1}$, satisfying $\langle T \Omega|T'\Omega'\rangle=\delta(T-T') \, \delta_{\Omega \Omega'}$ and $H|T \Omega\rangle = i \partial_T |T \Omega\rangle$.}  Then $H$ generates translations of $T$, so the wave functions of $|\omega\sigma\rangle$ in this basis are simply
\begin{align}
 \psi_{\omega\sigma}(T,\Omega) = \frac{1}{\sqrt{2\pi}} \, \delta_{\Omega,\sigma} \, e^{i \omega T} \, .
\end{align}
The cylinder is the conformal boundary of the future wedge, $F$ in fig.\ \ref{fig:penrose-app} (which actually splits in two wedges at $\Omega = \pm 1$ in the case of dS$_2$), and the $|\omega \pm\rangle$ are the states obtained by the usual free field quantization corresponding to the natural modes $\phi_{\omega\pm}(T,r)$ in this patch. 

It is now clear why the naive computation (\ref{nonsensechi}) of $\chi(t)$ in the basis $|\omega\sigma\rangle$  fails to produce the correct result: the wave functions $\psi_{\omega\sigma}(\varphi)$ are singular  precisely at the fixed points $\varphi=0,\pi$ of $H$ (top corners of Penrose diagram in fig.\ \ref{fig:penrose}), which are exactly the points at which the character trace computation of  section \ref{app:compchar} localizes. Closely related failure would be met in the basis $|T\Omega\rangle$: $H$ acts as by translating $T$, seemingly without fixed points, oblivious to their actual presence at $T = \pm \infty$. In other words, despite their  lure as being the bases in which the action of $H$ is maximally simple, $|T\Omega\rangle$ or its Fourier dual $|\omega\sigma\rangle$ are in fact the worst possible choice one could make to compute the trace.

Similar observations hold for in higher dimensions. The wave functions diagonalizing $H$ take the form $\psi_{\omega\sigma}(T,\Omega) \propto e^{i \omega T} Y_\sigma(\Omega)$ in $\IR \times S^{d-1}$ cylinder coordinates. Transformed to global $S^d$ coordinates, these are singular precisely at the fixed points of $H$, excluded from the cylinder, making this frame particularly ill-suited for computing ${\rm tr} \, e^{-i t H}$. 



\subsubsection*{Globally regular bases}

More generally, to ensure correct computation of the full Harish-Chandra group character $\chi_R(g) = {\rm tr} \, R(g)$, $g \in SO(1,d+1)$, we must use a basis on which finite $SO(1,d+1)$ transformations $g$ act in a globally nonsingular way. This is the case for a global dS$_{d+1}$ boundary basis $|\bar \Omega\rangle_{S^d}$, $\bar \Omega \in S^d$, generalizing the $d=1$ global $S^1$ basis $|\varphi\rangle_{S^1}$, but not for a planar basis $|x\rangle_{\IR^d}$ or a cylinder basis $|T \Omega\rangle_{\IR \times S^{d-1}}$. Indeed generic $SO(d+1)$ rotations of the global $S^d$ move the poles of the sphere, thus mapping finite points to infinity in planar or cylindrical coordinates. This singular behavior is inherited by the corresponding Fourier dual bases $|p\rangle \propto \int d^d x \, e^{i p x} |x\rangle$ and $|\omega\sigma\rangle \propto \int dT \, d\Omega \, e^{i \omega T} Y_\sigma(\Omega) \, |T\Omega\rangle$. From a bulk point of view these are the states obtained by standard mode quantization in the planar patch resp.\ future wedge. 
The singular behavior is evident here from the fact that these patches have horizons that are moved around by global $SO(d+1)$ rotations. Naively computing $\chi(g)$ in these frames will in general give incorrect results. More precisely the result will be wrong unless the fixed points of $g$ lie at finite position on the corresponding conformal boundary patch.  

On the other hand the normalizable dual basis $|\bar\sigma\rangle = \int d\bar \Omega \, Y_{\bar\sigma}(\bar\Omega) \, |\bar\Omega\rangle$ inherits the global regularity of $|\bar\Omega\rangle_{S^d}$. Here $Y_{\bar\sigma}(\Omega)$ is a basis of spherical harmonics on $S^d$, with $\bar\sigma$ labeling the global $SO(d+1)$ angular momentum quantum numbers, and $\langle \bar \sigma|\bar\sigma'\rangle = \delta_{\bar\sigma \bar\sigma'}$. (From the bulk point of view this is essentially the basis obtained by quantizing the natural mode functions of the global dS$_{d+1}$ metric in table \ref{dScoords}.) Although in practice much harder than computing $\chi(t)= \int d\bar\Omega \, \langle\bar\Omega|e^{-i Ht}|\bar\Omega\rangle$ as in section \ref{app:compchar}, computing 
\begin{align} \label{chiglobSO}
 \chi(t) = {\rm tr} \, e^{-i t H} = \sum_{\bar\sigma} \langle \bar\sigma|e^{-i t H}|\bar\sigma\rangle
\end{align}
gives in principle the correct result. Note that this suggests a natural UV regularization of $\chi(t)$ for $t \to 0$, by cutting off the global $SO(d+1)$ angular momentum. For example for a scalar on dS$_3$ with $SO(3)$ angular momentum cutoff $L$, this would be
\begin{align} \label{chiL}
 \chi_L(t) \equiv \sum_{\ell = 0}^L \langle \ell m|e^{-i t H}|\ell m \rangle \, .
\end{align}

\section{Density of states and quasinormal mode resonances} \label{sec:QNM}

The review in appendix \ref{app:chars} focuses mostly on mathematical and computational aspects of the Harish-Chandra  character $\chi(t) = {\rm tr} \, e^{-i t H}$. Here we focus on its physics interpretation, in particular the density of states $\rho(\omega)$ obtained as its Fourier transform. We define this in a general and manifestly covariant way using Pauli-Villars regularization in section \ref{sec:thermal}. Here we will not be particularly concerned with general definitions or manifest covariance, taking a more pedestrian approach. At the end we briefly comment on an ``S-matrix'' interpretation and a possible generalization of the formalism including interactions. 

In \ref{sec:dSvsAdSchar}, we contrast the spectral features encoded in the characters of unitary representations of the $\so(1,d+1)$ isometry algebra of global dS$_{d+1}$  with the perhaps more familiar characters of unitary representations of the $\so(2,d)$ isometry algebra of AdS$_{d+1}$: in a sentence, the latter encodes bound states, while the former encodes scattering states. In \ref{sec:DOS} we explicitly compare $\rho(\omega)$ obtained as the Fourier transform of $\chi(t)$  for dS$_2$ to the coarse-grained eigenvalue density obtained by numerical diagonalization of a model discretized by global angular momentum truncation, and confirm the results match at large $N$.
 In \ref{sec:resonances} we identify the poles of $\rho(\omega)$ in the complex $\omega$ plane as scattering resonances/quasinormal modes, counted by the power series expansion of the character. As a corollary this implies the relation $Z_{\rm PI} = Z_{\rm bulk}$ of (\ref{ZPIisZbulkss}) can be viewed as a precise version of the formal quasinormal mode expansion of $\log Z_{\rm PI}$ proposed in  \cite{Denef:2009kn}.

\subsection{Characters and the density of states: dS vs AdS} \label{sec:dSvsAdSchar}

We begin by highlight some important differences in the spectrum encoded in the characters of unitary $\so(1,d+1)$ representations furnished by global dS$_{d+1}$ single-particle Hilbert spaces and the characters of unitary $\so(2,d)$ representations furnished by global AdS$_{d+1}$ single-particle Hilbert spaces. 
 Although the discussion applies to arbitrary representations, for concreteness we  consider the example of a scalar of mass $m^2=(\frac{d}{2})^2+\nu^2$ on dS$_{d+1}$. Its character as computed in (\ref{chi0delta}) is   
\begin{align} \label{chids}
 \chi_{\rm dS}(t) \equiv {\rm tr} \, e^{-i t H} =  \frac{e^{-\Delta_+ t} + e^{-\Delta_- t}}{|1-e^{-t}|^d} \, , \qquad \Delta_\pm = \tfrac{d}{2} \pm i\nu  \, , \qquad t \in \IR .
\end{align} 
where ${\rm tr}$ traces over the {\it global} single-particle Hilbert space and we recall $H=M_{0,d+1}$ is a global $SO(1,1)$ boost generator, which becomes a spatial momentum operator in the future wedge and the energy operator in the southern static patch (cf.\ fig.\ \ref{fig:penrose-app}{\tt c}). 
This is to be contrasted with the familiar character of the unitary lowest-weight representation of a scalar of mass $m^2= -(\frac{d}{2})^2+\mu^2$ on global AdS$_{d+1}$ with standard boundary conditions:
\begin{align} \label{chiads}
  \chi_{\rm AdS}(t) \equiv {\rm tr} \, e^{-i t H} = \frac{e^{-i\Delta_+ t}}{(1-e^{-i t})^d} \, , \qquad \Delta_+ = \tfrac{d}{2} + \mu \, , \qquad {\rm Im} \, t < 0 \, .
\end{align}
Here the $\so(2)$ generator $H$ is the energy operator in global AdS$_{d+1}$. Besides the occurrence of both $\Delta_\pm$ in (\ref{chids}), another notable difference is the absence of factors of $i$ in the exponents.

The physics content of $\chi_{\rm AdS}$ is clear: $\chi_{\rm AdS}(-i \beta) = {\rm tr} \, e^{-\beta H}$ is the single-particle partition function at inverse temperature $\beta$ for a scalar particle trapped in the global AdS gravitational potential well. Equivalently for ${\rm Im} \, t < 0$, the expansion 
\begin{align} \label{NlamAdS}
 \chi_{\rm AdS}(t) = \sum_\lambda N_\lambda \, e^{-i t \lambda} \, , \qquad \lambda = \Delta_+ + n \, , \quad n \in \IN \, ,
\end{align}
counts normalizable single-particle states of energy $H=\lambda$, or equivalently global normal modes of frequency $\lambda$. The corresponding density of single-particle states is
\begin{align} \label{AdSDOS}
 \rho_{\rm AdS}(\omega) = \int_{-\infty}^\infty \frac{dt}{2\pi} \, \chi_{\rm AdS}(t) \, e^{i \omega t} = \sum_\lambda N_\lambda \, \delta(\omega-\lambda) \, .
\end{align}

\begin{figure}
   \includegraphics[height=3.5cm]{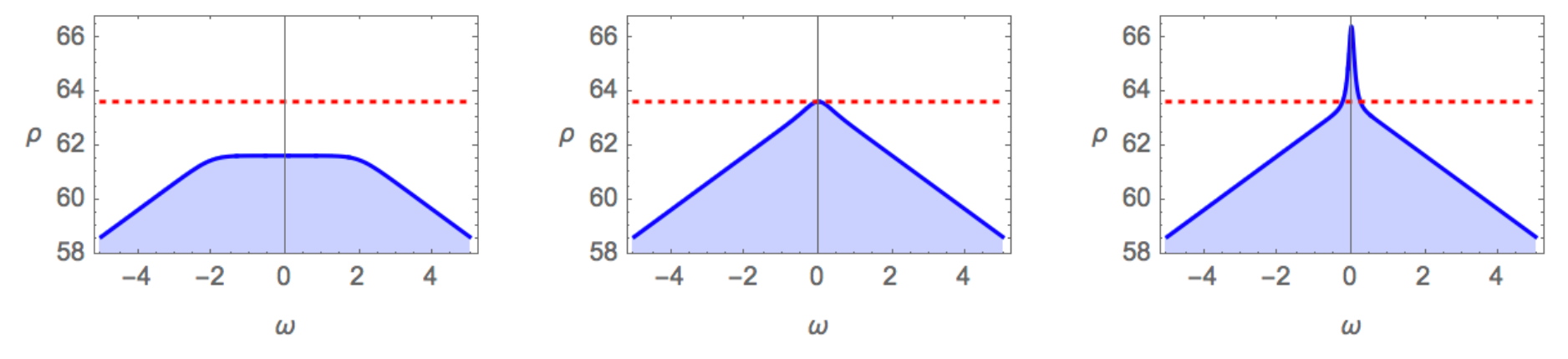} \vskip-5mm
 \caption{ \small Density of states $\rho_\Lambda(\omega)$ for dS$_3$ scalars with  $\Delta=1 + 2i$, $\Delta=\frac{1}{2}$, $\Delta=\frac{1}{10}$, and UV cutoff $\Lambda=100$, according to (\ref{rhoscds3}). The red dotted line represents the term $2 \Lambda/\pi $. The peak visible at $\Delta = \frac{1}{10}$ is due to a resonance  approaching the real axis, as explained in section  \ref{sec:resonances}.
 \label{fig:dosdS3}} 
\end{figure}

\noindent For dS, we can likewise expand the character as in (\ref{NlamAdS}). For $t>0$,
\begin{align} \label{qnmseries}
 \chi_{\rm dS}(t) = \sum_\lambda N_\lambda \, e^{-i t \lambda} \, , \qquad \lambda = -i(\Delta_\pm + n) =  -i(\tfrac{d}{2}+n) \pm \nu \, , \quad n \in \IN \, .
\end{align}
However  $\lambda$ is now {\it complex}, so evidently $N_\lambda$ does not count physical eigenstates of the hermitian operator $H$. Rather, as further discussed in section \ref{sec:resonances}, it counts {\it resonances}, or  {\it quasinormal} modes. 
The density of {\it physical} states with $H=\omega \in \IR$ is formally given by 
\begin{align} \label{rhofromchi}
 \rho_{\rm dS}(\omega) = \int_{-\infty}^\infty \frac{dt}{2\pi} \, \chi_{\rm dS}(t) \, e^{i \omega t} = \int_0^\infty \frac{dt}{2\pi} \, \chi_{\rm dS}(t) \bigl(e^{i \omega t} + e^{-i \omega t} \bigr) \, ,
\end{align} 
where $\omega$ can be interpreted as the momentum along the $T$-direction of the future wedge ($F$ in fig.\ \ref{fig:penrose-app} and table \ref{dScoords}).  Alternatively for $\omega >0$ it can be interpreted as the energy in the southern static patch, as discussed in section \ref{sec:thermalder}.    
A manifestly covariant Pauli-Villars regularization of the above integral is given by (\ref{PVregdost}). For our purposes here a simple $t>\Lambda^{-1}$ cutoff suffices.
For example for dS$_3$,
\begin{align} \label{rhoscds3}
 \rho_{\rm dS_3,\Lambda}(\omega) &\equiv \int_{\Lambda^{-1}}^\infty \frac{dt}{2\pi} \, \frac{e^{-(1+i\nu) t} + e^{-(1-i\nu) t}}{(1-e^{-t})^2}  \bigl(e^{i \omega t} + e^{-i \omega t} \bigr)  \\
  &=\frac{2 \Lambda}{\pi}  - \frac{1}{2} \sum_{\pm} (\omega \pm \nu) \coth\bigl(\pi(\omega \pm \nu)\bigr) \, . \nn
\end{align}
Some examples are illustrated in fig.\ \ref{fig:dosdS3}.   In contrast to AdS, $\rho_{\rm dS}(\omega)$ is continuous. Indeed energy eigenkets $|\omega\sigma\rangle$ of the static patch form a continuum of scattering states, coming out of and going into the horizon, instead of the discrete set of bound states one gets in the global AdS potential well.  
Note that although the above $\rho_{\rm dS_3,\Lambda}(\omega)$ formally goes negative in the large-$\omega$ limit, it is positive within its regime of validity, that is to say for $\omega,\nu \ll \Lambda$.

\subsection{Coarse-grained density of states in globally truncated model} \label{sec:DOS}

\begin{figure}
 \begin{center}
   \includegraphics[height=5cm]{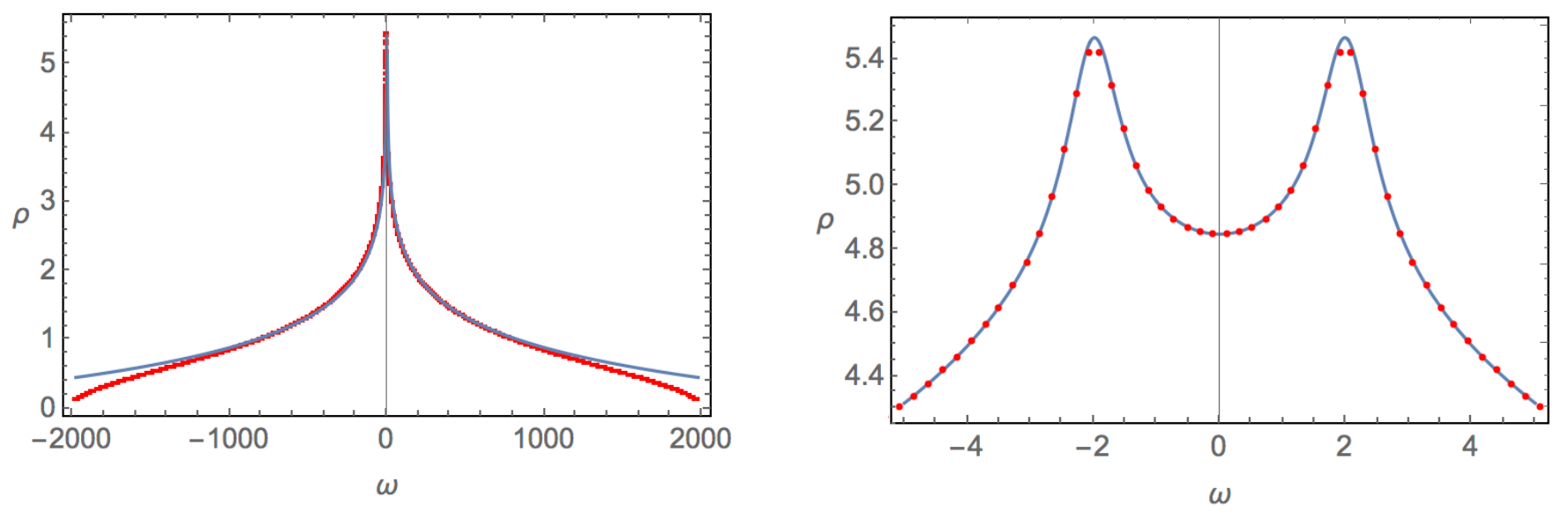}
 \caption{\small Density of states for a $\Delta=\frac{1}{2} + i \nu$ scalar with $\nu=2$  in dS$_2$. The red dots show the local eigenvalue density $\bar\rho_N(\omega)$, (\ref{rhondf}), of the truncated model with global angular momentum cutoff $N = 2000$, obtained by numerical diagonalization. The blue line shows $\rho(\omega)$ obtained as the Fourier transform of $\chi(t)$, explicitly (\ref{rholam}) with $e^{-\game} \Lambda \approx 4000$. The plot on the right zooms in on the IR region. The peaks are due to the proximity of quasinormal mode poles in $\rho(\omega)$, discussed in  \ref{sec:resonances}.
 \label{fig:DOScomparison}}
\end{center} \vskip-4mm
\end{figure}

For a $\Delta=\frac{1}{2}+i \nu$ scalar on dS$_2$, the density of states regularized by as in (\ref{rhoscds3}) is
\begin{align} \label{rholam}
 \rho(\omega) = \frac{2}{\pi} \log(e^{-\game} \Lambda) -\frac{1}{2\pi} \sum_{\pm,\pm}
 \psi\bigl(\tfrac{1}{2} \pm i \nu \pm i\omega) \bigr) \, ,
\end{align}
where $\game$ is the Euler constant, $\psi(x)=\Gamma'(x)/\Gamma(x)$ is the digamma function, and the sum is over the four different combinations of signs. 
To ascertain it makes physical sense  to identify this as the density of states, we would like to compare this to a model with discretized spectrum of eigenvalues $\omega$.

An efficient discretization --- which does not require solving bulk equations of motion and is quite natural from the point of view of dS-CFT approaches to de Sitter quantum gravity \cite{Maldacena:2002vr,Strominger:2001pn,Anninos:2011ui} --- 
is obtained by truncating the {\it global} dS$_{d+1}$ angular momentum $SO(d+1)$ of the single-particle Hilbert space, considering instead of $H$ a {\it finite}-dimensional matrix 
\begin{align}
  h_{\bar\sigma\bar\sigma'} \equiv \langle \bar\sigma|H|\bar\sigma'\rangle \, ,
\end{align}
where $\bar\sigma$ are $SO(d+1)$ quantum numbers, as in (\ref{chiglobSO}). For dS$_2$ this is $SO(2)$ and $\bar\sigma = n \in \IZ$, truncated e.g.\ by $|n| \leq N$. The matrix $h$ is sparse and can be computed either directly using $|n\rangle \propto \int d\varphi \, e^{i n \varphi} |\varphi\rangle$ and the explicit form of $H$ given in (\ref{globalHepxr}), or algebraically.

The algebraic way goes as follows. A normalizable basis $|n\rangle$ of the global dS$_2$ scalar single-particle Hilbert space can be constructed from the $SO(1,2)$ conformal algebra (\ref{SO12cofalg}), using a basis of generators $L_0$, $L_\pm$ related to $H$, $K$ and $P$ as $L_0 = \frac{1}{2}(P+K)$, $L_\pm = \frac{1}{2}(P-K) \pm i H$. Then $L_0$ is the global angular momentum generator $i \partial_\phi$ along the future boundary $S^1$ and $L_\pm$ are its raising and lowering operators. In some suitable normalization of the $L_0$ eigenstates $|n\rangle$, we have $L_0|n\rangle = n |n\rangle$, $L_\pm|n\rangle = (n \pm \Delta)|n \pm 1\rangle$. 
Cutting off the single-particle Hilbert space at $-N < n \leq N$,\footnote{The asymmetric choice here allows us to use the simple coarse graining prescription (\ref{rhondf}) and keep this discussion short. A symmetric choice $|n| \leq N$ would lead to an enhanced $\IZ_2$ and two families of eigenvalues distinguished by their $\IZ_2$ parity, inducing persistent microstructure in the level spacing. The most efficient way to proceed then is to compute $\bar\rho_{N,\pm}(\omega)$ as the inverse level spacing for these two families separately and then add the contributions together as interpolated functions. For dS$_3$ with $SO(3)$ cutoff $\ell \leq N$ one similarly gets $2N+1$ families of eigenvalues, labeled by $SO(2)$ angular momentum $m$, and one can proceed analogously. Alternatively, one can compute $\bar\rho_{N}(\omega)$ directly by binning and counting, but this requires larger $N$.}  the operator $H=\frac{i}{2}(L_--L_+)$ acts as a sparse $2N \times 2N$ matrix on the truncated basis $|n\rangle$.

A minimally coarse-grained density of states can then be defined as the inverse spacing of its eigenvalues $\omega_i$, $i=1,\ldots,2N$, obtained by numerical diagonalization: 
\begin{align} \label{rhondf}
 \bar\rho_N(\omega_i) \equiv \frac{2}{\omega_{i+1}-\omega_{i-1}}.
\end{align}
The continuum limit corresponds to $N \to \infty$ in the discretized model, and to $\Lambda \to \infty$ in (\ref{rholam}). To compare to (\ref{rholam}), we adjust $\Lambda$, in the spirit of renormalization, to match the density of states at some scale $\omega$, say $\omega=0$. The results of this comparison for $\nu = 2$, $N=2000$ are shown in fig.\ \ref{fig:DOScomparison}. Clearly they match remarkably well indeed  in the regime where they should, i.e.\ well below the UV cutoff scale.

\begin{figure}
 \begin{center}
   \includegraphics[height=3.5cm]{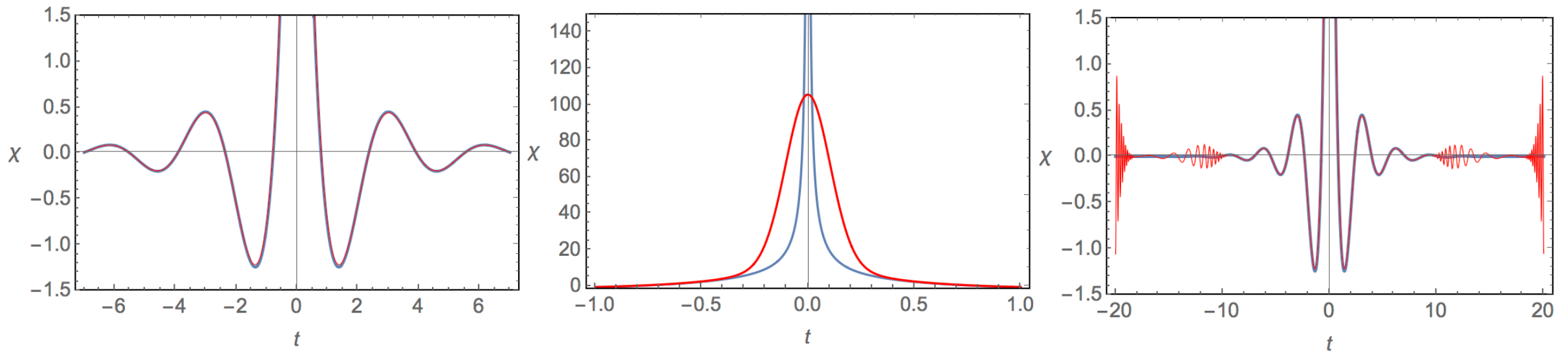}
 \caption{\small Comparison of $d=1$ character $\chi(t)$ defined in (\ref{chids}) (blue) to the coarse-grained discretized character $\bar\chi_{N,\delta}(t)$ defined in (\ref{chiNCG}) (red), with $\delta=0.1$ and other parameters as in fig.\ \ref{fig:DOScomparison}. Plot on the right shows wider range of $t$. Plot in the middle smaller range of $t$, but larger $\chi$.
 \label{fig:discretechar}}
\end{center} \vskip-4mm
\end{figure}

We can make a similar comparison directly at the (UV-finite) character level. The discrete character is $\sum_i e^{-i \omega_i t}$, which is a wildly oscillating function. At first sight this seems very different from the character $\chi(t) = {\rm tr} \, e^{-i H t}$ in (\ref{rhofromchi}). However to properly compare the two, we should coarse grain this at a small but finite resolution $\delta$. We do this by convolution with a Gaussian kernel, that is to say we consider 
\begin{align} \label{chiNCG}
 \bar\chi_{N,\delta}(t) \equiv \frac{1}{\sqrt{2\pi} \delta} \int_{-\infty}^\infty dt' \, e^{-(t-t')^2/2\delta^2} \, \sum_i e^{-i \omega_i t'} = \sum_i e^{-i t \omega_i - \delta^2 \omega_i^2/2} \, .
\end{align}
A comparison of $\bar\chi_{N,\delta}$ to $\chi$ is shown in fig.\ \ref{fig:discretechar} for $\delta=0.1$. The match is nearly perfect for $|t|$ not too large and not too small. For small $t$, the $\bar\chi_{N,\delta}(t)$ caps off at a finite value, the number of eigenvalues $|\omega_i| \lesssim 1/\delta$, while $\chi(t) \sim 1/|t| \to \infty$. The approximation gets better here when $\delta$ is made smaller.  For larger values of $t$, $\bar\chi_{N,\delta}(t)$ starts showing some oscillations again. These can be eliminated by increasing $\delta$, at the cost of accuracy at smaller $t$. In the $N \to \infty$ limit, the discretized approximation gets increasingly better over increasingly large intervals of $t$, with $\lim_{\delta \to 0} \lim_{N \to \infty} \bar\chi_{N,\delta}(t) = \chi(t)$. 


\vskip2mm
Note that there is no reason  to expect {\it any} discretization scheme will converge to $\chi(t)$ or $\rho(\omega)$. For example it is not clear a brick wall discretization along the lines described in section \ref{sec:brickwall} would. On the other hand, the convergence of the above global angular momentum cutoff scheme to the continuum $\chi(t)$ was perhaps to be expected, 
given (\ref{chiglobSO}) and the discussion preceding it. 

\subsection{Resonances and quasinormal mode expansion} \label{sec:resonances}

\begin{figure} 
 \begin{center}
   \includegraphics[height=5cm]{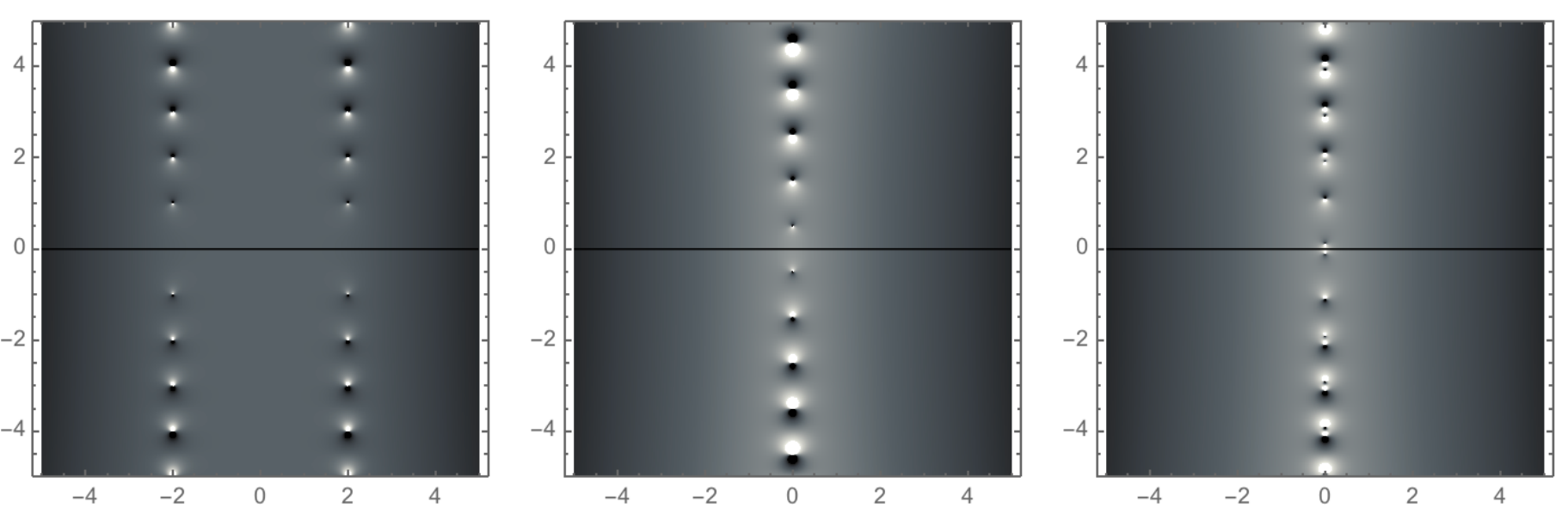}
 \caption{\small Plot of $|\rho(\omega)|$ in complex $\omega$-plane corresponding to the dS$_3$ examples of fig.\ \ref{fig:dosdS3}, that is $\Delta_\pm = \{1+2i,1-2i\}$, $\{\frac{1}{2},\frac{3}{2}\}$, $\{0.1,1.9\}$, and $2\Lambda/\pi \approx 64$. Lighter is larger with plot range $58 \text{ (black)}<|\rho|<67 \text{ (white)}$. Resonance poles are visible at $\omega = \mp i(\Delta_\pm + n)$, $n \in \IN$. 
 \label{fig:poles}}
\end{center}
\end{figure}
Substituting the expansion (\ref{qnmseries}) of the dS character,
\begin{align} \label{QNMexpchib}
 \chi(t) = \sum_\lambda N_\lambda \, e^{-i t \lambda} \qquad (t>0) \, ,
\end{align}
into (\ref{rhofromchi}), $\rho(\omega) = \frac{1}{2\pi} \int_0^\infty dt \, \chi(t) \, (e^{i\omega t}+e^{-i\omega t})$, we can formally express the density of states as  
\begin{align} \label{QNMexprho}
 \rho(\omega) = \frac{1}{2 \pi i} \sum_\lambda N_\lambda \Bigl( \frac{1}{\lambda-\omega} + \frac{1}{\lambda+\omega} \Bigr)   \, ,
\end{align}
From this we read off that $\rho(\omega)$ analytically continued to the complex plane has poles at $\omega = \pm \lambda$ which for massive representations means $\omega = \mp i(\Delta_\pm + n)$. This can also be checked from explicit expressions such as the dS$_3$ scalar density of states (\ref{rhoscds3}), illustrated in fig.\ \ref{fig:poles}. These values of $\omega$ are precisely the frequencies of the (anti-)quasinormal field modes in the static patch, that is to say modes with purely ingoing/outgoing boundary conditions at the horizon, regular in the interior. If we think of the normal modes as scattering states, the quasinormal modes are to be thought of as scattering resonances. Indeed the poles of $\rho(\omega)$ are related to the poles/zeros of the static patch $S$-matrix $S(\omega)$, cf.\ (\ref{rhominusrho0}) below. Thus we see the coefficients $N_\lambda$ in (\ref{QNMexpchib})
count resonances (or  quasinormal modes), rather than states (or normal modes) as in AdS. This expresses at the level of characters the observations made in \cite{Ng:2012xp}. It holds for any $SO(1,d+1)$ representation, including massless representations, as explored in more depth in \cite{Sun:2020sgn} (see also appendix \ref{sec:Zbulkmassless}). Some corresponding quasinormal mode expansions of bulk thermodynamic quantities are given in  (\ref{sofr}) and (\ref{CbulkQNM}), and related there to the quasinormal mode expansion of \cite{Denef:2009kn} for scalar and spinor path integrals.

\subsubsection*{``S-matrix'' formulation}

The appearance of resonance poles in the analytically continued density of states is well-known in quantum mechanical scattering off a fixed potential $V$. They are directly related to the poles/zeros in the S-matrix $S(\omega)$ at energy $\omega$ through the  relation\footnote{An elementary exposition can be found e.g.\ in chapter 39 of \cite{chaosbook}.}
\begin{align} \label{rhominusrho0}
\rho(\omega) - \rho_0(\omega)  = \frac{1}{2\pi i} \, \frac{d}{d\omega} \, {\rm tr} \log S(\omega) \, ,
\end{align}
where $\rho_0(\omega)$ is the density of states at $V=0$. 

Using the explicit form of the dS$_2$ dimension-$\Delta$ scalar static patch mode functions (\ref{SPmodes})  $\phi^{\Delta}_{\omega \ell}(r,T)$, expanding these for $r =: \tanh X \to 1$ as 
\begin{align}
 \phi^{\Delta}_{\omega \ell}(r) \to A^{\Delta}_\ell(\omega) \, e^{-i \omega (T+X)} + B^{\Delta}_\ell(\omega) \, e^{-i \omega(T-X)} \, ,
\end{align}
and defining $S^{\Delta}_\ell(\omega) \equiv B^{\Delta}_\ell(\omega)/A^{\Delta}_\ell(\omega)$, one can check that $\rho^\Delta(\omega)$ as obtained in (\ref{rholam}) satisfies
\begin{align}
 \rho^\Delta(\omega) - \rho_0(\omega) =  \frac{1}{2\pi i} \, \frac{d}{d\omega} \sum_{\ell=0,1} \log S^{\Delta}_\ell(\omega) \, ,
\end{align}
where $\rho_0(\omega)=\frac{1}{\pi}(\psi(i \omega)+\psi(-i\omega)) + {\rm const.}$ does not depend on $\Delta$. 
This can be viewed as a rough analog of (\ref{rhominusrho0}), although the interpretation of $\rho_0(\omega)$ in the present setting is not clear to us. Similar observations can be made in higher dimensions. 

In \cite{PhysRev.187.345}, a general (flat space) $S$-matrix formulation of statistical mechanics for interacting QFTs was developed. In this formulation, the canonical partition function is expressed as
\begin{align}
 \log Z - \log Z_0  = \frac{1}{2\pi i} \int dE \, e^{-\beta E} \, \frac{d}{d E} \, \bigl[{\rm Tr} \log S(E) \bigr]_c  \, ,
\end{align}
where the subscript $c$ indicates restriction to connected diagrams (where ``connected'' is defined with the rule  that particle permutations are interpreted as interactions \cite{PhysRev.187.345}). Combined with the above observations, this  hints at a possible generalization of our free QFT results to interacting theories.

\section{Evaluation of character integrals} \label{app:renpr}

The most straightforward way of UV-regularizing character integrals is to simply cut off the $t$-integral at some small $t=\epsilon$. However to compare to the standard heat kernel (or spectral zeta function) regularization for Gaussian Euclidean path integrals \cite{Vassilevich:2003xt}, it is useful to have explicit results in the latter scheme. In this appendix we give an efficient and general recipe to compute the exact heat kernel-regularized one-loop Euclidean path integral, with regulator $e^{-\epsilon^2/4\tau}$ as in (\ref{HKSP}), requiring only the unregulated character formula as input. For concreteness we consider the scalar case in the derivation, but because the scalar character $\chi_0(t)$ provides the basic building block for all other characters $\chi_S(t)$, the final result will be applicable in general. We spell out the derivation is some detail, and summarize the final result together with some examples in section \ref{sec:finresandex}. Application to the massless higher-spin case is discussed in section \ref{app:masslesssol}, where we work out the exact one-loop Euclidean path integral for Einstein gravity on $S^4$ as an example. In section \ref{sec:otherreg} we consider different regularizations, such as the simple $t>\epsilon$ cutoff.

\subsection{Derivation} \label{app:cherder}

As shown in section \ref{sec:scalars}, the scalar Euclidean path integral regularized as
\begin{align} 
 \log Z_\epsilon = \int_0^\infty \frac{d\tau}{2\tau} \, e^{-\frac{\epsilon^2}{4\tau}} \, F_D(\tau) \, , \qquad F_D(\tau) \equiv {\rm Tr} \, e^{-\tau D}  =  \sum_n D_n^{d+2} \, e^{-\left(n+\frac{d}{2}+i \nu \right)\left(n+\frac{d}{2}-i \nu \right)\tau} \, , \label{ZPIreg1}
\end{align}
where $D=-\nabla^2+\frac{d^2}{4}+\nu^2$, can be written in character integral form as
\begin{align} \label{origexpp}
 \log Z_\epsilon &= \int_\epsilon^\infty \frac{dt}{2 \sqrt{t^2-\epsilon^2}}  
 \sum_n D_n^{d+2} \Bigl( 
 e^{-(n+\frac{d}{2}) t - i \nu \sqrt{t^2-\epsilon^2}}
 + e^{-(n+\frac{d}{2}) t + i \nu \sqrt{t^2-\epsilon^2}} \Bigr)  \\
 &=\int_\epsilon^\infty \frac{dt}{2 \sqrt{t^2-\epsilon^2}} \, \, \frac{1+e^{-t}}{1-e^{-t}} \, \frac{e^{-\frac{d}{2} t - i \nu \sqrt{t^2-\epsilon^2}}
 +e^{-\frac{d}{2} t + i \nu \sqrt{t^2-\epsilon^2}}}{(1-e^{-t})^d} \, , \label{ZPIreg2}
\end{align}
Putting $\epsilon=0$ we recover the formal (UV-divergent) character formula
\begin{align} 
 \log Z_{\epsilon=0} &= \int_0^\infty \frac{dt}{2t}  \, F_\nu(t) \, , \nn \\
 F_\nu(t) &\equiv
  \sum_n D_n^{d+2} \Bigl( 
 e^{-(n+ \frac{d}{2} + i \nu) t}
 + e^{-(n+\frac{d}{2} - i \nu n) t} \Bigr) 
 =\frac{1+e^{-t}}{1-e^{-t}} \, 
 \frac{e^{-(\frac{d}{2} + i \nu) t}
 + e^{-(\frac{d}{2} - i \nu) t}}{(1-e^{-t})^d} \, . \label{Zformal}
\end{align}
To evaluate (\ref{ZPIreg2}), we split the integral into UV and IR parts, each of which can be evaluated in closed form in the limit $\epsilon \to 0$.

\subsubsection*{Separation into UV and IR parts}

The separation of the integral in UV and IR parts is analogous to the usual procedure in heat kernel regularization, where one similarly separates out the UV part of the $\tau$ integral by isolating the leading terms in the $\tau \to 0$ heat kernel expansion 
\begin{align} \label{HKESP}
 F_D(\tau) := {\rm Tr} \, e^{-\tau D} \, \to \,
 \sum_{k=0}^{d+1} \alpha_k \, \tau^{-(d+1-k)/2} =: F_D^{\rm uv}(\tau)  \, .
\end{align}
Introducing an infinitesimal IR cutoff $\mu \to 0$, we may write $\log Z_\epsilon = \log Z^{\rm uv}_\epsilon + \log Z^{\rm ir}$ where
\begin{align}
 \log Z^{\rm uv}_\epsilon \equiv \int_0^\infty \frac{d\tau}{2\tau} \, e^{-\frac{\epsilon^2}{4\tau}}  \, F_D^{\rm uv}(\tau) \,  e^{-\mu^2 \tau}  \, , \quad
 \log Z^{\rm ir} \equiv \int_0^\infty \frac{d\tau}{2\tau} \bigl( F_D(\tau) - F_D^{\rm uv}(\tau) \bigr) \, e^{-\mu^2 \tau} \, . 
\end{align}
Dropping the UV regulator in the IR integral is allowed because all UV divergences have been removed by the subtraction. The factor $e^{-\mu^2 \tau}$ serves as an IR regulator needed for the separate integrals when $F^{\rm uv}$ has a term $\frac{\alpha_{d+1}}{2\tau} \neq 0$, that is to say when $d+1$ is even. The resulting $\log\mu$ terms cancel out of the sum at the end. Evaluating this using the specific UV regulator of (\ref{ZPIreg1}) gives
\begin{align} \label{logZUVHK}
 \log Z_\epsilon =  \frac{1}{2}  \zeta_D'(0) + \alpha_{d+1} \log\bigl( \tfrac{2}{e^\game \epsilon} \bigr) + \frac{1}{2} \sum_{k=0}^{d} \alpha_k \, 
 \Gamma\bigl(\tfrac{d+1-k}{2}\bigr) \left(\tfrac{2}{\epsilon}\right)^{d+1-k} \, , 
\end{align}
where $\zeta_D(z) = {\rm Tr} \, D^{-z} = \frac{1}{\Gamma(z)} \int \frac{d\tau}{\tau} \, \tau^z \, {\rm Tr} \, e^{-\tau D}$ is the zeta function of $D$ and  $\alpha_{d+1}=\zeta_D(0)$. 
 
We can apply the same idea to the square-root regulated character formula (\ref{ZPIreg2}) for $Z_\epsilon$. The latter is obtained from the simpler integrand of the formal character formula (\ref{Zformal}) for $Z_{\epsilon=0}$ by dividing it by $r(\epsilon,t) \equiv \sqrt{t^2-\epsilon^2}/t$ and replacing $\nu$ by $\nu r(\epsilon,t)$:
\begin{align} \label{logZformal}
 \log Z_{\epsilon=0} = \int_0^\infty \frac{dt}{2t} \, F_\nu(t) 
 \qquad \Rightarrow \qquad
 \log Z_{\epsilon} = \int_\epsilon^\infty \frac{dt}{2rt} \, F_{r \nu}(t) \, ,
 \qquad r \equiv \frac{\sqrt{t^2-\epsilon^2}}{t} \, . 
\end{align}
Note that $0 < r<1$ for all $t>\epsilon$, $r \sim \CO(1)$ for $t \sim \epsilon$ and $r \to 1$ for $t \gg \epsilon$. Therefore, given the $t \to 0$ behavior of the integrand in the formal character formula for $Z_{\epsilon=0}$,
\begin{align} \label{Fnuuvdef}
 \frac{1}{2t} F_\nu(t) \to \frac{1}{t} \sum_{k=0}^{d+1} b_{k}(\nu) \, t^{-(d+1-k)} =: \frac{1}{2t} F_{\nu}^{\rm uv}(t) \, , \qquad b_k(\nu) = \sum_{\ell=0}^k b_{k\ell} \,\nu^\ell  \, ,
\end{align}
we get the $t \sim \epsilon \to 0$ behavior of the integrand for the exact $Z_{\epsilon}$:
\begin{align} \label{smalltexpF}
 \frac{1}{2rt} \, F_{r \nu}(t)  \to \frac{1}{2rt} \, F^{\rm uv}_{r\nu}(t) = \frac{1}{r t} \sum_{k,\ell} b_{k\ell} \, \nu^\ell \, r^\ell  \, t^{-(d+1-k)}  \, .
\end{align}
Thus we can separate $\log Z_\epsilon = \log \tilde Z^{\rm uv}_\epsilon + \log \tilde Z^{\rm ir}$, with
\begin{align}
 \log \tilde Z^{\rm uv}_\epsilon \equiv \int_\epsilon^\infty \frac{dt}{2rt}  \, F^{\rm uv}_{r \nu}(t)  \, e^{-\mu t} \, , \qquad
 \log \tilde Z^{\rm ir} \equiv \int_0^\infty \frac{dt}{2t} \, \bigl( F_\nu(t) - F^{\rm uv}_{\nu}(t) \bigr) e^{-\mu t} \, . 
\end{align}
Again the limit $\mu \to 0$ is understood. We were allowed to put $\epsilon = 0$ in the IR part because it is UV finite.

\subsubsection*{Evaluation of UV part}

Using the expansion (\ref{smalltexpF}), the UV part can be evaluated explicitly as 
\begin{align} \label{logZtildeuv}
 \log \tilde Z^{\rm uv}_\epsilon =  \frac{1}{2} \sum_{\ell,k\leq d} b_{k\ell} \,
 B\bigl(\tfrac{d+1-k}{2},\tfrac{\ell+1}{2} \bigr)
  \, \nu^\ell \, \epsilon^{-(d+1-k)} - \sum_\ell b_{d+1,\ell} \bigl(H_\ell - \tfrac{1}{2} H_{\ell/2} + \log( \tfrac{e^{\game} \, \epsilon \, \mu}{2}) \bigr) \nu^\ell 
\end{align}
where $B(x,y) = \frac{\Gamma(x) \Gamma(y)}{\Gamma(x+y)}$ is the Euler beta function and $H_x= \game+\frac{\Gamma'(1+x)}{\Gamma(1+x)}$ which for integer $x$ is the $x$-th harmonic number $H_x = 1+\frac{1}{2}+\cdots + \frac{1}{x}$. For example for $d=3$, we get
\begin{align} 
 \log \tilde Z^{\rm uv}_\epsilon = \tfrac{4}{3} \, \epsilon^{-4}  - \tfrac{4\nu^2+1}{12} \, \epsilon^{-2} - \bigl( \tfrac{\nu^4}{9} + \tfrac{\nu^2}{24} \bigr) - \bigl( \tfrac{\nu^4}{12} + \tfrac{\nu^2}{24} -\tfrac{17}{2880} \bigr) \log\bigl(\tfrac{e^\game \epsilon \mu}{2}\bigr) \, .
\end{align}
This gives an explicit expression for the part of $\log Z$ denoted ${\rm Pol}(\Delta)$ in \cite{Denef:2009kn}, without having to invoke an independent  computation of the heat kernel coefficients. 
Indeed, turning this around, by comparing (\ref{logZtildeuv}) to (\ref{logZUVHK}), we can express the heat kernel coefficients $\alpha_k$ explicitly in terms of the character coefficients $b_{k,\ell}$. In particular the Weyl anomaly coefficient is  simply given by the coefficient $b_{d+1}=\sum_\ell b_{d+1,\ell} \nu^\ell$ of the $1/t$ term in the integrand of the formal character formula (\ref{Zformal}). More generally, 
\begin{align} \label{HKCFORM0}
 \alpha_k = \sum_{\ell}  \frac{\Gamma(\frac{\ell+1}{2})}{2^{d+1-k} \Gamma(\frac{d+1-k+\ell+1}{2})} \, b_{k\ell} \, \nu^\ell \, .
\end{align}
For example for $d=3$, this becomes $\alpha_0=\frac{1}{12} b_{00}$, $\alpha_2 = \frac{1}{2} b_{20} + \frac{\nu^2}{6} b_{22}$ and $\alpha_4 = b_4$. From the small-$t$ expansion $\frac{1}{2t} F_\nu(t) \to \sum_k b_k t^{3-k}$ in (\ref{Zformal}) we read off $b_{0}=2$, $b_{2}=-\frac{1}{12}-\nu^2$ and
$b_4= -\frac{17}{2880}+\frac{1}{24} \nu^2  +\frac{1}{12}\nu^4$. Thus 
$\alpha_0 = \frac{1}{6}$, $\alpha_2 = -\frac{1}{24} - \frac{1}{6} \nu^2$ and $\alpha_4=-\frac{17}{2880}+\frac{1}{24} \nu^2  +\frac{1}{12}\nu^4$.

\subsubsection*{Evaluation of IR part}

As we explain momentarily, the IR part can be evaluated as
\begin{align} \label{ZIRzetanu}
 \log \tilde Z^{\rm ir} =  \frac{1}{2}  \zeta_\nu'(0) + b_{d+1} \log \mu \, , \qquad \zeta_\nu(z) \equiv \frac{1}{\Gamma(z)} \int_0^\infty \frac{dt}{t} \, t^z \, F_\nu(t) \, , 
\end{align}
where like for the spectral zeta function $\zeta_D(z)$, the ``character zeta function'' $\zeta_\nu(z)$ is defined by the above integral for $z$ sufficiently large and by analytic continuation for $z \to 0$. This zeta function representation of $\log Z^{\rm ir}$ follows from the following observations. If we define $\zeta^{\rm ir}_\nu(z) \equiv \frac{1}{\Gamma(z)} \int_0^\infty \frac{dt}{t} \, t^z \bigl(F_\nu(t) - F_\nu^{\rm uv}(t) \bigr) \, e^{-\mu t}$, then since the integral remains finite for $z \to 0$, while $\Gamma(z) \sim 1/z$ and $\partial_z(1/\Gamma(z)) \to 1$, we trivially have $\frac{1}{2} \partial_z \zeta^{\rm ir}_\nu(z)|_{z=0}=  {\log \tilde{Z}^{\rm ir}}$. Moreover for $z$ sufficiently large we have in the limit $\mu \to 0$ that $\frac{1}{2}\zeta^{\rm uv}(z) \equiv \frac{1}{\Gamma(z)} \int_0^\infty \frac{dt}{2t} \, t^z F^{\rm uv}_\nu(t) \, e^{-\mu t} = b_{d+1} \mu^{-z}$, so upon analytic continuation we have ${1/2} \partial_z \zeta^{\rm uv}(z)|_{z=0}=- b_{d+1} \log \mu$, and (\ref{ZIRzetanu}) follows. 

In contrast to the spectral zeta function, the character zeta function can straightforwardly be evaluated in terms of Hurwitz zeta functions. Indeed, denoting $\Delta_\pm=\frac{d}{2}\pm i\nu$, we have $F_D(t) = \sum_n Q(n) \, e^{-t(n+\Delta_+)(n+\Delta_-)}$ where the spectral degeneracy $Q(n)$ is some polynomial in $n$, and
$\zeta_D(z) = \sum_{n=0}^\infty Q(n) \, \bigl((n+\Delta_+)(n+\Delta_-)\bigr)^{-z}$,
which is quite tricky to evaluate, 
whereas $F_\nu(t) =\sum_n Q(n)  \bigl( e^{-t(n+\Delta_+)} + e^{-t(n+\Delta_-)} \bigr)$, and we can immediately express the associated character zeta function as a finite sum of Hurwitz zeta functions $\zeta(z,\Delta) = \sum_{n=0}^\infty (n+\Delta)^{-z}$:
\begin{align} \label{zetanuHur}
  \zeta_\nu(z) = \sum_{\pm} \sum_{n=0}^\infty Q(n) (n+\Delta_\pm)^{-z} = \sum_{\pm} Q(\hat\delta - \Delta_\pm) \, \zeta(z,\Delta_\pm) \, .
\end{align}
Here $\hat\delta$ is the unit $z$-shift operator acting as $\hat\delta^n \zeta(z,\Delta) = \zeta(z-n,\Delta)$; for example if $Q(n)=n^2$ we have $Q(\hat\delta - \Delta) \, \zeta(z,\Delta) = (\hat\delta^2 - 2 \Delta \hat\delta + \Delta^2) \zeta(z,\Delta) = \zeta(z-2,\Delta) - 2 \Delta \zeta(z-1,\Delta) + \Delta^2 \zeta(z,\Delta)$.

\subsection{Result and examples} \label{sec:finresandex}

\subsubsection*{Result}

Altogether we conclude that given a formal character integral formula 
\begin{align} \label{lozipinu}
 \log Z_{\rm PI} = \int_0^\infty \frac{dt}{2t} \, F_\nu(t)  \, , 
\end{align}
for a field corresponding to a dS$_{d+1}$ irrep of dimension $\frac{d}{2} + i \nu$, with IR and UV expansions 
\begin{align} \label{Fnuexpan}
  F_\nu(t) = \sum_{\Delta} \sum_{n=0}^\infty P_\Delta(n) \, e^{-(n+\Delta) t} \, ,
  \qquad \frac{1}{2t} F_\nu(t) =  \frac{1}{t} \sum_{k=0}^{d+1} b_k(\nu) \, t^{-(d+1-k)} \, + \CO(t^0) \, ,
\end{align}
where $b_k(\nu) = \sum_\ell b_{k\ell} \, \nu^\ell$, 
we obtain the exact $Z_{\rm PI}$ with  heat kernel regulator $e^{-\epsilon^2/4\tau}$ as
\begin{equation} \label{ZEXACT}
\boxed{\begin{aligned}
 \log Z_{{\rm PI},\epsilon} = &\frac{1}{2}\sum_\Delta P_\Delta(\hat\delta-\Delta) \, \zeta'(0,\Delta) 
 - \sum_{\ell=0}^{d+1} b_{d+1,\ell} \bigl(H_\ell - \tfrac{1}{2} H_{\ell/2} \bigr) \nu^\ell + b_{d+1}(\nu) \log (2 e^{-\game} / \epsilon) \!   \\
  &  +  \frac{1}{2} \sum_{k=0}^{d} \sum_{\ell=0}^k b_{k\ell} \,
 B\bigl(\tfrac{d+1-k}{2},\tfrac{\ell+1}{2} \bigr)
  \, \nu^\ell \, \epsilon^{-(d+1-k)}  \, .
\end{aligned}}
\end{equation}
Here $B(x,y) = \frac{\Gamma(x) \Gamma(y)}{\Gamma(x+y)}$, $H_x= \game+\frac{\Gamma'(1+x)}{\Gamma(1+x)}$, which for integer $x$ is the $x$-th harmonic number $H_x = 1+\frac{1}{2}+\cdots + \frac{1}{x}$, and 
$\hat\delta$ is the unit shift operator acting on the first argument of the Hurwitz zeta function $\zeta(z,\Delta)$: the polynomial $P_\Delta(\hat\delta - \Delta)$ is to be expanded in powers of $\hat\delta$, setting $\hat\delta^n \zeta'(0,\Delta) \equiv \zeta'(-n,\Delta)$. Finally the heat kernel coefficients are
\begin{align} \label{HKCFORM}
 \alpha_k = \sum_{\ell}  \frac{\Gamma(\frac{\ell+1}{2})}{2^{d+1-k} \Gamma(\frac{d+1-k+\ell+1}{2})} \, b_{k\ell} \, \nu^\ell \, .
\end{align}
If we are only interested in the finite part of $\log Z$, only the first three terms in (\ref{ZEXACT}) matter. Note that the third and the second term $\CM_\nu \equiv \sum_\ell b_{d+1,\ell} \bigl(H_\ell - \tfrac{1}{2} H_{\ell/2} \bigr)$ is in general nonvanishing for even $d+1$. By comparing (\ref{ZEXACT}) to (\ref{logZUVHK}), say in the scalar case discussed earlier, we see that $\zeta_D'(0) = \zeta_\nu'(0) + 2\CM_\nu$. Thus $2\CM_\nu$ can be thought of as correcting the formal factorization  $\sum_n \log (n+\Delta_+)(n+\Delta_-) = \sum_n \log(n+\Delta_+) + \sum_n \log(n+\Delta_-)$ in zeta function regularization. For this reason $\CM_\nu$ is called the multiplicative ``anomaly'', as reviewed in \cite{Dowker:2014xca}. The above thus generalizes the explicit formulae in \cite{Dowker:2014xca} for $\CM_\nu$ to fields of arbitrary representation content.

\subsubsection*{Examples}

\noindent {\bf 1}. A scalar on $S^2$ ($d=1$) with $\Delta_\pm = \frac{1}{2} \pm i\nu$ has $F_\nu(t) = \frac{1+e^{-t}}{1-e^{-t}} \frac{e^{- \Delta_+ t}+e^{-\Delta_- t}}{1-e^{-t}}$ so the IR and UV expansions are $F_\nu(t) = \sum_{\pm} \sum_{n=0}^\infty (2n+1) e^{-(\Delta_\pm + n)t}$ and $
 \frac{1}{2t} \, F_\nu(t) = \frac{2}{t^3} + \frac{\frac{1}{12}-\nu^2}{t} + \CO(t^0)$.
Therefore according to (\ref{ZEXACT})
\begin{align} \label{dS2exampleZPI}
 \log Z_{\rm PI,\epsilon} = &\sum_{\Delta=\frac{1}{2}\pm i \nu} \Bigl(\zeta'(-1,\Delta) - (\Delta-\tfrac{1}{2}) \zeta'(0,\Delta) \Bigr) + \nu^2 + \bigl(\tfrac{1}{12}-\nu^2\bigr) \log \bigl(2 \, e^{-\game}/\epsilon\bigr) + \frac{2}{\epsilon^2} \, . 
\end{align}
The heat kernel coefficients are obtained from (\ref{HKCFORM}) as $\alpha_0 = 1$ and $\alpha_2 = \frac{1}{12}-\nu^2$. \vskip3mm

\noindent {\bf 2}. For a scalar on $S^3$, $F_\nu(t) = \sum_{\pm} \sum_{n=0}^\infty (n+1)^2 e^{-(\Delta_\pm + n)t}$, $\frac{1}{2t} F_\nu(t) \to \frac{2}{t^4} - \frac{\nu^2}{t^2} + \CO(t^0)$, so 
\begin{align} \label{scalarS3}
 \log Z_{\rm PI,\epsilon} = &\sum_{\pm} \Bigl(\tfrac{1}{2} \zeta'(-2,1\pm i\nu) \mp i \nu \zeta'(-1,1\pm i\nu)  - \tfrac{1}{2}\nu^2 \zeta'(0,1\pm i \nu) \Bigr)  - \frac{\pi \nu^2}{4 \epsilon} + \frac{\pi}{2 \epsilon^3} \, . 
\end{align}
The heat kernel coefficients are $\alpha_0=\frac{\sqrt{\pi}}{4}$, $\alpha_2 = -\frac{\sqrt{\pi}}{4} \nu^2$. In particular for a conformally coupled scalar, i.e.\ $\Delta=\frac{1}{2},\frac{3}{2}$ or equivalently $\nu=i/2$, we get for the finite part the familiar result $\log Z_{\rm PI} = \frac{3 \zeta (3)}{16 \pi ^2}-\frac{\log (2)}{8}$. For $\Delta=1$, i.e.\ $\nu=0$, we get $\log Z_{\rm PI}=-\frac{\zeta(3)}{4\pi^2}$. Notice that the finite part looks quite different from (\ref{Zbubufi2}) obtained by contour integration. Nevertheless they are in fact the same function.  
\vskip3mm
\noindent {\bf 3}. A more interesting example is the massive spin-$s$ field on $S^4$ with $\Delta_{\pm} = \frac{3}{2} \pm i \nu$. In this case, (\ref{ZPItrueformula}) combined with (\ref{coin}) or equivalently (\ref{formuhs2}) gives $F_\nu = F_{\rm bulk} - F_{\rm edge}$ with
\begin{align}  \label{startcomp}
 F_{\rm bulk}(t) &=  \sum_{\Delta=\frac{3}{2} \pm i\nu} \sum_{n=-1}^\infty  D^{3}_s D^{5}_n  \, e^{-(n+\Delta) t} = D_s^{3} \, \frac{1+e^{-t}}{1-e^{-t}} \frac{e^{-(\frac{3}{2}+i\nu) t}+e^{-(\frac{3}{2}-i\nu) t}}{(1-e^{-t})^3}  \, , \qquad  \,   \\
 F_{\rm edge}(t) &=  \sum_{\Delta=\frac{1}{2} \pm i\nu} \sum_{n=-1}^\infty  D^{5}_{s-1} D^{3}_{n+1}  \, e^{-(n+\Delta) t} = D_{s-1}^{5} \, \frac{1+e^{-t}}{1-e^{-t}} \frac{e^{-(\frac{1}{2}+i\nu) t}+e^{-(\frac{1}{2}-i\nu) t}}{(1-e^{-t})} \, , 
\end{align}
where $D_p^{3}=2p+1$, $D^{5}_p = \frac{1}{6}(2p+3)(p+2)(p+1)$. In particular note that with $g_s \equiv D_s^{3} = 2s+1$, we have $D_{s-1}^{5}=\frac{1}{24}g_s(g_s^2-1)$. The small-$t$ expansions are
\begin{align}
 \tfrac{1}{2t} F_{\rm bulk}(t) &\to g_s \Bigl(2 \, t^{-5} - \bigl(\nu^2+\tfrac{1}{12}\bigr) t^{-3} + \bigl(\tfrac{\nu^4}{12}+\tfrac{\nu^2}{24}-\tfrac{17}{2880} \bigr) t^{-1} + \CO(t^0) \Bigr)  \\
 \tfrac{1}{2t} F_{\rm edge}(t) &\to \tfrac{1}{24} g_s(g_s^2-1) \Bigl( 2 \, t^{-3} + \bigl(\tfrac{1}{12}-\nu^2\bigr) t^{-1} + \CO(t^0) \Bigr) \, . 
\end{align}
Thus the exact partition function for a massive spin-$s$ field is
\begin{align} \label{ZPIexactmssd3}
 \log Z_{\rm PI,\epsilon} &= g_s \sum_{\Delta=\frac{3}{2}\pm i \nu} \Bigl( \tfrac{1}{6} \zeta'(-3,\Delta) \mp \tfrac{1}{2} i \nu \zeta'(-2,\Delta) - \bigl(\tfrac{1}{2}\nu^2 + \tfrac{1}{24} \bigr) \zeta'(-1,\Delta) 
 \pm i\bigl(\tfrac{1}{24} \nu + \tfrac{1}{6} \nu^3  \bigr) \zeta'(0,\Delta) \Bigr) \nn \\
 &\quad - \tfrac{1}{24} g_s(g_s^2-1)  \sum_{\Delta=\frac{1}{2}\pm i \nu} \Bigl( \zeta'(-1,\Delta) \mp i \nu \zeta'(0,\Delta) \Bigr) \, - \tfrac{1}{24} g_s^3 \nu^2 - \tfrac{1}{9} g_s \nu^4 \\
 &\quad + \Bigl( g_s^3 \bigl( \tfrac{1}{24} \nu^2 - \tfrac{1}{288} \bigr) + g_s \bigl( \tfrac{1}{12} \nu^4 - \tfrac{7}{2880} \bigr) \Bigr) \log(2 \, e^{-\game}/\epsilon)  - \bigl(\tfrac{1}{12} g_s^3 + \tfrac{1}{3} g_s \nu^2 \bigr) \epsilon^{-2} + \tfrac{4}{3} g_s \epsilon^{-4} \, . \nn
\end{align}
Finally the heat kernel coefficients are
\begin{align} \label{hkcex}
 \alpha_0 = \tfrac{1}{6} g_s \, , \qquad \alpha_2 = -\tfrac{1}{24} g_s^3 - \tfrac{1}{6} g_s \nu^2 \, , \qquad \alpha_4 = g_s^3 \bigl( \tfrac{1}{24} \nu^2 - \tfrac{1}{288} \bigr) + g_s \bigl( \tfrac{1}{12} \nu^4 - \tfrac{7}{2880} \bigr) \, .
\end{align} 

\subsubsection*{Single-mode contributions}

Contributions from single path integral modes and contributions of single quasinormal modes are of use in some of our derivations and applications. These are essentially special cases of the above general results, but for convenience we collect some explicit formulae here: \\
\noindent $\bullet$ {\bf Path integral single-mode contributions}: For our choice of heat-kernel regulator $e^{-\epsilon^2/4\tau}$, the contribution to $\log Z_{{\rm PI},\epsilon}$ from a single bosonic eigenmode with eigenvalue $\lambda$ is
\begin{align} \label{HKHKHKH}
 I_\lambda = \int_0^\infty \frac{d\tau}{2\tau} \, e^{-{\epsilon^2}/{4 \tau}} \, e^{-\tau \lambda} = K_0(\epsilon \sqrt{\lambda}) \to -\frac{1}{2} \log \frac{\lambda}{M^2} \,  , \qquad M \equiv \frac{2 e^{-\game}}{\epsilon} \, ,
\end{align}
Different regulator insertions lead to a similar result in the limit $\epsilon \to 0$, with $M=c/\epsilon$ for some regulator-dependent constant $c$.  A closely related formula is obtained for the contribution from an individual term in the  sum (\ref{origexpp}) or equivalently in the IR expansion of (\ref{Fnuexpan}), which amounts to computing (\ref{lozipinu}) with $F_\nu(t) \equiv e^{-\rho t}$, $\rho = a \pm i \nu$. The small-$t$ expansion is $\frac{1}{2t} F_\nu(t) = \frac{1}{2t} + \CO(t^0)$, so the UV part is given by the log term in (\ref{ZEXACT}) with coefficient $\frac{1}{2}$, and the IR part is $\frac{1}{2} \zeta_\nu'(0) = -\frac{1}{2} \log \rho$ as in (\ref{ZIRzetanu}). Thus   
\begin{align} \label{simplereg}
 I'_\rho = \int_0^\infty \frac{dt}{2t} \, e^{-\rho t} \to -\frac{1}{2} \log \frac{\rho}{M} \, , \qquad M = \frac{2 e^{-\game}}{\epsilon} \, ,
\end{align}
where the integral is understood to be regularized as in (\ref{origexpp}), $I'_\rho = \int_\epsilon^\infty \frac{dt}{2\sqrt{t^2-\epsilon^2}} \, e^{-t a -  i \nu \sqrt{t^2 - \epsilon^2}}$, left implicit here. 
The similarities  between (\ref{HKHKHKH}) and (\ref{simplereg}) are of course no accident, since in our setup, the former splits into the sum of two integrals of the latter type: writing $\lambda=a^2+\nu^2=(a+i\nu)(a-i\nu)$, we have $I_\lambda=I'_{a+i\nu} + I'_{a-i \nu}$. 
\vskip1mm 
\noindent $\bullet$ {\bf Quasinormal mode contributions}: Considering a character quasinormal mode expansion $\chi(t) = \sum_r N_r \, e^{-r|t|}$  as in (\ref{qnmexp}), the IR contribution from a single bosonic/fermionic QNM  is 
\begin{align} \label{QNMcontrib}
 \int_0^\infty \frac{dt}{2t}  \frac{1+e^{-t}}{1-e^{-t}} \, e^{-r \, t} 
 \biggr|_{\rm IR}   =  \log \frac{\Gamma(r+1)}{\mu^r \sqrt{2 \pi r}} \, , \qquad 
 -\int_0^\infty \frac{dt}{2t}   \frac{2 e^{-t/2}}{1-e^{-t}} \, e^{-r \, t} \biggr|_{\rm IR}   = - \log \frac{\Gamma(r+\frac{1}{2})}{\mu^r \sqrt{2 \pi}} 
\end{align} 
\vskip1mm 
\noindent $\bullet$ {\bf Harmonic oscillator}:
The character of a $d=0$ scalar of mass $\nu$ is $\chi(t) = e^{-i\nu t} + e^{i \nu t}$, hence  
\begin{align} \label{kzerocase}
 \log Z_{\rm PI,\epsilon} = \int_0^\infty \frac{dt}{2t} \,  \frac{1+e^{-t}}{1-e^{-t}} \,\bigl( e^{-i \nu t} + e^{i \nu t} \bigr) = \frac{\pi}{\epsilon} - \log \bigl(e^{\pi\nu} - e^{-\pi\nu} \bigr) \, .
\end{align}
The finite part gives the canonical bosonic harmonic oscillator thermal partition function  $ {\rm Tr} \, e^{-\beta H}= \sum_n e^{-\beta \nu(n+\frac{1}{2})} = \bigl(e^{\beta \nu/2}-e^{-\beta \nu/2}\bigr)^{-1}$ at  $\beta=2\pi$. 
The fermionic version is
\begin{align} \label{kzerocaseFer}
 \log Z_{\rm PI,\epsilon} = -\int_0^\infty \frac{dt}{2t} \,  \frac{2e^{-t/2}}{1-e^{-t}} \,\bigl( e^{-i \nu t} + e^{i \nu t} \bigr) = -\frac{\pi}{\epsilon} + \log \bigl(e^{\pi\nu} + e^{-\pi\nu}  \bigr) \, .
\end{align}

\subsection{Massless case} \label{app:masslesssol}

Here we give a few more details on how to use (\ref{ZEXACT}) to explicitly evaluate $Z_{\rm PI}$ in the massless case, and work out the exact $Z_{\rm PI}$ for Einstein gravity on $S^4$ as an example. 

Our final result for the massless one-loop $Z_{\rm PI}= \ZG \cdot \Zchar$ is given by (\ref{ZPIFINAL}):
\begin{align} \label{ZPIFINALapp}
 Z_{\rm PI}  = i^{-\pol} \, \frac{\gamma^{\rm dim G}}{\vc} \cdot \exp \inteps^{\noir} \frac{dt}{2t} \, F \, , \qquad F = \frac{1+q}{1-q} \Bigl( \bigl[\hat\chi_{\rm bulk}\bigr]_+ - \bigl[\hat\chi_{\rm edge}\bigr]_+ - 2 \dim G \Bigr) \, ,
\end{align}
where for $s=2$ gravity $\gamma = \sqrt{\frac{8 \pi G_{\rm N}}{\Ad}}$, $\pol=d+3$, $G=SO(d+2)$ and $\vc=(\ref{vcSO})$. 

\noindent $\bullet$ {\bf UV part:} As always, the coefficient of the log-divergent term simply equals the coefficient of the $1/t$ term in the small-$t$ expansion of the integrand in (\ref{ZPIFINALapp}). For the other UV terms in (\ref{ZEXACT}) (including the ``multiplicative anomaly''), a problem might seem to be that we need a continuously variable dimension parameter $\Delta = \frac{d}{2} + i \nu$, whereas massless fields, and our explicit formulae for $\hat \chi \to [\hat\chi]_+$, require fixed integer dimensions. This problem is easily solved, as the {\it UV part} can actually be computed from the original {\it naive} character formula (\ref{Znaive}):
\begin{align} \label{ZUVjustnaive}
  \log Z_{\rm PI}\bigr|_{\rm UV} = \inteps \frac{dt}{2t} \, \hat F \Bigr|_{\rm UV} \, , \qquad \hat F = \frac{1+q}{1-q} \bigl(\hat \chi_{\rm bulk} - \hat\chi_{\rm edge} \bigr) \, ,
\end{align}
Indeed since $\hat F \to F = \{\hat F\}_+$ in (\ref{Fsplusdef})  affects just a finite number of terms $c_k q^k \to c_k q^{-k}$, it does not alter the small-$t$ (UV) part of the integral. 
Moreover $\hat\chi_s = \hat\chi_{s,\nu_\phi} - \hat\chi_{s,\nu_\xi}$, where $\hat\chi_{s,\nu}$ is a massive spin-$s$ character. Thus the UV part may be obtained simply by combining the results of (\ref{ZEXACT}) for general $\nu$ and $s$, substituting the  values $\nu_\phi$, $\nu_\xi$ set by (\ref{phixidim}). 

\noindent $\bullet$ {\bf IR part:} The {\it IR part} is the $\zeta'$ part of (\ref{ZEXACT}), obtained from the $q$-expansion of $F(q)$ in (\ref{ZPIFINALapp}). This can be found in general by using 
\begin{align} \label{expansionrule}
 \frac{1+q}{1-q} \, \frac{q^\Delta}{(1-q)^k} = \sum_{n=0}^\infty P(n) \, q^{n+\Delta} \, , \qquad P(n) = D_n^{k+2} \, ,
\end{align}
with $D_n^{k+2}$ the polynomial given in (\ref{Dsods2}).
For $k=0$, (\ref{QNMcontrib}) is useful.
In particular, using the $\int^\noir$ prescription (\ref{noirdef}), the IR contribution from the last term in (\ref{ZPIFINALapp}) is obtained by considering the $r \to 0$ limit of the bosonic formula in (\ref{QNMcontrib}): 
\begin{align} \label{dimGcontrib}
  \int^\noir \frac{dt}{2t} \, \frac{1+q}{1-q} \bigl(-2 \dim G \bigr) \biggr|_{\rm IR} =  \dim G \cdot  \log (2\pi)    \, .
\end{align}

\subsubsection*{Example: Einstein gravity on $S^4$}

As a simple application, let us compute the exact one-loop Euclidean path integral for pure gravity on $S^4$. In this case $G=SO(5)$, $\dim G = 10$, $d=3$ and $s=2$. From (\ref{phixidim}) we read off $i\nu_\phi = \frac{3}{2}$, $i\nu_\xi = \frac{5}{2}$, and from (\ref{charexmpls}) we get 
\begin{align}
 \chi_{\rm bulk} = \bigl[\hat \chi_{\rm bulk} \bigr]_+ = \frac{10 \, q^3 - 6 \, q^4}{(1-q)^3} \, , \qquad \chi_{\rm edge} = \bigl[\hat \chi_{\rm edge} \bigr]_+ = \frac{10 \, q^2-2 \, q^3}{1-q} \, .
\end{align}
The small-$t$ expansion of the integrand in (\ref{ZPIFINALapp}) is $\frac{1}{2t} F = 4 \, t^{-5} -\frac{47}{3} \, t^{-3} -\frac{571}{45} \, t^{-1} +O(t^0)$.   
The coefficient of the log-divergent part of $\log Z_{\rm PI}$ is the coefficient of $t^{-1}$:
\begin{align} \label{logtermval}
 \log Z_{\rm PI}|_{\rm log \, div} = -\frac{571}{45} \, \log \bigl(2 e^{-\game} \epsilon^{-1} \bigr) \, ,
\end{align}
in agreement with \cite{Christensen:1979iy}. The complete heat-kernel regularized UV part of (\ref{ZEXACT}) can be read off directly from our earlier results for massive spin-$s$ in $d=3$ as
\begin{align}
 \log Z_{\rm PI} \bigr|_{\rm UV} &= \log Z_{\rm PI}(s=2,\nu=\tfrac{3}{2} i) \bigr|_{\rm UV} \, - \, 
 \log Z_{\rm PI}(s=1,\nu=\tfrac{5}{2} i) \bigr|_{\rm UV} \nn \\
 &= \frac{8}{3} \, \epsilon^{-4}-\frac{32}{3} \, \epsilon^{-2} - \frac{571}{45}  \log\bigl(2 e^{-\game} \epsilon^{-1} \bigr)   + \frac{715}{48} \, .
\end{align}
Here $\CM=\frac{715}{48}$ is the ``multiplicative anomaly'' term. The integrated heat kernel coefficients are similarly obtained from (\ref{hkcex}): $\alpha_0 = \frac{1}{3}$, $\alpha_2 = -\frac{16}{3}$, $\alpha_4 = - \frac{571}{45}$.

The IR ($\zeta'$) contributions from bulk and edge characters are obtained from the expansions
\begin{align}
 \frac{1+q}{1-q} \bigl(\chi_{\rm bulk} - \chi_{\rm edge} \bigr) = \sum_n P_{\rm b}(n) \, \bigl(10 \, q^{3+n} - 6 \, q^{4+n} \bigr) - \sum_n P_{\rm e}(n) \, \bigl( 10 \, q^{2+n} - 2 \, q^{3+n} \bigr) \, ,
\end{align}
where $P_{\rm b}(n) = D^{5}_n = \frac{1}{6} (n+1) (n+2) (2 n+3)$, $P_{\rm e}(n) = D^3_n = 2n+1$. According to (\ref{ZEXACT}) this gives a contribution to $\log \Zchar|_{\rm IR}$ equal to
\begin{align}
 &5 \, P_{\rm b}(\hat\delta-3) \, \zeta'(0,3) - 3 \, P_{\rm b}(\hat\delta-4) \, \zeta'(0,4) - 5 \, P_{\rm e}(\hat\delta-2) \, \zeta'(0,2) +  P_{\rm e}(\hat\delta-3) \, \zeta'(0,3) \, ,
\end{align}
where the polynomials are to be expanded in powers of $\hat\delta$, putting $\hat \delta^n \zeta'(0,\Delta) \equiv \zeta'(-n,\Delta)$. Working this out and adding the contribution (\ref{dimGcontrib}), we find
\begin{align}
 \log \Zchar\bigr|_{\rm IR} =    -\log 2 - \frac{47}{3} \, \zeta'(-1) +\frac{2}{3} \, \zeta'(-3) \, .
\end{align}
Combining this with the UV part and reinstating $\ell$, we get\footnote{This splits as $\log Z_{\rm char}=10 \, \log(2\pi) + \log Z_{\rm bulk} - \log Z_{\rm edge}$ where $\log Z_{\rm bulk}=\frac{8 \ell^4}{3 \epsilon ^4}-\frac{8 \ell^2}{3 \epsilon ^2}-\frac{331}{45} \log \frac{2 e^{-\gamma} \ell}{\epsilon}+\frac{475}{48}\\ -\frac{23}{3} \zeta'(-1)+\frac{2}{3} \zeta'(-3)-5 \log (2 \pi )$ and $\log Z_{\rm edge} = \frac{8 \ell^2}{\epsilon ^2}+\frac{16}{3} \log \frac{2 e^{-\gamma} \ell}{\epsilon} -5 + 8 \zeta'(-1)+\log 2+5 \log (2 \pi)$.}
\begin{align} \label{fullZchar}
 \log \Zchar = &\frac{8}{3} \, \frac{\ell^4}{\epsilon^4}-\frac{32}{3} \, \frac{\ell^2}{\epsilon^{2}} - \frac{571}{45}  \log \frac{2 e^{-\game} \LR}{\epsilon} \nn \\
  & - \frac{571}{45}  \log \frac{\ell}{L}
 +\frac{715}{48} -\log 2 - \frac{47}{3} \zeta'(-1) +\frac{2}{3} \zeta'(-3) \, ,
\end{align}
where $\LR$ is an arbitrary length scale introduced to split off a finite part: 
\begin{align} \label{renZchar}
 \log \Zchar^{\rm fin} =  - \frac{571}{45}  \log (\ell/\LR) 
  +\frac{715}{48} -\log 2 - \frac{47}{3} \zeta'(-1) +\frac{2}{3} \zeta'(-3) \, ,
\end{align} 
To compute the group volume factor $Z_G$ in (\ref{ZPIFINALapp}), we use (\ref{vcSO}) for $G=SO(5)$ to get $\vc = \frac{2}{3}(2\pi)^6$, and $\gamma = \sqrt{8 \pi G_{\rm N}/4 \pi \ell^2}$. Finally, $i^{-\pol} = i^{-(d+3)} = -1$. Thus we conclude that the one-loop Euclidean path integral for Einstein gravity on $S^4$  is 
\begin{align}
 Z_{\rm PI} = - \frac{(8 \pi G_{\rm N}/4 \pi \ell^2)^5 \, \Zchar}{\frac{2}{3}(2 \pi)^6} \, ,
\end{align}
where $\Zchar$ is given by (\ref{fullZchar}). 

\subsubsection*{Example: Einstein gravity on $S^5$}

For $S^5$ an analogous (actually simpler) computation gives $Z_{\rm PI} = i^{-7} \ZG \Zchar$ with
\begin{equation} \label{dS5Einst}
\begin{aligned}
 \log \Zchar &= \frac{15 \, \pi }{8} \, \frac{\ell^5}{\epsilon^{5}} -\frac{65 \, \pi}{24} \, \frac{\ell^3}{\epsilon^3} -\frac{105 \, \pi }{16} \, \frac{\ell}{\epsilon} + \frac{65 \, \zeta (3)}{48 \, \pi ^2}+\frac{5 \, \zeta (5)}{16 \, \pi ^4}+15 \log (2 \pi ) \\
 \log \ZG &= \frac{15}{2} \log \frac{8 \pi G_{\rm N}}{2\pi^2 \ell^3} 
 - \log \frac{(2\pi)^9}{12}  \, .
\end{aligned}
\end{equation}

\subsection{Different regularization schemes} \label{sec:otherreg}

If we simply cut off the character integral at $t = \epsilon$, we get the following instead of (\ref{ZEXACT}):
\begin{align} \label{ZEXACTcuteps}
 \log Z_{\epsilon} = &\frac{1}{2}\sum_\Delta P_\Delta(\hat\delta-\Delta) \, \zeta'(0,\Delta) 
 + b_{d+1}(\nu) \log (e^{-\game}/ \epsilon)  
    +  \sum_{k=0}^{d} \frac{b_k(\nu)}{d+1-k}  \, \epsilon^{-(d+1-k)}  \, ,
\end{align}
with $b_k(\nu)$ defined as before, $\frac{1}{2t} F_\nu(t) =  \sum_{k=0}^{d+1} b_k(\nu) \, t^{-(d+2-k)} \, + \CO(t^0)$. 
Unsurprisingly, this differs from (\ref{ZEXACT}) only in its UV part, more specifically in the terms polynomial in $\nu$, including the  ``multiplicative anomaly'' term discussed below (\ref{HKCFORM}). 
The transcendental ($\zeta'$) part and the $\log \epsilon$ coefficient remain unchanged. This remains true in any other regularization.  

If we stick with heat-kernel regularization but pick a different regulator $f(\tau/\epsilon^2)$ instead of $e^{-\epsilon^2/4\tau}$ (e.g.\ the $f=(1-e^{-\tau\Lambda^2})^k$ PV regularization of section \ref{sec:thermal}) 
or use zeta function regularization, more is true: the same finite part is obtained for any choice of $f$ provided logarithmically divergent terms (arising in even $d+1$) are expressed in terms of $M$ defined as in (\ref{HKHKHKH}) with $e^{-\epsilon^2/4\tau} \to f$. The relation $M(\epsilon)$ will depend on $f$, but nothing else. 

In dimensional regularization, some polynomial terms in $\nu$ will be different, including the ``multiplicative anomaly'' term. Of course no physical quantity will be affected by this, as long as self-consistency is maintained.     
In fact any regularization scheme (even (\ref{ZEXACTcuteps})) will lead to the same physically unambiguous part of the one-loop corrected dS entropy/sphere partition function of section \ref{sec:gravETD}. However to go beyond this, e.g.\ to extract more physically unambiguous data by comparing different saddles along the lines of (\ref{saddledif}) and (\ref{lincombinv}), a portable covariant regularization scheme, like heat-kernel regularization, must be applied consistently to each saddle. A sphere-specific ad-hoc regularization as in (\ref{ZEXACTcuteps}) is not suitable for such purposes.   


\section{Some useful dimensions, volumes and metrics}

\subsection{Dimensions of representations of $SO(K)$} \label{app:Weyldim}

General irreducible representations of $SO(K)$ with $K=2r$ or $K=2r+1$ are labeled by $r$-row Young diagrams or more precisely a set $S = (s_1,\ldots,s_r)$ of highest weights ordered from large to small, which are either all integer (bosons) or all half-integer (fermions). When $K=2r$, $s_r$ can be either positive of negative, distinguishing the chirality of the representation. For various applications in this paper we need the dimensions $D^{K}_S$ of these $SO(K)$ representations $S$.
The Weyl dimension formula gives a general expression for the dimensions of irreducible representations of simple Lie groups. For the $SO(K)$ this is

\noindent $\bullet$ $K=2r$:
\begin{align} \label{Weyldimeven}
 D^{K}_{S} = \CN_K^{-1} \prod_{1 \leq i < j \leq r} \bigl(\ell_i + \ell_j \bigr) \bigl( \ell_i - \ell_j \bigr)  \, , \qquad \ell_i \equiv s_i + \tfrac{K}{2} - i 
\end{align}
with $\CN_K$ independent of $S$, hence fixed by $D^K_0 = 1$, i.e.\ 
$\CN_K = \prod_{1 \leq i < j \leq r} (K - i - j ) ( j- i)$.

\noindent $\bullet$ $K=2r+1$:
\begin{align} \label{Weyldimodd}
 D^{K}_{S} = \CN_K^{-1} \prod_{1 \leq i \leq r} (2 \ell_i) \prod_{1 \leq i < j \leq r} \bigl(\ell_i + \ell_j \bigr) \bigl( \ell_i - \ell_j \bigr) \, ,
 \qquad \ell_i \equiv s_i + \tfrac{K}{2} - i  \, ,
\end{align}
where $\CN_K$ is fixed as above: $\CN_K = \prod_{1 \leq i \leq r} (K-2i) \prod_{1 \leq i < j \leq r} (K - i - j ) ( j- i)$.

\noindent For convenience we list here some low-dimensional explicit expressions:
\begin{equation} \label{SOKlowdim}  
\begin{array}{l|l|l|l}
 K & D_s^K & D_{n,s}^K & D^K_{k+\frac{1}{2},\bfhalf}  \\
 \hline
 2 & 1 & 
   & 1 \\
 3 & 2 s+1 & 
   & 2 {k+1 \choose 1} \\
 4 & (s+1)^2 & \left(n-s+1\right) \left(n+s+1\right) 
   & 2 {k +2 \choose 2} \\
 5 & \frac{(s+1) (s+2) (2 s+3)}{6} & \frac{\left(2 n+3\right)
   \left(n-s+1\right) \left(n+s+2\right) \left(2 s+1\right)}{6} 
   & 4 {k +3 \choose 3} \\
 6 & \frac{(s+1) (s+2)^2 (s+3)}{12} & \frac{ \left(n+2\right){}^2
   \left(n-s+1\right) \left(n+s+3\right) \left(s+1\right){}^2 }{12}
   & 4 {k + 4 \choose 4}  \\
 7 & \frac{ (s+1) (s+2) (s+3) (s+4) (2 s+5)}{120} & \frac{ \left(n+2\right)
   \left(n+3\right) \left(2 n+5\right) \left(n-s+1\right) \left(n+s+4\right) \left(s+1\right)
   \left(s+2\right)  \left(2 s+3\right)}{720}
    & 8 {k + 5 \choose 5} \\
 8 & \frac{ (s+1) (s+2) (s+3)^2 (s+4) (s+5)}{360} & \frac{\left(n+2\right)
   \left(n+3\right){}^2 \left(n+4\right) \left(n-s+1\right) \left(n+s+5\right) \left(s+1\right)
   \left(s+2\right){}^2 \left(s+3\right) }{4320} 
   & 8 {k + 6 \choose 6} \\
\end{array}
\end{equation}
Here $(k+\frac{1}{2},\bfhalf)$ means $(s_1,\ldots,s_r)=(k+\frac{1}{2},\frac{1}{2}, \ldots, \frac{1}{2})$, i.e.\ the spin $s=k+\frac{1}{2}$ representation. 

\vskip1mm \noindent For general $d \geq 3$, we can use (\ref{Dsods2}) and (\ref{coin}) to compute 
\begin{align} \label{coinapp}
  D^{K}_{s}=\mbox{\large$\binom{s+K-1}{K-1}-\binom{s+K-3}{K-1}$} \, , \qquad D^{K}_{n, s}=D^{K}_{n} D^{K-2}_{s}-D^{K}_{s-1}D^{K-2}_{n+1} \, .
\end{align}
Denoting 1 repeated $m$ times by $1^m$, e.g.\ $(5,1^2) = (5,1,1) = {\tiny \Yvcentermath1 \yng(5,1,1)}$, we furthermore have   
\begin{align} \label{hookmagic}
  D_{1^p}^d = \mbox{\large$\binom{d}{p}$} \quad (p<\tfrac{d}{2}), \qquad D^{2p}_{1^{p-1},\pm 1} = \mbox{\large $\tfrac{1}{2}  \binom{2p}{p}$} \, , \qquad D^{d+2}_{n, s,1^m}=D^{d+2}_{n}D^{d}_{s,1^m}-D^{d+2}_{s-1}D^{d}_{n+1,1^m} \, .
\end{align} 

\subsection{Volumes} \label{sec:volumes}

The volume of the unit sphere $S^n$ is
\begin{align} \label{volSn}
 \Omega_n \equiv {\rm vol}(S^n) = \frac{2 \, \pi^{\frac{n+1}{2}}}{\Gamma\bigl(\frac{n+1}{2}\bigr)} \, = \, \frac{2\pi}{n-1} \cdot \Omega_{n-2}
\end{align}
The volume of $SO(d+2)$ with respect to the invariant group metric normalized such that minimal $SO(2)$ orbits have length $2\pi$ is
\begin{align} \label{vcSO}
 {\rm \Vol}\bigl(SO(d+2)\bigr)_{\can}  = \prod_{k=2}^{d+2} {\rm \Vol}(S^{k-1}) = \prod_{k=2}^{d+2} 
 \frac{2 \pi^{\frac{k}{2}}}{\Gamma(\frac{k}{2})} \, .
\end{align} 
This follows from the fact that the unit sphere $S^{n-1} = SO(n)/SO(n-1)$, which implies ${\rm \Vol}(SO(n))_{\can} = {\rm \Vol}(S^{n-1}) \, {\rm \Vol}(SO(n-1))_{\can}$ in the assumed normalization. 

The volume of $SU(N)$ with respect to the invariant metric derived from the matrix trace norm on the Lie algebra $\su(N)$ viewed as traceless $N \times N$ matrices is (see e.g.\ \cite{Ooguri:2002gx})
\begin{align} \label{volsuN}
 {\rm vol}\bigl(SU(N)\bigr)_{{\rm Tr}_N} \, = \, \sqrt{N} \prod_{k=2}^N \frac{(2\pi)^k}{\Gamma(k)} \, = \, \sqrt{N} \, \frac{(2 \pi )^{\frac{1}{2} (N-1) (N+2)}}{{\tt G}(N+1)} \, . 
\end{align}

\subsection{de Sitter and its Wick rotations to the sphere} \label{app:dSWick}

\begin{figure}
 \begin{center}
   \includegraphics[height=4.7cm]{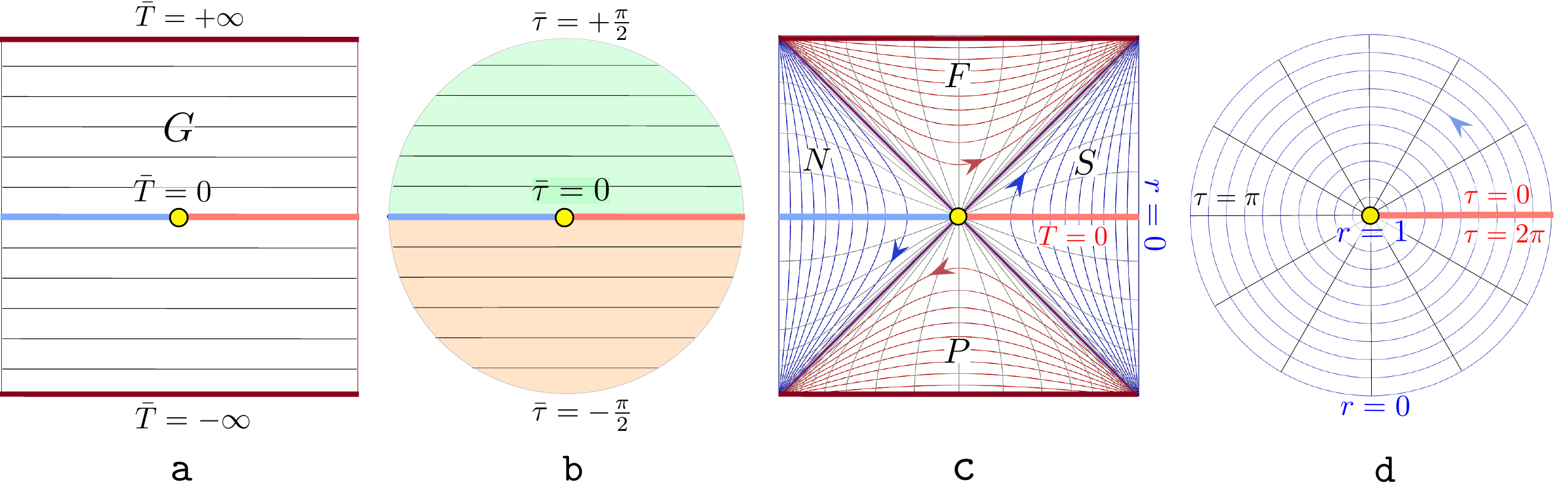}
 \caption{ \small \label{fig:penrose-app} Penrose diagrams of dS$_{d+1}$ and $S^{d+1}$ with coordinates \ref{dScoords}, \ref{Scoords}. Each point corresponds to an $S^{d-1}$, contracted to zero size at thin-line boundaries.    {\tt a:} Global dS$_{d+1}$ in slices of constant $\bar T$. {\tt b:} Wick rotation  of global dS$_{d+1}$ to $S^{d+1}$.  {\tt c:} S/N = southern/northern static patch, F/P = future/past wedge; slices of constant $T$ (gray) and $r$ (blue/red) = flows generated by $H$.  Yellow dot = horizon $r=1$. {\tt d:} Wick-rotation of static patch $S$ to $S^{d+1}$; slices of constant $\tau$ and constant $r$. 
  } 
 \end{center} \vskip-4mm
\end{figure}

Global dS$_{d+1}$ has a convenient description as a hyperboloid embedded in $\IR^{1,d+1}$,
\begin{align} \label{embeddingspacedS}
 X^I X_I \equiv \eta_{IJ} X^I X^J \equiv -X_0^2 + X_1^2 + \cdots + X_{d+1}^2  = \ell^2  \, , \qquad ds^2 = \eta_{IJ} dX^I dX^J   \, . 
\end{align}
Below we set $\boxed{\ell \equiv 1}$. The isometry group is $SO(1,d+1)$, with generators $M_{IJ}=X_I \partial_J-X_J \partial_I$. Various coordinate patches are shown in fig.\ \ref{fig:penrose-app}{\tt a,c}, with coordinates and metric given by
\begin{equation} \label{dScoords} \small
\begin{array}{|l|l|l|l|}
\hline
  \text{co} & \text{embedding } (X^0,\ldots,X^{d+1})   & \text{coordinate range} & \text{metric } ds^2 = \eta_{IJ} dX^I dX^J  \\
\hline 
 G &  (\sinh \bar T,\cosh \bar T \, \gom) &  \bar T \in \IR, \, \bar\Omega \in S^{d} & -d \bar T^2 + \cosh^2 \bar T \, d \gom^2  \\
 S & (\sqrt{1-r^2} \sinh T, r  \Omega,\sqrt{1-r^2} \cosh T)  & T \in \IR, 0 \leq r < 1, \, \Omega \in S^{d-1} &  -(1-r^2) dT^2 + \frac{dr^2}{1-r^2} + r^2 d\Omega^2 \\
 F & (\sqrt{r^2-1} \cosh T, r  \Omega, \sqrt{r^2-1} \sinh T) &  T \in \IR, \, r>1 , \, \Omega \in S^{d-1} & - \frac{dr^2}{r^2-1} + (r^2-1) dT^2 + r^2 d\Omega^2 \\
 \hline
\end{array}
\end{equation} 
illustrated in fig.\ \ref{fig:penrose-app}{\tt a,c}.
$N$ is obtained from $S$ by $X^{d+1} \to -X^{d+1}$, and $P$ from $F$ by $X^0 \to -X^0$. The southern static patch $S$ is the part of de Sitter causally accessible to an inertial observer at the south pole of the global spatial $S^d$. The metric in this patch is static, with the observer at $r=0$ and a horizon at $r=1$. The $SO(1,1)$ generator $H=M_{0,d+1}$ acts by translation of the coordinate $T$, which is timelike in $S,N$ and spacelike in $F,P$. From the direction of the flow lines in fig.\ \ref{fig:penrose-app}{\tt c}, it can be seen that the positive energy operator  is $H$ in $S$, whereas it is $-H$ in $N$. In $F/P$, $r$ is the time coordinate, and $H$ is the operator corresponding to spatial momentum along the $T$-axis of the $\IR \times S^{d-1}$ spatial slices. 

A Wick rotation $X^0 \to -i X^0$ maps (\ref{embeddingspacedS}) to the round sphere $S^{d+1}$:
\begin{align} \label{embeddingspaceSph}
 \delta_{IJ} X^I X^J = \ell^2 \, , \qquad ds^2 = \delta_{IJ} dX^I dX^J   \, . 
\end{align}
The full $S^{d+1}$ can be obtained either from global dS $G$ by Wick rotating global time $\bar T \to - i \bar \tau$, or from a single static patch $S$ by Wick rotating static time $T \to - i \tau$, as illustrated in fig.\ \ref{fig:penrose-app}{\tt b,d}. The corresponding sphere coordinates and metric are, again setting $\boxed{\ell \equiv 1}$ 
\begin{equation} \label{Scoords} \small
\begin{array}{|l|l|l|l|}
\hline
 \text{co}  &\text{embedding } (X^0,X^1,\ldots,X^{d+1}) & \text{coordinate range} & \text{metric } ds^2 = \delta_{IJ} dX^I dX^J \\
\hline 
 G &   (\sin \bar\tau,\cos\bar\tau 
 , \gom) &   -\frac{\pi}{2} \leq \bar\tau \leq \frac{\pi}{2} \, , \bar\Omega \in S^{d} & d\bar\tau^2 + \cos^2\bar\tau \, d \gom^2  \\
 S & (\sqrt{1-r^2} \sin \tau, r  \Omega,\sqrt{1-r^2} \cos \tau)  & 0 \leq r < 1, \tau \simeq \tau+2\pi , \, \Omega \in S^{d-1} &  (1-r^2) d\tau^2 + \frac{dr^2}{1-r^2} + r^2 d\Omega^2 \\
 \hline
\end{array}
\end{equation}

\section{
Euclidean vs canonical: formal \& physics expectations} \label{app:thermalZPI}

Given a QFT on a static spacetime $\IR \times M$ with  metric $ds^2 = -dt^2 + ds^2_M$, Wick rotating $t \to - i \tau$ yields a Euclidean QFT on a space with metric $ds^2 = d\tau^2 + ds^2_M$. 
The Euclidean path integral $Z_{\rm PI}(\beta)=\int \CD \Phi \, e^{-S[\Phi]}$ on $S^1 _\beta \times M$ obtained by identifying $\tau \simeq \tau+\beta$ equals the thermal partition function: $Z_{\rm PI}(\beta) = {\rm Tr} \, e^{-\beta H}$, as follows from cutting the path integral along constant-$\tau$ slices and viewing $e^{-\tau H}$ as the Euclidean time evolution operator. 

At least for noninteracting theories, it is in practice much more straightforward to compute the partition function as the state sum ${\rm Tr} \, e^{-\beta H}$ of an ideal gas in a box $M$ than as a one-loop path integral $Z_{\rm PI}=\int \CD \Phi \, e^{-S[\Phi]}$ on $S^1_\beta \times M$, in particular for higher-spin fields. 
In view of this, it is reasonable to wonder if a free QFT path integral on the sphere could perhaps similarly be computed as a simple state sum, by viewing the sphere as the Wick-rotated static patch (fig.\ \ref{fig:penrose-app}{\tt d}), with inverse temperature $\beta=2\pi$ given by the period of the angular coordinate $\tau$:
\begin{align} \label{ZPITreqq}
 Z_{\rm PI} \overset{?}{=} {\rm Tr}_S \, e^{-2 \pi H} \, .
\end{align} 
Below we review the formal path integral slicing argument suggesting the above relation and why it fails, emphasizing the culprit is the presence of a fixed-point locus of $H$, the yellow dot in fig.\ \ref{fig:penrose-app}. 
At the same formal level, we show the above relation is equivalent to $Z_{\rm PI} \overset{?}{=} Z_{\rm bulk}$, with $Z_{\rm bulk}$ defined as a character integral as in section \ref{sec:thermal}. This improves the situation, but is still incorrect for spin $s \geq 1$. 
In more detail, the content  is as follows:

In  \ref{sec:S1} we consider the $d=0$ case: a scalar of mass $\omega$ on dS$_1$ in its Euclidean vacuum state, i.e.\ an entangled pair of harmonic oscillators. 
Though surely superfluous to most readers, we use the occasion to provide a  pedagogical introduction to some standard constructions. 
 
In  \ref{sec:formform} we formally apply the same template to general $d$, ignoring yellow-dot issues, leading to the standard formal ``thermofield double'' description of the static patch of de Sitter \cite{Israel:1976ur}, and more specifically to $Z_{\rm PI} \simeq {\rm Tr} \, e^{-2 \pi H} \simeq Z_{\rm bulk}$. We review the pathological divergences that ensue when one attempts to evaluate the trace, and some of its proposed fixes such as the ``brick-wall'' cutoff \cite{tHooft:1984kcu} and refinements thereof. We contrast these to $Z_{\rm bulk}$ defined as a character integral.

In  \ref{sec:edgecor}, we turn to the edge corrections missed by such formal arguments, explaining from various points of view why they are to be expected.

\def\ham{H'}

\subsection{$S^1$} \label{sec:S1}

\begin{figure}
 \begin{center}
   \includegraphics[height=3.7cm]{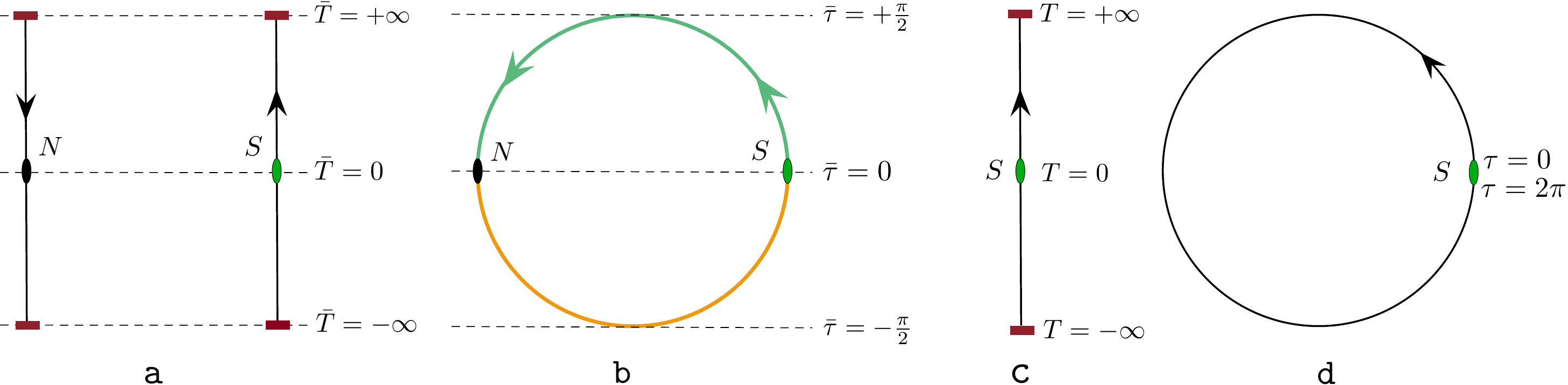}
 \caption{ \small \label{fig:dS1} dS$_1$ version of fig.\ \ref{fig:penrose-app} (in {\tt c} we only show $S$ here). Wick rotation of global time $\bar T \to -i \bar\tau$ maps ${\tt a} \to {\tt b}$ while wick rotation of static patch time $T \to - i \tau$ maps ${\tt c} \to {\tt d}$. Coordinates are as defined in tables \ref{dScoords} and \ref{Scoords} with $d=0$.}
 \end{center} \vskip-4mm
\end{figure}

Though slightly silly, it is instructive to first consider the $d=0$ case: a free scalar field of mass $\omega$ on dS$_1$ (fig.\ \ref{fig:dS1}). Global dS$_1$ is the hyperbola $X_0^2 - X_1^2 = 1$ according to (\ref{embeddingspacedS}), which consists of two causally disconnected lines,  globally parametrized according to table \ref{dScoords} by $(\bar T,\bar\Omega)$ where $\bar\Omega \in S^0 = \{-1,+1\} \equiv \{N,S\}$. The pictures of fig.\ \ref{fig:penrose-app} still apply, except there are no interior points, resulting in fig.\ \ref{fig:dS1}. Putting a free scalar of mass $\omega$ on this space just means we consider two harmonic oscillators $\phi_S$ and $\phi_N$, with action
\begin{align} \label{SLHO}
 S_L = \frac{1}{2} \int_{-\infty}^\infty d\bar T \bigl( \dot \phi_S^2 - \omega^2 \phi_S^2 + \dot \phi_N^2 - \omega^2 \phi_N^2  \bigr) \, .
\end{align}
The dS$_1$ isometry group is $SO(1,d+1)=SO(1,1)$, generated by $H \equiv M_{01}$, which acts as forward/backward time translations on $\phi_S$/$\phi_N$, to be contrasted with the global {\it Hamiltonian} $\ham$, which acts as forward time translations on both.  
The southern and northern static patch are parametrized by $T$, and each contains one harmonic oscillator, respectively $\phi_S$ and $\phi_N$. 
Introducing creation and annihilation operators $a_\omega^S,a_\omega^{S\dagger},a_{-\omega}^N,a_{-\omega}^{N\dagger}$ satisfying $[a,a^\dagger]=1$, we have
\begin{align} \label{HHSHN}
 H = H_S  -  H_N \, , \quad \ham = H_S + H_N \, , \qquad H_S = \omega \bigl( a_\omega^{S\dagger} a^S_\omega + \tfrac{1}{2} \bigr) \, , \quad
 H_N = \omega \bigl( a_{-\omega}^{N\dagger} a_{-\omega}^N + \tfrac{1}{2} \bigr) \, .
\end{align}
The subscript $\pm \omega$ refers to the $H$ eigenvalue: $[H,a^\dagger_{\pm\omega}] = \pm \omega \, a^\dagger_{\pm\omega}$, $[H,a_{\pm\omega}] = \mp \omega \, a_{\pm\omega}$.  The southern and northern Hilbert spaces $\CH_S$, $\CH_N$ each have a positive energy eigenbasis $|n)$ with energies $E_n = (n+\frac{1}{2}) \, \omega$. In QFT language, $|0)$ is the static patch ``vacuum'', and each patch has one ``single-particle'' state, $|1)=a^\dagger|0)$.  
The global Hilbert space is $\CH_G = \CH_S \otimes \CH_N$, with basis $|n_S,n_N\rangle=|n_S) \otimes |n_N)$ satisfying $H |n_S,n_N\rangle = \omega(n_S-n_N) |n_S,n_N\rangle$. 

\begin{figure}
 \begin{center}
   \includegraphics[height=2cm]{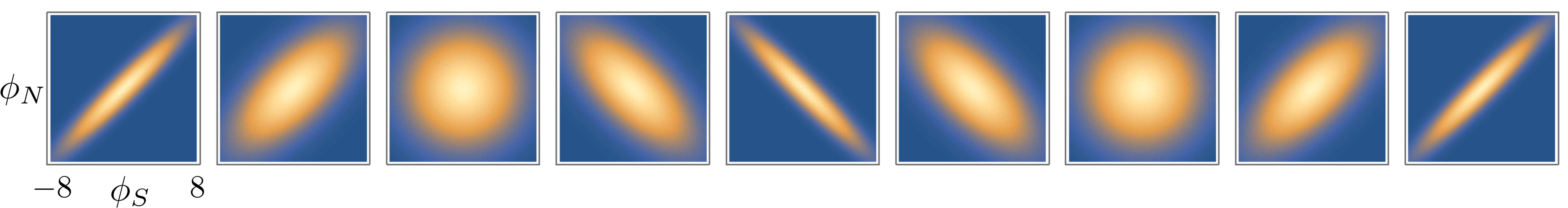}
 \caption{ \small \label{fig:HHosc} Global time evolution of  $P_{\bar T}(\phi_S,\phi_N)=\bigl|\langle \phi_S,\phi_N|e^{- i \ham \bar T}|O\rangle \bigr|^2$ for free $\omega=0.1$ scalar on dS$_1$, from $\bar T=0$ to $\bar T = \pi/\omega$. $P(\phi_S)=\int d\phi_N \, P_{\bar T}(\phi_S,\phi_N)$ is thermal and time-independent.
  }
 \end{center}
\end{figure}

Wick-rotating dS$_1$ produces an $S^1$ of radius $\ell=1$. If we consider this as the Wick rotation of the static patch as in fig.\ \ref{fig:penrose-app}{\tt d}/\ref{fig:dS1}{\tt d}, $S$ in table (\ref{Scoords}), the $S^1$ is parametrized by the periodic Euclidean time coordinate $\tau \simeq \tau + 2\pi$. The corresponding Euclidean action for the scalar is
\begin{align} \label{SES}
 S_E = \frac{1}{2} \int_0^{2\pi} d\tau \bigl( \dot \phi^2 + \omega^2 \phi^2 \bigr) \, \qquad \phi(2\pi)=\phi(0)  \, .
\end{align}  
The Euclidean path integral $Z_{\rm PI}$ on $S^1$ is most easily computed by reverting to the canonical formalism with $e^{-\tau H_S}=e^{-\tau H}$ as the Euclidean time evolution operator,  which maps it to the harmonic oscillator thermal partition function at inverse temperature $\beta=2\pi$:
\begin{align} \label{ZPIexpl}
 Z_{\rm PI} = \int \CD \phi \, e^{-S_E[\phi]} = {\rm Tr}_{\CH_S} \, e^{-2 \pi H} = \sum_n e^{-2 \pi \omega(n+\frac{1}{2})} = \frac{e^{-2\pi \omega/2}}{1-e^{-2\pi \omega}} \, .
\end{align}
We can alternatively consider the $S^1$ to be obtained as the Wick rotation of {\it global} dS$_1$ as in fig.\ \ref{fig:penrose-app}{\tt b}/\ref{fig:dS1}{\tt b}, $G$ in (\ref{Scoords}), parametrizing the $S^1$ by $(\bar\tau,\bar\Omega)$, $-\frac{\pi}{2} \leq \bar\tau \leq \frac{\pi}{2}$, $\bar\Omega \in S^0 = \{S,N\}$, identifying  $(\pm \frac{\pi}{2},S)=(\pm \frac{\pi}{2},N)$. The global action (\ref{SLHO}) then Wick rotates to  
\begin{align}
 S_E = \frac{1}{2} \int_{-\pi/2}^{\pi/2} d\bar\tau \bigl(\dot \phi_S^2 + \omega^2 \phi_S^2 + \dot \phi_N^2 + \omega^2 \phi_N^2 \bigr) \, , \qquad \phi_S(\pm\tfrac{\pi}{2}) = \phi_N(\pm\tfrac{\pi}{2}) \, , 
\end{align}
which is identical to (\ref{SES}), just written in a slightly more awkward form.  This  form naturally leads to an interpretation of $Z_{\rm PI}$ as computing the norm squared of the Euclidean vacuum state $|O\rangle$ of the scalar on the {\it global} dS$_1$ Hilbert space $\CH_G$, by cutting the path integral at the $S^0=\{N,S\}$ equator $\bar\tau=0$ of the $S^1$ (cf.\ fig.\ \ref{fig:dS1}{\tt b}):
\begin{align} \label{ZPIint1dS1}
 Z_{\rm PI} = \int d^2 \phi_{0} \, \langle O|\phi_0 \rangle \langle \phi_0 |O \rangle  \equiv \langle O|O\rangle \, , \qquad  \, \langle \phi_0 |O \rangle  \equiv \int_{\bar\tau \leq 0} \CD \phi|_{\phi_0} \, e^{-S_E[\phi]} \, ,
\end{align}
where $\phi_0=(\phi_{S,0},\phi_{N,0})$. The notation $\int_{\bar\tau \leq 0} \CD \phi|_{\phi_0}$ means the path integral of $\phi=(\phi_S,\phi_N)$ is performed on the lower hemicircle $\bar\tau \leq 0$ (orange part in fig.\ \ref{fig:dS1}{\tt b}), with boundary conditions  $\phi|_{\bar\tau = 0} = \phi_0$.
$\langle O|\phi_0 \rangle$ is similarly defined as a path integral on the upper hemicircle (green part). It is not too difficult to explicitly compute $|O\rangle$ in the $|\phi_{S,0},\phi_{N,0} \rangle$ basis, but it is easier to compute it in the oscillator basis $|n_S,n_N\rangle$, noticing that slicing the path integral defining $|O\rangle$    allows us to write it as $\langle n_S,n_N|O\rangle = ( n_S|e^{-\pi H}|n_N) = e^{-\pi \omega(n_S+\frac{1}{2})} \, \delta_{n_S,n_N}$. Thus
\begin{align} \label{Oeuclexpl}
  |O\rangle = \sum_n e^{-\pi \omega(n+\frac{1}{2})} |n,n\rangle  = e^{-\pi \omega/2} \exp\bigl(e^{-\pi \omega} a^{S\dagger}_\omega a_{-\omega}^{N\dagger} \bigr) |0,0\rangle \, . 
\end{align}
In the Schr\"odinger picture, $|O\rangle$ is to be thought of as an initial state at $\bar T=0$ for global dS$_1$: pictorially, we are gluing the bottom half of fig.\ \ref{fig:dS1}{\tt b} to the top half of fig.\ \ref{fig:dS1}{\tt a}. This state evolves nontrivially in global time $\bar T$: though invariant under $SO(1,1)$ generated by $H=H_S - H_N$, it is {\it not} invariant under forward global time translations generated by the global Hamiltonian $\ham = H_S + H_N$. For viewing pleasure this is illustrated in fig.\ \ref{fig:HHosc}, which also visually exhibits the north-south entangled nature of $|O\rangle$.

Note that $Z_{\rm PI} = \langle O|O\rangle = \sum_n e^{-2 \pi \omega (n+\frac{1}{2})}$, reproducing the dS$_1$ static patch thermal partition function (\ref{ZPIint1}). Indeed from the point of view of the static patch, the global Euclidean vacuum state looks thermal with inverse temperature $\beta=2\pi$: the 
southern reduced density matrix $\hat\varrho_S$ obtained by tracing out the northern degree of freedom $\phi_N$ in the global Euclidean vacuum $|O\rangle$ is $\hat\varrho_S = \sum_n e^{-2 \pi \omega(n+\frac{1}{2})} |n) ( n| = e^{-2 \pi H_S}$. In contrast to the global $|O\rangle$, the reduced density matrix is time-{\it independent}.  

The path integral slicing arguments we used did not rely on the precise form of the action. In particular the conclusions remain valid when we add interactions:
\begin{align} \label{thermalitydS1} 
  |O\rangle = \sum_n e^{-\beta E_n/2} |n,n\rangle  \, , 
  \qquad \hat\varrho_S = e^{-\beta H_S} \, , \qquad Z_{\rm PI} = \langle O|O\rangle = {\rm Tr}_S \, e^{- \beta H} \, \qquad (\beta=2\pi) 
\end{align}
Actually in the $d=0$ case at hand, we can generalize all of the above to arbitrary  values of $\beta$. (For $d>0$ this would create a conical singularity at $r=1$ on $S^{d+1}$, but for $S^1$ the point $r=1$ does not exist.) 
Note that since the reduced density matrix is thermal, the north-south entanglement entropy in the Euclidean vacuum $|O\rangle$ equals the thermal entropy: $S_{\rm ent} = - {\rm tr}_S \, \varrho_S \log \varrho_S = S_{\rm th} = (1-\beta \partial_\beta) \log Z$, where $\varrho_S \equiv \hat\varrho_S/Z$, $Z={\rm Tr}_S \, \hat\varrho_S$.

Despite appearing distinctly non-vacuous from the point of view of a local observer, and being globally time-dependent,  the state $|O\rangle$ does deserve its ``vacuum'' epithet. As already mentioned, it is invariant under the global $SO(1,1)$ isometry group: $H|O\rangle=0$. Moreover, for the free scalar, (\ref{Oeuclexpl}) implies  $|O\rangle$ is itself annihilated by a pair of {\it global} annihilation operators $a^G$ related related to $a^S$, $a^{S\dagger}$, $a^N$ and $a^{N\dagger}$ by a Bogoliubov transformation:
\begin{align} \label{bogoliubov}
 a^G_{\pm \omega} |O\rangle = 0 \, , \qquad 
 a^G_\omega \equiv 
 \frac{a^S_\omega - e^{-\pi \omega} a^{N\dagger}_{-\omega}}{\sqrt{1-e^{-2 \pi \omega}}} \, , \quad
 a^G_{-\omega} \equiv 
 \frac{a^N_{-\omega} - e^{-\pi \omega} a^{S\dagger}_\omega}{\sqrt{1-e^{-2 \pi \omega}}} \, ,
\end{align}
normalized such that $[a_{\pm\omega}^G,a_{\pm\omega}^{G\dagger}]= \delta_{\pm,\pm}$.  From (\ref{HHSHN}) we get $H=\omega \, a_\omega^{G\dagger}a_\omega^{G} - \omega \, a_{-\omega}^{G\dagger}a_{-\omega}^{G}$. Thus we can construct the global Hilbert space $\CH_G$ as a Fock space built on the Fock vacuum $|O\rangle$,  by acting with the global creation operators $a^{G\dagger}_{\pm \omega}$. The Hilbert space $\CH_G^{(1)}$ of ``single-particle'' excitations of the global Euclidean vacuum is two-dimensional, spanned by
\begin{align} \label{globSP}
 |\!\pm\! \omega\rangle \equiv a_{\pm \omega}^{G\dagger} |O\rangle, \qquad H|\!\pm \!\omega\rangle = \pm \omega\, |\!\pm\!\omega\rangle \, .
\end{align} 
The character $\chi(t)$ of the $SO(1,1)$ representation furnished by $\CH_G^{(1)}$ is 
\begin{align} \label{chitSO11}
 \chi(t) \equiv {\rm tr}_G \, e^{-i t H}  = e^{-i t \omega} + e^{i t \omega} \, .
\end{align}
The above constructions are straightforwardly generalized to fermionic oscillators. The character of a collection of bosonic and fermionic oscillators of frequencies $\omega_i$ and $\omega'_j$ is
\begin{align} \label{chigenbf}
 \chi(t) = {\rm tr}_G \, e^{-i H t} = \chi(t)_{\rm bos} + \chi(t)_{\rm fer} = \sum_{i,\pm} e^{ \pm i \omega_i} + 
 \sum_{j,\pm} e^{ \pm i \omega'_j} \, .
\end{align}

\noindent {\bf Character formula} 

\noindent For a single bosonic resp.\ fermionic oscillator of frequency $\omega$, $\log {\rm Tr} \, e^{-\beta H}$ has the following integral representation:\footnote{\label{iepspres} The $t^{-2}$ pole of the integrand is resolved by the $i\epsilon$-prescription $t^{-2} \to \frac{1}{2} \left((t-i\epsilon)^{-2} + (t+i\epsilon)^{-2}\right)$, left implicit here and in the formulae below. The integral formula can be checked by observing the integrand is even in $t$, extending the integration contour to the real line, closing the contour, and summing residues.}
\begin{equation} \label{integralformula} 
  \begin{aligned}
  \log\left( {e^{-\beta \omega/2}}{\bigl(1-e^{-\beta \omega}\bigr)^{-1}} \right) &=+ \int_0^\infty \frac{dt}{2t} \,  \frac{1+e^{-2 \pi t/\beta}}{1-e^{-2\pi t/\beta}} \, \bigl( e^{-i\omega t} + e^{i \omega t} \bigr) \\
  \log \Bigl(e^{+\beta \omega/2}\bigl(1 + e^{-\beta \omega} \bigr) \Bigr) &= -\int_0^\infty \frac{dt}{2t} \, \frac{2 \, e^{-\pi t/\beta}}{1-e^{-2 \pi t/\beta}} \, \bigl( e^{-i\omega t} + e^{i \omega t} \bigr) \, .
 \end{aligned}
\end{equation}
Combining this with (\ref{chigenbf}) expresses the thermal partition function of a collection of bosonic and harmonic oscillators as an integral transform of its $SO(1,1)$ character:
\begin{align} \label{generalbetacharS1}
  \log {\rm Tr} \, e^{-\beta H} = \int_{0}^\infty \frac{dt}{2t} \biggl( \frac{1+e^{-2 \pi t/\beta}}{1-e^{-2 \pi t/\beta}}   \, \, \chi(t)_{\rm bos}  \, -\, \frac{2 \,  e^{-\pi t/\beta}}{1-e^{-2 \pi t/\beta}}  \, \, \chi(t)_{\rm fer} \biggr)  \, .
\end{align}
The Euclidean path integral on an $S^1$ of radius $\ell=1$ for a collection of free bosons and fermions (the latter with thermal, i.e.\ antiperiodic, boundary conditions) is then given by putting $\beta=2\pi$ in the above:
\begin{align} \label{lalalili}
 \log Z_{\rm PI} = \int_{0}^\infty \frac{dt}{2t} \biggl( \frac{1+e^{- t}}{1-e^{- t}}   \, \, \chi(t)_{\rm bos}  \, -\, \frac{2 \,  e^{-t/2}}{1-e^{-t}}  \, \, \chi(t)_{\rm fer} \biggr) \, .
\end{align}

\subsection{$S^{d+1}$} \label{sec:formform}

The arguments in this section will be formal, following the template of section \ref{sec:S1} while glossing over some important subtleties, the consequence of which we discuss in section \ref{sec:edgecor}.

Wick-rotating a QFT on dS$_{d+1}$ to $S^{d+1}$, we get the Euclidean path integral 
\begin{align}
 Z_{\rm PI} = \int \CD \Phi \, e^{-S_E[\Phi]} \, ,
\end{align}
where $\Phi$ collects all fields in the theory.
Just like in the $d=0$ case, the two different paths from dS$_{d+1}$ to $S^{d+1}$, i.e.\ Wick-rotating global time $\bar T$ or static patch time $T$ (cf.\ fig.\ \ref{fig:penrose-app} and table \ref{Scoords}), naturally give rise to two different  dS Hilbert space interpretations: one involving the global Hilbert space $\CH_G$ and one involving the static patch Hilbert space $\CH_S$. 

The global Wick rotation of fig.\ \ref{fig:penrose-app}{\tt b} leads to an interpretation of $Z_{\rm PI}$ as computing $\langle O|O\rangle$, analogous to (\ref{ZPIint1dS1}), by cutting the path integral on the globlal $S^d$ equator $\bar\tau=0$:
\begin{align} \label{ZPIint1}
 Z_{\rm PI} = \int_{\bar \tau=0} d \Phi_0 \, \langle O|\Phi_0 \rangle \langle \Phi_0 |O \rangle  \equiv \langle O|O\rangle \, , \qquad \langle \Phi_0|O\rangle \equiv \int_{\bar\tau \leq 0} \CD \Phi|_{\Phi_0} \, e^{-S_E[\Phi]} \, ,
\end{align}
where $\int_{\bar\tau \leq 0} \CD \phi|_{\phi_0}$ means the path integral is performed on the lower hemisphere $\bar\tau \leq 0$ of $S^{d+1}$ (orange region in fig.\ \ref{fig:penrose-app}{\tt b}) with boundary conditions  $\Phi|_{\bar\tau = 0} = \Phi_0$. $\langle O|\Phi_0\rangle$ is similarly defined as a path integral on the upper hemisphere $\bar\tau \geq 0$ (green region). This defines the Hartle-Hawking/Euclidean vacuum state $|O\rangle$ \cite{PhysRevD.28.2960} of global dS$_{d+1}$, with $Z_{\rm PI}$ computing the natural pairing of $|O\rangle$ with  $\langle O|$.\footnote{For kind enough theories, such as a scalar field theory, this pairing can be identified with the Hilbert space inner product. However not all theories are kind enough, as is evident from the negative-mode rotation phase $i^{-(d+3)}$ in the one-loop graviton contribution to $Z_{\rm PI}=\langle O|O\rangle$ according to (\ref{ZPIFINAL}) and \cite{Polchinski:1988ua}. Indeed for gravity this pairing is  not in an obvious way related to the semiclassical inner product of \cite{PhysRev.160.1113}. On the other hand, in the CS formulation of 3D gravity it appears to be framing-dependent, vanishing in particular for canonical framing (cf.\ (\ref{Zrresult}) and discussion below it). The phase also drops out of  $\langle A \rangle \equiv \langle O|A|O\rangle/\langle O|O\rangle$.}


The static patch Wick rotation of fig.\ \ref{fig:penrose-app}{\tt d} on the other hand leads to an interpretation of $Z_{\rm PI}$ as a thermal partition function at inverse temperature $\beta=2\pi$, analogous to (\ref{ZPIexpl}): slicing the path integral along constant-$\tau$ slices as in fig.\ \ref{fig:penrose-app}{\tt d}, and viewing $e^{-\tau H}$ with $H=M_{0,d+1}$ as the Euclidean time evolution operator acting on  $\CH_S$, we formally get\footnote{\label{simeqdef} The notation $\simeq$ means ``equal according to these formal arguments''. Besides the default deferment of dealing with divergences, we are ignoring some additional important points here, including in particular the fixed points of $H$: the $S^{d-1}$ at $r=1$ (yellow dot in fig.\ \ref{fig:penrose-app}), where the equal-$\tau$ slicing of  (\ref{ZPIint2}) degenerates, and the $\CH_G = \CH_N \otimes \CH_S$ factorization implicit in (\ref{EuclvacPI}) breaks down. We return to these points in section \ref{sec:edgecor}.} 
\begin{align} \label{ZPIint2}
 Z_{\rm PI} \simeq {\rm Tr}_{\CH_S} \, e^{-\beta H}  \qquad (\beta = 2\pi) \, .
\end{align}    
Like in the $d=0$ case, this interpretation can be related to the global interpretation (\ref{ZPIint1}). Picking suitable bases of $\CH_S$ and $\CH_N$ diagonalizing $H$, and applying a similar slicing argument, we formally get the analog of (\ref{thermalitydS1}):
\begin{align} \label{EuclvacPI}
  |O\rangle \simeq \text{``}\sum_n\text{''} e^{-\beta E_n/2}|E_n,E_n\rangle \, , \qquad \hat\varrho_S \simeq e^{-\beta H_S} \,  \qquad (\beta = 2 \pi) \, ,
\end{align}
where we have put the sum in quotation marks because the spectrum is actually continuous, as we will describe more precisely for free QFTs below.  
Granting this, we conclude that an inertial observer in de Sitter space sees the global Euclidean vacuum as a thermal state at inverse temperature $\beta=2\pi$, the Hawking temperature of the observer's horizon \cite{PhysRevD.15.2738,PhysRevD.15.2752,Israel:1976ur}.  

Applying (\ref{ZPIint2}) to a free QFT on dS$_{d+1}$, we can write the corresponding  Gaussian $Z_{\rm PI}$ on $S^{d+1}$ as the thermal partition function of an ideal gas in the southern static patch:
\begin{align} \label{ZIGomega}
 \log Z_{\rm PI} \simeq \log {\rm Tr}_S \, e^{-2 \pi H} = \sum_{\pm} \mp \int_0^\infty d\omega \, \rho_S(
\omega)_\pm \bigl( \log(1 \mp \, e^{-2\pi\omega}) +  2\pi  {\omega}/2  \bigr)  \, ,
\end{align}
where $\rho_S(\omega) \equiv {\rm tr}_S \, \delta(\omega-H)$ is the density of single-particle states at energy $\omega>0$ above the vacuum energy in the static patch, split into bosonic and fermionic parts as $\rho_S = \rho_{S+} + \rho_{S-}$.
Using (\ref{chievenarg2}), we can write the character for arbitrary $SO(1,d+1)$ representations as
\begin{align}
 \chi(t) = {\rm tr}_G \, e^{-i t H} =  \int_0^\infty \rho_G(\omega) \, \bigl( e^{-i \omega t} + e^{i \omega t} \bigr) 
\end{align}
where $\rho_G(\omega) \equiv {\rm tr}_G \, \delta(\omega-H)$. The Bogoliubov map (\ref{bogoliubov}) formally implies $\rho_G(\omega) \simeq \rho_S(\omega)$ for $\omega > 0$, hence, following the reasoning leading to (\ref{lalalili}), 
\begin{align} \label{Zcharappth}
 \log Z_{\rm PI} \simeq  \log Z_{\rm bulk} \equiv \int_0^\infty \frac{dt}{2t} \biggl( \, \frac{1+e^{-t}}{1-e^{-t}}  \, \, \chi(t)_{\rm bos}  \, - \,  \frac{2 \, e^{-t/2}}{1-e^{-t}}  \, \, \chi(t)_{\rm fer} \biggr) \, . 
\end{align}

\subsection{Brick wall regularization}  
\label{sec:brickwall}

Here we review how attempts at evaluating the ideal gas partition function (\ref{ZIGomega}) directly hit a brick wall. Consider for example a scalar field of mass $m^2$ on dS$_{d+1}$. Denoting $\Delta_\pm = \frac{d}{2} \pm \bigl((\frac{d}{2})^2-m^2\bigr)^{1/2}$, the positive frequency solutions  on the static patch are of the form 
\begin{align} \label{SPmodes}
 \phi_{\omega \sigma}(T,\Omega,r) \propto e^{-i \omega T} \, Y_\sigma(\Omega) \,\, r^\ell \left(1-r^2\right)^{i \omega/2}  {}_2F_1\bigl(\tfrac{\ell+\Delta_+ +i \omega}{2},\tfrac{\ell+\Delta_- +i \omega }{2};\tfrac{d}{2}+\ell;r^2\bigr),
\end{align}
where $\omega>0$, and $Y_\sigma(\Omega)$ is a basis of spherical harmonics on $S^{d-1}$ labeled by $\sigma$, which includes the total $SO(d)$ angular momentum quantum number $\ell$. A basis of energy and $SO(d)$ angular momentum eigenkets is therefore given by $|\omega\sigma)$ satisfying $( \omega \sigma|\omega'\sigma')=\delta(\omega-\omega') \, \delta_{\sigma\sigma'}$. Naive evaluation of the density of states in this basis gives a pathologically divergent result $\rho_S(\omega) = \int d\omega' \sum_\sigma  ( \omega' \sigma| \delta(\omega-\omega') |\omega' \sigma) = \sum_\sigma \delta(0)$, and commensurate nonsense in (\ref{ZIGomega}). 

Pathological divergences of this type are generic in the presence of a horizon. Physically they can be thought of as arising from the fact that the infinite horizon redshift enables the existence of field modes with arbitrary angular momentum and energy localized in the vicinity of the horizon.
One way one therefore tries to deal with this is to replace the  horizon  by a ``brick wall'' at a distance $\delta$ away from the horizon  \cite{tHooft:1984kcu}, with some choice of boundary conditions, say  $\phi(T,\Omega,1-\frac{1}{2}\delta^2) = 0$ in the example above.  This discretizes the energy spectrum and lifts the infinite angular momentum degeneracy, allowing in principle to control the divergences as $\delta \to 0$. However, inserting a brick wall alters what one is actually computing, introduces  ambiguities (e.g.\ Dirichlet/Neumann), potentially leads to new pathologies (e.g.\ Dirichlet boundary conditions for the graviton are not elliptic \cite{Witten:2018lgb}), and breaks most of the symmetries in the problem.

A more refined version of the idea considers the QFT in Pauli-Villars regularization \cite{Demers:1995dq}. This eliminates the dependence on $\delta$ in the limit $\delta \to 0$ at fixed PV-regulator scale $\Lambda$. It was shown in \cite{Demers:1995dq} that for scalar fields the remaining divergences for $\Lambda \to \infty$ agree with those of the PV-regulated path integral.\footnote{This work directly inspired the use of Pauli-Villars regularization in section \ref{sec:thermal}.}  
A somewhat different approach, reviewed in \cite{Solodukhin:2011gn,Frolov:1998vs}, first maps the equations of motion in the metric $ds^2=g_{\mu \nu} dx^\mu dx^\nu$ 
by a (singular) Weyl transformation to formally equivalent equations of motion in the ``optical'' metric $d\bar s^2  = |g_{00}|^{-1} ds^2$. In the case at hand this would be $d\bar s^2 = -dT^2 + (1-r^2)^{-2} dr^2 + (1-r^2)^{-1} r^2 d\Omega^2$, corresponding to $\IR \times \mbox{hyperbolic $d$-ball}$. The thermal trace is then mapped to a path integral on the Euclidean optical geometry with an $S^1$ of constant radius $\beta$ and a Weyl-transformed action. (This is not a standard covariant path integral. In the case at hand, unless the theory happens to be conformal, non-metric $r$-dependent terms break the $SO(1,d)$ symmetry of the hyperbolic ball to $SO(d)$.) This path integral can be expressed in terms of a heat kernel trace $\int_x \langle x|e^{-\tau \bar D}|x\rangle$. The divergences encountered earlier now arise from the fact that the optical metric $d\bar s^2$ has infinite volume near $r=1$. This is regularized  by cutting the $\int_x$ integral off at $r=1-\delta$, analogous to the brick wall cutoff, though computationally more convenient. For scalars and spinors, Pauli-Villars or dimensional regularization again allows trading the $\delta \to 0$ divergences for the standard UV divergences \cite{Frolov:1998vs}.

Unfortunately, certainly for general field content and in the absence of conformal invariance, none of these variants offers any simplification compared to conventional Euclidean path integral methods. In the case of interest to us, the large underlying $SO(1,d+1)$ symmetry is broken, and with it one's hope for easy access to exact results. Generalization to higher-spin fields, or even just the graviton, appears challenging at best.   

\subsection{Character regularization}


\label{sec:evtoHC}

The character formula (\ref{Zcharappth}) is formally equivalent to the ideal gas partition function (\ref{ZIGomega}), and indeed at first sight, naive evaluation in a global single-particle basis $|\omega\sigma\rangle=a_{\omega\sigma}^{G\dagger}|0\rangle$ diagonalizing $H=\omega \in \IR$, obtained e.g.\ by quantization of the natural cylindrical mode functions of the future wedge ($F$ in fig.\ \ref{fig:penrose-app} and table \ref{dScoords}), gives a similarly pathological $\chi(t) = {\rm tr}_G \, e^{-i H t} = \int_{-\infty}^\infty d\omega \sum_\sigma \langle \omega \sigma|e^{-i \omega t}|\omega\sigma\rangle = 2\pi \delta(t) \sum_\sigma \delta(0)$; hardly a surprise in view of the Bogoliubov relation $\rho_G(\omega) \simeq \rho_S(\omega)$ and our earlier result $\rho_S(\omega)=\sum_\sigma \delta(0)$. Thus the conclusion would appear to be that the situation is as bad, if not worse, than it was before. 

However this is very much the wrong conclusion. As reviewed in appendix \ref{app:chars}, $\chi(t)$, properly defined as a Harish-Chandra character, is in fact rigorously well-defined, analytic in $t$ for $t \neq 0$, and moreover easily computed. For example for a scalar of mass $m^2$ on dS$_{d+1}$, we get (\ref{formulaforchi2}):
\begin{align} \label{scachia}
 \chi(t) =  \frac{e^{-t \Delta_+} + e^{-t \Delta_-}}{|1-e^{-t}|^d}   \qquad \Delta_{\pm} \equiv \tfrac{d}{2} \pm \sqrt{\bigl(\tfrac{d}{2}\bigr)^2-m^2} 
\end{align}
as explicitly computed in appendix \ref{app:compchar}. The reason why naive computation by diagonalization of $H$ fails so badly is explained in detail in appendix \ref{sec:traces}: it is not the trace itself that is sick, but rather the {\it basis} $|\omega\sigma\rangle$ used in the naive computation. 

Substituting the explicit $\chi(t)$ into the character integral (\ref{Zcharappth}), we still get a UV-divergent result, but this divergence is now easily regularized in a standard, manifestly covariant way, as explained 
in section \ref{sec:thermalder}. Keeping the large underlying symmetry manifest allows exact evaluation, for arbitrary particle content.


In section \ref{sec:scalars} we show that for scalars and spinors, the Euclidean path integral $Z_{\rm PI}$ on $S^{d+1}$, regularized as in (\ref{HKSP}), exactly equals $Z_{\rm bulk}$ as defined in (\ref{Zcharappth}), regularized as in (\ref{ZPIreg}):
\begin{align} \label{ZPIZbulkss}
 Z_{\rm PI,\epsilon} = Z_{\rm bulk,\epsilon} \qquad \mbox{(scalars and spinors)} \, ,
\end{align}


One might wonder how it is possible the switch to characters makes such a dramatic difference. After all, (\ref{ZIGomega}) and (\ref{Zcharappth}) are formally equal. Yet the former first evaluates to nonsense and then hits a brick wall, while the latter somehow ends up effortlessly producing sensible results upon standard UV regularization. The discussion in  \ref{sec:traces}, in particular (\ref{chiglobSO}), provides some clues: character regularization can be thought of, roughly speaking, as being akin to a regularization cutting off  {\it global} $SO(d+1)$ angular momentum.  

This goes some way towards explaining why the character formalism fits naturally with the Euclidean path integral formalism on $S^{d+1}$, as covariant (e.g.\ heat kernel) regularization of the latter effectively cuts off the $SO(d+2) \supset SO(d+1)$ angular momentum.

It also goes some way towards explaining what happened above. One way of thinking about the origin of the pathological divergences encountered in section \ref{sec:brickwall} is that, as mentioned in footnote \ref{simeqdef}, the formal argument implicitly starts from the premise that the QFT Hilbert space can be factorized as $\CH_G = \CH_S \otimes \CH_N$, like in the $S^1$ toy model. However this cannot be done in the continuum limit of QFT: locally factorized states, such as the formal state $|O) \otimes |O)$ in which both the southern and the northern static patch are in their minimal energy state, are violently singular objects \cite{Witten:2018zxz}. 
Cutting off the global $SO(d+1)$ angular momentum does indeed smooth out the sharp north-south divide: $SO(d+1)$ is the isometry group of the global spatial slice at $\bar T=0$ (fig.\ \ref{fig:penrose-app}{\tt b}). The angular momentum cutoff means we only have a finite number of spherical harmonics available to build our field modes. This makes it impossible in particular to build field modes sharply localized in the southern or northern hemisphere: the harmonic expansion of a localized mode always has infinitely many terms. Cutting off this expansion will necessarily leave some support on the other hemisphere. Quite similar in this way again to the Euclidean path integral, this offers some intuition on why the UV-regularized character integral avoids the pathological divergences induced by sharply cutting space.

\subsection{Edge corrections} \label{sec:edgecor}

In view of all this and (\ref{ZPIZbulkss}), one might be tempted at this point to jump to the conclusion that the arguments of section \ref{sec:formform}, while formal and glossing over some subtle points, are apparently good enough to give the right answer provided we use the character formulation, and that likewise $Z_{\rm PI}^{(1)}$ on the sphere for a field of arbitrary spin $s$, despite its off-shell baroqueness, is just the ideal gas partition function $Z_{\rm bulk}$ on the dS static patch, calculable with on-shell ease: mission accomplished. 
As further evidence in favor of declaring footnote \ref{simeqdef} overly cautious, one might point to the fact that in the context of theories of {\it quantum gravity}, identifying $Z_{\rm PI}^{\rm grav} = {\rm Tr}_\CH \, e^{-\beta H}$ elegantly reproduces the thermodynamics of horizons  inferred by other means \cite{PhysRevD.15.2752}, and that such identifications are moreover known to be valid in a quantitatively precise way in many well-understood cases in string theory and AdS-CFT. If the formal argument is good enough for quantum gravity, then surely it is good enough for field theory, one might think.

These naive considerations are wrong: the formal relation $Z_{\rm PI} \simeq Z_{\rm bulk}$ for fields of spin $s \geq 1$ receives ``edge'' corrections. In sections \ref{sec:mashsflds} and \ref{sec:massless}, we  determine these for massive resp.\ massless spin-$s$ fields on $S^{d+1}$ by direct computation. The results are eqs.\ (\ref{formuhs2}) and (\ref{ZPIFINAL}). The corrections we find exhibit a concise and suggestive structure: again taking the form of a character formula like (\ref{Zcharappth}), but encoding instead a path integral on a sphere in {\it two lower} dimensions, i.e.\ on $S^{d-1}$ rather than $S^{d+1}$. This $S^{d-1}$ is naturally identified with the horizon $r=1$, i.e.\ the edge of the static patch hemisphere, the yellow dot in fig.\ \ref{fig:penrose-app}. The results of section \ref{sec:all-loop} then imply $S_{\rm PI} \simeq S_{\rm bulk}$ likewise receives edge corrections (besides corrections due to nonminimal coupling to curvature, which arise already for scalars). 

Similar edge corrections, to the entropy $S_{\rm PI} \simeq S_{\rm bulk}$ in the conceptually analogous case of Rindler space, were anticipated long ago in \cite{Susskind:1994sm} and explicitly computed shortly thereafter for massless spin-1 fields in \cite{Kabat:1995eq}. The result of \cite{Kabat:1995eq} was more recently revisited in several works including \cite{Donnelly:2014fua,Donnelly:2015hxa,Blommaert:2018rsf}, relating it to the local factorization problem of constrained QFT Hilbert spaces \cite{Buividovich:2008gq, Donnelly:2011hn, Casini:2013rba, Soni:2015yga, Dong:2018seb, Jafferis:2015del} and given an interpretation in terms of the edge modes arising in this context. Given the sensitivity of the edge contributions to ultraviolet physics, it is natural to try and pose the edge contribution problem in terms of less UV sensitive quantities. A recent approach to do so by studying the mutual information of operator algebras for theories with global symmetries can be found in \cite{Casini:2019kex}.

We leave the precise physical interpretation of the explicit edge corrections we obtain in this paper to future work. Below we will review why they were to be expected, and how related corrections can be interpreted in analogous, better-understood contexts in quantum gravity and QFT. We begin by explaining why the quantum gravity argument was misleading and what its correct version actually suggests, first from a boundary CFT point of view in the precise framework of AdS-CFT, then from a bulk point of view in a qualitative picture based on string theory on Rindler space. Finally we return to interpretations within QFT itself, clarifying more directly why the caution expressed in footnote \ref{simeqdef} was warranted indeed.

\subsubsection{AdS-CFT considerations} \label{app:adscft}
 
As mentioned above, there are reasons to believe that in theories of quantum gravity, the identification $Z_{\rm PI}^{\rm grav} = {\rm Tr}_{\CH} \, e^{-\beta H}$ is exact as a semiclassical (small-$G_{\rm N}$) expansion.
 
However, the key point here is that $\CH$ is the Hilbert space of the {\it fundamental} microscopic degrees of freedom, {\it not} the Hilbert space of the low energy effective field theory. This can be made very concrete in the context of AdS-CFT, where $\CH$ has a precise boundary CFT definition. For example for asymptotically Euclidean AdS$_{d+1}$ geometries with $S^1_\beta \times S^{d-1}$ conformal boundary, certain analogs of the formal relations (\ref{ZPIint2}) and (\ref{EuclvacPI}) then become exact in the semiclassical/large-$N$ expansion \cite{Witten:1998zw,Maldacena:2001kr}:
\begin{align} \label{thermalAdSCFT}
 Z_{\rm PI}^{\rm grav}  = {\rm Tr}_\CH \, e^{-\beta H} = \langle O|O\rangle  \, , \qquad |O\rangle = \sum_n e^{-\beta E_n/2} |E_n)_{\CH} \otimes |E_n)_{\CH} \, .
\end{align}
Crucially, $\CH$ here is the complete boundary CFT Hilbert space, and $|O\rangle$ is the Euclidean vacuum state of two disconnected copies of the boundary CFT, constructed exactly like in the dS$_1$ toy model of section \ref{sec:S1}, but with the hemicircle $\frac{1}{2} S^1$ replaced by $\frac{1}{2} S^1 \times S^{d-1}$. From a semiclassical bulk dual point of view this can be viewed as the Euclidean vacuum of two disconnected copies of global AdS or of the eternal AdS-Schwarzchild geometry \cite{Maldacena:2001kr}, depending on whether $\beta$ lies above or below the Hawking-Page phase transition point $\beta_c$ \cite{Hawking:1982dh}. 

\begin{figure} 
 \begin{center}
   \includegraphics[height=4.1cm]{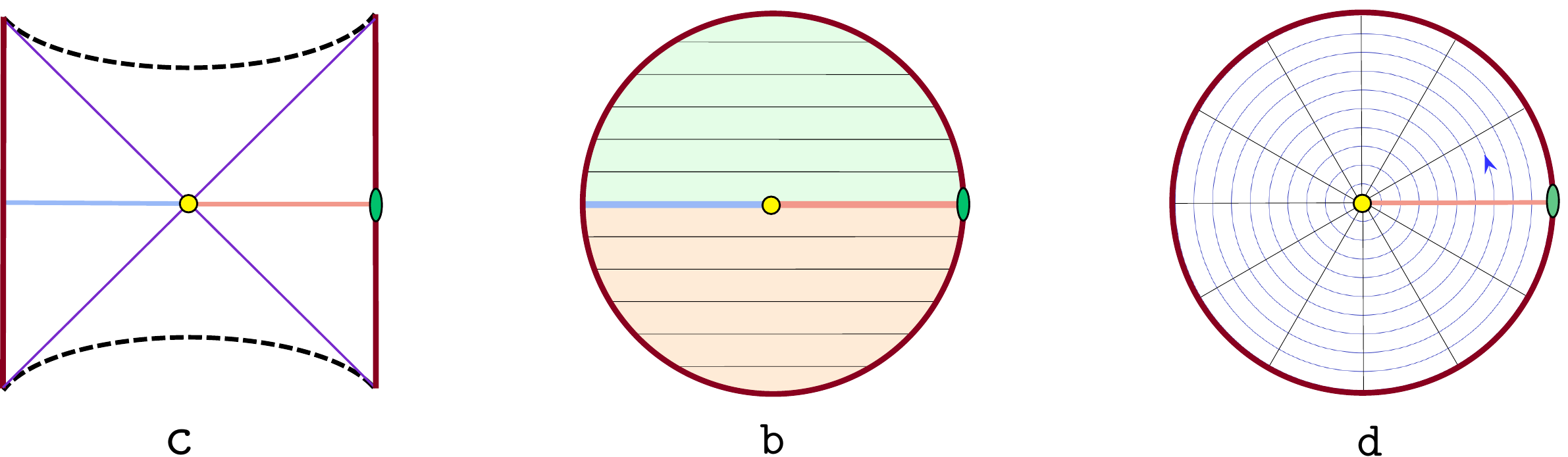}
 \caption{\small AdS-Schwarzschild analogs of {\tt c,b,d} in fig.\  \ref{fig:penrose-app}. Black dotted line = singularity. Thick brown line = conformal boundary.  
 \label{fig:AdS}}
 \end{center} \vskip-4mm
\end{figure}

When $\beta > \beta_c$, where $\beta_c \sim O(1)$ assuming the low-energy gravity theory is approximately Einstein with $G_{\rm N} \ll \ell^{d-1} = 1$, $Z_{\rm PI}^{\rm grav}$ is dominated by the thermal EAdS saddle \cite{Hawking:1982dh}, with on-shell action $S_E \equiv 0$, so in the limit $G_{\rm N} \to 0$,
$Z_{\rm PI}^{\rm grav} = Z_{\rm PI}^{(1)}$. Thus in this case, the relation $Z_{\rm PI}^{\rm grav} = {\rm Tr}_\CH \, e^{-\beta H}$ of (\ref{thermalAdSCFT}) indeed implies $Z_{\rm PI}^{(1)}$ equals a statistical mechanical partition function. There is no need to invoke quantum gravity to see this, of course: the thermal $S^1$ is noncontractible in the bulk geometry, so the bulk path integral slicing argument is free of subtleties, directly implying $Z_{\rm PI}^{(1)}$ equals the partition function ${\rm Tr} \, e^{-\beta H}$ of an ideal gas in global AdS.

On the other hand if $\beta < \beta_{\rm crit}$, the dominant saddle is the  Euclidean Schwarzschild geometry (fig.\ \ref{fig:AdS}), with on-shell action $\tilde S_E \propto - \frac{1}{G_{\rm N}}$, so in the limit $G_{\rm N} \to 0$, $Z_{\rm PI}^{\rm grav} =  Z_{\rm PI}^{(0)}  = e^{-\tilde S_E}$. In this case the identification $Z_{\rm PI}^{\rm grav} = {\rm Tr}_\CH \, e^{-\beta H}$ of (\ref{thermalAdSCFT}) no longer implies the one-loop correction $ Z_{\rm PI}^{(1)}$ can be identified as a statistical mechanical partition function. In particular the bulk one-loop contributions $S^{(1)} = (1-\beta\partial_\beta) \log  Z_{\rm PI}^{(1)}$ to the entropy need not be positive. (More specifically its leading divergent term, which in a UV-complete description of the bulk theory would become finite but generically still dominant, need not be positive.)  
From the CFT point of view, these are just O(1) corrections in the large-$N$ expansion of the statistical entropy. Although the total entropy must of course be positive, corrections can come with either sign. From the bulk point of view, since the Euclidean geometry is the Wick-rotated exterior of a black hole, the thermal circle is contractible, shrinking to a point analogous to the yellow dot in fig.\ \ref{fig:penrose-app}{\tt d}, leading to the same issues as those mentioned in footnote \ref{simeqdef}.

\subsubsection{Strings on Rindler considerations} \label{app:rindler}

To gain some insight from a bulk point of view, we consider the simplest example of a spacetime with a horizon: the Rindler wedge $ds^2=-\rho^2 dt^2 + d\rho^2 + dx_\perp^2$ of Minkowski space. While not quite at the level of AdS-CFT, we do have a perturbative theory of quantum gravity in Minkowski space: string theory. In fact, that $Z_{\rm PI}^{(1)}$ on a Euclidean geometry with a contractible thermal circle cannot be interpreted as a statistical mechanical partition function in general, even if  the full $Z_{\rm PI}^{\rm grav}$ has such an interpretation, was anticipated long ago in \cite{Susskind:1994sm}, in an influential attempt at developing a string theoretic understanding of the thermodynamics of the Rindler horizon. Rindler space Wick rotates to 
\begin{align} \label{ERindmet}
 ds^2=\rho^2 d\tau^2 + d\rho^2 + dx_\perp^2, \qquad \tau \simeq \tau + \beta, \qquad \beta=2\pi-\epsilon \, .
\end{align}   
with the conical defect $\epsilon=0$ on-shell.
The argument given in \cite{Susskind:1994sm} is based on the point of view developed in their work that loop corrections in the semiclassical expansion of the Rindler entropy $S_{\rm PI} \equiv (1-\beta \partial_\beta) \log Z^{\rm grav}_{\rm PI}|_{\beta=2\pi}$ are equivalent to loop corrections to the Newton constant, ensuring the entropy $S=A/4 G_{\rm N}$ involves the physically measured $G_{\rm N}$ rather than than the bare $G_{\rm N}$. In $\CN = 4$ compactifications of string theory to 4D Minkowski space (and in $\CN=4$ supergravity theories more generally), loop corrections to the Newton constant vanish. By the above observation, this implies loop corrections to $S_{\rm PI}$ vanish as well. Hence there must be cancelations between different particle species, and in particular the one-loop contribution to the entropy of some fields in the supergravity theory must be {\it negative}. Since statistical entropy is always positive, the one-loop $Z_{\rm PI}^{(1)}$ of such fields  cannot be  equal to a statistical mechanical partition function. 

\begin{figure} 
 \begin{center}
   \includegraphics[height=5cm]{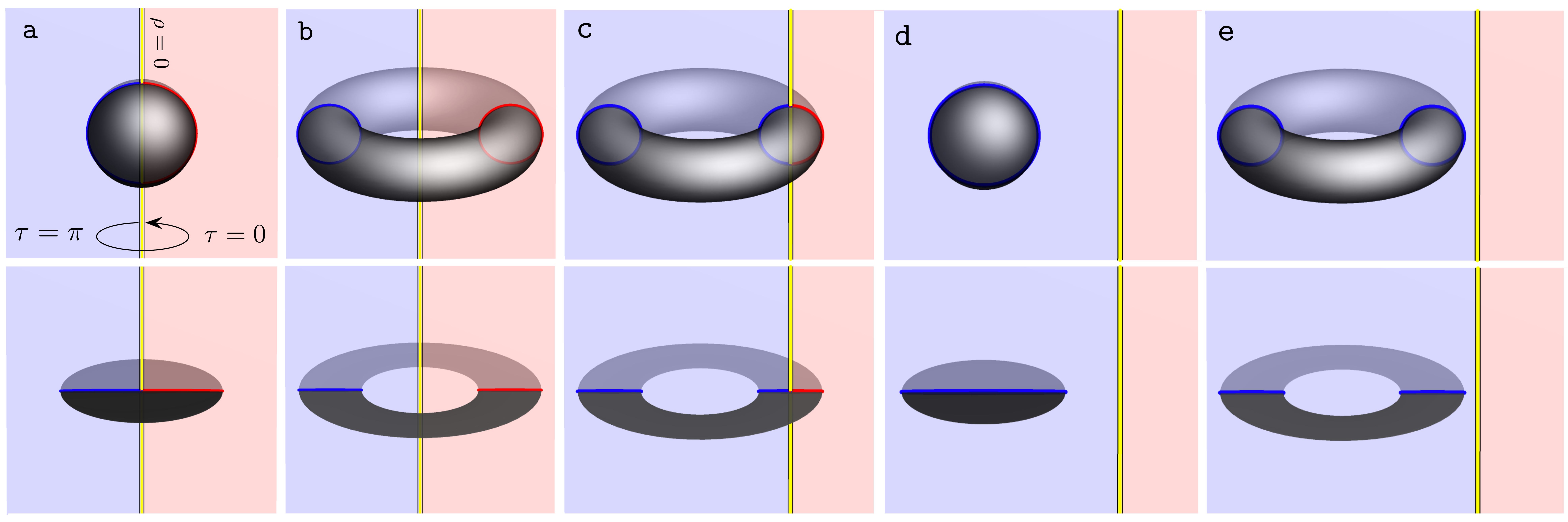}
 \caption{\small \label{fig:worldsheets} Closed/open string contributions to the total Euclidean Rindler ($ds^2=\rho^2 d\tau^2 + d\rho^2 + dx^2$) partition function according to the picture of \cite{Susskind:1994sm}. $\tau$ = angle around yellow axis $\rho=0$; blue$|$red plane is $\tau=\pi|0$. {\tt a,b,c} contribute to the entropy. 
Sliced along Euclidean time $\tau$, {\tt a} and {\tt b} can be viewed as free bulk resp.\ edge  string  thermal traces  contributing positively to the entropy, 
while $\tt c$ can be viewed as an edge string emitting and reabsorbing a bulk string, contributing a (negative) interaction term. 
  }
 \end{center} \vskip-4mm
\end{figure}

In the same work \cite{Susskind:1994sm}, a qualitative stringy picture was sketched giving some bulk intuition about the nature of such negative contributions to $S_{\rm PI}$ when the total $S_{\rm PI}$ is a statistical entropy.
In this picture, all relevant microscopic fundamental degrees of freedom are presumed to be realized in the bulk quantum gravity theory as weakly coupled strings. More specifically it is presumed that $Z_{\rm PI}^{\rm grav} = {\rm Tr}_{\CH} \, e^{-\beta H}$ where $\CH$ is the string Hilbert space on Rindler space and $H$ is the Rindler Hamiltonian, so $S_{\rm PI}=S$, the statistical entropy. Tree level and one-loop contributions to $\log Z_{\rm PI}$ are shown in fig.\ \ref{fig:worldsheets}. Diagrams {\tt d,e} do not contribute to the entropy $S_{\rm PI}=(1-\beta \partial_\beta) \log Z_{\rm PI}$ as their $\log Z_{\rm PI} \propto \beta$. Cutting {\tt b} along constant-$\tau$ slices gives it an interpretation as a thermal trace over ``bulk''  string states away from $\rho=0$ (closed strings in top row).\footnote{As a simple analog of what is meant here, consider a free scalar field on $S^1$ parametized by $\tau \simeq \tau + \beta$. Then $\log Z_{\rm PI} = \int_0^\infty \frac{ds}{2s} \, {\rm Tr} \, e^{-s \frac{1}{2}(-\partial_\tau^2+m^2)} = \sum_n \int \frac{ds}{2s}  \int \CD \tau|_n \, \exp[-\frac{1}{2} \int_0^s (\dot \tau^2 + m^2)] = \sum_n \frac{1}{2 |n|} \, e^{-|n| \beta m}$. Here $n$ labels the winding number sector of the particle worldline path integral with target space $S^1$. Discarding the UV-divergent $\frac{1}{0}$ term, this sums to $\log Z_{\rm PI} = -\log(1-e^{-\beta m}) - \frac{1}{2}\beta m=\log {\rm Tr} \, e^{-\beta H}$ as in (\ref{ZPIexpl}). {\tt b} is analogous to the $|n|=1$ contribution $e^{-\beta m}$, {\tt e} is analogous to $n=0$, and higher winding versions of {\tt b} correspond to $|n|>1$.} Similarly, {\tt a} can be viewed as a thermal trace over ``edge'' string states stuck to $\rho=0$ ({\it open} strings in top row). On the other hand {\tt c} represents an {\it interaction} between bulk and edge strings, with no thermal or state counting interpretation on its own.
Being statistical mechanical partition functions, $\tt a$ and $\tt b$ contribute positively to $S_{\rm PI}$, whereas $\tt c$ may contribute negatively. In fact in the $\CN=4$ case discussed above, {\tt c} {\it must} be negative, canceling {\tt b} to render $S^{(1)}_{\rm PI}=0$. From an effective field theory point of view,  {\tt b} and {\tt e} correspond to the bulk ideal gas partition function inferred from formal arguments along the lines of section \ref{sec:formform}, while {\it {\tt c} represents ``edge'' corrections missed by such arguments}. 

This picture is qualitative, as the individual contributions corresponding to a sharp split of the worldsheet path integrals along these lines are likely ill-defined/divergent  \cite{Mertens:2015adr}. Moreover, even without any splitting, an actual string theory calculation of $S_{\rm PI}=(1-\beta\partial_\beta) \log Z_{\rm PI}|_{\beta=2\pi}$ is problematic, as Euclidean Rindler with a generic conical defect $\epsilon=2\pi-\beta$ is off-shell. Shortly after \cite{Susskind:1994sm}, \cite{Dabholkar:1994ai} proposed to compute $Z_{\rm PI}$ on the orbifold $\IR^2/\IZ_N$ for general integer $N$ and then analytically continue the result to $N \to 1 + \epsilon$. Unfortunately such orbifolds have closed string tachyons leading to befuddling IR-divergences \cite{Dabholkar:1994ai,Lowe:1994ah}. Recently, progress was made in resolving some of these issues: in an open string version of the idea, arranged in type II string theory by adding a sufficiently low-dimensional $D$-brane, it was shown in \cite{Witten:2018xfj} that upon careful analytic continuation, the tachyon appears to disappear at $N = 1 + \epsilon$.

\subsubsection{QFT considerations} \label{sec:QFTcons}

The problem of interest to us is really just a problem involving Gaussian path integrals in free quantum field theory, so there should be no need to invoke quantum gravity to gain some insight in what kind of corrections we should expect to the naive $Z_{\rm PI} \simeq Z_{\rm bulk}$. Indeed the above stringy Rindler considerations have much more straightforwardly computable low-energy counterparts in QFT. 

\begin{figure}
 \begin{center}
   \includegraphics[height=2.8cm]{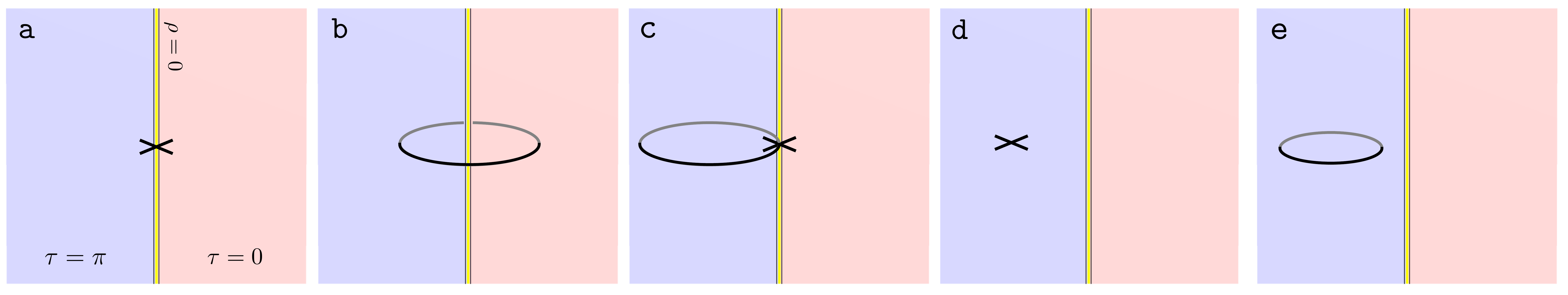}
 \caption{\small \label{fig:worldlines} Tree-level and one-loop contributions to $\log Z_{\rm PI}$ for massless vector field in Euclidean Rindler (\ref{ERindmet}). These can be viewed as field theory limits of fig.\ \ref{fig:worldsheets}, with verbatim the same comments applicable to {\tt a}-{\tt e}. The worldline path integral {\tt c} appears with a sign opposite to {\tt b} in $\log Z^{(1)}_{\rm PI}$ \cite{Kabat:1995eq}. }
 \end{center}\vskip-3mm
\end{figure}

Motivated by \cite{Susskind:1994sm}, \cite{Kabat:1995eq} computed $Z_{\rm PI}^{(1)}$ for scalars, spinors and Maxwell fields on Rindler space. For scalars and spinors, this was found to coincide with the ideal gas partition function, whereas for Maxwell an additional contact term was found, expressible in terms of a ``edge'' worldline path integral with coincident start and end points at $\rho=0$, fig.\  \ref{fig:worldlines}{\tt c}.  This term contributes negatively to $S_{\rm PI}^{(1)} = (1-\beta \partial_\beta)\log Z_{\rm PI}|_{\beta=2\pi}$ and thus has no thermal interpretation on its own. In fact it causes the total $S_{\rm PI}^{(1)}$ to be negative in less than 8 dimensions. The results of \cite{Kabat:1995eq} and more generally the picture of \cite{Susskind:1994sm} were further clarified by low-energy effective field theory analogs in \cite{Kabat:1995jq}, emphasizing in particular that whereas $S_{\rm PI}$ remains invariant under Wilsonian RG, the division between contributions with or without a low-energy statistical interpretation does not, the former gradually turning into the latter as the UV-cutoff $\Lambda$ is lowered. At $\Lambda = 0$, only the tree-level contribution $S = A/4 G_{\rm N}$ of fig.\ \ref{fig:worldlines}{\rm a} is left.    

The contact/edge correction of fig.\ \ref{fig:worldlines}{\tt c} to $\log Z_{\rm PI}$ can be traced to the presence of a curvature coupling $X$ linear in the Riemann tensor in $S_E=\int A (-\nabla^2 + X) A + \cdots$ \cite{Solodukhin:2011gn,Larsen:1995ax,Kabat:1995jq}. Such terms appear for any spin $s \geq 1$ field, massless or not. Hence, as one might have anticipated from the stringy picture of fig.\ \ref{fig:worldsheets}, they are the norm rather than the exception.

\begin{figure} 
 \begin{center}
   \includegraphics[height=3cm]{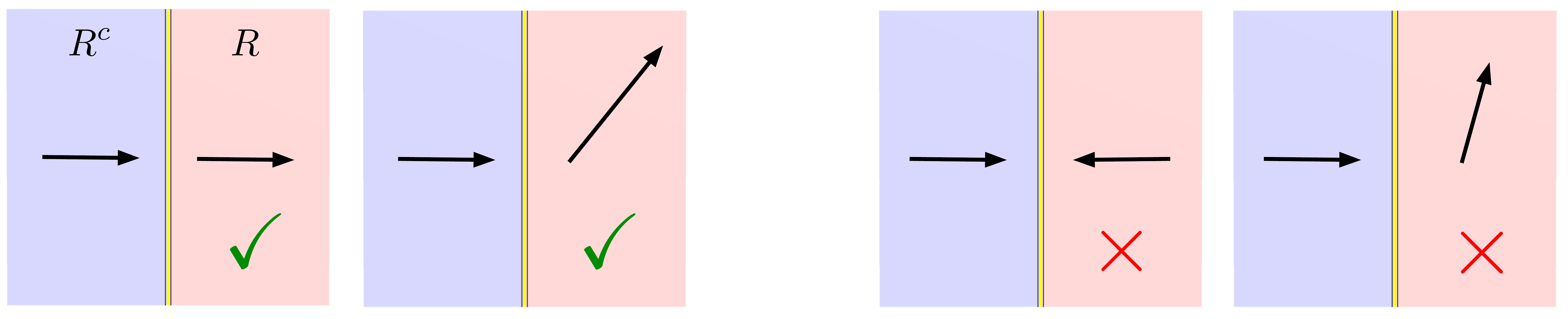}
 \caption{\small Candidate classical initial electromagnetic field configurations (phase space points), with $A_0 = 0$, $A_i=0$, showing electric field $E_i = \Pi_i = \dot{A}_i$. Gauss' law requires continuity $E_\perp$ across the boundary, disqualifying the two candidates on the right.  
 \label{fig:gauss}}
 \end{center} \vskip-4mm
\end{figure}

The result of \cite{Kabat:1995eq} was more recently revisited in \cite{Donnelly:2014fua,Donnelly:2015hxa}, relating the appearance of edge corrections to the 
local factorization problem of QFT Hilbert spaces with gauge constraints \cite{Buividovich:2008gq, Donnelly:2011hn, Casini:2013rba, Soni:2015yga, Dong:2018seb, Jafferis:2015del} like Gauss' law $\nabla \cdot E = 0$ in Maxwell theory. This problem arises more generally when contemplating the definition of entanglement entropy $S_R = - {\rm Tr}\,  \varrho_R \log \varrho_R$ of a spatial subregion $R$ in gauge theories. In principle $\varrho_R$ is obtained by factoring the global Hilbert space $\CH_G = \CH_R \otimes  \CH_{R^c}$ and tracing out $\CH_{R^c}$. As mentioned at the end of \ref{sec:evtoHC}, local factorization is impossible in the continuum limit of any QFT, including scalar field theories, but the issue raised there can be dealt with by a suitable regularization.  
However for a gauge theory such as free Maxwell theory, there is an additional obstruction to local factorization, which persists after regularization, and indeed is present already at the classical phase space level: the Gauss law constraint $\nabla \cdot E = 0$ prevents us from picking independent initial conditions in both $R$ and $R^c$ (fig.\ \ref{fig:gauss}), unless the boundary is a physical object that can accommodate compensating surface charges --- but this is not the case here. One way to resolve this is to decompose the global phase space into sectors labeled by ``center'' variables located at the boundary surface \cite{Buividovich:2008gq, Donnelly:2011hn, Casini:2013rba, Soni:2015yga, Dong:2018seb, Jafferis:2015del}, for example the normal component $E_\perp$ of the electric field. The center variables Poisson-commute with all local observables inside $R$ and $R^c$. In any given sector, factorization then becomes possible.  

Building on this framework it was shown in \cite{Donnelly:2014fua,Donnelly:2015hxa} that in a suitable brick wall-like regularization scheme and  for some choice of measure $\CD E_\perp$, the edge correction of \cite{Kabat:1995eq} arises as a classical contribution $\int \CD E_\perp \, e^{-S_E[E_\perp]}$ to the thermal statistical partition function. Here $S_E[E_\perp]$ is the on-shell action for static electromagnetic field  modes in Euclidean Rindler space with prescribed $E_\perp$, localized vanishingly close to $\rho=0$ when the brick-wall cutoff is taken to zero, and thus interpreted as edge modes.   They also find a more precise form for the result of \cite{Kabat:1995eq} for Rindler with its transverse dimensions compactified on a torus, which is  identical in form to our de Sitter result (\ref{ZPIFINAL}) for $s=1$, $G=U(1)$.

Similar results for massive vector fields were obtained in \cite{Blommaert:2018rsf}. (The Stueckelberg action for a massive vector has a $U(1)$ gauge symmetry, so from that point of view it may fit into the above considerations.) 
An open string realization of the above ideas was proposed in \cite{Balasubramanian:2018axm}. It has been suggested that edge modes and ``soft hair'' might be related \cite{Strominger:2017zoo}.

\section{Derivations 
for massive higher spins 
}

\subsection{Massive spin-$s$ fields} \label{sec:EPIspins}

Here we derive (\ref{formuhs2}) and (\ref{ZPItrueformula}). The starting point is the path integral (\ref{massiveZPIdefin}). To get a result guaranteed to be consistent with QFT locality and general covariance, we should in principle start with the full off-shell system \cite{Zinoviev:2001dt} involving auxiliary Stueckelberg fields of all spin $s'<s$. 

\subsubsection*{Transverse-traceless part $Z_{\rm TT}$}

One's initial hope might be that $Z_{\rm PI}$ ends up being equal to the path integral $Z_{\rm TT}$ restricted to the propagating degrees of freedom, the transverse traceless modes of $\phi$, with kinetic operator given by the second-order equation of motion in (\ref{FPeom}). Regularized as in (\ref{HKSP}), this is
\begin{align} \label{ZPITTmassive}
 \log Z_{\rm TT} \equiv   \int_0^\infty \frac{d\tau}{2\tau} \,  e^{-\epsilon^2/4\tau}  \, {\rm Tr}_{\rm TT} \, e^{-\tau \left(-\nabla^2_{s,\rm TT}+\ms^2 \right)} \, .
\end{align}
The index ${\rm TT}$ indicates the object is defined on the restricted space of transverse traceless modes. This turns out to be correct for Euclidean AdS with standard boundary conditions \cite{Giombi:2013yva}. However, this is not quite true for the sphere, related to the presence of normalizable tensor decomposition zeromodes.

The easiest way to convince oneself that $Z_{\rm PI} \neq Z_{\rm TT}$ on the sphere is to just compute $Z_{\rm TT}$ and observe it is inconsistent with locality, in a sense made clear below.
To evaluate $Z_{\rm TT}$, all we need is the spectrum of $-\nabla^2_{\rm TT}+\ms^2$ \cite{Rubin:1983be,doi:10.1063/1.530850}. The eigenvalues are $\lambda_n = \left(n+\tfrac{d}{2}\right)^2+\nu^2$, $n \geq s$ with degeneracy  given by  the dimension $D^{d+2}_{n, s}$ of the $\text{so}(d+2)$ representation corresponding to the two-row Young diagram $(n,s)$, for example for $d=3$, $(n,s)=(7,3)$, \Yvcentermath1
\begin{align}
 D_{\rm 7,3}^{5} = \dim^{\so(5)} \, \yng(7,3)  \quad  = 1190 \, .
\end{align}
Explicit dimension formulae and tables can be found in appendix \ref{app:Weyldim}.

Following the same steps as for the scalar case in section \ref{sec:scalars}, we end up with    
\begin{align} \label{ZPIperpformula}
  \log Z_{\rm TT} =   \int_0^\infty \frac{dt}{2t} \, \bigl( q^{i\nu} +q^{-i\nu}  \bigr)  \, f_{\rm TT}(q) \, ,\qquad  f_{\rm TT}(q) \equiv \sum_{n \geq s} D^{d+2}_{n, s} \, q^{\frac{d}{2}+n} \, . 
\end{align}
Now let us evaluate this explicitly for the example of a massive vector on $S^5$, i.e.\  $d=4$, $s=1$. From (\ref{SOKlowdim}) we read off $D_{n,1}^6 = \frac{1}{3} n (n+2)^2 (n+4)$. Performing the sum we end up with 
\begin{align} \label{dis4sis1}
  f_{\rm TT}(q) = \frac{1+q}{1-q} \biggl( \frac{4 \, q^2}{(1-q)^4}-  \frac{q}{(1-q)^2} \biggr) \, + q \, \qquad (d=4,s=1) \, .
\end{align}
The first term inside the brackets can be recognized as the $d=4$ massive spin-1 bulk character. 

The small-$t$ expansion of the integrand in (\ref{ZPIperpformula}) contains a term $1/t$. This term arises from the term $+q$ in the above expression, as the other parts give contributions to the integrand that are manifestly even under $t \to -t$. The presence of this $1/t$ term in the small-$t$ expansion implies $\log Z_{\rm TT}$ has a logarithmic UV divergence $\log Z_{\rm TT}|_{\rm log \, div} = \log M$ where $M$ is the UV cutoff scale. More precisely in the heat-kernel regularization under consideration, the contribution of the term $+q$ to $\log Z_{\rm TT}$ is, according to (\ref{simplereg}),
\begin{align} \label{explicitfinite}
 \int_0^\infty \frac{dt}{2t} \bigl( q^{1+i\nu} +q^{1-i\nu}  \bigr) = \log \frac{M}{\sqrt{1+\nu^2}} \, ,  \qquad M \equiv \frac{2 e^{-\game}}{\epsilon} \, .
\end{align}
Note that $m=\sqrt{1+\nu^2}$ is the Proca mass (\ref{convmdef}) of the vector field. 
The presence of a logarithmic divergence means $Z_{\rm PI} \neq Z_{\rm TT}$, for $\log Z_{\rm PI}$ itself is defined as a manifestly covariant, local QFT path integral on $S^5$, which cannot have any logarithmic UV divergences, as there are no local curvature invariants of mass dimension 5. 

For $s=2$ and $d=4$ we get similarly
\begin{align}
 f_{\rm TT}(q) =   \frac{1+q}{1-q} \biggl( \frac{9 \, q^2}{(1-q)^4}-\frac{6 \, q}{(1-q)^2} \biggr) \, + 6 \, q + 15 \, q^2   \qquad (d=4,s=2) \, .
\end{align}
The terms $6 \, q + 15 \, q^2$ produce a nonlocal logarithmic divergence $\log Z_{\rm TT}|_{\rm log \, div} = c \, \log M$, where $c=6+15=21$, so again $Z_{\rm PI} \neq Z_{\rm TT}$. Note that $21 = \frac{7 \times 6}{2} = \dim\so(1,6)$, the number of conformal Killing vectors on $S^5$. That this is no coincidence can be ascertained by repeating the same exercise for general $d \geq 3$ and $s=2$:
\begin{align}
 f^{(s=2)}_{\rm TT}(q) &=   \frac{1+q}{1-q} \biggl( D_2^d \cdot \frac{ q^\frac{d}{2}}{(1-q)^d}-D_1^{d+2} \cdot \frac{ q^{\frac{d-2}{2}}}{(1-q)^{d-2}} \biggr) \, + D_1^{d+2} \, q + D_{1,1}^{d+2} \, q^2 \,  \, .
\end{align}
The $q,q^2$ terms generate a log-divergence $c_2 \log M$, $c_2=D_1^{d+2} + D_{1,1}^{d+2} = (d+2) + \frac{1}{2}(d+2)(d+1) = \frac{1}{2}(d+3)(d+2) = D_{1,1}^{d+3} = \dim \so(1,d+2)$, the number of conformal Killing vectors on $S^{d+1}$. The identity $N_{\rm CKV} = D_{1,1}^{d+3} =  D_{1,1}^{d+2} + D_{1,0}^{d+2}$ and its generalization to the spin-$s$ case will be a crucial ingredient in establishing our claims. It has a simple group theoretic origin. As a complex Lie algebra, the conformal algebra $\so(1,d+2)$ generated by the conformal Killing vectors is the same as $\so(d+3)$, which is generated by antisymmetric matrices and therefore forms the irreducible representation with Young diagram ${\footnotesize \yng(1,1)}$ of $\so(d+3)$. This decomposes into irreps of $\so(d+2)$ by the branching rule
\begin{align} \label{CKVbranching}
 \yng(1,1) \quad \to \quad  \yng(1,1) \,\, + \,\,  \yng(1) \, ,
\end{align}
implying in particular $D_{1,1}^{d+3} = D_{1,1}^{d+2} + D_{1,0}^{d+2}$. 
Geometrically this reflects the fact that the conformal Killing modes split into two types:  $(i)$ transversal vector modes $\varphi^{i_1}_\mu$, ${i_1}=1,\ldots,D^{d+2}_{1,1}$, satisfying the ordinary Killing equation $\nabla_{(\mu} \varphi^{i_1}_{\nu)} = 0$, spanning the ${\footnotesize \yng(1,1)}$ eigenspace of the transversal vector Laplacian, and $(ii)$ longitudinal modes  $\varphi^{i_0}_\mu = \nabla_\mu \varphi^i$, $i_0=1,\ldots,D^{d+2}_1$, satisfying $\nabla_\mu \nabla_\nu \varphi^{i_0} + g_{\mu\nu} \varphi^{i_0} = 0$, with the scalar $\varphi^{i_0}$ modes spanning the  ${\footnotesize \yng(1)}$ eigenspace of the scalar Laplacian on $S^{d+2}$.  

We can extend the above to general $s,d$ by observing the following key relation:\footnote{This can be checked for any given $d$ from the Weyl dimension formula of appendix \ref{app:Weyldim}, or from the general $d$ formula given in e.g.\ \cite{Rubin:1983be,doi:10.1063/1.530850}, or proven directly (with some effort) by an $SO(d+2) \to SO(2) \times SO(d)$ reduction. It has a stronger version as an $\so(d+2)$ character relation: $\chi^{\so(d+2)}_{n, s}(x)= D^{d}_{s} \, \chi^{\so(d+2)}_{n}(x)-D^{d}_{n+1} \, \chi^{\so(d+2)}_{s-1}(x)$. 
} 
\begin{align} \label{coin}
 \boxed{D^{d+2}_{n, s}=D^{d+2}_{n} D^{d}_{s}-D^{d+2}_{s-1}D^{d}_{n+1}} 
\end{align}
which together with the explicit expression (\ref{Dsods2}) for $D_s^d$ with $d \geq 3$ 
immediately leads to 
\begin{align}
 f_{\rm TT}(q) &= \sum_{n \geq -1} D^{d+2}_{n,s} \, q^{\frac{d}{2}+n} - \sum_{n=-1}^{s-1} D^{d+2}_{n,s} \, q^{\frac{d}{2}+n} \label{fTT1} \\
 &=\frac{1+q}{1-q} \biggl( D^d_s \cdot \frac{ q^{\frac{d}{2}}}{(1-q)^d}- D^{d+2}_{s-1} \cdot \frac{ q^{\frac{d-2}{2}}}{(1-q)^{d-2}} \biggr) + \sum_{n=-1}^{s-2} D^{d+2}_{s-1,n+1} \, q^{\frac{d}{2}+n}  \, .
\end{align} 
To rewrite the finite sum we used $D_{n,s}^{d+2}=-D_{s-1,n+1}^{d+2}$ and $D_{s-1,s}^{d+2}=0$, both of which follow from (\ref{coin}). Substituting this into the integral (\ref{ZPIperpformula}), we get
\begin{align}
 \log Z_{\rm TT} = \log Z_{\rm bulk} - \log Z_{\rm edge} + \log Z_{\rm res} \, , 
\end{align} 
where $\log Z_{\rm bulk}$ and $\log Z_{\rm edge}$ are the character integrals  defined in (\ref{formuhs2})-(\ref{chibulkedgeprev}), and, evaluating the integral of the remaining finite sum as in (\ref{explicitfinite}),
\begin{align} \label{Zres}
 \log Z_{\rm res} = \sum_{n=-1}^{s-2} D^{d+2}_{s-1,n+1} \!\int\! \frac{dt}{2t} \, \bigl( q^{\frac{d}{2}+n + i \nu} +q^{\frac{d}{2}+n - i \nu}  \bigr)
     =
   \sum_{n=-1}^{s-2} D^{d+2}_{s-1,n+1} \log \frac{M}{\sqrt{(\frac{d}{2} + n)^2+\nu^2}} 
\end{align}
The term $\log Z_{\rm res}$ has a logarithmic UV-divergence: 
\begin{align} \label{logZreslogdiv}
 \log Z_{\rm res} = c_s \log M + \cdots, \qquad  c_s = \sum_{n=-1}^{s-2} D^{d+2}_{s-1,n+1} = D^{d+3}_{s-1,s-1} = N_{\rm CKT} \, ,
\end{align} 
where $N_{\rm CKT} = D^{d+3}_{s-1,s-1}$ is the number of rank $s-1$ conformal Killing tensors on $S^{d+2}$ \cite{Eastwood:2002su}. This identity has a group theoretic origin as an $\so(d+3) \to \so(d+2)$ branching rule generalizing (\ref{CKVbranching}). For example for $s=4$:
\begin{align} \label{CKTbranching}
 \yng(3,3) \quad \to \quad  \yng(3,3) \,\, + \,\,  \yng(3,2) \,\, + \,\,  \yng(3,1)
 \,\, + \,\,  \yng(3)  \, .
\end{align}
Geometrically this reflects the fact that the rank $s-1=3$ Killing tensor modes split up into $4$ types: Schematically $\varphi^{i_3}_{\mu_1 \mu_2 \mu_3}$, $i_3=1,\ldots,D_{3,3}$; $\varphi^{i_2}_{\mu_1 \mu_2 \mu_3} \sim \nabla_{(\mu_1} \varphi^{i_2}_{\mu_2 \mu_3)}$, $i_2 = 1, \ldots, D_{3,2}$; $\varphi^{i_1}_{\mu_1 \mu_2 \mu_3} \sim \nabla_{(\mu_1} \nabla_{\mu_2} \varphi^{i_1}_{\mu_3)}$, $i_1=1,\ldots,D_{3,1}$; $\varphi^{i_0}_{\mu_1 \mu_2 \mu_3} \sim \nabla_{(\mu_1} \nabla_{\mu_2} \nabla_{\mu_3)} \varphi^{i_0}$, $i_0=1,\ldots,D_{3}$, where the $\varphi^{i_r}_{\mu_1 \cdots \mu_r}$ span the eigenspace of the TT spin-$r$ Laplacian labeled by the above Young diagrams.   

As pointed out in examples above and discussed in more detail below, the log-divergence of $\log Z_{\rm res}$ is inconsistent with locality, hence $Z_{\rm PI} \neq Z_{\rm TT}$: locality must be restored by the non-TT part of the path integral. Below we argue this part in fact {\it exactly cancels} the $\log Z_{\rm res}$ term, thus ending up with $\log Z_{\rm PI} = \log Z_{\rm bulk} - \log Z_{\rm edge}$, i.e.\  the character formula (\ref{formuhs2}). 

\subsubsection*{Full path integral $Z_{\rm PI}$: locality constraints}

The full, manifestly covariant, local path integral takes the form (a simple example is (\ref{ZTTnonTTexample})):
\begin{align}
 Z_{\rm PI} = Z_{\rm TT} \cdot Z_{\text{non-TT}} = Z_{\rm bulk} \cdot Z_{\rm edge}^{-1} \, \cdot \,  Z_{\rm res} \cdot Z_{\text{non-TT}} \, .
\end{align}
All UV-divergences of $\log Z_{\rm PI}$ are {\it local}, in the sense they can be canceled by local counterterms, more specifically local curvature invariants of the background metric. In particular for odd $d+1$, this implies there cannot be any logarithmic divergences at all, as there are no curvature invariants of odd mass dimension. 
Recall from (\ref{logZreslogdiv}) that the term $\log Z_{\rm res}$ is logarithmically divergent. For odd $d+1$, this is clearly the only log-divergent contribution to $\log Z_{\rm TT}$, as the integrands of both $\log Z_{\rm bulk}$ and $\log Z_{\rm edge}$ are even in $t$ in this case. More generally, for even or odd $d+1$, $\log Z_{\rm res}$ is the only {\it nonlocal} log-divergent contribution to $\log Z_{\rm TT}$, as follows from the result of \cite{Fradkin:1983mq,Tseytlin:2013fca} mentioned below (\ref{massiveZPIdefin}), combined with the observation in (\ref{logZreslogdiv}) that $c_s = N_{\rm CKT}$. Therefore the log-divergence of $\log Z_{\rm res}$ must be canceled by an equal log-divergence in $\log Z_{\text{non-TT}}$ of the opposite sign.

The simplest way this could come about is if $Z_{\text{non-TT}}$ exactly cancels $Z_{\rm res}$, that is if
\begin{align} \label{conj}
 Z_{\text{non-TT}} = Z_{\rm res}^{-1} = \prod_{n=-1}^{s-2}\biggl(M^{-1}\sqrt{(\tfrac{d}{2}+n)^2 + \nu^2} \, \biggr)^{D_{s-1,n+1}^{d+2}}  \qquad 
 \Rightarrow \qquad Z_{\rm PI}  = \frac{Z_{\rm bulk}}{Z_{\rm edge}} \, .
\end{align}
Note furthermore that
from (\ref{fTT1}), or from (\ref{Zres}) and $D^{d+2}_{s-1,n+1}=-D^{d+2}_{n,s}$, it follows this identification is equivalent to the following simple prescription: {\it The full $Z_{\rm PI}$ is obtained from $Z_{\rm TT}$ by extending the TT eigenvalue sum $\sum_{n \geq s}$ in (\ref{ZPIperpformula}) down to $\sum_{n \geq -1}$}:
\begin{align} \label{ZPItrueformulaB}
 \log Z_{\rm PI} = \int_0^\infty \frac{dt}{2t} \, \bigl( q^{i\nu} +q^{-i\nu}  \bigr) \sum_{n \geq -1} D^{d+2}_{n, s} \, q^{\frac{d}{2}+n} \, ,
\end{align} 
i.e.\ (\ref{ZPItrueformula}).
In what follows we establish this is indeed the correct identification. We start by showing it precisely leads to the correct spin-$s$ unitarity bound, and that it moreover exactly reproduces the critical mass (``partially massless'') thresholds at which a new set of terms in the action defining the path integral $Z_{\text{non-TT}}$ fails to be positive definite. Assisted by those insights, it will then be rather clear how (\ref{conj}) arises from explicit path integral computations.       

\subsubsection*{Unitarity constraints}

A significant additional piece of evidence beyond consistency with locality is consistency with unitarity. It is clear that both the above integral (\ref{ZPItrueformulaB}) for $\log Z_{\rm PI}$ and the integral (\ref{ZPIperpformula}) for $\log Z_{\rm TT}$ are real provided $\nu$ is either real or imaginary. Real $\nu$ corresponds to the principal series $\Delta=\frac{d}{2} + i \nu$, while imaginary $\nu = i \mu$ corresponds to the complementary series $\Delta = \frac{d}{2} - \mu \in \IR$. In the latter case there is in addition a bound on $|\mu|$ beyond which the   
integrals cease to make sense, due to the appearance of negative powers of $q=e^{-t}$ and the integrand blowing up at $t \to \infty$. The bound can be read off from the term with the smallest value of $n$ in the sum. In the $Z_{\rm TT}$ integral (\ref{ZPIperpformula}) this is the $n=s$ term $\propto q^{\frac{d}{2}+s \pm \mu}$, yielding a bound $|\mu|<\frac{d}{2}+s$. 
In the $Z_{\rm PI}$ integral (\ref{ZPItrueformulaB}), assuming $s \geq 1$, this is the $n=-1$ term $\propto q^{\frac{d}{2}-1 \pm \mu}$, so the bound becomes much tighter:  
\begin{align} \label{Unbounds}
   |\mu| < \frac{d}{2} - 1 \,  \qquad (s \geq 1) \, .
\end{align}
This is exactly the correct unitarity bound for the spin-$s \geq 1$ complementary series representations of $SO(1,d+1)$ \cite{10.3792/pja/1195522333,10.3792/pja/1195523378,10.3792/pja/1195523460,Basile:2016aen}. In terms of the mass $m^2=(\frac{d}{2}+s-2)^2-\mu^2$ in (\ref{convmdef}), this becomes $m^2> (s-1)(d-3+s)$, also known as the Higuchi bound \cite{Higuchi:1986py} (a convenient concise summary is given in \cite{Lust:2019lmq} s.a.\ \cite{Noumi:2019ohm}). From a path integral perspective, this bound can be understood as the requirement that the full off-shell action is positive definite \cite{Zinoviev:2001dt}, so indeed $\log Z_{\rm PI}$ should diverge exactly when the bound is violated. Moreover, we get new divergences in the integral formula for $\log Z_{\text{non-TT}}$, according to the above identifications, each time $|\mu|$ crosses a critical value $\mu_{*n}=\frac{d}{2} + n$, where $n=-1,0,1,2,\ldots,s-2$. These correspond to critical masses $m^2_{*n} = (\frac{d}{2}+s-2)^2-(\frac{d}{2} + n)^2 = (s-2-n)(d+s-2+n)$, which on the path integral side precisely correspond to the points where a new set of terms in the action fails to be positive definite.    
\cite{Zinoviev:2001dt}. 

This establishes the terms in the integrand of (\ref{Zres}), or equivalently the extra terms $n=-1,\ldots,s-2$ in (\ref{ZPItrueformulaB}), have exactly the correct powers of $q$ to match with $\log Z_{\text{non-TT}}$. It does not yet confirm the precise values of the coefficients $D_{n,s}^{d+2}$ --- except for their sum (\ref{logZreslogdiv}), which was fixed earlier by the locality constraint. To complete the argument, we determine the origin of these coefficients from the path integral point of view in what follows.         

\subsubsection*{Explicit path integral considerations}

Complementary to but guided by the above general considerations, we now turn to more concrete path integral calculations to confirm the expression (\ref{conj}) for $Z_{\rm non-{\rm TT}}$, focusing in particular on the origin of the coefficients $D_{n,s}^{d+2}$.

\vskip3mm\noindent {\it Spin 1:}\\
We first consider the familar $s=1$ case, a  vector field of mass $m$, related to $\nu$ by (\ref{convmdef}) as $m=\sqrt{(\frac{d}{2}-1)^2+\nu^2}$. The local field content in the Stueckelberg description consists of a vector $\phi_\mu$ and a scalar $\chi$, with action and gauge symmetry given by
\begin{align} \label{SStueck1}
 S_0 = \int \nabla_{[\mu} \phi_{\nu]} \nabla^{[\mu} \phi^{\nu]}  + \tfrac{1}{2}  (\nabla_\mu \chi - m \, \phi_\mu)(\nabla^\mu \chi - m \, \phi^\mu)  \, ;  \qquad 
  \delta \chi = m \xi \, , \quad \delta \phi_\mu = \nabla_\mu \xi \, .
\end{align}
Gauge fixing the path integral by putting $\chi \equiv 0$, we get the gauge-fixed action
\begin{align}
 S =  \int \nabla_{[\mu} \phi_{\nu]} \nabla^{[\mu} \phi^{\nu]} + \tfrac{1}{2}  m^2 \phi_\mu \phi^\mu  +  m \bar c c \, ,
\end{align}
with BRST ghosts $c$, $\bar c$. Decomposing $\phi_\mu$ into a transversal and longitudinal part, $\phi_\mu = \phi^T_{\mu} + \phi'_{\mu}$, we can decompose the path integral as $Z_{\rm PI} = Z_{\rm TT} \cdot Z_{\text{non-TT}}$ with
\begin{align} \label{ZTTnonTTexample}
 Z_{\rm TT} = \int \CD \phi^T \, e^{-\frac{1}{2}\int \phi^T (-\nabla^2+\m1^2)\phi^T} \, , \qquad 
 Z_{\text{non-TT}} = \int \CD \phi' \CD c \CD \bar c \, e^{-\int \frac{1}{2} m^2 \phi^{\prime 2} + m \bar c c} \, ,
\end{align}
Both the ghosts and the longitudinal vectors $\phi'_\mu = \nabla_\mu \varphi$ have an    mode decomposition in terms of orthonormal real scalar spherical harmonics $Y_i$.\footnote{Explicitly, $c=\sum_i c_i Y_i$, $\bar c=\sum_i \bar c_i Y_i$, $\phi'_\mu = \sum_{i:\lambda_i \neq 0} \phi'_i \nabla_\mu Y_i/\sqrt{\lambda_i}$ where $\nabla^2 Y_i = -\lambda_i Y_i$, $\int Y_i Y_j = \delta_{ij}$.} In our heat kernel regularization scheme, each longitudinal vector mode integral gives a factor $M/m$, which is exactly canceled by a factor $m/M$ from integrating out the corresponding ghost mode.\footnote{A priori there might be a relative numerical factor $\kappa$ between ghost and longitudinal factors, depending on the so far unspecified normalization of the measure $\CD c$. But because $c$ is local, {\it unconstrained}, rescaling $\CD c = \prod_i dc_i \to \prod_i (\lambda dc_i)$ merely amounts to a trivial constant shift of the bare cc. So we are free to take $\kappa = 1$.} 
 However there is one ghost mode which remains unmatched: the constant mode. A constant scalar does not map to a longitudinal vector mode, because $\phi'_{\mu} = \nabla_\mu \varphi = 0$ for constant $\varphi$. Thus we end up with a ghost factor $m/M$ in excess, and 
\begin{align}
 Z_{\text{non-TT}} = m/M = M^{-1} \sqrt{(\tfrac{d}{2}-1)^2+\nu^2} \, ,
\end{align}
in agreement with (\ref{conj}) for $s=1$. 

\vskip3mm \noindent {\it Spin 2:}\\
For $s=2$, the analogous Stueckelberg action involves a symmetric tensor $\phi_{\mu\nu}$, a vector $\chi_\mu$, and a scalar $\chi$, subject to the gauge transformations \cite{Zinoviev:2001dt}
\begin{align}
 \delta \chi = a_{-1} \, \xi \, , \qquad 
 \delta \chi_\mu = a_0 \, \xi_\mu + \sqrt{\tfrac{d-1}{2 d}} \, \nabla_\mu \xi \, , \qquad
 \delta \phi_{\mu\nu} = \nabla_\mu \xi_\nu + \nabla_\nu \xi_\mu +\sqrt{\tfrac{2}{d(d-1)}} \,  a_0 \, \xi \, ,
\end{align}
where $a_0 \equiv m$ and $a_{-1} \equiv \sqrt{m^2-(d-1)}$. Equivalently, recalling (\ref{convmdef}), $a_n = \sqrt{(\tfrac{d}{2}+n)^2+\nu ^2}$. Gauge fixing by putting $\chi=0$, $\chi_\mu=0$, we get a ghost action 
\begin{align}
 S_{\rm gh} = \int a_{-1}  \, \bar c  c + a_0 \, \bar c^\mu  c_\mu \, .
\end{align}
We can decompose $\phi_{\mu\nu}$ into a TT part and a non-TT part orthogonal to it as $\phi_{\mu\nu} = \phi_{\mu\nu}^{\rm TT} + \phi'_{\mu\nu}$, where $\phi'_{\mu\nu}$ can be decomposed into vector and scalar  modes  as $\phi'_{\mu\nu} = \nabla_{(\mu} \varphi_{\nu)} + g_{\mu\nu} \varphi$. Analogous to the $s=1$ example, we should expect that integrating out $\phi'$ cancels against integrating out the ghosts, up to unmatched modes of the latter. The unmatched modes correspond to mixed vector-scalar modes  solving $\nabla_{(\mu} \varphi_{\nu)} + g_{\mu\nu} \varphi = 0$. This is equivalent to the conformal Killing equation. Hence the unmatched modes are the conformal Killing modes. As discussed below (\ref{CKVbranching}), the conformal Killing modes split according to ${\tiny \yng(1,1) \to \yng(1,1) + \yng(1)}$ into $D_{1,1}$ vector ${\tiny \yng(1,1)}$-modes and $D_{1,0}$ scalar ${\tiny \yng(1)}$-modes. Integrating out the ${\tiny \yng(1,1)}$-modes of the  vector ghost $c_\mu$ then yields an unmatched factor $(a_0/M)^{D_{1,1}}$, while integrating out the ${\tiny \yng(1)}$-modes of the scalar ghost $c$ yields an unmatched factor $(a_{-1}/M)^{D_1}$. 
All in all, we get
\begin{align}
 Z_{\text{non-TT}} = (a_{-1}/M)^{D_{1,0}} (a_0/M)^{D_{1,1}}  = \Bigl(M^{-1}\sqrt{(\tfrac{d}{2}-1)^2+\nu ^2} \Bigr)^{D_{1,0}}\Bigl(M^{-1}\sqrt{(\tfrac{d}{2})^2+\nu ^2} \Bigr)^{D_{1,1}}  
\end{align}
in agreement with (\ref{conj}) for $s=2$. 

\vskip3mm \noindent {\it Spin $s$:}\\
The pattern is now clear: according to \cite{Zinoviev:2001dt}, the Stueckelberg system for a massive spin-$s$ field consists of an unconstrained symmetric $s$-index tensor $\phi^{(s)}$ and of a tower of unconstrained symmetric $s'$-index auxiliary Stueckelberg fields $\chi^{(s')}$ with $s'=0,1,\ldots,s-1$, with gauge symmetries of the form
\begin{align}
 \delta \chi^{(s')} = a_{s'-1} \, \xi^{(s')} + \cdots \, , \qquad \delta \phi^{(s)} = \cdots \, , \qquad a_n \equiv \sqrt{(\tfrac{d}{2}+n)^2+\nu ^2} \, ,
\end{align}
where the dots indicate terms we won't technically need --- which is to say, as transpired from $s=1,2$ already, we need very little indeed. The ghost action is $S = \sum_{s'=0}^{s-1} a_{s'-1} \bar c^{(s')} c^{(s')}$. The unmatched modes correspond to the conformal Killing tensors modes on $S^{d+1}$, decomposed for say $s=4$ as in (\ref{CKTbranching}) into $D_{3,3}$ ${\tiny \yng(3,3)}$-modes, $D_{3,2}$ ${\tiny \yng(3,2)}$-modes, $D_{3,1}$ ${\tiny \yng(3,1)}$-modes, and $D_{3,0}$ ${\tiny \yng(3)}$-modes. The corresponding unmatched modes of respectively $c^{(3)}$, $c^{(2)}$, $c^{(1)}$ and $c^{(0)}$ then integrate to unmatched factors $(a_2/M)^{D_{3,3}} (a_1/M)^{D_{3,2}} (a_0/M)^{D_{3,1}} (a_{-1}/M)^{D_{3,0}}$. For general $s$:
\begin{align}
 Z_{\text{non-TT}}  = \prod_{s'=0}^{s-1} \bigl(a_{s'-1}/M \bigr)^{D_{s-1,s'}}  = \prod_{n=-1}^{s-2}\biggl(M^{-1}\sqrt{(\tfrac{d}{2}+n)^2 + \nu^2} \, \biggr)^{D_{s-1,n+1}^{d+2}} \, ,
\end{align}
in agreement with (\ref{conj}) for general $s$. This establishes our claims.

The above computation was somewhat schematic of course, and one could perhaps still worry about missed purely numerical factors independent of $\nu$, perhaps leading to an additional finite constant term being added to our final formulae (\ref{formuhs2}) -(\ref{ZPItrueformula}) for $\log Z_{\rm PI}$. However at fixed UV-regulator scale, the limit $\nu \to \infty$ of these final expressions manifestly approaches zero, as should be the case for particles much heavier than the UV cutoff scale. This would not be true if there was an additional constant term. Finally, we carefully checked the analogous result in the massless case (which has a more compact off-shell formulation \cite{Fronsdal:1978rb}),  discussed in section \ref{sec:massless},  by direct path integral computations in complete gory detail \cite{Law:2020cpj}, for all $s$. 

Also, the result is pretty.

\subsection{General massive representations} \label{sec:EPIC}

Here we give a generalization of (\ref{ZPItrueformula})  for arbitrary massive representations of the dS$_{d+1}$ isometry group $SO(1,d+1)$.

Massive irreducible representations of $SO(1,d+1)$ are labeled by a dimension $\Delta=\frac{d}{2} + i \nu$ and an $\so(d)$ highest weight $S=(s_1,\ldots,s_r)$ \cite{10.3792/pja/1195522333,10.3792/pja/1195523378,10.3792/pja/1195523460}. The massive spin-$s$ case considered in (\ref{ZPItrueformula}) corresponds to $S=(s,0,\ldots,0)$, a totally symmetric tensor field. More general irreps correspond to more general mixed-symmetry fields. The analog of (\ref{ZPITTmassive}) in this generalized setup is 
\begin{align} \label{ZnaivePIgen}
 \log Z^{\text{``TT''}}_{\rm PI} = \pm \int_0^\infty \frac{d\tau}{2 \tau} \, e^{-\epsilon^2/4\tau} \, \sum_{n \geq s_1} \, D_{n,S}^{d+2} \, e^{-\tau ( (n+\frac{d}{2})^2 + \nu^2 )} \, ,
\end{align}
where for bosons the sum runs over integer $n$ with an overall $+$ sign and for fermions the sum runs over half-integer $n$ with an overall $-$ sign.  The dimensions of the $\so(d+2)$ irreps $(n,S)$ are given explicitly as polynomials in $(n,s_1,\ldots,s_r)$ by the Weyl dimension formulae (\ref{Weyldimeven})-(\ref{Weyldimodd}). From this it can be seen that $D^{d+2}_{n,S}$ is (anti-)symmetric under reflections about $n=-\frac{d}{2}$, more precisely 
\begin{align} \label{DnSsymm}
 D^{d+2}_{n,S} = (-1)^d D^{d+2}_{-d-n,S}.
\end{align}
Moreover the exponent in (\ref{ZnaivePIgen}) is symmetric under the same reflection. The most natural extension of the sum is therefore to {\it all} (half-)integer $n$, taking into account the sign in (\ref{DnSsymm}) for odd $d$, and adding an overall factor $\frac{1}{2}$ to correct for double counting, suggesting 
\begin{align} \label{ZPIgenconj1}
 \log Z_{\rm PI} = \pm \frac{1}{2} \int_0^\infty \frac{d\tau}{2 \tau} \, e^{-\epsilon^2/4\tau} \,\sum_n \, \sigma_d(\tfrac{d}{2}+n) \, D_{n,S}^{d+2} \, e^{-\tau ( (n+\frac{d}{2})^2 + \nu^2 )} \, ,
\end{align}
where $\sigma_d(x) \equiv 1$ for even $d$ and $\sigma_d(x) \equiv \mbox{sign}(x)$ for odd $d$. Equivalently, in view of (\ref{DnSsymm})
\begin{align}  \label{ZPIgenconj2}
 \boxed{\log Z_{\rm PI} = \pm \int_0^\infty \frac{d\tau}{2 \tau} \, e^{-\epsilon^2/4\tau} \, \sum_n \, \Theta(\tfrac{d}{2}+n) \, D_{n,S}^{d+2} \, e^{-\tau ( (n+\frac{d}{2})^2 + \nu^2 )}} 
\end{align}
where $n \in \IZ$ for bosons and $n \in \frac{1}{2} + \IZ$ for fermions, and
\begin{align}
 \mbox{$\Theta(x) = 1$ for $x>0$, \quad  $\Theta(0) = {\displaystyle{\frac{1}{2}}}$, \quad $\Theta(x) = 0$ for $x<0$.}
\end{align}
At first sight this seems to be different from the extension to $n \geq -1$ in (\ref{ZPItrueformula}) for the spin-$s$ case $S=(s,0,\ldots,0)$. However it is actually the same, as (\ref{Weyldimeven})-(\ref{Weyldimodd}) imply that 
$D_{n,S}$ vanishes for $2-d \leq n \leq -2$ when $S=(s,0,\ldots,0)$. 

The obvious conjecture is then that (\ref{ZPIgenconj2}) is true for general massive representations. Here are some consistency checks, which are satisfied precisely for the sum range in (\ref{ZPIgenconj2}):

\noindent $\bullet$ {\it Locality:} For even $d$, the summand in (\ref{ZPIgenconj1}) is analytic in $n$. Applying the Euler-Maclaurin formula to extract the $\tau \to 0$ asymptotic expansion of the sum gives in this case
\begin{align}
 \sum_n \,  D_{n,S}^{d+2} \, e^{-\tau (n+\frac{d}{2})^2 } \sim \int_{-\infty}^\infty dn \,  D_{n,S}^{d+2} \, e^{-\tau (n+\frac{d}{2})^2 } \, . 
\end{align} 
The symmetry (\ref{DnSsymm}) tells us that the integrand on the right hand side is even in $x \equiv n + \frac{d}{2}$. Since $\int dx \, x^{2k} \, e^{-\tau x^2} \propto \tau^{-k-1/2}$, this implies the absence of $1/\tau$ terms in the $\tau \to 0$ expansion of the integrand in  (\ref{ZPIgenconj1}), and therefore, in contrast to (\ref{ZnaivePIgen}), the absence of nonlocal log-divergences, as required by locality of $Z_{\rm PI}$ in odd spacetime dimension $d+1$. 

\noindent $\bullet$ {\it Bulk $-$ edge structure:} By following the usual steps, we can rewrite (\ref{ZPIgenconj2}) as
\begin{align} \label{ZPItrueformulaGeneral}
  \boxed{\log Z_{\rm PI} =  \inteps \frac{dt}{2t} \, F(e^{-t}) \, , \qquad  \, 
 F(q) = \pm \bigl( q^{i\nu} +q^{-i\nu}  \bigr) \sum_n \Theta\bigl(\tfrac{d}{2}+n \bigr) \, D^{d+2}_{n,S} \, q^{\frac{d}{2}+ n} } 
\end{align}
Using (\ref{Weyldimeven})-(\ref{Weyldimodd}), this can be seen to sum up to the form $\log Z_{\rm PI} = \log Z_{\rm bulk} - \log Z_{\rm edge}$, where $Z_{\rm bulk}$ is the physically expected bulk character formula for an ideal gas in the dS$_{d+1}$ static patch consisting of massive particles in the $(\Delta,S)$ UIR of $SO(1,d+1)$, and $Z_{\rm edge}$  can be interpreted as a Euclidean path integral of local fields living on the $S^{d-1}$ edge/horizon. 

\noindent $\bullet$ {\it Unitarity:} Note that for $\Delta = \frac{d}{2} + \mu$ with $\mu \equiv i\nu$ real, we get a bound on $\mu$ from requiring $t \to \infty$ (IR) convergence of the integral (\ref{ZPItrueformulaGeneral}),  generalizing (\ref{Unbounds}), namely  
\begin{align}
 |\mu| < \frac{d}{2} + n_*(S)  \, , 
\end{align}
where  $n_*(S)$ is the lowest value of $n$ in the sum for which $D_{n,S}^{d+2}$ is nonvanishing. This coincides again with the unitarity bound on $\mu$ for massive representations of $SO(1,d+1)$ \cite{10.3792/pja/1195522333,10.3792/pja/1195523378,10.3792/pja/1195523460,Basile:2016aen}.  Recalling the discussion below (\ref{Unbounds}), this can be viewed as a generalization of the Higuchi bound to arbitrary representations.

\vskip3mm

Combining (\ref{ZPIgenconj2}) with (\ref{ZEXACT}), we thus arrive at an exact closed-form solution for the Euclidean path integral on the sphere for arbitrary massive field content.

\section{Derivations 
for massless higher spins} \label{app:ZPImassless}

In this appendix we derive (\ref{ZPIFINAL}) and provide details of various other points summarized in section \ref{sec:massless}. 

\subsection{Bulk partition function: $Z_{\rm bulk}$} \label{sec:Zbulkmassless}

The bulk partition function $Z_{\rm bulk}$ as defined in (\ref{Zbulkdef}) for a massless spin-$s$ field is given by
\begin{align} \label{Zbulkfirst}
 Z_{\rm bulk} = \int \frac{dt}{2t} \, \frac{1+q}{1-q} \, \chi_{{\rm bulk},s}(q) \, ,
\end{align} 
where $q=e^{-t}$, and $\chi_{{\rm bulk},s}(q) = {\rm tr} \, q^{iH}$ in the case at hand is the (restricted) $q$-character of the massless spin-$s$  $SO(1,d+1)$ UIR. For generic $d$, this UIR is part of the exceptional series \cite{Basile:2016aen}.  More precisely in the notation of \cite{10.3792/pja/1195522333,10.3792/pja/1195523378,10.3792/pja/1195523460} it is the $D^j_{S;p}$ representation, with $p=0$, $j=(d-4)/2$ for even $d$, $j=(d-3)/2$ for odd $d$, and $S=(s,s,0,\ldots,0)$. 
In the notation of \cite{Basile:2016aen} this is the exceptional series with $\Delta=p=2$, $S=(s,s,0,\ldots,0)$.\footnote{For $d=3$, it is in the discrete series, but (\ref{excserieschi}) still applies. 
} The characters $\chi_{{\rm bulk},s}(q)$ for these irreps are quite a bit more intricate than their massive counterparts. The full $SO(1,d+1)$ characters $\tilde \chi(g)$ were obtained in \cite{10.3792/pja/1195522333,10.3792/pja/1195523378,10.3792/pja/1195523460}. Restricting to $g=e^{-i t H}$ gives $\chi_{{\rm bulk},s}(t)$,\footnote{Actually we obtained the formula from (\ref{flippedchargenPM}), then Mathematica checked agreement with \cite{10.3792/pja/1195522333,10.3792/pja/1195523378,10.3792/pja/1195523460}.} 
\begin{align} \label{excserieschi}
 (1-q)^d \, \chi_{{\rm bulk},s}(q) =& 
 \bigl(1-(-1)^d \bigr) \bigl( D_s^d \, q^{s+d-2} - D_{s-1}^d \, q^{s+d-1} \bigr)  \\
 &  + \sum_{m=0}^{r-2} (-1)^{m} D^d_{s,s,1^{m}} \bigl( q^{2+m} + (-1)^d q^{d-2-m} \bigr) \, , \qquad r \equiv \mbox{rank $\so(d)$} = \lfloor \tfrac{d}{2} \rfloor \, , \nn
\end{align}
Here we used the notation of  \cite{Basile:2016aen}: the $\so(d)$ irrep $(s,s,1^m) \equiv (s,s,1,\ldots,1,0,\ldots,0)$ with $1$ repeated $m$ times.
The degeneracies $D^d_{s,s,1^m}$ can be read off from (\ref{Weyldimeven})-(\ref{Weyldimodd}).   Some explicit low-dimensional examples are
\begin{equation} \label{chiexcex}
\begin{array}{ll|l}
 d & r & (1-q)^d \, \chi_{{\rm bulk},s}(q)  \\
 \hline
 3 & 1 & 2  D_{s}^3 \, q^{s+1} -2  D_{s-1}^3 \, q^{s+2} \\
 4 & 2 & D_{s,s}^4 \, 2 \, q^2  \\
 5 & 2 & D_{s,s}^5 \left(q^2-q^3\right)  + 2 D_{s}^5 q^{s+3} -2 D_{s-1}^5 q^{s+4}  \\
 6 & 3 & D_{s,s}^6 \left(q^2+q^4\right) - D_{s,s,1}^6 \, 2 \, q^3  \\
 7 & 3 & D_{s,s}^7 \left(q^2-q^5\right) - D_{s,s,1}^7 \left(q^3-q^4\right) 
 + 2 D_{s}^7 \, q^{s+5}   -2 D_{s-1}^7 \, q^{s+6}  \, , \\
\end{array}
\end{equation}
where $D_s^3 = 2s+1$, $D_{s,s}^4 = 2s+1$, $D_s^5=\frac{1}{6} (s+1) (s+2) (2 s+3)$, $D_{s,s}^5 = \frac{1}{3} (2 s+1) (s+1) (2 s+3)$, $D_{s,s}^6=\frac{1}{12} (s+1)^2 (s+2)^2 (2 s+3)$, $D_{s,s,1}^6=\frac{1}{12} s (s+1) (s+2) (s+3) (2 s+3)$, etc.      
For $s=1$, the character can be expressed more succinctly as
\begin{align} \label{spin1casechis}
 \chi_{{\rm bulk},1}(q) = d \cdot \frac{q^{d-1}+q}{(1-q)^d}  - \frac{q^d+1}{(1-q)^d} + 1 \, .
\end{align}
With the exception of the $d=3$ case, the above $\so(1,d+1)$ $q$-characters encoding the $H$-spectrum of massless spin-$s$ fields in dS$_{d+1}$ are very different from the $\so(2,d)$   characters encoding the energy spectrum of massless spin-$s$ fields in AdS$_{d+1}$ with standard boundary conditions, the latter being $\chi_{{\rm bulk},s}^{{\rm AdS}_{d+1}} = (D_s^d \, q^{s+d-2} - D_{s-1}^d \, q^{s+d-1})/(1-q)^d$.     In particular for $d \geq 4$, the lowest power $q^\Delta$ appearing in the $q$-expansion of the character is $\Delta=2$, and is associated with the $\so(d)$ representation $S=(s,s)$, i.e.\ ${\tiny \Yvcentermath1 \yng(1,1),\yng(2,2),\yng(3,3),\ldots}$ for $s=1,2,3,\ldots$, whereas for the $\so(2,d)$  character this is $\Delta=s+d-2$ and $S=(s)$.  An explanation for this was given in \cite{Basile:2016aen}: in dS, $S$ should be thought of as associated with the higher-spin Weyl curvature tensor of the gauge field rather than the gauge field itself. 

This fits well with the interpretation of the expansion 
\begin{align} \label{QNMcountml}
 \chi(q) = \sum_r N_r q^r \, ,
\end{align}
as counting the number $N_r$ of {\it physical} static patch quasinormal modes decaying as $e^{-r T}$ (cf.\ section \ref{sec:thermal} and appendix \ref{sec:resonances}).
Indeed for $d \geq 4$, the longest-lived physical quasinormal modes of a massless spin-$s$ field in the static patch of dS$_{d+1}$ always decay as $e^{-2 T}$ \cite{Sun:2020sgn}, which can be understood as follows.   Physical quasinormal modes of the southern static patch can be thought of as sourced by insertions of gauge-invariant\footnote{More precisely, invariant under linearized gauge transformations acting on the conformal boundary.} local operators on the past conformal boundary $T = -\infty$ of the static patch, or equivalently at the south pole of the past conformal boundary (or alternatively the north pole of future boundary) of global dS$_{d+1}$ \cite{Sun:2020sgn,Ng:2012xp,Jafferis:2013qia}. By construction, the dimension $r$ of the operator maps to the decay rate $r$ of the quasinormal mode  $\propto e^{-r T}$. For $s=1$, the gauge-invariant operator with the smallest dimension $r=\Delta$ is the magnetic field strength $F_{ij} = \partial_i A_j - \partial_j A_i$ of the  boundary gauge field $A_i$, which has $\Delta =  \dim \partial + \dim A = 1+1=2$.    
For $s=2$ in $d \geq 4$, the gauge-invariant operator with smallest dimension is the Weyl tensor of the boundary metric: $\Delta=2+0=2$. Similarly for higher-spin fields we get the spin-$s$ Weyl tensor, with $\Delta = s + 2-s = 2$. The reason $d=3$ is special is that the Weyl tensor vanishes identically in this case. To get a nonvanishing gauge-invariant tensor, one has to act with at least $2s-1$ derivatives (spin-$s$ Cotton tensor), yielding $\Delta=(2s-1)+(2-s)=s+1$. An extensive analysis is given in \cite{Sun:2020sgn}.   

\vskip2mm \noindent {\it Note on a literature disagreements:} The characters (\ref{excserieschi}) agree with the characters listed in the original work \cite{10.3792/pja/1195522333,10.3792/pja/1195523378,10.3792/pja/1195523460}, 
computed by undisclosed methods. They do not agree with those listed in the more recent work \cite{Basile:2016aen}, computed by Bernstein-Gelfand-Gelfand resolutions. Indeed \cite{Basile:2016aen} emphasized they disagreed with \cite{10.3792/pja/1195522333,10.3792/pja/1195523378,10.3792/pja/1195523460} for even $d$. More precisely, in their eq.\ (2.14) applied to $p=2$, $Y_p=(s,s,0,\ldots,0)$, $\vec x=0$, they find a factor $2=(1+(-1)^d)$ instead of the factor $(1-(-1)^d)=0$ in (\ref{excserieschi}). It is stated in \cite{Basile:2016aen} that on the other hand their results do agree with \cite{10.3792/pja/1195522333,10.3792/pja/1195523378,10.3792/pja/1195523460} for {\it odd} $d$. Actually we find this is not quite true either, as in that case eq.\ (2.13) in \cite{Basile:2016aen} applied to $p=2$, $Y_p=(s,s,0,\ldots,0)$, $\vec x=0$) has a factor $1$ instead of the factor $(1-(-1)^d)=2$ in (\ref{excserieschi}). Our Euclidean path integral result (\ref{flippedchargenPM}) coupled with the physics of section \ref{sec:thermal} strongly suggests the original results in \cite{10.3792/pja/1195522333,10.3792/pja/1195523378,10.3792/pja/1195523460} and (\ref{excserieschi}) are the correct versions. Further support is provided in \cite{Sun:2020sgn} by direct construction of higher-spin quasinormal modes. 

\subsection{Euclidean path integral: $Z_{\rm PI} = \ZG \Zchar$} \label{sec:masslessZPI}


The Euclidean path integral of a collection of gauge fields $\phi$ on $S^{d+1}$ is formally given by
\begin{align}  \label{formalZPI}
 Z_{\rm PI} = \frac{\int \CD \phi \, e^{-S[\phi]}}{{\rm \Vol}(\CG)}
\end{align}
where $\CG$ is the local gauge group generated by the local field $\xi$ appearing in (\ref{linGT}). This ill-defined formal expression is turned into something well-defined by BRST gauge fixing. A convenient gauge for higher-spin fields is the de Donder gauge. At the Gaussian level, the resulting analog of (\ref{ZPITTmassive}) is\footnote{A detailed discussion of normalization conventions left implicit here is given above and below (\ref{canonnorm}).}   
\begin{align} \label{ZPITTmassless}
 \log Z_{\rm TT} \equiv 
 \sum_{s} \int_0^\infty\frac{d\tau}{2\tau}  \, e^{-\epsilon^2/4\tau}  \Bigl(  \Tr_{\rm TT} \, e^{-\tau\left(-\nabla^2_{s,\rm TT} + m_{{\phi,s}}^2\right)} - 
  \Tr_{\rm TT} \, e^{-\tau\left(-\nabla^2_{s-1,\rm TT} + m_{\xi,s}^2\right)} \Bigr) \, ,
\end{align}
where $\sum_s$ sums over the spin-$s$ gauge fields in the theory (possibly with multiplicities) and
$m_{\phi,s}^2$ and $m_{\xi,s}^2$ are obtained from the relations below (\ref{FPeom}) using (\ref{phixidim}). The first term arises from the path integral over the TT modes of $\phi$, while the second arises from the TT part of the gauge fixing sector in de Donder gauge --- a combination of integrating out the TT part of the spin-$(s-1)$ ghost fields and the corresponding longitudinal degrees of freedom of the spin-$s$ gauge fields. The above (\ref{ZPITTmassless}) is the difference of two expressions of the form (\ref{ZPITTmassive}). Naively applying  the formula (\ref{ZPItrueformulaGeneral}) or (\ref{ZPItrueformula}) for the corresponding full $Z_{\rm PI}$, we get
\begin{align} \label{Znaive}
 Z_{\rm PI} = \exp \sum_s  \inteps \frac{dt}{2t} \, \hat F_s(e^{-t}) \, \qquad \mbox{(naive)}
\end{align}
where (assuming $d \geq 3$)
\begin{align}   \label{masslessF}
 \hat F_s(q) &= \sum_{n \geq -1} D^{d+2}_{n, s} \bigl( q^{s+d-2+n} +q^{2-s+n}  \bigr)    - \sum_{n \geq -1} D^{d+2}_{n, s-1}  \bigl( q^{s+d-1+n} +q^{1-s+n}  \bigr)    
\end{align}
However this is clearly problematic. One problem is that for $s \geq 2$, the above $\hat F_s(q)$ contains {\it negative} powers of $q=e^{-t}$, making (\ref{Znaive}) exponentially divergent at $t \to \infty$. The appearance of such ``wrong-sign'' powers of $q$ is directly related to the appearance of ``wrong-sign'' Gaussian integrals in the path integral, as can be seen for instance from the relation between (\ref{ZPItrueformulaGeneral}) and the heat kernel integral (\ref{ZPIgenconj2}). In the path integral framework, one deals with this problem by analytic continuation, generalizing the familiar contour rotation prescription for negative modes in the gravitational Euclidean path integral \cite{Hawking:1976ja}. Thus one defines $\int dx \, e^{-\lambda x^2/2}$ for $\lambda < 0$ by rotating $x \to i x$, or equivalently by rotating $\tau \to -\tau$ in the heat kernel integral. Essentially this just boils down to flipping any $\lambda<0$ to $-\lambda>0$. Since the Laplacian eigenvalues are equal to the products of the exponents appearing in the pairs $\bigl(q^{\Delta + n} + q^{d-\Delta+n} \bigl)$ in (\ref{masslessF}), the implementation of this prescription in our setup is to flip the negative powers $q^k$ in $\hat F_s(q) = \sum_k c_k \, q^k$ to positive powers $q^{-k}$, that is to say  replace
\begin{align} \label{Fsplusdef}
 \hat F_s(q) \to  F_s(q) \equiv \bigl\{ \hat F_s(q) \bigr\}_+ =  \Bigl\{ \sum_k c_k \, q^k \Bigr\}_+ \equiv \sum_{k < 0} c_k \, q^{-k} + \sum_{k \geq 0} c_k \, q^{k} \, .
\end{align}
In addition,  each negative mode path integral contour rotation produces a phase $\pm i$, resulting in a definite, finite overall phase in $Z_{\rm PI}$ \cite{Polchinski:1988ua}. The analysis of \cite{Polchinski:1988ua} translates to each corresponding flip in (\ref{Fsplusdef}) contributing with the {\it same} sign,\footnote{This can be seen in a more careful path integral analysis \cite{Law:2020cpj}.} hence to an overall phase $i^{-\pol_s}$ with $\pol_s$ the total degeneracy of negative modes in (\ref{masslessF}). Using $D^{d+2}_{n,s}= - D_{s-1,n+1}^{d+2}$:
\begin{align} \label{polphase}
 Z_{\rm PI} \to i^{-\pol_s} Z_{\rm PI} \, , \qquad  \pol_s 
 = \sum_{n'=0}^{s-2} D_{s-1,n'}^{d+2} + \sum_{n'=0}^{s-2} D_{s-2,n'}^{d+2} 
 =  D_{s-1,s-1}^{d+3} - D_{s-1,s-1}^{d+2} + D^{d+3}_{s-2,s-2}
\end{align}
In particular this implies $\pol_1=0$ and $\pol_2=d+3$ in agreement with \cite{Polchinski:1988ua}.

After having taken care of the {\it negative} powers of $q$, the resulting amended formula $Z_{\rm PI} = \inteps \frac{dt}{2t} \, F_s(q)$ is still problematic, however, as $F_s(q)$ still contains terms proportional to  $q^0$, causing the integral to diverges (logarithmically) for $t \to \infty$. These correspond to zeromodes in the original path integral. Indeed such zermodes were to be expected: they are due to the existence of normalizable rank $s-1$ traceless Killing tensors $\bar \xi^{(s-1)}$, which by definition satisfy $\nabla_{(\mu_1} \bar\xi_{\mu_2 \cdots \mu_{s})} = 0$, and therefore correspond to vanishing gauge transformations (\ref{linGT}), leading in particular to ghost zeromodes. Zeromodes of this kind must be omitted from the Gaussian path integral. They are easily identified in (\ref{masslessF}) as the values of $n$ for which a term proportional to $q^0$ appears. Since we are assuming $d > 2$, this is $n=s-2$ in the first sum and $n=s-1$ in the second. Thus we should refine (\ref{Fsplusdef}) to
\begin{align} \label{FplusminF0}
 \hat F_s \to F_s - F_s^0  \, ,
\end{align}
where $F_s^0 = D_{s-2,s}^{d+2} ( q^{2s+d-4} + 1 ) - D_{s-1,s-1}^{d+2} ( q^{2s+d-2} + 1 )$. Noting that $D^{d+2}_{s-2,s}=-D^{d+2}_{s-1,s-1}$ and $D_{s-1,s-1}^{d+2}$ is the number $N^{\rm KT}_{s-1}$ of rank $s-1$ traceless Killing tensors on $S^{d+1}$, we can rewrite this as
\begin{align} \label{F0def}
 F_s^0 &= - N^{\rm KT}_{s-1} \bigl( q^{2s+d-4} + 1 + q^{2s+d-2} + 1 \bigr) \, , \qquad N^{\rm KT}_{s-1} = D_{s-1,s-1}^{d+2} \, ,
\end{align}
making the relation to the existence of normalizable Killing tensors manifest.
For example $N^{\rm KT}_0 = 1$, corresponding to constant $U(1)$ gauge transformations; $N^{\rm KT}_1 = \frac{1}{2} (d+2)(d+1) = {\rm dim} \, SO(d+2)$, corresponding to the Killing vectors of the sphere; and $N^{\rm KT}_{s-1} \propto s^{2d-3}$ for $s \to \infty$, corresponding to large-spin generalizations thereof.


We cannot just drop the zeromodes and move on, however. 
The original formal path integral expression (\ref{formalZPI}) is local by construction, as both numerator and denominator are defined with a local measure on local fields. In principle BRST gauge fixing is designed to maintain manifest locality, but if we remove any finite subset of modes by hand, including in particular zeromodes, locality is lost. Indeed the $- F^{(0)}_s$ subtraction results in nonlocal log-divergences in the character integral, i.e.\ divergences which cannot be canceled by local counterterms. From the point of view of (\ref{formalZPI}), the loss of locality is due the fact that we are no longer dividing by the volume of the local gauge group $\CG$, since we are effectively omitting the subgroup $G$ generated by the Killing tensors. To restore locality, and to correctly implement the idea embodied in (\ref{formalZPI}), we must divide by the volume of $G$ by hand. This volume must be computed using the same local measure defining ${\rm \Vol}(\CG)$, i.e.\ the invariant measure on $\CG$ normalized such that integrating the gauge fixing insertion in the path integral over the gauge orbits results in a factor 1. Hence the appropriate measure defining the volume of $G$ in this context is inherited from the BRST path integral measure. As such we will denote it by $\vg$. A detailed general discussion of the importance of these specifications for consistency with locality and unitarity in the case of Maxwell theory can be found in \cite{Donnelly:2013tia}. 
Relating $\vg$ to a ``canonical'', theory-independent definition of the group volume $\vc$ (such as for example ${\rm \Vol}(U(1))_{\rm c} \equiv 2\pi$) is not trivial, requiring considerable care in keeping track of various normalization factors and conventions. Moreover $\vg$ depends on the nonlinear interaction structure of the theory, as this determines the Lie algebra of $G$. We postpone further analysis of $\vg$ to section \ref{sec:Gvol}.



\subsubsection*{Conclusion}

To summarize, instead of the naive (\ref{Znaive}), we get the following formula for the 1-loop Euclidean path integral on $S^{d+1}$ for a collection of massless spin-$s$ gauge fields:
\begin{align} \label{ZPIM0summary1}
Z_{\rm PI} = i^{-\pol_s} \,\bigl(\vg\bigr)^{-1} \, \exp \sum_s  \inteps \frac{dt}{2t} \bigl(  F_s - F_s^0 \bigr)  \, ,
\end{align}
where $F_s = \{\hat F_s\}_+$ and $F_s^0$ were defined in (\ref{masslessF}), (\ref{Fsplusdef}) and (\ref{F0def}); $G$ is the subgroup of gauge transformations generated by the Killing tensors $\bar\xi^{(s-1)}$, i.e.\ the zeromodes of  (\ref{linGT}); and $i^{-\pol_s}$ is the phase (\ref{polphase}).
We can split up the integrals by introducing an IR regulator:
\begin{align} \label{ZPIM0summary2}
 \boxed{Z_{\rm PI} =  i^{-\pol_s} \, \ZG \, \Zchar \, , \qquad \ZG \equiv \frac{\exp \bigl(-\sum_s  \inteps^{\noir} \frac{dt}{2t} F_s^0\bigr)}{\vg} \, , \qquad 
  \Zchar \equiv  \exp \sum_s  \inteps^\noir \frac{dt}{2t} \, F_s} 
\end{align}
where the notation $\int^\noir$ means we IR regulate by introducing a factor $e^{-\mu t}$, take $\mu \to 0$, and subtract the $\log \mu$ divergent term. For a function $f(t) $ such that $\lim_{t \to \infty} f(t) = c$, this means
\begin{align} \label{noirdef}
 \int^\noir \frac{dt}{t} \, f(t)  \equiv \lim_{\mu \to 0} \Bigl( c \log \mu + \int \frac{dt}{t} \, f(t) \, e^{-\mu t}  \Bigr)
\end{align}
For example for $f(t) = \frac{t}{t+1}$ this gives $\int_0^\noir \frac{dt}{t} \, \frac{t}{t+1} = \log\mu -\log(e^{\game} \mu) = - \game$, and for $f(t) = 1$ with the integral UV-regularized as in (\ref{simplereg}) we get $\inteps^\noir \frac{dt}{t} =  \log(2 e^{-\game}/\epsilon)$.  

In section \ref{sec:CFmasslessX} we recast $\Zchar$ as a character integral formula. In section \ref{sec:Gvol} we express $\ZG$ in terms of the canonical group volume $\vc$ and the coupling constant of the theory.

\subsection{Character formula: $\Zchar = Z_{\rm bulk}/Z_{\rm edge} Z_{\rm KT}$} \label{sec:CFmasslessX}

In this section we derive a character formula for $\Zchar$ in (\ref{ZPIM0summary2}). If we start from the naive $\hat F_s$ given by (\ref{masslessF}) and follow the same steps as those bringing (\ref{ZPItrueformula}) to the form  (\ref{formuhs2}), we get  
\begin{align} \label{hatFchar}
 \hat F_s = \frac{1+q}{1-q} \, \chin_s \, , \qquad \chin_s =  \chin_{{\rm bulk},s} - \chin_{{\rm edge},s}   \, ,
\end{align}
where
\begin{align} \label{bulknaive}
 \chin_{{\rm bulk},s}  &=  D_s^d \, \frac{q^{s+d-2} + q^{2-s}}{(1-q)^d} - D_{s-1}^d \, \frac{q^{s+d-1} + q^{1-s}}{(1-q)^d} \, \\
 \chin_{{\rm edge},s} &=  D_{s-1}^{d+2} \, \frac{q^{s+d-3}+q^{1-s}}{(1-q)^{d-2}} - D_{s-2}^{d+2} \, \frac{q^{s+d-2}+q^{-s}}{(1-q)^{d-2}} \, . \label{edgenaive}
\end{align} 
Note that these take the form of ``field $-$ ghost'' characters obtained respectively by substituting the values of $\nu_\phi$ and $\nu_\xi$ given by (\ref{phixidim}) into the massive spin $s$ and spin $s-1$ characters (\ref{chibulkedgeprev}). The naive bulk characters $\hat\chi_{\rm bulk,s}$ thus obtained cannot possibly be the character of any UIR of $SO(1,d+1)$, as is obvious from the presence of negative powers of $q$. In particular it is certainly not equal to the physical exceptional series bulk character (\ref{excserieschi}). Now let us consider the actual $F_s=\{\hat F_s\}_+$ appearing in (\ref{ZPIM0summary2}). 
Then we find\footnote{To check (\ref{Fplustochiplus}) starting from (\ref{Fsplusdef}), observe that $\bigl\{ \frac{1+q}{1-q} \bigl( q^k + q^{-k} - 2 \bigr) \bigr\}_+ = 0$ for any integer $k$, so $\bigl\{ \frac{1+q}{1-q} \, q^k \bigr\}_+ = \frac{1+q}{1-q} \bigl(-q^{-k} + 2 \bigr)$ for $k < 0$, while of course $\bigl\{ \frac{1+q}{1-q} \, q^k \bigr\}_+ =  \frac{1+q}{1-q} \, q^k$ for $k \geq 0$, . This accounts for the $k<0$ and $k>0$ terms in the expansion $\sum_k c_k \, q^k$ of (\ref{Fplustochiplus}). The coefficient $2 N_{s-1}^{\rm KT}$ of the $q^0$ term is most easily checked by comparing the $q^0$ terms on the left and right hand sides of (\ref{Fplustochiplus}), taking into account that, by definition, $[\chi]_+$ has no $q^0$ term, and that the $q^0$ terms of the left hand side are given by (\ref{F0def}).}    
\begin{align} \label{Fplustochiplus}
 F_s =  \Bigl\{ \frac{1+q}{1-q} \, \chin_s \Bigr\}_+ = \frac{1+q}{1-q} \Bigl( \bigl[ \chin_s \bigr]_+ 
 - 2 \, N^{\rm KT}_{s-1} \Bigr)   \, ,
\end{align}
where 
the ``flipped'' character $[\hat \chi]_+$ is obtained from  
 $\hat \chi=\sum_k c_k q^k$ by flipping $c_k q^k \to -c_k q^{-k}$ for $k<0$ and dropping the $k=0$ terms:
\begin{align} \label{sqbrackplusop}
 \bigl[\hat\chi \bigr]_+ = \Bigl[ \sum_k c_k \, q^k \Bigr]_+ \equiv \sum_{k<0} (-c_k) \, q^{-k} + \sum_{k>0} c_k \, q^k \, = \hat \chi - c_0 - \sum_{k<0} c_k \bigl(q^{k} + q^{-k} \bigr) \, .
\end{align}
%
Thus this flipping prescription can be thought of as the character analog of contour rotations for ``wrong-sign'' Gaussians in the path integral.
Notice the slight differences in the map $\hat \chi \to [\hat \chi]_+$ and the related but different map $\hat F \to \{\hat F\}_+$ defined in (\ref{Fsplusdef}).

Substituting (\ref{Fplustochiplus}) into (\ref{ZPIM0summary2}), we conclude
\begin{align} 
 \boxed{\log \Zchar = \sum_s  \inteps^{\noir} \frac{dt}{2t} \, \frac{1+q}{1-q} \Bigl( \bigl[  \chin_{{\rm bulk},s} \bigr]_+ - \bigl[  \chin_{{\rm edge},s} \bigr]_+ - 2 N_{s-1}^{\rm KT} \Bigr)}  \label{newZchardef}
\end{align}

\subsubsection*{Consistency with  ideal gas bulk character formula} 

Consistency with the physical ideal gas picture used to derive $Z_{\rm bulk}$ in section \ref{sec:Zbulkmassless} requires the bulk part of $\log \Zchar$ as given in (\ref{newZchardef}) agrees with (\ref{Zbulkfirst}), that is to say it requires 
\begin{align} \label{flipeqexc}
 \chi_{{\rm bulk},s} = [\chin_{{\rm bulk},s}]_+ \,  \qquad (?)
\end{align}
where $\chi_{{\rm bulk},s}$ is one of the intricate exceptional series characters (\ref{excserieschi}), while $[\hat\chi_{{\rm bulk},s}]_+$ is obtained from the simple naive bulk character (\ref{bulknaive}) just by flipping some polar terms as in (\ref{sqbrackplusop}). At first sight this might seems rather unlikely. Nevertheless, quite remarkably, it turns out to be true. 
Let us first check this in some simple examples:
\begin{itemize}
\item {\it $s=2$ in $d=3$}: The naive character (\ref{bulknaive}) is 
 \begin{align}
    \chin_{\rm bulk}  &=  5 \cdot \frac{q^{3} + 1}{(1-q)^3} - 3 \cdot \frac{q^{4} + q^{-1}}{(1-q)^3} \, .
 \end{align}
The polar and $q^0$ terms are obtained by expanding $\chin_{\rm bulk} = -\frac{3}{q} - 4 + \CO(q)$. Thus 
 \begin{align}
  \bigl[\chin_{\rm bulk} \bigr]_+ = \chin_{\rm bulk} + 4 +\frac{3}{q} + 3 \, q =  \frac{2 \cdot 5 \cdot q^{3} - 2 \cdot 3 \cdot q^4}{(1-q)^3}  \, ,
 \end{align}
correctly reproducing the $d=3,s=2$ character in (\ref{chiexcex}). 

\item {\it $s=1$, general $d \geq 3$}: In this case the map $[...]_+$ merely eliminates the $q^0$ term in the naive character (\ref{bulknaive}) by adding $+1$:
\begin{align}
 \bigl[\chin_{\rm bulk} \bigr]_+ = \chin_{\rm bulk} + 1 = d \cdot \frac{q^{d-1}+q}{(1-q)^d}  - \frac{q^d+1}{(1-q)^d} + 1 \, ,
\end{align}
correctly reproducing (\ref{spin1casechis}).

\end{itemize} 
Using Mathematica, it is straightforward to check an arbitrary large number of examples in this way. 
Below we will derive a general explicit formula for the character flip map $\hat \chi \to [\hat \chi]_+$. This will provide a general proof of (\ref{flipeqexc}), and more generally will enable efficient closed-form computation of the proper  bulk and edge characters for general $s$ and $d$. Generalizations are implemented with equal ease: as an illustration thereof we compute the bulk and edge characters for partially massless fields.

\subsubsection*{Flipping formula}

We wish to derive a general explicit formula for $\bigl[ \frac{q^\Delta}{(1-q)^d} \bigr]_+$ defined in (\ref{sqbrackplusop}), for 
\begin{align}
  \Delta \in \{ 0,-1,-2,-3, \ldots \}
\end{align}
which suffices to obtain an explicit expression for $[\hat\chi]_+$ for any bosonic character of interest, including in particular (\ref{bulknaive})-(\ref{edgenaive}).
This is achieved by expanding $(1-q)^{-d}=\sum_k {d-1+k \choose k} q^k$, splitting the sum in its polar and nonpolar part, incorporating the appropriate sign flips, and resumming in terms of hypergeometric functions $_2F_1(a,b,c;q) = \sum_{n=0}^\infty \frac{(a)_n (b)_n}{(c)_n} \frac{q^n}{n!}$. After a bit more sanding and polishing we find
\begin{align} \label{flip1}
 \biggl[ \frac{q^\Delta}{(1-q)^d} \biggr]_+ &= 
 \frac{(-1)^{d+1} q^{d-\Delta} + p_\Delta(q) + (-1)^d p_{d-\Delta}(q)}{(1-q)^d} \, ,
\end{align}
where $p_\Delta(q) = \tbinom{d-\Delta}{d-1} \cdot q \cdot {}_2F_1\bigl(1-d,1-\Delta,2-\Delta,q\bigr) = \tbinom{d-\Delta}{d-1} \cdot q \cdot 
 \sum_{k=0}^{d-1} (-1)^k \tbinom{d - 1}{k }
   \tfrac{1 - \Delta}{k +1 - \Delta} \, q^k$.
The hypergeometric series terminates because $1-d \leq 0$. A more interesting version is 
\begin{align} \label{flip2}
 &\biggl[ \frac{q^\Delta}{(1-q)^d} \biggr]_+ = \frac{\CP_\Delta(q)}{(1-q)^d} \, \nn \\
 & \CP_\Delta(q) \equiv (-1)^{d+1} q^{d-\Delta} + \sum_{m=0}^{r-1} (-1)^m \, D^d_{1-\Delta,1^m} \bigl( q^{1+m} + (-1)^d q^{d-1-m} \bigr) \qquad r = \left\lfloor \tfrac{d}{2} \right\rfloor \, .
\end{align} 
Here $D^d_{1-\Delta,1^m}$ is  the dimension of the irrep of $\so(d)$ corresponding to the Young diagram $S=(1-\Delta,1,\ldots,1,0,\ldots,0)$, with $1$ repeated $m$ times, explicitly given in (\ref{Weyldimeven})-(\ref{Weyldimodd}). 
We obtained this formula using Mathematica and also obtained a proof of (\ref{flip2}) by expressing (\ref{Weyldimeven})-(\ref{Weyldimodd}) in terms of gamma functions and comparing to (\ref{flip1}). This is somewhat tedious and not especially illuminating, so we omit it here. 


\def\sp{{s'}}

\subsubsection*{Bulk characters for (partially) massless fields}

Now let us apply this to a slight generalization of the massless $\hat\chi_{\rm bulk,s}$ given in (\ref{bulknaive}), 
\begin{align} \label{PMchinaive}
 \chin^{\rm bulk}_{s\sp}(q) \equiv D_s^d \, \frac{q^{1-\sp}+q^{d-1+\sp}}{(1-q)^d} - D_\sp^d \, \frac{q^{1-s}+q^{d-1+s}}{(1-q)^d} \, .
\end{align}
This is the naive bulk character for a {\it partially massless} spin-$s$ field $\phi_{\mu_1 \cdots \mu_s}$ with a spin-$s'$ ($0 \leq \sp \leq s-1$) gauge parameter field $\xi_{\mu_1 \cdots \mu_\sp}$ \cite{Bekaert:2013zya}. The massless case (\ref{bulknaive}) corresponds to 
\begin{align}
 \sp = s-1 \, \qquad \mbox{(massless case)}.
\end{align}
We consider the more general partially massless case here to illustrate the versatility of (\ref{flip2}), and because in a sense the resulting formulae are more elegant than in the massless case, due to the neat $s \leftrightarrow \sp$ symmetry already evident in (\ref{PMchinaive}).
Applying (\ref{flip2}), still with $r = \left\lfloor \tfrac{d}{2} \right\rfloor$,
\begin{equation} \label{flippedchargenPM}
\boxed{
 \begin{aligned}
 (1-q)^d \bigl[\chin^{\rm bulk}_{s\sp}(q)\bigr]_+ &= \bigl( 1 + (-1)^{d+1} \bigr) \bigl(D_s^d \, q^{d-1+\sp} - D_\sp^d \, q^{d-1+s} \bigr)   \\
 &  \quad + \sum_{m=1}^{r-1} (-1)^{m-1} \, D^d_{s,\sp+1,1^{m-1}} \bigl( q^{1+m} + (-1)^d q^{d-1-m} \bigr)  \, .
\end{aligned}
}
\end{equation}
We used $D_s^d \,  \CP_{1-\sp}(q) - D_\sp^d  \CP_{1-s}(q)  
 = \sum_{m=0}^{r-1} (-1)^m \, \bigl( D_s^d \, D^d_{\sp,1^m} - D_\sp^d \, D^d_{s,1^m} \bigr) \bigl( q^{1+m} + (-1)^d q^{d-1-m} \bigr) \\
 = \sum_{m=1}^{r-1} (-1)^{m-1} \, D^d_{s,\sp+1,1^{m-1}} \bigl( q^{1+m} + (-1)^d q^{d-1-m} \bigr)$ with $\CP_\Delta$ as defined in (\ref{flip2}) to get the second term. Like for (\ref{flip2}), we obtained this formula using Mathematica.  
It can be proven starting from (\ref{Weyldimeven})-(\ref{Weyldimodd}).
 
Remarkably, (\ref{flippedchargenPM}) precisely reproduces the massless exceptional series characters (\ref{excserieschi}) for $\sp=s-1$, further strengthening our physical picture, adding evidence for (\ref{ZPItrueformulaGeneral}), proving (\ref{flipeqexc}), and generalizing it moreover to partially massless gauge fields. Comparing to \cite{10.3792/pja/1195522333,10.3792/pja/1195523378,10.3792/pja/1195523460}, the partially massless gauge field characters we find here coincide with those of the unitary exceptional series $D^j_{S,p}$ with $p=0$, $S=(s,\sp+1)$, and $j=(d-3)/2$ for odd $d$, $j=(d-4)/2$ for even $d$. In the notation of \cite{Basile:2016aen} this is the  exceptional series with $\Delta=p=2$, $S=(s,\sp+1)$, which was indeed identified in  \cite{Basile:2016aen} as the $\so(1,d+1)$ UIR for partially massless fields.


\subsubsection*{Edge characters for (partially) massless fields}

For the edge correction we proceed analogously. The naive PM edge character is
\begin{align}
 \chin_{s\sp}^{\rm edge}(q) = D_{s-1}^{d+2} \, \frac{q^{-\sp}+q^{d-2+\sp}}{(1-q)^{d-2}} - D_{\sp-1}^{d+2} \, \frac{q^{-s}+q^{d-2+s}}{(1-q)^{d-2}} \, ,
\end{align}
reducing to the massless case (\ref{edgenaive}) for $\sp=s-1$. 
Applying (\ref{flip2}) gives, still with $r=\lfloor \frac{d}{2} \rfloor$,
\begin{equation} \label{flippedchargenPMedge}
\boxed{
\begin{aligned} 
 (1-q)^{d-2} \bigl[ \chin^{\rm edge}_{s\sp}(q) \bigr]_+ &= \bigl( 1 + (-1)^{d+1} \bigr) \bigl( D_{s-1}^{d+2} \, q^{d-2+\sp} - D_{\sp-1}^{d+2} \, q^{d-2+s} \bigr)  \\
 & \quad + \sum_{m=0}^{r-2} (-1)^m \, \tilde D_m \bigl( q^{1+m} + (-1)^d q^{d-3-m} \bigr) 
\end{aligned}
}
\end{equation}
where $\tilde D_m \equiv D_{s-1}^{d+2} \, D^{d-2}_{\sp+1,1^m} - D_{\sp-1}^{d+2} \, D^{d-2}_{s+1,1^m}$.

Note that in the massless spin-1 case
\begin{align} \label{spin1edgem0}
 \bigl[\chin_{1}^{\rm edge}(q) \bigr]_+ = \frac{q^{d-2}+1}{(1-q)^{d-2}} - 1 \, .
\end{align}
In the notation of \cite{10.3792/pja/1195522333,10.3792/pja/1195523378,10.3792/pja/1195523460}, this equals the character for the unitary $SO(1,d-1)$ irrep in the exceptional series $D^j_{S;p=0}$ with $S=(1)$ and $j=(d-4)/2$ for even $d$ and $j=(d-3)/2$ for odd $d$  --- the irreducible representation indeed of a massless scalar on dS$_{d-1}$ with its zeromode removed. The fact that $S=(1)$ is analogous to what happens in the 2D CFT of a massless free scalar $X$: the actual CFT primary operators are the spin $\pm 1$ derivatives $\partial_\pm X(0)$.

In contrast to (\ref{flippedchargenPM}), we did not find a way of rewriting $\tilde D_m$ for general spin to suggest an interpretation along these lines in general. Indeed unlike (\ref{flippedchargenPM}), $\bigl[ \chin^{\rm edge}_{s\sp}(q) \bigr]_+$ in general does not appear to be proportional to the character of a {\it single} exceptional series irrep of $SO(1,d-1)$. This is not in conflict with the picture of edge corrections as a Euclidean path integral of some collection of local fields on $S^{d-1}$, since if the fields have nontrivial spin / $\so(d-2)$ weights, the corresponding character integrals will have a complicated structure, involving sums of iterations of $SO(1,d-1-2k)$ characters with $k=0,1,2,\ldots$, exhibiting patterns that might be hard to discern without knowing what to look for.  It should also be kept in mind we have not identified a reason the edge correction {\it must} have a local QFT path integral interpretation. On the other hand, 
the coefficients of the $q$-expansion of the effective edge character {\it do} turn out to be positive, consistent with an interpretation in terms of some collection of fields corresponding to unitary representations of dS$_{d-1}$. A more fundamental group-theoretic or physics understanding of the edge correction would evidently be desirable.   

For practical purposes, the interpretation does not matter of course. The formula (\ref{flippedchargenPMedge}) gives a general formula for $\chi_{\rm edge}$, which is all we need. For example for $d=3$, this gives $\bigl[\chin_{s}^{\rm edge}(q)\bigr]_+ = 2 \, \frac{ D_{s-1}^5 \, q^{s} - D_{s-2}^5 \, q^{s+1}}{1-q} = 2 D_{s-1}^5 \, q^s + 2 D_{s-1}^4 \frac{q^{s+1}}{1-q}$, where $D_{s-1}^5=\frac{1}{6} s (s+1) (2 s+1)$ and $D_{s-1}^4=D_{s-1}^5-D_{s-2}^5=s^2$. The second form makes positivity of coefficients manifest. For $d=4$ we get $\bigl[\chin_{s}^{\rm edge}(q)\bigr]_+ = D^5_{s-1} \frac{2 \, q}{(1-q)^2}$.


\subsubsection*{Conclusion}

We conclude that (\ref{newZchardef}) can be written as 
\begin{align}
 \log \Zchar = \log Z_{\rm bulk} - \log Z_{\rm edge} - \log Z_{\rm KT} \, ,
\end{align}
where the bulk and edge contributions are explicitly given by (\ref{flippedchargenPM})-(\ref{flippedchargenPMedge}) with $\sp=s-1$,\footnote{In the partially massless case $\log Z_{\rm KT}$ takes the same form, but with $D^{d+2}_{s-1,s-1}$ replaced by $D^{d+2}_{s-1,s'}$.} and 
\begin{align}  \label{ZKTIR}
 \log Z_{\rm KT} = \dim G \, \inteps^{\noir} \frac{dt}{2t} \, \frac{1+q}{1-q} \cdot 2 \, , \qquad \dim G = \sum_s N^{\rm KT}_{s-1} = \sum_s D_{s-1,s-1}^{d+2} \, .
\end{align}
The finite (IR) part of $Z_{\rm KT}$ is given by (\ref{dimGcontrib}): $Z_{\rm KT}|_{\rm IR} = (2\pi)^{-\dim G}$. 
 


\subsection{Group volume factor: $\ZG$} \label{sec:Gvol}

The remaining task is to compute the factor $\ZG$ defined in (\ref{ZPIM0summary2}), that is
\begin{align} \label{ZG1b}
 \ZG &= \bigl(\vg \bigr)^{-1} \exp \sum_s N^{\rm KT}_{s-1} \inteps^{\noir} \frac{dt}{2t} \bigl(q^{2s+d-4}+ q^{2s+d-2} + 2 \bigr) 
\end{align}
We imagine the spin range to be finite, or cut off in some way. (The infinite spin range case is discussed in section \ref{sec:CHSgr}.) 
In the heat kernel regularization scheme of appendix \ref{app:renpr}, we can then evaluate the integral using (\ref{simplereg}):
\begin{align} \label{ZGhalf}
 \ZG &= \bigl(\vg \bigr)^{-1} \prod_s \biggl( \frac{M^4}{(2s+d-4)(2s+d-2)} \biggr)^{\frac{1}{2} N_{s-1}^{\rm KT}}  \, , \qquad M \equiv \frac{2 e^{-\game}}{\epsilon} \, ,
\end{align} 
On general grounds, the nonlocal UV-divergent factors $M$ appearing here in $\ZG$ should cancel against factors of $M$ in $\vg$, as we will explicitly confirm below.

\subsubsection*{Generalities}

Recall that $G$ is the group of gauge transformations generated by the Killing tensors. Equivalently it is the subgroup of gauge transformations leaving the background invariant. $\vg$ is the volume of $G$ with respect to the path integral induced measure. This is different from what we shall call the ``canonical'' volume $\vc$, defined  with respect to the invariant metric normalized such that the generators of some standard basis of the Lie algebra have unit norm. (In the case of Yang-Mills, this coincides with the metric defined by the canonically normalized Yang-Mills action, providing some justification for the (ab)use of the word canonical.) In particular, in contrast to $\vc$, $\vg$ depends on the coupling constants and UV cutoff of the field theory. 

As mentioned at the end of section \ref{sec:masslessZPI}, the computation of $\ZG$ brings in a series of new complications. One reason is that the Lie algebra structure constants defining $G$ are not determined by the free part of the action, but by its interactions, thus requiring data going beyond the usual one-loop Gaussian level. Another reason is that due to the omission of zeromodes and the ensuing loss of locality in the path integral, a precise computation of $\vg$ requires keeping track of an unpleasantly large number of normalization factors, such as for instance constants multiplying kinetic operators, as these can no longer be automatically discarded by adjusting local counterterms. Consequently, exact, direct path integral computationz of $\ZG$ for general higher-spin theories requires great care and considerable persistence, although it can be done  \cite{Law:2020cpj}. Below we obtain an exact expression for $\ZG$ in terms of $\vc$ and the Newton constant in a comparatively painless way, by combining results and ideas from \cite{Joung:2013nma,Joung:2014qya,Sleight:2016xqq,Witten:1995gf,Giombi:2013yva,Donnelly:2013tia,Sleight:2016dba}, together with the observation that the form of (\ref{ZPITTmassless}) actually determines all the normalization factors we need. Although the expressions at intermediate stages are still a bit unpleasant, the end result takes a strikingly simple and universal form.

If $G$ is finite-dimensional, as is the case for example for Yang-Mills, Einstein gravity and certain (topological) higher-spin theories \cite{Boulanger:2011qt, Joung:2015jza, Manvelyan:2013oua, Brust:2016gjy,Blencowe:1988gj,Bergshoeff:1989ns,Castro:2011fm,Perlmutter:2012ds,Campoleoni:2010zq} including the dS$_3$ higher-spin theory analyzed in section \ref{sec:HSCS}, we can then proceed to compute $\vc$, and we are done. If $G$ is infinite-dimensional, as is the case in generic higher-spin theories, one faces the remaining problem of making sense of $\vc$ itself. Glossing over the already nontrivial problem of exponentiating the higher-spin algebra to an actual group \cite{Monnier:2014tfa}, the obvious issue is that $\vc$ is going to be divergent.  
We discuss and interpret this and other infinite spin range issues in section \ref{sec:CHSgr}. In what follows we will continue to assume the spin range is finite or cut off in some way as before, so $G$ is finite-dimensional.

We begin by determining the path integral measure to be used to compute $\vg$ in (\ref{ZGhalf}). Then we compute $\ZG$ in terms of  $\vc$ and the coupling constant of the theory, first for Yang-Mills, then for Einstein gravity, and finally for general higher-spin theories.

\subsubsection*{Path integral measure}

To determine $\vg$ we have to take a quick look at the path integral measure. This is fixed by locality and consistency with the regularized heat kernel definition of Gaussian path integrals we have been using throughout. For example for a scalar field as in (\ref{HKSP}), we have $\int \CD \phi \, e^{-\frac{1}{2} \phi(-\nabla^2+m^2) \phi} \equiv \exp \int \frac{d\tau}{2 \tau} \, e^{-\epsilon^2/4\tau} \, {\rm Tr} \, e^{-\tau (-\nabla^2+m^2)}$. An eigenmode of $-\nabla^2+m^2$ with eigenvalue $\lambda_i$ contributes a factor $M/\sqrt{\lambda_i}$ to the right hand side of this equation, with $M= \exp \int \frac{d\tau}{2\tau} \, e^{-\epsilon^2/4\tau} e^{-\tau} =  2 e^{-\game}/\epsilon$, the same parameter as in (\ref{ZGhalf}) (essentially by definition). To ensure the left hand side matches this, we must use a path integral measure derived from the local metric $ds^2_\phi = \frac{M^2}{2\pi} \int (\delta \phi)^2$. To see this, expand $\phi(x)=\sum_i \varphi_i \psi_i(x)$ with $\psi_i(x)$ an orthonormal basis of eigenmodes of $-\nabla^2+m^2$ on $S^{d+1}$. The metric in this basis becomes $ds^2 = \sum_i \frac{M^2}{2\pi} d\varphi_i^2$, so a mode with eigenvalue $\lambda_i$ contributes a factor $\int d\varphi_i \, \frac{M}{\sqrt{2\pi}} \, e^{-\frac{1}{2} \lambda_i \varphi_i^2} = M/\sqrt{\lambda_i}$ to the left hand side, as required. 

We work with canonically normalized fields. For a spin-$s$ field $\phi$ this means  the quadratic part of the action evaluated on its transverse-traceless part $\phi^{\rm TT}$ takes the form
\begin{align} \label{canonnorm}
  S\bigl[\phi^{\rm TT}\bigr] = \frac{1}{2} \int \phi^{\rm TT} (-\nabla^2 + \overline m^2) \phi^{\rm TT}.
\end{align}
Consistency with (\ref{ZPITTmassive}) or  (\ref{ZPITTmassless}) then requires the measure for $\phi$ to be derived again from the metric $ds^2_\phi = \frac{M^2}{2\pi} \int (\delta \phi)^2$. If $\phi$ has a gauge symmetry, the formal division by the volume of the gauge group $\CG$ is conveniently implemented by BRST gauge fixing. For example for a spin-1 field with gauge symmetry $\delta \phi_\mu = \partial_\mu \xi$,  we can gauge fix in Lorenz gauge by adding 
the BRST-exact action $S_{\rm BRST}= \int i B \, \nabla^\mu \phi_\mu - \bar c \nabla^2 c$. This requires specifying a measure for the Lagrange multiplier field $B$ and the ghosts $c,\bar c$. It is straightforward to check that a ghost measure derived from $ds^2_{\bar c c} = M^2 \int \delta \bar c \, \delta c$ (which translates to a mode measure $\prod_i \frac{1}{M^2} d \bar c_i \, dc_i$) combined with a $B$-measure derived from $ds^2_B = \frac{1}{2\pi} \int (\delta B)^2$, reproduces precisely the second term in  (\ref{ZPITTmassless}) upon integrating out $B$, $c$, $\bar c$ and the longitudinal modes of $\phi$. It is likewise straightforward to check that BRST gauge fixing is then formally equivalent to dividing by the volume of the local gauge group $\CG$ with respect to the measure derived from the following metric on the algebra of local gauge transformations: 
\begin{align} \label{dsxigenf}
 ds^2_\xi = \frac{M^4}{2\pi} \int (\delta \xi)^2 \, .
\end{align} 
Note that all of these metrics take the same form, with the powers of $M$ fixed by dimensional analysis. An important constraint in the above was that the second term in (\ref{ZPITTmassless}) is exactly reproduced, without some extra factor multiplying the Laplacian. This matters when we omit zeromodes. For this to be the case with the above measure prescriptions, it was important that the gauge transformation took the form $\delta \phi_\mu = \alpha_1 \partial_\mu \xi$ with $\alpha_1 = 1$ as opposed to some different value of $\alpha_1$, as we a priori allowed in (\ref{linGT}). For a general $\alpha_1$, we would have obtained an additional factor $\alpha_1$ in the ghost action, and a corresponding factor $\alpha_1^2$ in the kinetic term in the second term of (\ref{ZPITTmassless}). To avoid having to keep track of this, we picked $\alpha_1 \equiv 1$. For Yang-Mills theories, everything remains the same, with internal index contractions understood, e.g.\ $S[\phi^{\rm TT}] = \frac{1}{2} \int \phi^{a {\rm TT}} (-\nabla^2+\overline m^2)\phi^{a {\rm TT}}$,   $ds^2_\phi = \frac{M^2}{2\pi} \int (\delta \phi^a)^2$, $ds^2_\xi = \frac{M^4}{2\pi} \int (\delta \xi^a)^2$. 

For higher-spin fields, we gauge fix in the de Donder gauge. All metrics remain unchanged, except for the obvious additional spacetime index contractions. The second term of (\ref{ZPITTmassless}) is exactly reproduced upon integrating out the ${\rm TT}$ sector of the BRST fields together with the corresponding longitudinal modes of $\phi$, provided we pick
\begin{align} \label{aplphasval}
 \alpha_s = \sqrt{s} 
\end{align} 
in (\ref{linGT}), with symmetrization conventions such that $\phi_{(\mu_1 \cdots \mu_s)} = \phi_{\mu_1 \cdots \mu_s}$. (Technically the origin of the factor $s$ can be traced to the fact that if $\phi_{\mu_1 \cdots \mu_s} = \nabla_{(\mu_1} \xi_{\mu_2 \cdots \mu_s)}$ for a TT $\xi$, we have $\int \phi^2 = s^{-1} \int \xi (-\nabla^2 + c_s) \xi$.) Equation (\ref{aplphasval}) fixes the normalization of $\xi$, and (\ref{dsxigenf}) then determines unambiguously the measure to be used to compute $\vg$ in (\ref{ZGhalf}).  
We will see more concretely how this works in what follows, first spelling out the basic idea in detail in the familiar YM and GR examples, and then moving on to the  general higher-spin gauge theory case considered in \cite{Joung:2013nma}.
 

\subsubsection*{Yang-Mills}

Consider a Yang-Mills theory with with a simple Lie algebra  
\begin{align} \label{LieAlgDef}
  [ L^a, L^b]=f^{abc}  L^c \, ,
\end{align}
with the $L^a$ some standard basis of anti-hermitian matrices and $f^{abc}$ real and totally antisymmetric. For example for $\su(2)$ Yang-Mills, $L^a = -\frac{1}{2} i \sigma^a$ and $[L^a,L^b] = \epsilon^{abc} L^c$. Consistent with our general conventions, we take the gauge fields $\phi_\mu=\phi_\mu^a L^a$ to be canonically normalized: the curvature takes the form $F_{\mu\nu}^a L^a = F_{\mu\nu} = \partial_\mu \phi_\nu - \partial_\nu \phi_\mu + g [\phi_\mu,\phi_\nu]$, and the  action is
\begin{align} \label{ScanYM}
 S = \frac{1}{4} \int  F^a \cdot F^a  
\end{align}
The quadratic part of $S$ is invariant under the linearized gauge transformations $\delta^{(0)}_{\xi} \phi_\mu =  \partial_\mu \xi$, where $\xi = \xi^a L^a$, taking  the form (\ref{linGT}) with $\alpha_1=1$ as required. The full $S$ is invariant under local gauge transformations  $\delta_{\xi} \phi_\mu = \partial_\mu \xi  + g [\phi_\mu,\xi]$, generating the local gauge algebra 
\begin{align} \label{locGA}
[\delta_{\xi},\delta_{\xi'}] = \delta_{g[\xi',\xi]} \, .  
\end{align}
The rank-0 Killing tensors $\bar \xi$ satisfy $\partial_\mu \bar\xi = 0$: they are  the constant gauge transformations $\bar\xi = \bar\xi^a L^a$ on the sphere, forming the subalgebra $\lieg$ of local gauge transformations acting trivially on the background $\phi_\mu=0$, generating the group $G$ whose volume we have to divide by.  The  bracket of $\lieg$, denoted  $[\![\cdot,\cdot]\!]$ in \cite{Joung:2013nma}, is inherited from the  local gauge algebra (\ref{locGA}): 
\begin{align} \label{KSbrack}
 [\![\bar\xi,\bar\xi']\!] = g [\bar\xi',\bar\xi] \, .
\end{align}
Evidently this is isomorphic to the original YM Lie algebra. Being a simple Lie algebra, $\lieg$ has an up to normalization unique invariant bilinear form/metric. The path integral metric  $ds^2_{\rm PI}$ of (\ref{dsxigenf}) corresponds to such an invariant bilinear form with a specific normalization:
\begin{align} \label{pipi}
 \langle \bar\xi|\bar\xi'\rangle_{\rm PI} = \frac{M^4}{2\pi} \int \bar\xi^a \bar\xi^{\prime a}  =  \frac{M^4}{2\pi} \, {\rm \Vol}(S^{d+1}) \, \bar\xi^a \bar\xi^{\prime a} \, .
\end{align}
We define the theory-independent ``canonical'' invariant bilinear form $\langle \cdot|\cdot \rangle_\can$ on $\lieg$ as follows. First pick a ``standard'' basis $M^a$ of $\lieg$, i.e.\ a basis satisfying the same commutation relations as (\ref{LieAlgDef}):  $[\![M^a,M^b]\!] = f^{abc} M^c$. This fixes the normalization of the $M^a$. Then we fix the normalization of $\langle \cdot|\cdot \rangle_\can$ by requiring these standard generators have unit norm, i.e.\ 
\begin{align}
 \langle M^a|M^b \rangle_c \equiv \delta^{ab} \, .
\end{align}
The explicit form of (\ref{KSbrack}) implies such a basis is given by the constant functions $M^a = -L^a/g$ on the sphere. Thus we have $\langle L^a|L^b \rangle_\can = g^2 \delta^{ab}$ and 
\begin{align} \label{kaka}
  \langle \bar\xi|\bar\xi'\rangle_{\rm c} = g^2 \bar\xi^a \bar\xi^{\prime a}
\end{align}
Comparing (\ref{kaka}) and (\ref{pipi}), we see the path integral and canonical metrics on $G$ and their corresponding volumes are related by
\begin{align} \label{dsPIdscYM}
 ds^2_{\rm PI} = \frac{M^4}{2\pi} \frac{{\rm \Vol}(S^{d+1})}{g^2} \, ds^2_c
 \qquad \Rightarrow \qquad \frac{\vg}{\vc} = \biggl(\frac{M^4}{2\pi} \frac{{\rm \Vol}(S^{d+1})}{g^2} \biggr)^{\frac{1}{2}\dim G}  \, .
\end{align}
From (\ref{ZGhalf}), we get $\ZG = \vg^{-1} \bigl(\frac{M^4}{(d-2)d}\bigr)^{\frac{1}{2}\dim G}$, hence 
\begin{align} \label{ZGYM}
 \ZG  =   \frac{\gamma^{\dim G}}{\vc}  \, , \qquad  \gamma 
 \equiv  \frac{g}{\sqrt{(d-2)\Ad}} \, , \qquad  \Ad \equiv {\rm \Vol}(S^{d-1}) = \frac{2 \pi^{\frac{d}{2}}}{\Gamma(\frac{d}{2})} \, ,
\end{align} 
where we used ${\rm \Vol}(S^{d+1}) = \frac{2\pi}{d} {\rm \Vol}(S^{d-1})$. (Recall  we have been assuming $d>2$. The case $d=2$ is discussed in appendix \ref{app:dS3charform}.) The quantity $\gamma$ may look familiar: the Coulomb potential energy for two unit charges at a distance $r$ in flat space is $V(r)=\gamma^2/r^{d-2}$. 

\vskip3mm

Practically speaking, the upshot is that $Z_G$ is given by (\ref{ZGYM}), with $\vc$ the volume of the Yang-Mills gauge group with respect to the metric defined by the Yang-Mills action (\ref{ScanYM}). For example for $ G=SU(2)$ with $f^{abc} = \epsilon^{abc}$ as before, ${\rm \Vol}( G)_{\rm c} = 16 \pi^2$, because $SU(2)$ in this metric is the round $S^3$ with circumference $4\pi$, hence radius $2$. 

The relation (\ref{KSbrack}) can be viewed as defining the coupling constant $g$ given our normalization conventions for the kinetic terms and linearized gauge transformations. Of course the final result is independent of these conventions. Conventions without explicit factors of $g$ in the curvature and gauge transformations are obtained by rescaling $\phi \to \phi/g$, $\xi \to \xi/g$. Then there won't be a factor $g$ in (\ref{KSbrack}), but instead $g$ is read off from the action $S=\frac{1}{4g^2} \int (F^a)^2$. We could also write this without explicit reference to a basis as $S=\frac{1}{4g^2} \int {\rm Tr} \, F^2$, where the trace ``${\rm Tr}$'' is normalized such that ${\rm Tr}(L^a L^b) \equiv \delta^{ab}$. Then we can say the canonical bilinear/metric/volume is defined by the trace norm appearing in the YM action. We could choose a differently normalized trace ${\rm Tr}' = \lambda^2 {\rm Tr}$. The physics remains unchanged provided $g'=\lambda g$. Then $\vc'=\lambda^{\dim G} \vc$, hence, consistently, $Z_G'=Z_G$. 

As a final example, for $SU(N)$ Yang-Mills with $\su(N)$ viewed as anti-hermitian $N \times N$ matrices, $S=-\frac{1}{4g^2} \int {\rm Tr}_N F^2$ in conventions without a factor $g$ in the gauge algebra, and ${\rm Tr}_N$ the ordinary $N \times N$ matrix trace, ${\rm vol}(SU(N))_\can = (\ref{volsuN})$.
  

\subsubsection*{Einstein gravity}

The Einstein gravity case proceeds analogously.  Now we have single massless spin-2 field $\phi_{\mu\nu}$. The gauge transformations are diffeomorphisms generated by vector fields $\xi_\mu$. The subgroup $G$ of diffeomorphisms leaving the background $S^{d+1}$ invariant is $SO(d+2)$, generated by Killing vectors $\bar\xi_\mu$. The usual standard basis $M_{IJ} = - M_{JI}$, $I=1,\ldots,d+2$ of the $\so(d+2)$ Lie algebra satisfies $[M_{12},M_{23}]=M_{13}$ etc. We define the canonical bilinear $\langle \cdot|\cdot\rangle_c$ to be the unique invariant form normalized such that the $M_{IJ}$ have unit norm:
\begin{align} \label{cannorm}
 \langle M_{12}|M_{12}\rangle_c = 1 \, .
\end{align}
With respect to the corresponding metric $ds^2_c$, orbits $g(\varphi) = e^{\varphi M_{12}}$ with $\varphi$ ranging from $0$ to $2\pi$ have length $2\pi$. The canonical volume is then given by (\ref{vcSO}).


To identify the standard generators $M_{IJ}$ more precisely in our normalization conventions for $\bar\xi$, we need to look at the field theory realization in more detail.  The $\so(d+2)$ algebra generated by the Killing vectors $\bar\xi$ is realized in the interacting Einstein gravity theory as a subalgebra of the gauge (diffeomorphism) algebra. As in the Yang-Mills case (\ref{KSbrack}), the bracket $[\![\cdot,\cdot]\!]$ of this subalgebra is inherited from the gauge algebra.
Writing the Killing vectors as $\bar\xi = \bar\xi^\mu \partial_\mu$, the standard Lie bracket is
$
 [\bar\xi,\bar\xi']_{\rm L} = \bigl(\bar\xi^\mu \partial_\mu \bar\xi^{\prime \nu}  - \bar\xi^{\prime\mu} \partial_\mu \bar\xi^\nu \bigr) \partial_\nu
$. 
If we had normalized $\phi_{\mu\nu}$ as $\phi_{\mu\nu} \equiv g_{\mu\nu} - g_{\mu\nu}^0$ with $g_{\mu\nu}^0$ the background sphere metric, and if we had normalized $\xi_\mu$ by putting $\alpha_2 \equiv 1$ in (\ref{linGT}), the bracket $[\![\cdot,\cdot]\!]$ would have coincided with the Lie bracket $[\cdot,\cdot]_{\rm L}$. However, we are working in different normalization conventions, in which $\phi_{\mu\nu}$ is canonically normalized and $\alpha_2 = \sqrt{2}$ according to (\ref{aplphasval}). In these conventions we have instead
\begin{align} \label{Gnewt}
 [\![\bar \xi,\bar\xi']\!] = \sqrt{16 \pi G_{\rm N}} \, [\bar\xi',\bar\xi ]_{\rm L} \, ,
\end{align} 
where $G_{\rm N}$ is the Newton constant. This can be checked by starting from the Einstein-Hilbert action, expanding to quadratic order (see e.g.\ \cite{Goon:2018fyu} for convenient and reliable explicit expressions in dS$_{d+1}$), and making the appropriate convention rescalings. This is the Einstein gravity analog of (\ref{KSbrack}). To be more concrete, let us consider the ambient space description of the sphere $S^{d+1}$, i.e.\ $X^I X_I = 1$ with $X \in \IR^{d+2}$. Then the basis of Killing vectors $M_{IJ} \equiv -(X_I \partial_J - X_J \partial_I)/\sqrt{16 \pi G_{\rm N}}$ satisfy our standard $\so(d+2)$ commutation relations $[\![M_{12},M_{23}]\!] = M_{13}$ etc, hence by (\ref{cannorm}), $\langle M_{12}|M_{12} \rangle_c = 1$.  
The path integral metric (\ref{dsxigenf}) on the other hand corresponds to the invariant bilinear $\langle \bar\xi | \bar\xi' \rangle_{\rm PI} = \frac{M^4}{2\pi} \, \int \bar\xi \cdot \bar\xi'$, so $\langle M_{12}|M_{12} \rangle_{\rm PI}=\frac{M^4}{2\pi} \frac{1}{16 \pi G_{\rm N}} \int_{S^{d+1}} (X_1^2 + X_2^2) = \frac{M^4}{2\pi} \frac{1}{16 \pi G_{\rm N}} \frac{2}{d+2} {\rm \Vol}(S^{d+1})$. Thus we obtain the following relation between PI and canonical metrics and volumes for $G=SO(d+2)$: 
\begin{align} \label{ds2PItocgrav}
  ds^2_{\rm PI} = \frac{\Ad}{4G_{\rm N}} \frac{1}{d(d+2)} \frac{M^4}{2\pi} \,  ds^2_{\rm c} \qquad \Rightarrow \qquad \frac{\vg}{\vc} = \left(\frac{\Ad}{4G_{\rm N}} \frac{1}{d(d+2)} \frac{M^4}{2\pi} \right)^{\frac{1}{2}\dim G}  
\end{align}
where $\dim G = \frac{1}{2}(d+2)(d+1)$, $\Ad={\rm \Vol}(S^{d-1})$ as in (\ref{ZGYM}), and we again used ${\rm \Vol}(S^{d+1}) = \frac{2\pi}{d} {\rm \Vol}(S^{d-1})$. Combining this with (\ref{ZGhalf}), we get our desired result:
\begin{align} \label{ZGEinst}
  \ZG  =  \frac{\gamma^{\dim G}}{\vc}  \, , \qquad\qquad  \gamma \equiv   \sqrt{\frac{8 \pi G_{\rm N}}{\Ad}} \, .
\end{align}

\subsubsection*{Higher-spin gravity}

We follow the same template for the higher-spin case. 
In the interacting higher-spin theory, the Killing tensors generate a subalgebra of the nonlinear gauge algebra, with bracket $[\![ \cdot,\cdot ]\!]$ inherited from the gauge algebra, just like in the Yang-Mills and Einstein examples, except the gauge algebra is much more complicated in the higher-spin case. Fortunately it is not necessary to construct the exact gauge algebra to determine the Killing tensor algebra: it suffices to determine the lowest order deformation of the linearized gauge transformation (\ref{linGT}) fixed by the transverse-traceless cubic couplings of the theory \cite{Joung:2013nma}. The Killing tensor algebra includes in particular an $\so(d+2)$ subalgebra, that is to say an algebra of the same general form (\ref{Gnewt}) as in Einstein gravity, with some constant appearing on the right-hand side determined by the spin-2 cubic coupling in the TT action. We {\it define} the ``Newton constant'' $G_{\rm N}$ of the higher-spin theory to be this constant, that is to say we read off $G_{\rm N}$ from the $\so(d+2)$ Killing vector subalgebra by writing it as
\begin{align} \label{GnewtHS}
 [\![\bar \xi,\bar\xi']\!] = \sqrt{16 \pi G_{\rm N}} \, [\bar\xi',\bar\xi ]_{\rm L} \, .
\end{align}
The standard Killing vector basis is then again given by $M_{IJ} \equiv -(X_I \partial_J - X_J \partial_I)/\sqrt{16 \pi G_{\rm N}}$, satisfying $[\![M_{12},M_{23}]\!]=M_{13}$ etc. 

It was argued in  \cite{Joung:2013nma} that for the most general set of consistent parity-preserving cubic interactions, assuming the algebra  does not split as a direct sum of subalgebras, i.e.\ assuming the algebra is simple, there exists an up to normalization unique invariant bilinear form $\langle \cdot|\cdot \rangle_c$ on the Killing tensor algebra. We fix its normalization again by requiring the standard $\so(d+2)$ Killing vectors $M_{IJ}$ have unit norm,
\begin{align} \label{M12M12nnn}
 \langle M_{12}|M_{12} \rangle_c \equiv 1 \, .
\end{align} 
Expressed in terms of the bilinears $\langle \bar\xi_{s-1} | \bar\xi_{s-1} \rangle_{\rm PI} = \frac{M^4}{2\pi} \, \int \bar\xi_{s-1} \cdot \bar\xi_{(s-1)}$ corresponding to (\ref{dsxigenf}), the invariant bilinear on the Killing tensor algebra takes the general form
\begin{align} \label{Bscoeff}
 \langle \bar\xi | \bar\xi' \rangle_c = \sum_s B_s \, \langle \bar\xi_{s-1} | \bar\xi_{s-1}' \rangle_{\rm PI} \, , 
\end{align}   
where $B_s$ are certain constants fixed in principle by the algebra. The arguments given in \cite{Joung:2013nma} moreover imply that up to overall normalization, the coefficients $B_s$ are independent of the coupling constants in the theory. 
 More specifically, adapted (with some work, as described below) to our setting and conventions, and correcting for what we believe is a typo in \cite{Joung:2013nma}, the coefficients are $B_s \propto (2s+d-4)(2s+d-2)$. We confirmed this by comparison to  \cite{Sleight:2016xqq}, where the invariant bilinear form for minimal Vasiliev gravity in AdS$_{d+1}$, dual to the free $O(N)$ model, was spelled out in detail, building on \cite{Joung:2013nma,Joung:2014qya,Sleight:2016dba}. Analytically continuing to positive cosmological constant, implementing their ambient space $X$-contractions by a Gaussian integral, and reducing this integral to the sphere by switching to spherical coordinates, the expression in \cite{Sleight:2016xqq} can be brought to the form (\ref{Bscoeff}). This transformation almost completely cancels the factorials in the analogous coefficients $b_s$ in \cite{Sleight:2016xqq}, reducing to the simple $B_s \propto (2s+d-4)(2s+d-2)$. (The alternating signs of \cite{Sleight:2016xqq} are absent here due to the analytic continuation to positive cc.) Taking into account our normalization prescription (\ref{M12M12nnn}) (which is different from the normalization chosen in \cite{Sleight:2016xqq}), we thus get
\begin{align} \label{dsPIdscHS}
 \bigl\langle \bar\xi | \bar\xi \bigr\rangle_{\can} = \frac{2\pi}{M^4} \cdot \frac{4G_{\rm N}}{\Ad} \sum_s (2s+d-4)(2s+d-2) \, \bigl\langle\bar\xi^{(s-1)} \bigl|\bar\xi^{(s-1)}\bigr\rangle_{\rm PI} \, ,
\end{align}
with $\Ad = {\rm \Vol}(S^{d-1})$ as before.
In view of the independence of the coefficients $B_s$ of the couplings within the class of theories considered in \cite{Joung:2013nma}, i.e.\ all parity-invariant massless higher-spin gravity theories consistent to cubic order, this result is universal, valid for this entire class. 

As before for Einstein gravity and Yang Mills, from (\ref{dsPIdscHS}) we get the ratio
\begin{align} \label{groupvol2}
 \frac{\vg}{\vc} = \prod_s \left(\frac{M^4}{2\pi} \cdot \frac{\Ad}{4 G_{\rm N}} \cdot \frac{1}{ \bigl(2s+d-4 \bigr) \bigl(2s+d-2 \bigr)} \right)^{\frac{1}{2} N^{\rm KT}_{s-1}} \, .
\end{align}
Combining this with (\ref{ZGhalf}) we see that, rather delightfully, all the unpleasant-looking factors cancel, leaving us with
\begin{align} \label{cancel}
 \boxed{\quad \ZG  =  \frac{\gamma^{\dim G}}{\vc} \,   \, , \qquad  \gamma \equiv   \sqrt{\frac{8 \pi G_{\rm N}}{\Ad}} \quad }
\end{align}
This takes exactly the same form as the Einstein gravity result (\ref{ZGEinst}) except $G$ is now the higher-spin symmetry group rather than the $SO(d+2)$ spin-$2$ symmetry group. 

The cancelation of the UV divergent factors $M$ is as expected from consistency with locality. The cancelation of the $s$-dependent factors on the other hand seems surprising, in view of the different origin of the numerator (spectrum of quadratic action) and the denominator (invariant bilinear form on  higher spin algebra of interactions). Apparently the former somehow knows about the latter. We do not see an obvious reason why this is the case, although the simplicity and universality of the result suggests we should, and that this entire section should be replaceable by a one-line argument. Perhaps it is obvious in a frame-like formalism.


\subsubsection*{Newton constant from central charge}

Recall that the Newton constant $G_{\rm N}$ appearing in (\ref{cancel}) was defined by the $\so(d+2)$ algebra (\ref{GnewtHS}) in our normalization conventions. An analogous definition can be given in dS$_{d+1}$ or AdS$_{d+1}$ where the algebra becomes $\so(1,d+1)$ resp.\ $\so(2,d)$. Starting from this definition,  $G_{\rm N}$ can also be formally related to the Cardy central charge $C$ of a putative\footnote{There is no assumption whatsoever this CFT actually exists. One just imagines it exists and uses the formal holographic dictionary to infer the two-point function of this imaginary CFT's stress tensor. In dS, this ``dual CFT'' can be thought of as computing the Hartle-Hawking wave function of the universe \cite{Maldacena:2002vr}.}  boundary CFT for AdS or dS, defined as the coefficient of the CFT 2-point function of the putative energy-momentum tensor. With our definition of $G_{\rm N}$, the computation of \cite{Kovtun:2008kw} remains unchanged, so we can just copy the result obtained there:
\begin{align} \label{CGNewt}
  \boxed{\quad C =   \frac{(\pm 1)^{\frac{d-1}{2}} \, \Gamma(d+2)}{(d-1) \, \Gamma\bigl(\frac{d}{2}\bigr)^2} \cdot \frac{A_{d-1}}{8 \pi G_{\rm N}} \quad }  
\end{align}
where as before $A_{d-1}= 2 \pi^{d/2} \ell^{d-1}/\Gamma(\frac{d}{2})$, and $\pm 1$ = $+1$ for AdS and $-1$ for dS. The central charge of $N$ free real scalars equals $C = \frac{d}{2(d-1)} \, N$ in the conventions used here.  
Note that (\ref{CGNewt}) reduces to the Brown-Henneaux formula $C=3 \ell/2G_{\rm N}$ for $d=2$. In \cite{Anninos:2011ui} it was argued that the Hartle-Hawking wave function of minimal Vasiliev gravity in dS$_4$ is perturbatively computed by a $d=3$ CFT of $N$ free Grassmann scalars. This CFT has central charge $C = -\frac{3}{4} N$, hence according to (\ref{CGNewt}), $G_{\rm N} = 2^5/ \pi N$ and $\gamma = \sqrt{2 G_{\rm N}} = 8/\sqrt{\pi N}$. 


 
\vskip6mm 
 
\noindent The final result of this appendix, putting everything together, is stated in (\ref{ZPIFINAL}).

\section{One-loop and exact results for 3D theories} \label{app:dS3}

\subsection{Character formula for $Z_{\rm PI}^{(1)}$} \label{app:dS3charform}

For $d = 2$, i.e.\ dS$_3$ / $S^3$, some of the generic-$d$ formulae in sections  \ref{sec:mashsflds} and  \ref{sec:massless} become  a bit degenerate, requiring separate discussion. 
One reason $d=2$ is a bit more subtle is that the spin-$s$ irreducible representation of $SO(2)$ actually comes in two distinct chiral versions $\pm s$, as do the corresponding $SO(1,3)$ irreducible representations $(\Delta,\pm s)$. Likewise the field modes of a spin $s$ field in the path integral on $S^3$ split into chiral irreps $(n,\pm s)$ of $SO(4)$. The dimensions $D^2_s=D^2_{-s}=1$ and $D^4_{n,s}=D^4_{n,-s}=(1+n-s)(1+n+s)$ of the $SO(2)$ and $SO(4)$ irreps are correctly reproduced by the Weyl dimension formula (\ref{Weyldimeven}), rather than (\ref{Dsods2}). It should however be kept in mind that the single-particle Hilbert space of for instance a massive spin-$s \geq 1$ Pauli-Fierz field on dS$_3$ carries {\it both} helicity versions $(\Delta,\pm s)$ of the massive spin-$s$ SO(1,3) irrep, hence the character $\chi$ to be used in expressions for $Z_{\rm PI}$ in this case is $\chi = \chi_{+s} + \chi_{-s} = 2 \chi_{+s} = 2 (q^\Delta+q^{2-\Delta})/(1-q)^2$. On the other hand for a real scalar field, we just have $\chi = \chi_0 = (q^\Delta+q^{2-\Delta})/(1-q)^2$. 

For massless higher-spin gauge fields of spin $s \geq 2$, a similar reasoning implies we should include an overall factor of 2 in (\ref{bulknaive})-(\ref{edgenaive}). For an $s=1$ Maxwell field on the other hand, we get a factor of 2 in the first term but not in the second term (since the gauge parameter/ghost field is a scalar). The proper massless spin-$s$ bulk and edge characters are then obtained from these by the polar term flip (\ref{sqbrackplusop}) as usual. This results in
\begin{align}
 \chi_{\rm bulk,s} = 0 \quad (s \geq 2) \, , 
 \qquad \chi^{(s=1)}_{\rm bulk} =  \frac{2q}{(1-q)^2} \, , \qquad \chi_{\rm edge,s} = 0 \quad (\mbox{all } s) \, ,
\end{align}
expressing the absence of propagating degrees of freedom (i.e.\ particles) for massless spin-$s\geq 2$ fields on dS$_3$. 

This can also be derived more directly from the general path integral formula (\ref{ZPItrueformulaGeneral}), taking into account the $\pm s$ doubling. In particular for massless $s \geq 2$, (\ref{masslessF}) gets replaced by
\begin{align}   \label{masslessFdS3}
 \hat F_s &= \sum_{n \geq -1} \Theta(1+n) \, 2 D^{4}_{n, s} \bigl( q^{s+n} +q^{2-s+n}  \bigr)    - \sum_{n \geq -1} \Theta(1+n) \, 2 D^{4}_{n, s-1}  \bigl( q^{s+1+n} +q^{1-s+n}  \bigr),   
\end{align}
which matters for the $n=-1$ term because $\Theta(0) \equiv \frac{1}{2}$.
For $s=1$, we get instead
\begin{align} \label{F1YM}
 \hat F_1 &= \sum_{n \geq -1} \Theta(1+n) \, 2 D^4_{n,1} \bigl( q^{1+n} +q^{1+n}  \bigr)  -\sum_{n \geq 0} D^{4}_{n, 0}  \bigl( q^{2+n} +q^{n}  \bigr) \, .
\end{align}
For $s \geq 2$, the computation of $Z_{\rm char}$ and $Z_G$ remains essentially unchanged.   For $s=1$ there are some minor changes. The edge character in (\ref{edgenaive}) acquires an extra $q^0$ term in $d=2$ because $q^{s+d-3} = q^0$, so the map $\hat\chi_{\rm edge} \to [\hat \chi_{\rm edge}]_+$ gets an extra $-1$ subtraction, as a result of which the factor $-2$ in (\ref{Fplustochiplus}) becomes a $-3$. Relatedly we get an extra $q^0$ term in $q^{2s+d-4}+ q^{2s+d-2} + 2 = q^{2} + 3$ in (\ref{ZG1b}), and we end up with $\ZG = \tilde\gamma^{\dim G}/\vc$ with $\tilde\gamma = g \ell /\sqrt{A_1} = g \sqrt{\ell}/\sqrt{2\pi}$ instead of (\ref{ZGYM}). Everything else remains the same.

Finally, the phase $i^{-\pol_s}$  (\ref{polphase}) is somewhat modified. For $s \geq 2$, from (\ref{masslessFdS3}), 
\begin{align} \label{pols3D}
 \pol_s = -\sum_{n=-1}^{s-3} \Theta(1+n) \, 2 D_{n,s}^{4} - \sum_{n=-1}^{s-2} \Theta(1+n) \, 2 D_{n,s-1}^{4} = \frac{1}{3} (2 s-3) (2 s-1) (2 s+1)
\end{align}
Note that $\pol_2=5$, in agreement with \cite{Polchinski:1988ua}. $\pol_1=0$ as before, since there are no negative modes.  

\subsubsection*{Conclusion}

The final result for $Z^{(1)}_{\rm PI} = \ZG \Zchar$ in dS$_3$ replacing (\ref{ZPIFINAL})-(\ref{gammadefs}) is:
\begin{itemize}
\item For Einstein and HS gravity theories with $s \geq 2$, 
\begin{align} \label{HSdS3reg}
 Z_{\rm PI}^{(1)} = i^{- \pol} \, \frac{\gamma^{\dim G}}{\vc} \cdot \Zchar \, , \qquad \Zchar = e^{-2 \dim G \inteps^\noir \frac{dt}{2t} \, \frac{1+q}{1-q}} = (2\pi)^{\dim G} e^{-\dim G \cdot c \, \ell \, \epsilon^{-1}} \, ,
\end{align}
where as before $\gamma = \sqrt{8 \pi G_{\rm N}/A_1}=\sqrt{4 G_{\rm N}/\ell}$, $\pol=\sum_s \pol_s$, and $\vc$ is the volume with respect to the metric for which the standard $\rm so(4)$ generators $M_{IJ}$ have norm 1. We used (\ref{dimGcontrib}) to evaluate $\Zchar$.  The coefficient $c$ of the linearly divergent term is an order 1 constant depending on the regularization scheme. (For the heat kernel regularization of appendix \ref{app:renpr}, following section \ref{app:masslesssol}, $c=\frac{3 \pi}{4}$. For a simple cutoff at $t=\epsilon$ as in section \ref{sec:otherreg},  $c=2$.) 
The finite part is 
\begin{align} \label{HSdS3}
 \boxed{\quad Z^{(1)}_{\rm PI,fin} = i^{-\pol} \, \frac{(2\pi \gamma)^{\dim G}}{\vc} \, , \qquad \gamma =  \sqrt{\frac{8 \pi G_{\rm N}}{2 \pi \ell}} \quad }
\end{align}  
For example for Einstein gravity with $G=SO(4)$, we get 
\begin{align} \label{ZPIS3Einst}
 Z^{(1)}_{\rm PI,fin} = i^{- 5} \, \frac{(2\pi \gamma)^6}{(2\pi)^4} = - i \, 4 \pi^2 \gamma^6 \, .
\end{align} 

\item For Yang-Mills theories with gauge group $G$ and coupling constant $g$, we get
\begin{align} \label{ZPIYMd3}
 Z_{\rm PI}^{(1)} = \frac{\tilde\gamma^{\dim G}}{\vc} \cdot e^{ 
 \dim G \inteps^\noir \frac{dt}{2t} \, \frac{1+q}{1-q} 
 \left( \frac{2  q}{(1-q)^2} - 3 \right) } \, , \qquad \tilde \gamma = \frac{g \sqrt{\ell}}{\sqrt{2\pi}} \, .
\end{align}
Using (\ref{ZEXACT}), (\ref{dimGcontrib}), the finite part evaluates to 
\begin{align} \label{YM3D}
 Z^{(1)}_{\rm PI,fin} = \frac{(2\pi g \sqrt{\ell} \, Z_1)^{\dim G}}{\vc} \, , \qquad 
 Z_1 = 
  e^{- \frac{\zeta(3)}{4 \pi^2}} \, .
\end{align}
As in (\ref{ZPIFINAL}), $\vc$ is the volume of $G$ with respect to the metric defined by the trace appearing in the Yang-Mills action.
As a check, for $G=U(1)$ we have $\vc = 2\pi$, so $Z = g \, e^{-\zeta(3)/4\pi^2} \sqrt{\ell} $ in agreement with \cite{Klebanov:2011td}  eq.\ (3.25). 

\item We could also consider the Chern-Simons partition function on $S^3$,
\begin{align} \label{csac}
  Z_{k} = \int \CD A \, e^{i \, k \, S_{\rm CS}[A]} \, , \qquad S_{\rm CS}[A] \equiv \frac{1}{4 \pi} \int {\rm Tr} (A \wedge dA + \frac{2}{3} A \wedge A \wedge A) \, ,
\end{align}
with $k>0$ suitably quantized ($k \in \IZ$ for $G=SU(N)$ with ${\rm Tr}$ the trace in the $N$-dimensional representation). Because in this case the action is first order in the derivatives and not parity-invariant, it falls outside the class of theories we have focused on in this paper. It is not too hard though to generalize the analysis to this case. The main difference with Yang-Mills is that $\chi_{\rm bulk} = 0 = \chi_{\rm edge}$: like in the $s \geq 2$ case, the $s=1$ Chern-Simons theory has no particles. 
The function $\hat F_{1}$ is no longer given by the Maxwell version  (\ref{F1YM}), but rather by (\ref{masslessFdS3}), except without the factors of $2$, related to the fact that the CS action is first order in the derivatives. This immediately gives $F_1 = \hat F_1  = -2 \, \frac{1+q}{1-q}$.
The computation of the volume factor is analogous to our earlier discussions. The result  (in canonical framing \cite{Witten:1989ti}) is 
\begin{align} \label{Z1LCS}
 Z^{(1)}_{k} = \frac{\tilde\gamma^{\dim G}}{{\rm vol}(G)_{\rm Tr}} \, e^{-2 \dim G \inteps^\noir \frac{dt}{2t}  \frac{1+q}{1-q}} \, , \qquad
  Z^{(1)}_{k,\rm fin} = \frac{(2 \pi \tilde\gamma)^{\dim G}}{{\rm vol}(G)_{\rm Tr}} \, , \qquad \tilde \gamma = \frac{1}{\sqrt{k}} \, ,
\end{align}
where ${\rm vol}(G)_{\rm Tr}$ is the volume with respect to the metric defined by the trace appearing in the Chern-Simons action (\ref{csac}). This agrees with the standard results in the literature, nicely reviewed in section 4 of \cite{Marino:2011nm}.   


\end{itemize}

\subsection{Chern-Simons formulation of Einstein gravity} \label{app:EinstCS}

3D Einstein gravity can be reformulated as a Chern-Simons theory \cite{Witten:1988hc,Achucarro:1986uwr}. Although well-known, we briefly review some of the basic ingredients and  conceptual points here to facilitate the discussion of the higher-spin generalization in section \ref{app:HSCS}. A more detailed review of certain aspects, including more explicit solutions, can be found in section 4 of \cite{Cotler:2019nbi}. Explicit computations using the Chern-Simons formulation of $\Lambda>0$ Euclidean quantum gravity with emphasis on topologies more sophisticated than the sphere can be found in \cite{Carlip:1992wg,Guadagnini:1994ahx,Castro:2011xb}.        

\subsubsection*{Lorentzian gravity}

For the Lorentzian theory with positive cosmological constant, amplitudes are computed by path integrals $\int \CD A \, e^{i S_L}$  with real Lorentzian $SL(2,\IC)$ Chern-Simons action \cite{Witten:1989ip}
\begin{align} \label{SLCSdS}
 S_L = (\lcs + i \kappa)   S_{\rm CS}[A_+] + (\lcs - i \kappa)  S_{\rm CS}[A_-] \, ,  \quad\qquad A_+^* = A_- , 
\end{align} 
where $S_{\rm CS}$ is as in (\ref{csac}) with
$A_\pm$ an ${\rm sl}(2,\IC)$-valued connection and ${\rm Tr} ={\rm Tr}_2$. 
The vielbein $e$ and spin connection $\omega$ are the real and imaginary parts of the connection:  
\begin{align} \label{ApmsigmaetcSL}
 A_\pm = \omega \pm i   e/\ell  \, , \qquad ds^2 = 2 \, {\rm Tr}_2 \, e^2 = \eta_{ij} e^i e^j \, .
\end{align}
For the last equality we decomposed $e=e^i L_i$ in a basis $L_i$ of ${\rm sl}(2,\IR)$, say 
\begin{align} \label{Lbasis}
  (L_1,L_2,L_3) \equiv (\tfrac{1}{2} \sigma_1, \tfrac{1}{2} i \sigma_2, \tfrac{1}{2} \sigma_3) \, \qquad \Rightarrow \qquad \eta_{ij} \equiv 2 \, {\rm Tr}_2(L_i L_j) = {\rm diag}(1,-1,1) \, .
\end{align}
Note that $[L_i,L_j]=-\epsilon_{ijk} L^k$ with $L^k \equiv \eta^{kk'} L_{k'}$. When $\lcs=0$, the action reduces to the firs-order form of the Einstein action with Newton constant $G_{\rm N}= \ell/4\kappa$ and cosmological constant $\Lambda=1/\ell^2$. The equations of motion stipulate $A_\pm$ must be flat connections:
\begin{align}  \label{CSGEOM}
  dA_\pm + A_\pm \wedge A_\pm= 0 \,, 
\end{align}
equivalent with the Einstein gravity torsion constraint (with ${\omega^i}_j \equiv \eta^{i l} \epsilon_{ljk} \omega^k$) and the Einstein equations of motion \cite{Witten:1988hc}.
Turning on $\lcs$ deforms the action by parity-odd terms of gravitational Chern-Simons type. This does not affect the equations of motion (\ref{CSGEOM}).  We can take $\lcs \geq 0$ without loss of generality. The part of the action multiplied by $\lcs$ has a discrete ambiguity forcing $\lcs$ to be integrally quantized, like $k$ in (\ref{csac}).   Summarizing,     
\begin{align} \label{kaplam}
   0\leq \lcs \in \IZ \,  \, , \qquad 0 < \kappa = \frac{2 \pi \ell}{8 \pi G_{\rm N}}   \in \IR \,  , 
\end{align}

\subsubsection*{dS$_3$ vacuum solution}

A flat connection corresponding to the de Sitter metric can be obtained as follows. (We will be brief because the analog for the sphere below will be simpler and make this more clear.) Define $Q(X) \equiv 2 \, (X^4 L_4 + i X^i L_i)$ with $L_4 \equiv \frac{1}{2} {\bf 1}$ and note that $\det Q = X_4^2 + \eta_{ij} X^i X^j =: \eta_{IJ} X^I X^J$, so $\CM \equiv \{X|\det Q(X) =1 \}$ is the dS$_3$ hyperboloid, and $Q$ is a map from $\CM$ into $SL(2,\IC)$. Its square root $h \equiv Q^{1/2}$ is then a map from $\CM$ into $SL(2,\IC)/\IZ_2 \simeq SO(1,3)$, so $A_+ \equiv h^{-1} dh$ is a flat ${\rm sl}(2,\IC)$-valued connection on $\CM$. Moreover on $\CM$ we have $Q^* = Q^{-1}$, so $h^* = h^{-1}$, $A_- = A_+^* = -(dh)h^{-1}$, and $ds^2 = -\frac{1}{2} \ell^2 \, {\rm Tr} (A_+-A_-)^2=-\frac{1}{2} \ell^2 \, {\rm Tr} \, (Q^{-1}dQ)^2 = \ell^2 \eta_{IJ} dX^I dX^J$, which is the de Sitter metric of radius $\ell$ on $\CM$. 

\subsubsection*{Euclidean gravity}

Like the Einstein-Hilbert action --- or any other action for that matter ---  (\ref{SLCSdS}) may have complex saddle points, that is to say flat connections $A_\pm$ which do not satisfy the reality constraint (\ref{SLCSdS}), or equivalently solutions for which some components of the vielbein and spin connection are not real. Of particular interest for our purposes is the solution corresponding to the round metric on $S^3$. This can be obtained from the dS$_3$ solution as usual by a Wick rotation of the time coordinate. Given our choice of ${\rm sl}(2,\IR)$ basis (\ref{Lbasis}), this means $X^2 \to -i X^2$. At the level of the vielbein $e=e^i L_i$ such a Wick rotation is implemented as $e^2 \to -i e^2$. Similarly, recalling $\omega_{ij} = \epsilon_{ijk} \omega^k$, the spin connection $\omega=\omega^i L_i$ rotates as $\omega^1 \to i \omega^1$, $\omega^3 \to i \omega^3$. Equivalently, $A_\pm \to (\omega^i \pm e^i/\ell) S_i$ where $S_i \equiv \frac{1}{2} i \sigma_i$. Notice the $S_i$ are the generators of $\su(2)$, satisfying $[S_i,S_j]=-\epsilon_{ijk} S_k$, $-2 \, {\rm Tr}_2(S_i S_j) = \delta_{ij}$ and $S_i^\dagger = - S_i$. Thus the Lorentzian metric $\eta_{ij}$ gets replaced by the Euclidean metric $\delta_{ij}$, the Lorentzian  ${\rm sl}(2,\IC) = \so(1,3)$ reality condition gets replaced by the Euclidean $\su(2) \oplus \su(2) = \so(4)$ reality condition,
and the Lorentzian path integral $\int \CD A \, e^{i S_L}$ becomes a Euclidean path integral $\int \CD A \, e^{-S_E}$, where $S_E \equiv -i S_L$ is the Euclidean action: 
\begin{align} \label{Seucl}
 S_E =  (\kappa-i\lcs) \, S_{\rm CS}[A_+] - (\kappa + i \lcs) \, S_{\rm CS}[A_-] \, , \qquad A_\pm^\dagger = - A_\pm \, .
\end{align}
This can be interpreted as the Chern-Simons formulation of Euclidean Einstein gravity with positive cosmological constant. The $\su(2) \oplus \su(2)$-valued connection $(A_+,A_-)$ encodes the Euclidean vielbein, spin connection and metric as
\begin{align} \label{Apmsigmaetc}
 A_\pm = \omega \pm   e/\ell = (\omega^i \pm e^i/\ell)S_i \, , \qquad S_i \equiv  \tfrac{1}{2} i \sigma_i  \, , \qquad ds^2 =  -2 \, {\rm Tr}_2 \, e^2 = \delta_{ij} e^i e^j \, .
\end{align}
The Euclidean counterpart of the reality condition of the Lorentzian action is that $S_E$ gets mapped to $S_E^*$ under reversal of orientation. Reversal of orientation maps $S_{\rm CS}[A] \to - S_{\rm CS}[A]$, and in addition here it also exchanges the $\pm$ parts of the decomposition $\so(4) = \su(2)_+ \oplus \su(2)_-$ into self-dual and anti-self-dual parts, that is to say it exchanges $A_+ \leftrightarrow A_-$. Thus orientation reversal maps $S_E \to  -(\kappa-i\lcs) \, S_{\rm CS}[A_-] + (\kappa + i \lcs) \, S_{\rm CS}[A_+]  = S_E^*$, as required.

\subsubsection*{Round sphere solutions}

Parametrizing $S^3$ by $g \in SU(2) \simeq S^3$, it is easy to write down a flat $\su(2) \oplus \su(2)$ connection yielding the round metric of radius $\ell$:
\begin{align} \label{Asol1}
  (A_+,A_-) = (g^{-1}d g,0)  \qquad \Rightarrow \qquad e/\ell = \tfrac{1}{2}  g^{-1} dg = \omega \, , \quad ds^2 = -\tfrac{1}{2} \ell^2 \, {\rm Tr}(g^{-1} dg)^2 \, .
\end{align} 
The radius can be checked by observing that along an orbit $g(\varphi)=e^{\varphi S_3}$, we get $g^{-1} dg=d\varphi \, S_3$ so $ds=\frac{1}{2} \ell d\varphi$ and the orbit length is $ \int_0^{4\pi} ds = 2 \pi \ell$. The on-shell action is $S_E = -\frac{\kappa-i \lcs}{12\pi} \int_{S^3} {\rm Tr}_2 (g^{-1} dg)^3 = - \frac{2 (\kappa-i \lcs)}{3 \pi  \ell^3} \int e^i e^j e^k \, {\rm Tr}_2 (S_i S_j S_k) = -  \frac{\kappa-i \lcs}{6 \pi \ell^3} \int e^i e^j e^k \, \epsilon_{ijk} 
= - 2 \pi (\kappa-i \lcs)$, so 
\begin{align} \label{treelevelGCS}
  \exp(-S_E) = \exp\bigl(2 \pi \kappa + 2 \pi i \lcs \bigr) = \exp\Bigl(\frac{2\pi\ell}{4G_{\rm N}} \Bigr) \, ,
\end{align}
where we used (\ref{kaplam}). This
reproduces the standard Gibbons-Hawking result \cite{PhysRevD.15.2752} for dS$_3$. 

More generally we can consider flat connections of the form $(A_+,A_-)=(h_+^{-1} d h_+,h_-^{-1} dh_-)$ with $h_\pm = g^{n_{\pm}}$, where $n_\pm \in \IZ$ if we take the gauge group to be $G = SU(2) \times SU(2)$. 
These are all related to the trivial connection $(0,0)$ by a {\it large} gauge transformation $g \in S^3 \to (h_+,h_-) \in G$. All other flat connections on $S^3$ are obtained from these by gauge transformations continuously connected to the identity, which are equivalent to diffeomorphisms and vielbein rotations continuously connected to the identity in the metric description \cite{Witten:1988hc}. 
Large gauge transformations on the other hand are in general {\it not} equivalent to large diffeomorphisms. Indeed, 
\begin{align}
 e^{-S_E} = e^{2 \pi  n \kappa+ 2 \pi i  \tilde n \lcs} = e^{2 \pi n \kappa} \, , \qquad n \equiv n_+ - n_-, \quad \tilde n \equiv n_+ + n_- \, ,
\end{align}
so evidently different values of $n$ are physically inequivalent. Conversely, for a fixed value of $n$ but different values of $\tilde n$, we get the same metric, so these solutions are geometrically equivalent. 
In particular the $n  = 1$ solutions all produce the same round metric (\ref{Asol1}). For $n = 0$, the metric vanishes. For $n < 0$, we get a vielbein with negative determinant. Only vielbeins with positive determinant reproduce the Einstein-Hilbert action with the correct sign, so from the point of view of gravity we should discard the $n < 0$ solutions. Finally the cases $n > 1$ correspond to a metric describing a chain of $n$ spheres connected by throats of zero size, presumably more appropriately thought of as $n$ disconnected spheres. 


\subsubsection*{Euclidean path integral}


The object of interest to us is the Euclidean path integral $Z=\int \CD A \, e^{-S_E[A]}$, defined perturbatively around an $n=n_+-n_-=1$ round sphere solution $(\bar A_+,\bar A_-)=(g^{-n_+} d g^{n_+},g^{-n_-} d g^{n_-})$, such as the $(1,0)$ solution (\ref{Asol1}). Physically, this can be interpreted as the all-loop quantum-corrected Euclidean partition function of the dS$_3$ static patch.  For simplicity we take $G=SU(2) \times SU(2)$, so $n_\pm \in \IZ$ and we can formally factorize $Z$ as an $SU(2)_{k_+} \times SU(2)_{k_-}$ CS partition function where $k_\pm = \lcs \pm i \kappa$, with $\lcs \in \IZ^+$ and $\kappa \in \IR^+$:
\begin{align} \label{GCSZ}
 Z = \int_{(n_+,n_-)} \!\!\! \CD A \, e^{i k_+  S_{\rm CS}[A_+] + i k_-  S_{\rm CS}[A_-]} = Z_{\rm CS}\bigl(SU(2)_{k_+}|\bar A_{n_+} \bigr) \, Z_{\rm CS}\bigl(SU(2)_{k_-}|\bar A_{n_-} \bigr) \, , 
\end{align}
Here the complex-$k$ CS partition function $Z_{\rm CS}(SU(2)_k|\bar A_m) \equiv \int_m \CD A \, e^{i k S_{\rm CS}[A]}$ is defined perturbatively around the critical point $\bar A=g^{-m} d g^m$.  It is possible, though quite nontrivial in general, to define Chern-Simons theories at complex level $k$ on general 3-manifolds $M_3$ \cite{Witten:2010cx,Gukov:2016njj}. Our goal is less ambitious, since we only require a perturbative expansion of $Z$ around a given saddle, and moreover we restrict to $M_3=S^3$. In contrast to generic $M_3$, at least for integer $k$, the CS action on $S^3$ has a unique critical point modulo gauge transformations, and its associated perturbative large-$k$ expansion is not just asymptotic, but actually converges to a simple, explicitly known function: in canonical framing \cite{Witten:1989ti},
\begin{align} \label{ZCSkm}
 Z_{\rm CS}\bigl(SU(2)_k|\bar A \bigr)_0 = \sqrt{\tfrac{2}{2 + k}} \, \sin\bigl( \tfrac{\pi}{2  + k} \bigr) \, e^{ i (2 + k)  S_{\rm CS}[\bar A]}  \qquad (k \in \IZ^+) \, .
\end{align}
The dependence on the choice of critical point $\bar A=g^{-m} d g^m$ actually drops out for integer $k$, as $S_{\rm CS}[\bar A] = - 2 \pi m \in 2 \pi \IZ$. We have kept it in the above expression to because this is no longer the case for complex $k$. Analytic continuation to $k_\pm=\lcs \pm i \kappa$ with $\lcs \in \IZ^+$ and $\kappa \in \IR^+$ 
in (\ref{GCSZ}) then gives:
\begin{align} \label{Z0canfram}
 Z_{0} &=  \Bigl| \sqrt{\tfrac{2}{2+\lcs+  i \kappa}} \, \sin\bigl( \tfrac{\pi}{2+\lcs+  i \kappa} \bigr) \Bigr|^2
   \, e^{  2 \pi n  \kappa - 2 \pi i \, \tilde n (2+\lcs)  } 
=     
 \Bigl| \sqrt{\tfrac{2}{2+\lcs+  i \kappa}} \sin\bigl( \tfrac{\pi}{2+\lcs + i \kappa} \bigr) \cdot e^{\pi \kappa} \Bigr|^2   \, .
\end{align}

\subsubsection*{Framing dependence of phase and one-loop check}

For a general choice of $S^3$ framing with $SO(3)$ spin connection $\hat\omega$, (\ref{ZCSkm}) gets replaced by \cite{Witten:1989ti}
\begin{align} \label{ZCSframed}
 Z_{\rm CS}\bigl(SU(2)_k|\bar A \bigr) = \exp \bigl(  \tfrac{i}{24} c(k)  I(\hat\omega) \bigr) \, Z_{\rm CS}\bigl(SU(2)_k|\bar A \bigr)_0  \, , \qquad c(k) = 3  \bigl(1-\tfrac{2}{2+k} \bigr) \, ,
\end{align}
where $I(\hat\omega) = \frac{1}{4\pi} \int {\rm Tr}_3 (\hat \omega \wedge d\hat \omega + \frac{2}{3} \hat \omega \wedge \hat \omega \wedge \hat \omega)$ is the gravitational Chern-Simons action. The action $I(\hat\omega)$ can be defined more precisely as explained under (2.22) of \cite{Witten:2007kt}, by picking a 4-manifold $M$ with boundary $\partial M = S^3$ and putting 
\begin{align}
  I(\hat\omega) \equiv I_M \equiv \frac{1}{4 \pi} \int_M {\rm Tr} (R \wedge R) \, ,
\end{align}
where $R$ is the curvature form of $M$, ${R^\mu}_\nu=\frac{1}{2} {R^\mu}_{\nu\rho\sigma} dx^\rho \wedge dx^\sigma$. Taking $M$ to be a flat 4-ball $B$, the curvature vanishes so $I_B=0$, corresponding to canonical framing. Viewing $B$ as a 4-hemisphere with round metric has ${\rm Tr}(R \wedge R)=0$ pointwise so again $I_B=0$. Gluing any other 4-manifold $M$ with boundary $S^3$ to $B$, we get a closed 4-manifold $X=M-B$, with $I_M - I_B = \frac{1}{4 \pi} \int_X {\rm Tr} (R \wedge R) = 2 \pi p_1(X)$, where $p_1(X)$ is the Pontryagin number of $X$. According to the Hirzebruch signature theorem, the signature $\sigma(X)=b_2^+ - b_2^-$ of the intersection form of the middle cohomology of $X$ equals $\frac{1}{3} p_1(X)$. Therefore, for any choice of $M$,
\begin{align}
 I_M = 6 \pi r \, , \qquad r = \sigma(X) \in \IZ \, .
\end{align}  
For example $r=1$ for $X=\ICP^2$ and $r=p-q$ for $X=p\ICP^2 \# q \overline{\ICP}^2$.  Thus for general framing,
(\ref{ZCSframed}) becomes $Z_{\rm CS}(k|m) = Z_{\rm CS}(k|m)_0 \, \exp \bigl(  r \, c(k) \, \tfrac{i \pi}{4}   \bigr)$ and (\ref{Z0canfram}) becomes 
\begin{align} \label{Zrresult}
 Z_{r} = e^{i r \phi} \, \Bigl| \sqrt{\tfrac{2}{2+\lcs+  i \kappa}} \sin\bigl( \tfrac{\pi}{2+\lcs + i \kappa} \bigr) e^{\pi \kappa} \Bigr|^2  \, , \qquad r \in \IZ \, ,
\end{align}
where, using $c(k) = 3  \bigl(1-\tfrac{2}{2+k} \bigr)$, the phase is given by 
\begin{align}
 r\phi = r\bigl(c(\lcs+i\kappa) + c(\lcs-i\kappa)\bigr) \frac{\pi}{4}  = r\bigl(1 - \tfrac{2 (2+\lcs)}{(2+\lcs)^2+\kappa^2}\bigr) \frac{3 \pi}{2} \, .
\end{align}
In the  weak-coupling limit $\kappa \to \infty$,
\begin{align} \label{Zrrrrrrr}
 Z_{r} \to (-i)^{r} \, \frac{2 \pi^2}{\kappa^3} \cdot e^{2 \pi \kappa} \, .
\end{align}
Using (\ref{kaplam}) and taking into account that we took $G=SU(2) \times SU(2)$ here, the absolute value agrees with our general one-loop result (\ref{HSdS3}) in the metric formulation, with the phase $(-i)^r$ matching Polchinski's phase $i^{-P} = i^{-5} = - i$ in (\ref{ZPIFINAL}) for odd framing $r$.\footnote{Strictly speaking for $r = 1$ mod 4, but $i^P$ vs $i^{-P}$ in (\ref{ZPIFINAL}) is a matter of conventions,  so there is no meaningful distinction we can make here.} We do not have any useful insights into why (or whether) CS framing and the phase $i^{-P}$ might have anything to do with each other, let alone why odd but not even framing should reproduce the phase of \cite{Polchinski:1988ua}. Perhaps different contour rotation prescriptions as those assumed in \cite{Polchinski:1988ua} might reproduce the canonically framed ($r=0$) result in the metric formulation of Euclidean gravity. We  leave these questions open.

\vskip1mm \noindent {\it Comparison to previous results:}
The Chern-Simons formulation of gravity was applied to calculate Euclidean $\Lambda>0$ partition functions in \cite{Carlip:1992wg,Guadagnini:1994ahx,Castro:2011xb}. The focus of these works was on summing different topologies. Our one-loop (\ref{Zrrrrrrr}) in canonical framing agrees with \cite{Carlip:1992wg} up to an unspecified overall normalization constant in the latter, agrees with $Z(S^3)/Z(S^1 \times S^2)$ in \cite{Guadagnini:1994ahx} combining their eqs.\ (13),(32), and
disagrees with eq.\ (4.39) in \cite{Castro:2011xb}, $Z^{(1)}(S^3) = \pi^3/(2^5 \kappa)$.  

\def\CF{{\cal F}}

\subsection{Chern-Simons formulation of higher-spin gravity} \label{app:HSCS}

The $SL(2,\IC)$ Chern-Simons formulation of Einstein gravity (\ref{SLCSdS}) has a natural extension to an $SL(n,\IC)$ Chern-Simons formulation of higher-spin gravity --- the positive cosmological constant analog of the theories studied e.g.\ in \cite{Blencowe:1988gj,Bergshoeff:1989ns,Castro:2011fm,Perlmutter:2012ds,Campoleoni:2010zq,Ammon:2011nk}. The Lorentzian action is
\begin{align} \label{SCSGHSLor}
 S_L = (\lcs + i \kappa) \, S_{\rm CS}[\CA_+] + (\lcs - i \kappa) \, S_{\rm CS}[\CA_-]  \, , \qquad \CA_+^* = \CA_- \, ,
\end{align}
where $S_{\rm CS}[\CA] =  \frac{1}{4 \pi} \int {\rm Tr}_n\bigl( \CA \wedge d \CA  + \tfrac{2}{3} \CA \wedge \CA \wedge \CA \bigr)$, an $\CA$ is an ${\rm sl}(n,\IC)$-valued connection, $\kappa \in \IR^+$ and $\lcs \in \IZ^+$. The corresponding Euclidean action $S_E = - i S_L$ extending (\ref{Seucl}) is given by
\begin{align} \label{SCSGHS}
 S_E = (\kappa-i\lcs) \, S_{\rm CS}[\CA_+] - (\kappa + i \lcs) \, S_{\rm CS}[\CA_-]  \, , \qquad \CA_\pm^\dagger = - \CA_\pm \, ,
\end{align} 
where $\CA_\pm$ are now independent $\su(n)$-valued connections.  

\def\sl{{\rm sl}}

\subsubsection{Landscape of dS$_3$ vacua} \label{sec:landscape}

The solutions $A$ of the original ($n=2$) Einstein gravity theory can be lifted to solutions $\CA=R(A)$ of the extended ($n>2$) theory by choosing an embedding $R$ of $\sl(2)$ into $\sl(n)$. More concretely, such lifts are specified by picking an $n$-dimensional representation $R$ of $\su(2)$, 
\begin{align}
 R = \oplus_a {\bf m}_a \, , \qquad \sum_a m_a = n \, , \qquad \CS_i = R(S_i) = \oplus_a J^{(m_a)}_i \, .
\end{align}
Here $J^{(m)}_i$ are the standard anti-hermitian spin $j=\frac{m-1}{2}$ representation matrices of $\su(2)$, satisfying the same commutation relations and reality properties as the spin-$\frac{1}{2}$ generators $S_i$  in (\ref{Apmsigmaetc}). 
Then the matrices $\CL_i \equiv R(L_i)$ with the $L_i$ as in (\ref{Lbasis}) 
are real, generating the corresponding $n$-dimensional representation of $\sl(2,\IR)$. 
The Casimir eigenvalue of the spin $j=\frac{m-1}{2}$ irrep is $j(j+1)=\frac{1}{4}(m^2-1)$, so
\begin{align} \label{TRval}
 {\rm Tr}_n (\CS_i \CS_j) = -\tfrac{1}{2} T_R \, \delta_{ij} \, , \qquad {\rm Tr}_n(\CL_i \CL_j) = \tfrac{1}{2} T_R \, \eta_{ij} \, ,  \qquad T_R \equiv \frac{1}{6} \sum_a m_a(m_a^2-1) \, ;
\end{align}
A general $SL(2,\IC)$ connection $A=A^i L_i$ has curvature $dA + A \wedge A = \bigl(d A^i - \frac{1}{2} {\epsilon^i}_{jk} A^j A^k \bigr) L_i$, and an $SL(n,\IC)$ connection of the form $\CA=R(A) = A^i \CL_i$ has curvature $d\CA + \CA \wedge \CA = \bigl(d A^i - \frac{1}{2} {\epsilon^i}_{jk} A^j A^k \bigr) \CL_i$, hence $\CA=R(A)$ solves the equations of motion of the extended $SL(n,\IC)$ theory iff $A$ solves the equations of motion of the original Einstein $SL(2,\IC)$ theory. In other words, restricting to connections $\CA = A^i \CL_i$ amounts to a consistent truncation, which may be interpreted as the gravitational subsector of the $n>2$ theory. Substituting $\CA=R(A)$ into the action (\ref{SCSGHSLor}) gives the consistently truncated action
\begin{align} \label{TRdef}
 S_L = (\lcs + i \kappa) T_R \, S_{\rm CS}[A_+] + (\lcs - i \kappa) T_R \, S_{\rm CS}[A_-]  \, , \qquad A_+^* = A_- \, ,
\end{align}
which is of the exact same form as the original Einstein CS gravity theory (\ref{SLCSdS}), except $\lcs+i \kappa$ is replaced by $ (\lcs+i \kappa) T_R$. Thus we can naturally interpret the components $A^i_\pm$ again as metric/vielbein/spin connection degrees of freedom, just like in (\ref{ApmsigmaetcSL}), i.e.\ $A^i_{\pm} = \omega^i \pm i e^i/\ell$, $ds^2 = \eta_{ij} e^i e^j$, and the lift $\CA=R(A)$ of the original solution $A$ corresponding to the dS$_3$ metric again as a solution corresponding to the dS$_3$ metric. The difference is that the original relation (\ref{kaplam}) between $\kappa$ and $\ell/G_{\rm N}$ gets modified  
to 
\begin{align} \label{gamkapn}
 \kappa \, T_R = \frac{2 \pi \ell}{8 \pi G_{\rm N}}   \, .
\end{align}
Since $\kappa$ is fixed, this means the dimensionless ratio $\ell/G_{\rm N}$ depends on the choice of $R$. Thus the different solutions $\CA = R(A)$ of the $SL(n,\IC)$ theory can be thought of as different de Sitter vacua of the theory, labeled by $R$, with different values of the curvature radius in Planck units $\ell/G_{\rm N}$. These are the dS analog of the AdS vacua discussed in \cite{Ammon:2011nk}. The total number of vacua labeled by $R=\oplus_a {\bf m}_a$ equals the number of partitions of $n=\sum_a m_a$, 
\begin{align}
 \CN_{\rm vac} \sim e^{2 \pi \sqrt{n/6}} \, \qquad (n \gg 1) \, .
\end{align} 
For, say, $n \sim 2 \times 10^5$, this gives $\CN_{\rm vac} \sim 10^{500}$. 

Analogous considerations hold for the Euclidean version of the theory.  For example the round sphere solution (\ref{Asol1}) is lifted to 
\begin{align} \label{lifedAsol}
 (\CA_+,\CA_-) = \bigl(R(A_+),0\bigr) = \bigl(R(g)^{-1} d R(g),0 \bigr), \qquad R(e^{\alpha^i S_i} ) \equiv e^{\alpha^i \CS_i} \, ,
\end{align}
with the sphere radius $\ell$ in Planck units given again by (\ref{gamkapn}). The tree-level contribution of the solution (\ref{lifedAsol}) to the  Euclidean path integral is
\begin{align} \label{treelevelsun}
 \exp(-S_E) = \exp\bigl(2 \pi \kappa \, T_R \bigr) = \exp\Bigl( \frac{2 \pi \ell}{4 \, G_{\rm N}} \Bigr) \, .
\end{align}
Note that $\CS^{(0)} \equiv -S_E = \frac{2 \pi \ell}{4  G_{\rm N}}$ is the usual dS$_3$ Gibbons-Hawking horizon entropy \cite{PhysRevD.15.2752}. Its value $\CS^{(0)} = 2 \pi \kappa \, T_R$ depends on the vacuum $R = \oplus_a {\bf m}_a$ through $T_R$ as given by (\ref{TRval}). The vacuum $R$ maximizing $e^{-S_E}$ corresponds to the partition of $n=\sum_a m_a$ maximizing $T_R$. Clearly the maximum is achieved for $R={\bf n}$:
\begin{align} \label{TRprinc}
 \max_R \, T_R = T_{\bf n} = \frac{n (n^2-1)}{6}  \, .
\end{align}
The corresponding embedding of $\su(2)$ into $\su(n)$ is called the ``principal embedding''. Thus the ``principal vacuum'' maximizes the entropy at $S_{\rm GH, {\bf n}} =  \frac{1}{6} n (n^2-1) \, 2 \pi \kappa$, exponentially dominating the Euclidean path integral in the semiclassical (large-$\kappa$) regime.  In the remainder we focus on the Euclidean version of the theory.

\subsubsection{Higher-spin field spectrum and  algebra} \label{sec:hsa}

Of course for $n>2$, there are more degrees of freedom in the $2(n^2-1)$ independent  components of $\CA_\pm$ than just the 3+3 vielbein and spin connection degrees of freedom $\CA^i_\pm \CS_i$. The full set of fluctuations around the vacuum solution can be interpreted in a metric-like formalism as higher-spin field degrees of freedom. The precise spectrum depends on the vacuum $R$. For the principal vacuum $R={\bf n}$, we get the higher-spin vielbein and spin connections of a set of massless spin-$s$  fields of $s=2,3,\ldots,n$, as was worked out in detail for the AdS analog in \cite{Campoleoni:2010zq}. Indeed $\su(n)$ decomposes under  the principally embedded $\su(2)$ subalgebra into spin-$r$ irreps, $r=1,2,\ldots,n-1$, generated by the traceless symmetric products $\CS_{i_1 \cdots i_r}$ of the generators $\CS_i$. As reviewed in \cite{Joung:2014qya}, this means we can identify the $\su(n)_+ \oplus \su(n)_-$ Lie algebra of the theory (\ref{SCSGHS}) with the higher-spin algebra 
$\hs_n(\su(2))_+ \oplus \hs_n(\su(2))_-$, where $\su(2)_+ \oplus \su(2)_- = \so(4)$ is the principally embedded gravitational subalgebra. In the metric-like formalism the spin-$r$ generators correspond to (anti-)self-dual Killing tensors of rank $r$.  These are the Killing tensors of massless symmetric spin-$s$ fields with $s=r+1$. As a check, recall the number of (anti-)self-dual rank $r$ Killing tensors  on $S^3$ equals $D^4_{r,\pm r}=2r+1$, correctly adding up to \vskip-7mm
\begin{align} \label{dimdecomp}
 \sum_\pm \sum_{r=1}^{n-1} D^4_{r,\pm r} = 2 \sum_{r=1}^{n-1} (2r+1) = 2(n^2-1) \, . 
\end{align} \vskip-1mm \noindent
For different choices of embedding $R$, we get different $\su(2)$ decompositions of $\su(n)$. For example for $n=12$, while the principal embedding  $R={\bf 12}$ considered above gives the $\su(2)$ decomposition ${\bf 143}_{\su(12)} =  
{\bf 3}+{\bf 5}+{\bf 7}+{\bf 9}+{\bf 11}+{\bf 13}+{\bf 15}+{\bf 17}+{\bf 19}+{\bf 21}+{\bf 23}$, taking $R={\bf 6} \oplus {\bf 4} \oplus {\bf 2}$ gives ${\bf 143}_{\su(12)} = 2\cdot{\bf 1}+7\cdot{\bf 3}+8\cdot{\bf 5}+6\cdot{\bf 7}+3\cdot{\bf 9}+{\bf 11}$. Interpreting these as Killing tensors for $n_s$ massless spin-$s$ fields, we get for the former $n_2=1,n_3=1,\ldots,n_{12}=1$, and for the latter $n_1=2,n_2=7,n_3=8,n_4=6,n_5=3,n_6=1$. The tree-level entropy $\CS^{(0)}=2\pi \ell/4 G_{\rm N}$ for $R={\bf 12}$ is $\CS^{(0)}=286 \cdot 2 \pi \kappa$, and for $R={\bf 6} \oplus {\bf 4} \oplus {\bf 2}$ it is $\CS^{(0)}= 46 \cdot 2 \pi \kappa$.

\subsubsection{One-loop Euclidean path integral from metric-like formulation} \label{sec:zpim} 

In view of the above higher-spin interpretation of the theory, we can apply our general massless HS formula (\ref{HSdS3}) with $G = SU(n) \times SU(n)$ to obtain the one-loop contribution to the Euclidean path integral (for $\lcs=0$). In combination with (\ref{gamkapn}) this takes the form
\begin{align} \label{HSdS3bis}
 Z^{(1)}_{\rm PI} = i^{-\pol} \, \frac{(2\pi \gamma)^{\dim G}}{\vc} \, , \qquad \gamma = \sqrt{\frac{8 \pi G_{\rm N}}{2 \pi \ell}} = \frac{1}{\sqrt{\kappa \, T_R}} \, .
\end{align} 
Recall that $\vc$ is the volume of $G$ with respect to the metric normalized such that $\langle M|M \rangle_{\rm c} = 1$, where $M$ is one of the standard $\so(4) = \su(2) \oplus \su(2)$ generators, which we can for instance take to be the rotation generator $M= \CS_{3} \oplus \CS_{3}$. In the context of Chern-Simons theory, it is more natural to consider the volume ${\rm vol}(G)_{{\rm Tr}_n}$ with respect to the metric defined by the trace appearing in the Chern-Simons action (\ref{SCSGHS}). Using the definition of $T_R$ in (\ref{TRdef}), we see the trace norm of $M$ is $\langle M|M\rangle_{{\rm Tr}_n} = -2 \, {\rm Tr}_n(\CS_3 \CS_3)  = T_R = T_R \langle M|M\rangle_{\rm c}$, hence ${\rm vol}(G)_{{\rm Tr}_n} = (\sqrt{T_R})^{\dim G} \, \vc$. Note that upon substituting this in (\ref{HSdS3bis}), the $T_R$-dependent factors cancel out. Finally, using (\ref{pols3D}), we get $\pol = \sum_{s=2}^n \pol_s = \frac{1}{3} (2 s-3) (2 s-1) (2 s+1) = \frac{2}{3} n^2 (n-1)(n+1)- (n^2-1)$. Because $(n-1) \cdot n \cdot (n+1)$ is divisible by 3, the first term is an integer, and moreover a multiple of $8$ because either $n^2$ or $(n+1)(n-1)$ is a multiple of $4$. Hence $i^{-P} = i^{(n^2-1)}$, which equals $-i$ for even $n$ and $+1$ for odd $n$. Thus we get
\begin{align} \label{ZPIkappa}
 Z^{(1)}_{\rm PI} =  i^{n^2-1} \, \frac{(2 \pi \tilde \gamma)^{\dim G}}{{\rm vol}(G)_{{\rm Tr}_n}}  \, , \qquad \quad \tilde\gamma \equiv \frac{1}{\sqrt{\kappa}} \, .
\end{align}

\subsubsection{Euclidean path integral from CS formulation } \label{sec:epics}

As in the $SU(2) \times SU(2)$ Einstein gravity case, we can derive an all-loop expression for the Euclidean partition function $Z(R)$ of the $SU(n) \times SU(n)$ higher-spin gravity theory (\ref{SCSGHS}) expanded around a lifted round sphere solution $\bar\CA=R(\bar A)$ such as (\ref{lifedAsol}), by naive analytic continuation of the exact $SU(n)_{k_+} \times SU(n)_{k_-}$ partition function on $S^3$ to $k_{\pm} = \lcs \pm i \kappa$, paralleling (\ref{GCSZ}) and the subsequent discussion there. The $SU(n)_k$ generalization of the canonically framed $SU(2)_k$ result (\ref{ZCSkm}) as spelled out e.g.\ in \cite{Marino:2004uf,Periwal:1993yu} is
\begin{align}
 Z_{\rm CS}(SU(n)_k|\bar \CA)_0 = \frac{1}{\sqrt{n}}  \frac{1}{(n+k)^{\frac{n-1}{2}}} \prod_{p=1}^{n-1} \Bigl( 2  \sin \frac{\pi p}{n + k} \Bigr)^{(n-p)} \cdot e^{i(n+k) S_{\rm CS}[\bar \CA]} \, .
\end{align}
The corresponding higher-spin generalization of (\ref{Z0canfram}) is therefore
\begin{align} \label{ZR0cs2}
 Z(R)_{0} = \biggl| \frac{1}{\sqrt{n}}  \frac{1}{(n+\lcs+i\kappa)^{\frac{n-1}{2}}} \prod_{p=1}^{n-1} \Bigl( 2  \sin \frac{\pi p}{n + \lcs + i \kappa} \Bigr)^{(n-p)} \,  \biggr|^2 \, \cdot e^{2 \pi \kappa T_R} \, .
\end{align}
Physically this can be interpreted as the all-loop quantum-corrected Euclidean partition function of the dS$_3$ static patch in the vacuum labeled by $R$. 
The analog of the result (\ref{ZCSframed}) for more general framing $I_M$ is
\begin{align} \label{ZCSframedgenn}
 Z_{\rm CS}(SU(n)_k|\bar \CA) =  \exp \bigl(  \tfrac{i}{24} c(k)  I_M \bigr) \, Z_{\rm CS}(SU(n)_k|\bar \CA)_0 \, , \qquad c(k) = (n^2-1)  \bigl(1-\tfrac{n}{n+k} \bigr) \, ,
\end{align}
hence the generalization of (\ref{Zrresult}) for arbitrary framing $I_M=6 \pi r$, $r \in \IZ$, is
\begin{align} \label{ZRframed}
 Z(R)_{r} = e^{i r \phi} \, Z(R)_0 \, ,
\end{align}
where $\phi = \bigl(c(\lcs+i\kappa) + c(\lcs-i\kappa)\bigr) \frac{\pi}{4}  = \bigl(1 - \frac{2 (n+\lcs)}{(n+\lcs)^2+\kappa^2}\bigr) (n^2-1) \frac{\pi}{2}$. In the limit $\kappa \to \infty$, 
\begin{align}
 Z(R)_r \to i^{r(n^2-1)} \, \frac{1}{n}  \frac{1}{\kappa^{n-1}} \prod_{r=1}^{n-1} \Bigl(\frac{2 \pi r}{\kappa} \Bigr)^{2(n-r)} = i^{r(n^2-1)} \left(\frac{2\pi}{\sqrt{\kappa}}\right)^{2(n^2-1)}  \biggl( \frac{1}{\sqrt{n}} \prod_{s=2}^n \frac{\Gamma(s)}{(2\pi)^s}\biggr)^2 \, . 
  \label{ZoneloopCSgenn}
\end{align}
Recognizing $n^2-1 = \dim SU(n)$ and $\sqrt{n} \prod_{s=2}^n (2\pi)^s/\Gamma(s) = {\rm vol}(SU(n))_{{\rm Tr}_n}$ (\ref{volsuN}),  
we see this precisely reproduces the one-loop result (\ref{ZPIkappa}). Like in the original $n=2$ case, the phase again matches for odd framing $r$. (The agreement at one loop can also be seen more directly by a slight variation of the computation leading to (\ref{Z1LCS}).) This provides a nontrivial check of our higher-spin gravity formula (\ref{HSdS3}) and more generally (\ref{ZPIFINAL}).

\subsubsection{Large-$n$ limit and topological string description} \label{sec:top}

In generic dS$_{d+1}$ higher-spin theories, $\dim G = \infty$. To mimic this case, consider the $n \to \infty$ limit of $SU(n) \times SU(n)$ dS$_3$ higher-spin theory with $\lcs=0$. 
A basic observation is that the loop expansion is only reliable then if $n/\kappa \ll 1$. Using (\ref{gamkapn}), this translates to $T_R \, n \ll \frac{\ell}{G_{\rm N}}$
For the exponentially dominant principal vacuum $R={\bf n}$, this becomes $n^4 \ll \ell/G_{\rm N}$ while at the other extreme, for the nearly-trivial $R={\bf 2} \oplus {\bf 1} \oplus \cdots \oplus {\bf 1}$, this becomes $n \ll \ell/G_{\rm N}$. Either way,  for   fixed $\ell/G_N$, the large-$n$ limit is necessarily strongly coupled, and the one-loop formula (\ref{HSdS3bis}), or equivalently (\ref{ZPIkappa}) or (\ref{ZoneloopCSgenn}), becomes unreliable. Indeed, according to this formula, $\log Z^{(1)} \sim \log\left(\frac{n}{\kappa} \right) \cdot n^2$ in this limit, whereas the exact expression (\ref{ZR0cs2}) actually implies {$\log Z^{\rm{(loops)}} \to 0$}. 

In fact, the partition function {\it does} have a natural weak coupling expansion in the $n \to \infty$ limit --- not as a 3D higher-spin gravity theory, but rather as a topological string theory. $U(n)_k$ Chern-Simons theory on $S^3$ has a description \cite{Witten:1992fb} as an open topological string theory on the deformed conifold $T^* S^3$ with $n$ topological D-branes wrapped on the $S^3$, and a large-$n$ 't Hooft dual description \cite{Gopakumar:1998ki} as a {\it closed} string theory on the {\it resolved} conifold. Both descriptions are reviewed in \cite{Marino:2004uf}, whose notation we follow here. The string coupling constant is $g_s = 2\pi/(n+k)$ and the K\"ahler modulus of the resolved conifold is $t =  \int_{S^2} J + i B  = i g_s n = 2 \pi i n/(n+k)$. Under this identification, 
\begin{align}
 Z_{\rm CS}(SU(n)_k)_0 =\sqrt{\tfrac{n+k}{n}}  \, Z_{\rm CS}(U(n)_k)_0 = \sqrt{\tfrac{2 \pi i}{t}} \, Z_{\rm top}(g_s,t) \equiv \tilde Z_{\rm top}(g_s,t) \, .
\end{align}
Thus we can write the $SU(n)_{\lcs+i\kappa} \times SU(n)_{\lcs-i\kappa}$ higher-spin Euclidean gravity partition function (\ref{ZR0cs2}) expanded around the round $S^3$ solution $\bar\CA=R(\bar A)$ as
\begin{align} \label{ZRisZtopsquared}
 \boxed{Z(R)_{0} = \left| \tilde Z_{\rm top}(g_s,t) \,
  e^{-\pi T_R \cdot 2\pi i/g_s} \right|^2 } 
\end{align}
where $T_R$ was defined in (\ref{TRval}), maximized for $R={\bf n}$ at $T_{\bf n} = \frac{1}{6} n(n^2-1)$, and 
\begin{align}
 g_{s}  
  = \frac{2\pi}{n + \lcs + i\kappa} \, , \qquad 
 t = i g_s n = \frac{2 \pi i n}{n + \lcs + i\kappa} 
  \, .
\end{align}
Note that $t$ takes values inside a half-disk of radius $\frac{1}{2}$ centered at $t=i \pi$, with ${\rm Re} \, t >0$. The higher-spin gravity theory (or the open string theory description on the deformed conifold) is weakly coupled when $\kappa \gg n$, which implies $|t| \ll 1$. In the free field theory limit $\kappa \to \infty$, we get $g_s \sim - 2 \pi i/\kappa \to 0$ and $t \sim 2\pi n/\kappa \to 0$, which is singular from the closed string point of view.  
In the 't Hooft limit $n \to \infty$ with $t$ kept finite, the closed string  is weakly coupled and sees a smooth geometry. The earlier discussed Vasiliev-like limit $n \to \infty$ with $\lcs=0$ and $\ell/G_{\rm N} \sim T_{\bf n} \kappa \sim n^3 \kappa$ fixed, infinitely strongly coupled from the 3D field theory point of view, maps to $g_s \sim 2\pi/n \to 0$ and $t \sim 2 \pi i + 2 \pi \kappa/n \to 2\pi i$, which is again singular from the closed string point of view, differing from the 3D free field theory singularity by a mere $B$-field monodromy, reflecting the more general $n \leftrightarrow \lcs + i \kappa$, $t \leftrightarrow 2 \pi i - t$ level-rank symmetry.

\section{Quantum dS entropy: computations and examples} \label{app:rendS}

Here we provide the details for section \ref{sec:gravETD}.

\subsection{Classical gravitational dS thermodynamics} \label{sec:classquantthermo}


\subsubsection{3D Einstein gravity example} \label{sec:cgt}

For concreteness we start with pure 3D Einstein gravity as a guiding example, but we will phrase the discussion so generalization will be clear. The Euclidean action in this case is 
\begin{equation} \label{action3Dgrav}
S_E[g] = \frac{1}{8\pi G} \int d^3 x \sqrt{g}\Bigl(\Lambda  -\frac{1}{2} R \Bigr)\, ,
\end{equation}
with $\Lambda>0$. The tree-level contribution to the entropy (\ref{UPISPIgrav}) is 
\begin{align} \label{CZtree}
 \CS^{(0)} = \log \CZ^{(0)}  \, , \qquad \CZ^{(0)} =  \lint_{\!\!\!\text{tree}} \, \CD g \, e^{-S_E[g]} \, . 
\end{align}
The dominant saddle of (\ref{CZtree}) is a round $S^3$ metric $g_\ell$ of radius $\ell\!=\!\ell_0$ minimizing $S_E(\ell) \equiv S_E[g_\ell]$:
\begin{align} \label{SEelldef}
 \CZ^{(0)} = \lint_{\!\!\!\text{tree}} \, d\ell \, e^{-S_E(\ell)} \, , \qquad S_E(\ell) = \frac{2\pi^2}{8 \pi G} \bigl( \Lambda \ell^3 - 3 \ell \bigr) \, ,
\end{align}
where $\int_{\rm tree}$ means evaluation at the saddle point, here at the on-shell radius $\ell=\ell_0$: 
\begin{align} \label{CZtree2}
  \partial_\ell S_E(\ell_0)=0  \qquad \Rightarrow \qquad \Lambda = \frac{1}{\ell_0^2} \, , \qquad \CS^{(0)}   = -S_E(\ell_0) = \frac{2\pi\ell_0}{4G} \, ,
\end{align}
reproducing the familiar area law  $\CS^{(0)}=A/4G$ for the horizon entropy. 


We now recast the above in a way that will allow us to make contact with the formulae of section \ref{sec:ETDfix} and will naturally generalize beyond tree level in a diffeomorphism-invariant way. To this end we define an ``off-shell'' tree-level partition function at fixed (off-shell) volume $V$:  
\begin{align} \label{defZ0V}
  Z^{(0)}(V)   \equiv \lint_{\!\!\!\text{tree}} \, d\sigma \, \lint_{\!\!\!\text{tree}} \, \CD g \, e^{-S_E[g] + \sigma(\int \! \sqrt{g} - V)} \, .
\end{align}
Evaluating the integral is equivalent to a constrained extremization problem with Lagrange multiplier $\sigma$ enforcing the constraint $\int\! \sqrt{g} = V$. 
The dominant saddle is the round sphere $g=g_\ell$ of radius $\ell(V)$ fixed by the  volume constraint:  
\begin{align} \label{ellVrel}
 Z^{(0)}(V) = e^{-S_E(\ell)} \, , \qquad  
 2 \pi^2 \ell^3= V \, .
\end{align}  
Paralleling (\ref{UPIdef}) and (\ref{SPIdef}), we define from this an off-shell energy density and entropy,
\begin{equation} \label{deftho0S0}
\begin{aligned} 
  \rho^{(0)} &\equiv -\partial_V \log Z^{(0)} = - {\tfrac{1}{3} \ell \partial_\ell \log Z^{(0)}} \, / \, {V} = \bigl( \Lambda - \ell^{-2} \bigr)/8\pi G  \\
  S^{(0)} 
  &\equiv \bigl(1-V \partial_V \bigr) \log Z^{(0)} = \bigl(1-\tfrac{1}{3} \ell \partial_\ell \bigr) \log Z^{(0)} = \frac{2 \pi \ell}{4 G} \, .
\end{aligned}
\end{equation}
$\rho^{(0)}$ is the sum of the positive cosmological constant and negative curvature energy densities. $S^{(0)}$ is independent of $\Lambda$. 
It is the Legendre transform of  $\log Z^{(0)}$: 
\begin{align} \label{LegRel0}
   S^{(0)} = \log Z^{(0)} + V \rho^{(0)} \, , \qquad d \log Z^{(0)} = - \rho^{(0)} dV \, , \qquad    dS^{(0)} = V d \rho^{(0)} \, ,
\end{align}  
Note that evaluating $\int_{\rm tree} d\sigma$ in (\ref{defZ0V}) sets $\sigma=-\partial_V \log Z^{(0)} =\rho^{(0)}(V)$.
On shell, 
\begin{align} \label{onshell3DE}
  \rho^{(0)}(\ell_0) = 0 \, , \qquad \CS^{(0)} = \log Z^{(0)}(\ell_0) = S^{(0)}(\ell_0) = \frac{2 \pi \ell_0}{4G} \, .
\end{align}
Paralleling (\ref{LegendrelZtoS}), the differential relations in (\ref{LegRel0})  can be viewed as the first law of tree-level de Sitter thermodynamics. We can also consider variations of coupling constants such as $\Lambda$. Then $d\log Z^{(0)} = -\rho^{(0)} dV - \frac{1}{8\pi G} V d\Lambda$, $dS^{(0)}=V d\rho^{(0)} - \frac{1}{8\pi G} V d\Lambda$.  On shell, $d\CS^{(0)} = -\frac{V_0}{8 \pi G} \, d\Lambda$. 






\subsubsection{General $d$ and higher-order curvature corrections} \label{sec:gdhoc}

The above formulae readily extend to general dimensions and to gravitational actions $S_E[g]$ with general higher-order curvature corrections.  
Using that $R_{\mu\nu\rho\sigma} = (g_{\mu\rho} g_{\nu\sigma} - g_{\mu\sigma}g_{\nu\rho})/\ell^2$ 
for the round\footnote{By virtue of its $SO(d+2)$ symmetry, the round sphere metric $g_\ell$ with $\Omega_{d+1} \ell^{d+1} = V$ is a saddle of (\ref{defZ0V}). Spheres of dimension $\geq 5$ admit a plentitude of Einstein metrics that are {\it not} round \cite{10.4310/jdg/1214431962,Bohm:1998vc,Gibbons:2002th,Boyer:2003pe}, but as explained e.g.\ in \cite{Gibbons:2011dh}, by Bishop's theorem \cite{bishop}, these saddles are subdominant in Einstein gravity. In the large-size limit, higher-order curvature corrections are small, hence the round sphere dominates in this regime. \label{manyEinst}} $S^{d+1}$, $Z^{(0)}(V)$ (\ref{defZ0V}) can be evaluated explicitly for any action. It takes the form   
\begin{align} \label{logZ0Vgend}
 \log Z^{(0)}(V) = -S_E[g_\ell] = \frac{\Omega_{d+1}}{8 \pi G} \bigl( -\Lambda \ell^{d+1} + \tfrac{d(d+1)}{2} \, \ell^{d-1} +  \cdots \bigr) \, , \qquad \Omega_{d+1} \ell^{d+1} = V \, ,
\end{align}
where $+\cdots$ is a sum of $R^n$ higher-order curvature corrections $\propto \ell^{-2n}$  and $\Omega_{d+1}=(\ref{volSn})$.
  The off-shell energy density and entropy are defined as in (\ref{deftho0S0})  
\begin{equation} \label{rho0S0gend}
\begin{aligned}
 \rho^{(0)} &= {-\tfrac{1}{d+1} \ell \partial_\ell \log Z^{(0)}}\,/ \, {V} 
   =  \bigl( \Lambda - \tfrac{d(d-1)}{2} \, \ell^{-2}   + \cdots  \bigr) /8\pi G  \\
 S^{(0)} &= \bigl(1-\tfrac{1}{d+1} \ell \partial_\ell \bigr) \log Z^{(0)} 
    = \frac{A}{4G} \bigl(1+\cdots\bigr) \, .  
 \end{aligned}
\end{equation}             
where 
$A=\Omega_{d-1} \ell^{d-1}$ 
and $+\cdots$ are $1/\ell^{2n}$ curvature corrections. 
The on-shell radius $\ell_0$ solves $\rho^{(0)}(\ell_0)=0$, most conveniently viewed as giving a parametrization $\Lambda(\ell_0)$. 

\def\g{\lambda}

As an example, consider the general action up to order $R^2$ written as  
\begin{align} \label{SERiem2}
S_E=\frac{1}{8 \pi G}\int \!\! \sqrt{g}\Bigl(\Lambda - \tfrac{1}{2} R - \ls^2  \bigl( \g_{C^2}  C^{\mu\nu\rho\sigma} C_{\mu\nu\rho\sigma} + \g_{R^2}  R^2 + \g_{E^2}  E^{\mu\nu} E_{\mu\nu} \bigr) \Bigr) \, ,
\end{align}
where $E_{\mu\nu} \equiv R_{\mu\nu} - \frac{1}{d+1} R \, g_{\mu\nu}$, $C_{\mu\nu\rho\sigma}$ is the Weyl tensor, $\ls$ is a length scale and the $\g_i$ are dimensionless. The Weyl tensor vanishes on the round sphere and $R_{\mu\nu}=d \, g_{\mu\nu}/\ell^2$, hence 
\begin{align} \label{Z0exhcurvgen}
 \log Z^{(0)} = \frac{\Omega_{d+1}}{8 \pi  G} \bigl(-\Lambda \, \ell^{d+1} + \tfrac{1}{2} d(d+1) \, \ell^{d-1} +    \g_{R^2}  \, d^2(d+1)^2 \, \ls^2 \, \ell^{d-3} \,  \bigr) \, ,
\end{align}
For example for $d=2$, 
\begin{align} \label{Z0exhcurv}
 \log Z^{(0)} = \frac{\pi}{4  G} \Bigl(-\Lambda \, \ell^3 + 3 \, \ell + \frac{36  \,\ls^2 \g_{R^2}}{\ell} \Bigr) \, ,
\end{align}
hence, using (\ref{rho0S0gend}) and $\rho^{(0)}(\ell_0)=0$, 
\begin{align} \label{rho0S0ex}
 \CS^{(0)} = S^{(0)}(\ell_0) = \frac{2 \pi \ell_0}{4G} \Bigl(1 + \frac{24 \, \ls^2 \g_{R^2}}{\ell_0^2} \Bigr)  \, , \qquad \Lambda = \frac{1}{\ell_0^2} \Bigl(1 - \frac{12 \,  \ls^2 \g_{R^2}}{\ell_0^2} \Bigr) \, . 
\end{align}

\subsubsection{Effective field theory expansion and field redefinitions} \label{sec:eftfrd}

Curvature corrections such as those considered above naturally appear as terms in the derivative expansion of low-energy effective field theories of quantum gravity, with $\ls$ the characteristic length scale of UV-completing physics and higher-order curvature corrections terms suppressed by higher powers of $\ls^2/\ell^2 \ll 1$. 
The action (\ref{SERiem2}) is then viewed as a truncation at order $\ls^2$, and  
 (\ref{rho0S0ex}) can be solved perturbatively to obtain $\ell_0$ and $\CS^{(0)}$ as a function of $\Lambda$. 

Suppose someone came up with some fundamental theory of de Sitter quantum gravity,  
producing both a precise microscopic computation of the entropy and a precise low-energy effective action, 
with the large-$\ell_0/\ls$ expansion reproduced as some large-$N$ expansion.  
At least superficially, the higher-order curvature-corrected entropy obtained above looks like a Wald entropy \cite{Wald:1993nt,Iyer:1994ys}.
 In the spirit of for instance the nontrivial matching of $R^2$ corrections to the macroscopic BPS black hole entropy computed in \cite{LopesCardoso:1998tkj} and the microscopic entropy computed from M-theory  in \cite{Maldacena:1997de}, it might seem then that matching microscopic $1/N$-corrections and  macroscopic  $\ls^2/\ell_0^2$-corrections to the entropy such as those in (\ref{rho0S0ex}) could offer a nontrivial way of testing such a hypothetical theory. 

However, this is not the case. Unlike the Wald entropy, there are no charges $Q$ (such as energy, angular momentum or gauge charges) available here to give these corrections physical meaning as corrections in the large-$Q$ expansion of a function $S(Q)$. Indeed, the detailed structure of the $\ls/\ell_0$ expansion of $\CS^{(0)}=S^{(0)}(\ell_0)$ has no intrinsic physical meaning at all, because all of it can be wiped out by a local metric field redefinition, order by order in $\ls/\ell_0$, bringing the entropy to pure Einstein area law form, and leaving only the value of $\CS^{(0)}$ itself as a physically meaningful, field-redefinition invariant, dimensionless quantity. 

This is essentially a trivial consequence of the fact that in perturbation theory about the round sphere, the round sphere itself is the unique solution to the equations of motion. Let us however recall in more detail how this works at the level of local field redefinitions, and show how this is expressed at the level of  $\log Z^{(0)}(\ell)$, as this will be useful later in interpreting quantum corrections. 
For concreteness, consider again (\ref{SERiem2}) viewed as a gravitational effective field theory action expanded to order $\ls^2 R^2$. Under a local metric field redefinition  
\begin{align} \label{fieldredef}
 g_{\mu\nu} \to  g_{\mu\nu} + \delta g_{\mu\nu} + O(\ls^4)  \, , \qquad \delta g_{\mu\nu} \equiv \ls^2 \bigl(u_0 \, \Lambda \,  g_{\mu\nu} + u_1 \, R  \, g_{\mu\nu} + u_2 \,   R_{\mu\nu} \bigr)  \, ,
\end{align}
where the $u_i$ are dimensionless constants, 
the action transforms as 
\begin{align}
 S_E \to S_E +  \frac{1}{16 \pi G} \int \!\! \sqrt{g} \,\bigl( R^{\mu\nu} -\tfrac{1}{2}R g^{\mu\nu} +  \Lambda g^{\mu\nu} \bigr) \,   \delta g_{\mu\nu}  \, + \,  O(\ls^4) \, ,
\end{align}
shifting $\g_{R^2},\g_{E^2}$ and rescaling $G,\Lambda$ in (\ref{SERiem2}). A suitable choice of $u_i$ brings $S_E$ to the form
\begin{align} \label{SEred}
 S_E = \frac{1}{8 \pi G}\int \!\! \sqrt{g}\Bigl(\!\Lambda' - \tfrac{1}{2} R - \ls^2  \g_{C^2} C^{\mu\nu\rho\sigma} C_{\mu\nu\rho\sigma} + O(\ls^4)  \!\Bigr)  , \qquad \Lambda' = \Lambda \bigl(1 -\tfrac{4(d+1)^2}{(d-1)^2} \, \g_{R^2}  \ls^2 \Lambda \bigr).
\end{align}
Equivalently, this is obtained by using the $O(\ls^0)$ equations of motion $R_{\mu\nu} = \frac{2}{d-1} \Lambda g_{\mu\nu}$ in the $O(\ls^2)$ part of the action. Since $\lambda'_{R^2}=0$, the entropy computed from this equivalent action takes a pure Einstein area law form $\CS^{(0)}=\Omega_{d+1} \ell_0^{\prime \, d-1}/4G$, with $\ell_0'=\sqrt{d(d-1)/2\Lambda'}$. The on-shell  value $\CS^{(0)}$ itself remains unchanged of course under this change of variables. 

In the above we picked a field redefinition keeping $G'=G$. Further redefining $g_{\mu\nu} \to \alpha \, g_{\mu\nu}$ leads to another equivalent set of couplings $G'',\Lambda'',\ldots$ rescaled with powers of $\alpha$ according to their mass dimension. We could then pick $\alpha$  such that instead $\Lambda''=\Lambda$, or such that $\ell_0''=\ell_0$, now with $G'' \neq G$. If we keep $\ell_0''=\ell_0$, we get
\begin{align} \label{Gprimeprime}
 \CS^{(0)} = \frac{\Omega_{d+1} \ell_0^{d-1}}{4 G''} \, , \qquad \Lambda'' = \frac{d(d-1)}{2\ell_0^2} \, ,
\end{align}   
where for example in $d=2$ starting from (\ref{rho0S0ex}), $G''=G\bigl(1 -24 \,\g_{R^2} \ls^2 /\ell_0^2 + O(\ls^4) \bigr)$. 


At the level of $\log Z^{(0)}(\ell)$ in (\ref{Z0exhcurvgen}) 
the metric redefinition (\ref{fieldredef}) amounts to a radius redefininition $\ell \to \ell \, f(\ell)$ with $f(\ell) = 1 +  \ls^2\bigl( v_{10} \Lambda + v_{11} \ell^{-2} \bigr) + O(\ls^4)$. 
For suitable $v_i$ this brings $\log Z^{(0)}$ and therefore $\CS^{(0)}$ to pure Einstein form. E.g.\ for the $d=2$ example (\ref{Z0exhcurv}),  
\begin{align} \label{ltolprimetr}
  \ell = \bigl( 1 - 12\, \g_{R^2} \ls^2\bigl(\Lambda + \ell^{\prime \, -2} \bigr)  \bigr)  \ell' 
  \qquad \Rightarrow \qquad \log Z^{(0)} = \frac{\pi}{4  G} \bigl(-\Lambda' \, \ell^{\prime \, 3} + 3 \, \ell' + O(\ls^4) \bigr)  \, .
\end{align}
The above considerations generalize to all orders in the $\ls$ expansion. 
$R^n$ corrections  to $\log Z^{(0)}$ are $\propto (\ls/\ell)^{2n}$ and can be removed order by order by a local metric/radius redefinition 
\begin{equation} \label{ellredef}
  \ell \to  \alpha \, f(\ell) \, \ell \, , \qquad f(\ell) = 1 + \ls^2 \bigl(v_{10} \Lambda + v_{11} \ell^{-2} \bigr) +
 \ls^4 \bigl(v_{20} \Lambda^2 + v_{21} \Lambda \ell^{-2} + v_{22} \ell^{-4} \bigl) \, + \cdots
\end{equation}
bringing $\log Z^{(0)}$ and thus $\CS^{(0)}$ to Einstein form to any order in the $\ls$ expansion.

In $d=2$, the Weyl tensor vanishes identically. The remaining higher-order curvature invariants involve the Ricci tensor only, so can be removed by field redefinitions, reducing the action to Einstein form in general. Thus in $d=2$, $\CS^{(0)}$ is the only tree-level invariant in the theory, i.e.\ the only physical coupling constant. In the Chern-Simons formulation of \ref{app:EinstCS}, $\CS^{(0)}=2\pi\kappa$.  In $d \geq 3$, there are infinitely many independent coupling constants, such as the Weyl-squared $\lambda_{C^2}$ in (\ref{SERiem2}), which are not picked up by $\CS^{(0)}$, but are analogously probed by invariants $\CS^{(0)}_M = \log \CZ^{(0)}[g_M] = -S_E[g_M]$ for saddle geometries $g_M$ different from the round sphere. We comment on those and their role in the bigger picture in section \ref{sec:other}.      

The point of considering quantum corrections to the entropy $\CS$ is that these include nonlocal contributions, not removable by local redefinitions,  and thus, unlike the tree-level entropy $\CS^{(0)}$, offering actual data quantitatively constraining candidate microscopic models.

\def\Vb{\bar{V}}

\subsection{Quantum gravitational thermodynamics} \label{sec:qgttd}

The {\it quantum} off-shell partition function $Z(V)$ generalizing the tree-level $Z^{(0)}(V)$  (\ref{defZ0V})  is defined by replacing $\int_{\text{tree}} \CD g \to \int \CD g$ in that expression:\footnote{$Z(V)$ is reminiscent of but different from the fixed-volume partition function considered in the 2D quantum gravity literature, e.g.\ (2.20) in \cite{Ginsparg:1993is}. The latter would be defined as above but with $\frac{1}{2\pi i} \int_{i  \IR} d\sigma$ instead of $\int_{\text{tree}} d\sigma$, constraining the volume to $V$, whereas $Z(V)$ constrains the {\it expectation value} of the volume to $V$.}  
\begin{align} \label{ZVdef}
  Z(V) \equiv    \lint_{\!\!\!\text{tree}} \, d\sigma \, \lint \CD g \, e^{-S_E[g] + \sigma(\int \! \sqrt{g} - V)} \, .
\end{align} 
The quantum off-shell energy density and entropy generalizing (\ref{deftho0S0}) are
\begin{align} \label{rhoSVdef}
 \rho(V) \equiv -\partial_V \log Z  \, , \qquad S(V) \equiv \bigl(1-V\partial_V) \log Z \, . 
\end{align}
$S$ is the Legendre transform of $\log Z$: 
\begin{align} \label{LegRel}
   S = \log Z + V \rho \, , \qquad d \log Z = - \rho \, dV \,  , \qquad  dS = V d \rho \, .
\end{align} 
Writing $e^{-\Gamma(V)} \equiv Z(V)$, the above definitions imply that as a function of $\rho$,
\begin{align}
 S(\rho) = \log \lint_{\!\!\!\text{tree}} \, dV \, e^{-\Gamma(V) + \rho V} = \log \lint \CD g \, e^{-S_E[g] + \rho \int \! \sqrt{g}} 
\end{align}
hence $S(\rho)$ is the generating function for moments of the volume. In particular
\begin{align}
  V = \bigl\langle \smint\!\sqrt{g} \bigr\rangle_{\rho} = \partial_\rho S(\rho)   
\end{align}
is the expectation value of the volume in the presence of a source $\rho$ shifting the cosmological constant $\frac{\Lambda}{8\pi G} \to \frac{\Lambda}{8\pi G} - \rho$. 
$\Gamma(V)$ can be viewed as a quantum effective action for the volume, in the spirit of the QFT 1PI effective action \cite{weinberg_1996,Vilkovisky:1984st,Barvinsky:1985an} but taking only the volume off-shell. At tree level it reduces to $S_E[g_\ell]$ appearing in (\ref{logZ0Vgend}). 
At the quantum on-shell value $V=\bar V = \bigl\langle \smint\!\sqrt{g} \bigr\rangle_0$,  
\begin{align} \label{onshellentropy}
  \rho(\bar V) = 0 \, ,  \qquad \CS = \log Z(\bar V) = S(\bar V) \,.
\end{align}
It will again be convenient to work with a linear scale variable $\ell$ instead of $V$, {\it defined} by  
\begin{align} \label{ellVdef}
 \Omega_{d+1} \ell^{d+1} \equiv V  \, ,
\end{align} 
Since the mean volume $V = \bigl\langle \int\!\! \sqrt{g} \bigr\rangle_\rho$ is  diffeomorphism invariant, (\ref{ellVdef}) gives a manifestly diffeomorphism-invariant definition of the ``mean radius'' $\ell$ of the fluctuating geometry. 
Given $Z(\ell) \equiv Z(V(\ell))$, the off-shell energy density and entropy are then computed as  
\begin{align} \label{rhoSelldef}
 \rho(\ell) = -\frac{\frac{1}{d+1}\ell \partial_\ell \log Z}{V} \, , \qquad S(\ell) = \bigl(1-\tfrac{1}{d+1} \ell \partial_\ell \bigr) \log Z \, . 
\end{align}
The quantum on-shell value of $\ell$ is denoted by $\bar \ell$ and satisfies $\rho(\bar\ell)=0$. 

The magnitude of quantum fluctuations of the volume about its mean value is given by $\delta V^2 \equiv \bigl\langle  \bigl(\int \!\! \sqrt{g} - V \bigr)^2 \bigr\rangle_\rho = S''(\rho) = 1/\Gamma''(V)=1/\rho'(V)=V/S'(V)$. 
 At large $V$, $\delta V/V \propto 1/\sqrt{S}$.  

\subsection{One-loop corrected de Sitter entropy} \label{sec:olrgt}


The path integral (\ref{ZVdef}) for $\log Z$ can be computed perturbatively about its round sphere saddle in a semiclassical expansion in powers of $G$. To leading order it reduces to $\log Z^{(0)}$ defined in (\ref{defZ0V}). 
For 3D Einstein gravity, 
\begin{align} \label{logZelltreeap}
 \log Z(V) = \log Z^{(0)}(V)  +  O(G^0)   = \frac{\Omega_3}{8 \pi G} \bigl(-\Lambda \, \ell^3 + 3  \, \ell \bigr) \, + \, O(G^0) \, , 
\end{align} 
To compute the one-loop $O(G^0)$ correction, recall that evaluation of $\int_{\rm tree} d\sigma$ in (\ref{ZVdef}) is equivalent to extremization with respect to $\sigma$, which sets $\sigma=-\partial_V \log Z(V)=\rho(V)$ and
\begin{align} \label{ZVdefnosigma}
  \log Z(V) =  \log \lint \CD g \, e^{-S_E[g] + \rho(V) (\int \! \sqrt{g} - V)} \, .
\end{align} 
To one-loop order, we may replace $\rho$ by its tree-level approximation $\rho^{(0)} = - \partial_V \log Z^{(0)}$. By construction this ensures the round sphere metric $g=g_\ell$ of radius $\ell(V)$ given by (\ref{ellVdef}) is a saddle. Expanding the action to quadratic order in fluctuations about this saddle then gives a massless spin-2 Gaussian path integral of the type solved in general by (\ref{ZPIFINAL}), or more explicitly in (\ref{ZPIgraviton})-(\ref{Cdcoeff}). For  3D Einstein gravity, using (\ref{Cdcoeff}), 
\begin{align} \label{ZVrhoVe} 
 \log Z = -\Bigl( \frac{\Lambda}{8 \pi G} + c_0' \Bigr)  \Omega_3 \ell^3 + \Bigl( \frac{1}{8 \pi G} + c_2' - \frac{3}{4 \pi \epsilon}  \Bigr)  3  \Omega_3 \ell     -3 \log \frac{2\pi \ell}{4G}  + 5 \log(2\pi)  +   O(G) 
\end{align}   
Here $c_0'$ and $c_2'$ arise from $O(G^0)$ local counterterms 
\begin{align} \label{SEct3D}
  S_{E, \rm ct} = \int \!\! \sqrt{g} \Bigl( c_0' - \frac{c_2'}{2} R \Bigr) \, ,
\end{align}
split off from the bare action (\ref{action3Dgrav}) to keep the tree-level couplngs $\Lambda$ and $G$ equal to their ``physical'' (renormalized) values 
to this order. We define these physical values as 
 the coefficients of the local terms $\propto \ell^3,\ell$ in the $V \to \infty$ asymptotic expansion of the {\it quantum} $\log Z(V)$. 
That is to say, we fix $c_0'$ and $c_2'$ by imposing the renormalization condition  
\begin{align} \label{logZVasphys}
 \log Z(V) = \frac{\Omega_3}{8 \pi G} \bigl(-\Lambda \, \ell^3 + 3  \, \ell \, \bigr) + \cdots
  \qquad (V \to \infty) \, . 
\end{align} 
This renormalization prescription is diffeomorphism invariant, since $Z(V)$, $V$ and $\ell$ were all defined in a manifestly diffeomorphism-invariant way.   
In (\ref{ZVrhoVe}) it fixes $c_0' = 0$, $c_2' = \frac{3}{4 \pi \epsilon}$, 
hence $\log Z(\ell) = \log Z^{(0)} + \log Z^{(1)} + O(G)$, where 
\begin{align} \label{lZ0lZ1Einst}
 \log Z^{(1)} =   -3 \log \frac{2\pi \ell}{4G}  + 5 \log(2\pi)  \, .
\end{align}
We can express the renormalization condition (\ref{logZVasphys}) equivalently as
\begin{align} \label{renprescr}
  \boxed{\log Z^{(1)} = \log Z_{\rm PI}^{(1)} + \log Z_{\rm ct} \, , \qquad \lim_{\ell \to \infty} \partial_\ell  \log Z^{(1)}  = 0} 
\end{align} 
where $\log Z_{\rm ct} = -S_{E,\rm ct}[g_\ell]$ with $g_\ell$ the round sphere metric of volume $V$. On $S^3$ we have $\log Z_{\rm ct} = c_0 \ell^3 + c_2 \ell$, and the $\ell \to \infty$ condition fixes $c_0$ and $c_2$. Recalling (\ref{UPIdef}), we can physically interpret this as requiring the renormalized one-loop Euclidean energy $U^{(1)}$ of the static patch vanishes in the $\ell \to \infty$ limit.   
   
For general $d$, the UV-divergent terms in $\log Z_{\rm PI}^{(1)}$ come with non-negative powers $\propto \ell^{d+1-2n}$, canceled by counterterms consisting of $n$-th order curvature invariants. 
For example on $S^5$, $\log Z_{\rm ct} = c_0 \ell^5 + c_2 \ell^3 + c_4 \ell$. In odd $d+1$, the renormalization prescription (\ref{renprescr}) then fixes the $c_{2n}$. In even $d+1$, $\log Z_{\rm ct}$ has a constant term $c_{d+1}$, which is not fixed by (\ref{renprescr}). As we will make explicit in examples later, it can be fixed by $\lim_{\ell \to \infty} Z^{(1)} = 0$ for massive field contributions, and for massless field contributions by minimal subtraction at scale $L$, $c_{d+1} = -\alpha_{d+1}\log(M_\epsilon L)$, $M_\epsilon=2e^{-\gamma}/\epsilon$ (\ref{HKHKHKH}), with $L \partial_L \log Z = 0$, i.e.\ $L \partial_L \log Z^{(0)} = \alpha_{d+1}$. 

The renormalized off-shell $\rho$ and $S$ 
are obtained from $\log Z$ as in (\ref{rhoSelldef}). For 3D Einstein, 
\begin{align}
 \rho^{(1)} = \frac{1}{2 \pi^2 \ell^3} \, , \qquad 
 S^{(1)} = -3 \log \frac{2\pi \ell}{4G}  + 5 \log(2\pi) + 1 \, .
\end{align}
The  on-shell quantum dS entropy $\CS=\log Z(\bar\ell) = S(\bar\ell)$ (\ref{onshellentropy}) is     
\begin{align} \label{CS1ex}
 \boxed{\CS = S(\bar\ell) = S^{(0)}(\bar\ell) + S^{(1)}(\bar\ell) \, + \, O(G)} 
\end{align} 
where $\bar\ell$ is the quantum mean radius satisfying $\rho(\bar\ell) \propto \partial_\ell \log Z(\bar\ell) = 0$. 
For 3D Einstein,
\begin{align} \label{CSEinstf1}
  \CS = \frac{2\pi \bar\ell}{4G} -3 \log \frac{2\pi \bar\ell}{4G}  + 5 \log(2\pi) + 1 \, + \, O(G) \, , \qquad \Lambda = \frac{1}{\bar \ell^2} - \frac{4 G}{\pi\bar\ell^3} + O(G^2) \, .
\end{align}
Alternatively, $\CS$ can be expressed in terms of the tree-level $\ell_0$, $\rho^{(0)}(\ell_0) \propto \partial_\ell \log Z^{(0)}(\ell_0)=0$, using $\CS=\log Z^{(0)}(\bar\ell) + \log Z^{(1)}(\bar\ell) + O(G)$, $\bar\ell = \ell_0 + O(G)$ and Taylor expanding in $G$: 
\begin{align} \label{CS2ex}
 \boxed{\CS = \log Z(\bar\ell) = S^{(0)}(\ell_0) + \log Z^{(1)}(\ell_0) \, + \, O(G)} 
\end{align}
This form would be obtained from (\ref{UPISPIgrav}) by a more standard computation. For 3D Einstein,
\begin{align} \label{CSEinstf2}
 \CS = \frac{2\pi\ell_0}{4G} -3 \log \frac{2\pi \ell_0}{4G}  + 5 \log(2\pi) \, + \, O(G)   \, , \qquad \Lambda = \frac{1}{\ell_0^2} \, .
\end{align}
The equivalence of (\ref{CS1ex}) and (\ref{CS2ex}) can be checked directly here noting $\bar\ell = \ell_0 -\frac{2}{\pi}  G \, + \, O(G^2)$, so $\frac{2 \pi \bar\ell}{4G} = \frac{2\pi\ell_0}{4G} - 1 + O(G)$. The $-1$ cancels the $+1$ in (\ref{CSEinstf1}), reproducing (\ref{CSEinstf2}). 

More generally and more physically, the relation between these two expressions can be understood as follows. At tree level, the entropy equals the geometric horizon entropy $S^{(0)}(\ell_0)$, with radius $\ell_0$ such that the geometric energy density $\rho^{(0)}$ vanishes. At one loop, we get additional contributions from quantum field fluctuations. The UV contributions are absorbed into the gravitational coupling constants. The remaining IR contributions shift the entropy by $S^{(1)}$ and the energy density by $\rho^{(1)}$. 
The added energy backreacts on the fluctuating geometry: its mean radius changes from $\ell_0$ to $\bar \ell$ such that the geometric energy density changes by  $\delta \rho^{(0)} = - \rho^{(1)}$, ensuring the total energy density vanishes. This in turn changes the geometric horizon entropy by an amount dictated by the first law (\ref{LegRel0}), 
\begin{align}
  \delta S^{(0)} = V_0\, \delta \rho^{(0)} =  -V_0\, \rho^{(1)} \, .
\end{align}
We end up with a total entropy $\CS=S^{(0)}(\bar\ell) + S^{(1)} = S^{(0)}(\ell_0) - V_0 \, \rho^{(1)} + S^{(1)} = S^{(0)}(\ell_0) + \log Z^{(1)}$, up to $O(G)$ corrections, relating (\ref{CS1ex}) to (\ref{CS2ex}). (See also fig.\ \ref{fig:Z1plot}.)

More succinctly, obtaining (\ref{CS2ex}) from (\ref{CS1ex}) is akin to obtaining the canonical description of a thermodynamic system from the microcanonical description of system + reservoir. 
The analog of the canonical partition function is $Z^{(1)}=e^{S^{(1)}-V_0 \, \rho^{(1)}}$, with $-V_0 \, \rho^{(1)}$ capturing the reservoir (horizon) entropy change due to energy transfer to the system. 

\def\ns{N}

\subsection{Examples} \label{sec:exampilezz}

\subsubsection{3D scalar} \label{sec:3dscipip}

\begin{figure}
  \includegraphics[height=4.1cm]{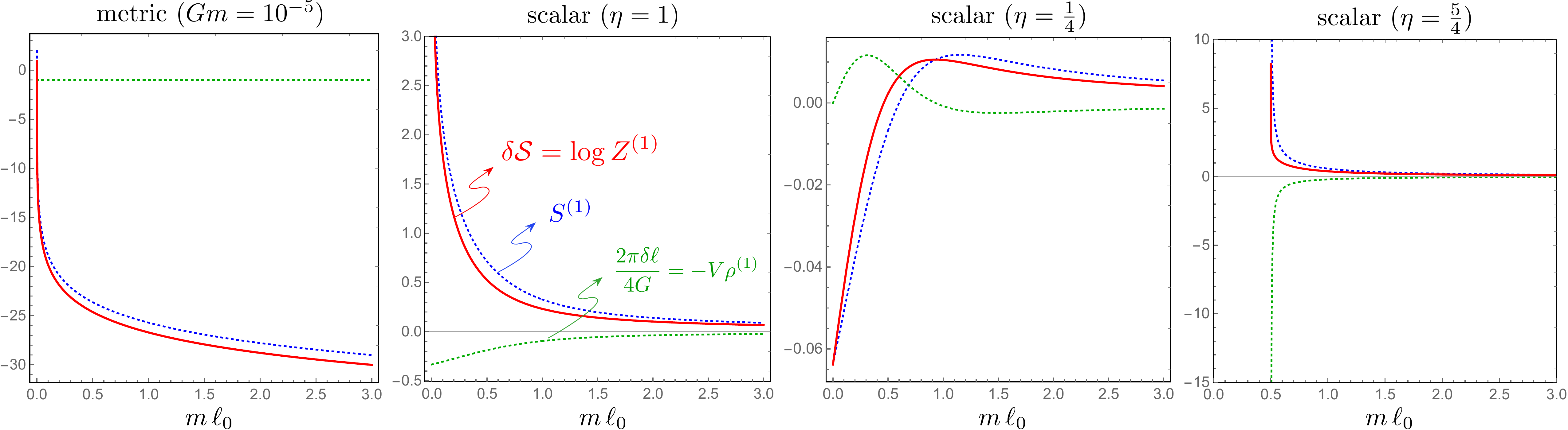} \vskip-2mm
 \caption{ \small \label{fig:Z1plot} One-loop contributions to the dS$_3$ entropy from metric and scalars with $\eta = 1,\frac{1}{4},\frac{5}{4}$, i.e.\ $\xi=0,\frac{1}{8},-\frac{1}{24}$. 
 Blue dotted line = renormalized entropy $S^{(1)}$. Green dotted line = horizon entropy change $\delta S^{(0)}=2\pi \delta \ell/4G = - V \rho^{(1)}$ due to quantum backreaction $\ell_0 \to \bar\ell = \ell_0 + \delta \ell$, as dictated by first law.      
Solid red line = total $\delta \CS  = S^{(1)} -V \rho^{(1)} = \log Z^{(1)}$. The metric contribution is negative within the semiclassical regime of validity $\ell \gg G$. 
The renormalized scalar entropy and energy density are positive for $m\ell \gg 1$, and for all $m\ell$ if $\eta=1$. If $\eta\!>\!1$ and $\ell_0 \to \ell_* \equiv \frac{\sqrt{\eta-1}}{m}$, the  correction $\delta \ell \sim  -\frac{G}{3\pi} \frac{\ell_*}{\ell_0-\ell_*} \to -\infty$, meaning the one-loop approximation breaks down. The scalar becomes tachyonic beyond this point. If a $\phi^4$ term is included in the action, two new dominant saddles emerge with $\phi \neq 0$.          
  }
\end{figure} 

An example with matter is 3D Einstein gravity + scalar $\phi$ as in (\ref{scalaractionfull}). Putting $\xi\equiv\frac{1-\eta}{6}$,
\begin{align} \label{EinstplusScalar}
 S_E[g,\phi] = \frac{1}{8\pi G} \int \!\!\sqrt{g} \, \bigl(\Lambda - \tfrac{1}{2} R \bigr) \, + \,  \frac{1}{2} \int \!\! \sqrt{g} \, \phi \bigl(-\nabla^2 + \mo^2 + \tfrac{1-\eta}{6}R  \,\bigr)  \phi \, ,  
\end{align}
The metric contribution to $\log Z^{(1)}$ remains $\log Z^{(1)}_{\rm metric} = -3 \log \frac{2\pi \ell}{4G}  + 5 \log(2\pi)$ as in (\ref{lZ0lZ1Einst}). 
The scalar $Z_{\rm PI}^{(1)}$ was given in  (\ref{Zbubufi3}). 
Its finite part is
\begin{align} \label{Zpifnscalar}
 \log Z^{(1)}_{\rm PI,fin,scalar} =  
 \frac{\pi \nu^3}{6} 
 -\sum_{k=0}^{2} \frac{\nu^k}{k!} \, \frac{{\rm Li}_{3-k}(e^{-2 \pi \nu})}{(2\pi)^{2-k}} \, , \qquad \nu \equiv \sqrt{\mo^2 \ell^2-\eta} \, .
\end{align}
The polynomial 
 $\log Z_{\rm ct}(\ell)=c_0 \ell^3 + c_2\ell$ corresponding to the counterterm action (\ref{SEct3D}) is  
fixed by the renormalization condition (\ref{renprescr}), 
resulting in 
\begin{align} \label{logZ1renscal}
 \log Z^{(1)}_{\rm scalar} = \log Z^{(1)}_{\rm PI,fin,scalar} - \frac{\pi}{6} \,  \mo^3 \ell^3 + \frac{\pi \eta }{4} \,  \mo \ell \, .
\end{align}
The finite polynomial cancels the local terms $\propto \ell^3,\ell$ in the large-$\ell$ asymptotic expansion of the finite part: 
 $\log Z_{\rm scalar}^{(1)} =  \frac{\pi  \eta^2}{16} (\mo \ell)^{-1} +\frac{\pi  \eta ^3}{96 } (\mo \ell)^{-3} + \cdots$ when $m\ell \to \infty$. 
The $(m\ell)^{-{2n-1}}$ terms have the $\ell$-dependence of $R^n$ terms in the  action and can effectively be thought of as finite shifts of higher-order curvature couplings in the $\mo \ell \gg 1$ regime. In the opposite regime $\mo \ell \ll 1$, IR bulk modes of the scalar becomes thermally activated and $\log Z^{(1)}_{\rm scalar}$ ceases to have a local expansion. In particular in the minimally-coupled case $\eta=1$,  
\begin{align} \label{smallmlexpsc3d}
 \log Z^{(1)}_{\rm scalar} \simeq  
  -  \log (\mo \ell) 
  \qquad (m\ell \to 0) \, .
\end{align} 
The total energy density is $\rho=-\frac{1}{3} \ell \partial_\ell \log Z / V  = \frac{1}{8 \pi G} \bigl(\Lambda - \ell^{-2} \bigr) + 1 + \rho^{(1)}_{\rm scalar}$ where 
\begin{align}
 V  \rho^{(1)}_{\rm scalar} =   -\frac{\pi}{6} (m\ell)^2 \nu \coth (\pi  \nu )  + \frac{\pi}{6} (m\ell)^3- \frac{\pi \eta}{12}  \, m \ell  \, .
\end{align}
The on-shell quantum dS entropy is given to this order by (\ref{CS2ex}) or by (\ref{CS1ex}) as  
\begin{align} \label{Snr1}
 \CS = \CS^{(0)} + \CS^{(1)} = S^{(0)}(\ell_0) + \log Z^{(1)} = S^{(0)}(\ell_0) - V \rho^{(1)} +  S^{(1)} = S^{(0)}(\bar\ell) + S^{(1)} \, ,  
\end{align}
where 
$\ell_0^{-2} = \Lambda = \bar\ell^{-2} - 8 \pi G \, \rho^{(1)}(\bar\ell)$ and $S^{(1)} 
= S_{\rm PI,fin}^{(1)}+ \frac{1}{6} \pi \eta \, m \ell$, with the scalar contribution to $S_{\rm PI,fin}^{(1)}$ given by the finite part of (\ref{SPIscalar}). 
Some examples are shown in fig.\ \ref{fig:Z1plot}.   


For a massless scalar, $m=0$, the renormalized scalar one-loop correction to $\CS$ is a constant independent of $\ell_0$ given by (\ref{Zpifnscalar}) evaluated at $\nu=\sqrt{-\eta}$, and $\rho^{(1)}_{\rm scalar}=0$.  For example for a massless conformally coupled scalar, $\eta=\frac{1}{4}$, $Z^{(1)}_{\rm scalar} = \frac{3 \, \zeta (3)}{16 \pi ^2}-\frac{\log (2)}{8}$. 

\subsubsection{3D massive spin $s$} \label{sec:3dmsps}

\begin{figure}
  \includegraphics[height=8.8cm]{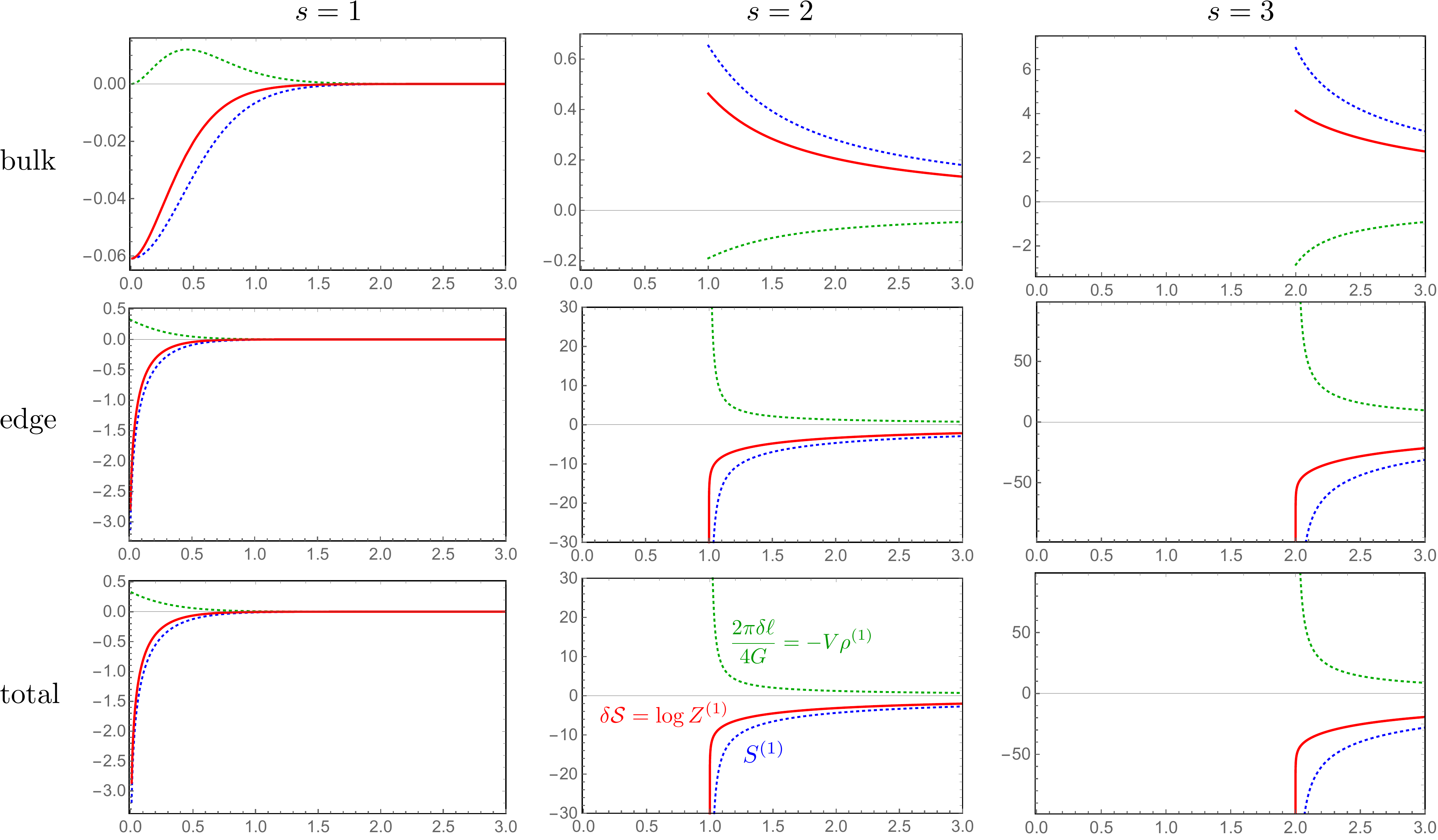} \vskip-2mm
 \caption{ \small \label{fig:Z1plotHS} Contributions to the dS$_3$ entropy from massive spin $s=1,2,3$ fields, as a function of $m \ell_0$, with coloring as in fig.\ \ref{fig:Z1plot}. Singularities = Higuchi bound, as discussed under (\ref{SPIscalar}).          
  }
\end{figure}

The renormalized one-loop correction $\CS^{(1)}_s = \log Z^{(1)}_s$ to the dS$_3$ entropy from a massive spin-$s$ field is obtained similarly from (\ref{exmasssd3}):  
\begin{align} \label{deltaCsll}
 \CS^{(1)}_s = \log Z^{(1)}_{s,\rm bulk} - \log Z^{(1)}_{s,\rm edge} \, ,
\end{align}
where $\log Z^{(1)}_{s,\rm bulk}$ equals twice the contribution of an $\eta = (s-1)^2$ scalar as given in (\ref{logZ1renscal}), while the edge contribution is, putting $\nu \equiv \sqrt{m^2\ell^2-(s-1)^2}$,
\begin{align}
 \log Z^{(1)}_{s,\rm edge} = s^2 \bigl(\pi (m\ell-\nu) - \log (1-e^{-2 \pi \nu})\bigr) \, .
\end{align}
The edge contribution to (\ref{deltaCsll}) is manifestly negative. It dominates the bulk part, and increasingly so as $s$ grows. Examples are shown in fig.\ \ref{fig:Z1plotHS}.

\def\F{{\tt f}}

\subsubsection{2D scalar} \label{sec:2dsc}

As mentioned below (\ref{renprescr}), the counterterm polynomial  $\log Z_{\rm ct}^{(0)}$ has a constant term in even spacetime dimensions $d+1$, which is not fixed yet by the renormalization prescription given there. Let us consider the simplest example: a $d=1$ scalar with action (\ref{scalaractionfull}).
Denoting  
\begin{align}
 \F(\nu) \equiv \sum_{\pm} \zeta'(-1,\tfrac{1}{2}\pm i \nu) \mp i \nu \zeta'(0,\tfrac{1}{2}\pm i \nu)  \, ,    
 \, ,
\end{align}
and $M_{\epsilon} = {2 e^{-\game}}/{\epsilon}$ as in (\ref{HKHKHKH}),
we get from (\ref{dS2exampleZPI}) with $\nu \equiv \sqrt{m^2\ell^2-\eta}$, $\eta \equiv \tfrac{1}{4} - 2 \xi$, 
\begin{align} \label{logZPI2dscalarfull}
 \log Z_{\rm PI}^{(1)} =    \bigl(2 \epsilon^{-2}  - m^2 \log (M_\epsilon \ell) + m^2 \bigr) \, \ell^2  + \bigl(\eta + \tfrac{1}{12} \bigr) \log \bigl(M_\epsilon \ell \bigr) -\eta  + \F(\nu) \, ,
\end{align}
In the limit $m \ell \to \infty$, using the asymptotic expansion of the Hurwitz zeta function\footnote{See e.g.\  appendix A of \cite{Dib:2000hd}}, 
\begin{align} \label{largemlasd1}
 \log Z_{\rm PI}^{(1)} = \bigl(2 \epsilon^{-2} - m^2 \log (M_\epsilon/m) -\tfrac{1}{2} m^2 \bigr) \, \ell^2 +(\eta+\tfrac{1}{12}) \log (M_\epsilon/m)  \, + \, O((m\ell)^{-2}) \, .
\end{align}
Notice the $\log \ell$ dependence apparent in (\ref{logZPI2dscalarfull}) has canceled out. The counterterm action to this order is again of the form (\ref{SEct3D}), corresponding to $\log Z_{\rm ct} = 4 \pi (-c_0' \ell^2 + c_2')$. The renormalization condition (\ref{renprescr}) fixes $c_0'$ 
but leaves $c_2'$ undetermined. Its natural extension here is to pick $c_2=4\pi c_2'$ to cancel off the constant term as well, that is
\begin{align} \label{massivescalarconstc}
  c_2 = -(\eta+\tfrac{1}{12}) \log(M_\epsilon/m) \qquad \Rightarrow \qquad  \lim_{\ell \to \infty} \log Z^{(1)} = 0
   \, , 
\end{align}
ensuring the tree-level $G$ equals the renormalized Newton constant to this order, as in (\ref{logZVasphys}). 
The renormalized scalar one-loop contribution to the off-shell partition function is then
\begin{equation} \label{rensc4dZ}
 \log Z^{(1)} =   
  \bigl(\tfrac{3}{2}- \log ( m \ell) \bigr) (m\ell)^2  
 + (\eta +\tfrac{1}{12}) \log ( m \ell)    -\eta + \F(\nu) \, .
\end{equation}
In the large-$m\ell$ limit, $\log Z^{(1)} = \frac{240 \, \eta ^2+40 \, \eta +7}{960}  (m\ell)^{-2} + \cdots > 0$, while in the small-$m\ell$ limit 
\begin{align}
 \log Z^{(1)} \simeq \bigl(\eta+\tfrac{1}{12}\bigr) \log(m\ell) \quad  (\eta < \tfrac{1}{4}) \, , \qquad \log Z^{(1)} \simeq \bigl(\tfrac{1}{4}+\tfrac{1}{12}-1\bigr) \log(m\ell) \quad  (\eta = \tfrac{1}{4}) \, .
\end{align}
The extra $-\log(m\ell)$ in the minimally-coupled case $\eta=\frac{1}{4}$ is the same as in (\ref{smallmlexpsc3d}) and has the same thermal interpretation. The energy density is $\rho^{(1)}=-\frac{1}{2} \ell\partial_\ell \log Z^{(1)}/V$ with $V=4\pi\ell^2$:  
\begin{align} \label{Vrho1scalar}
  V \rho^{(1)} = -\tfrac{1}{2}(\eta+\tfrac{1}{12}) 
  +\tfrac{1}{2} (m\ell)^2 \bigl(2\log (m \ell) - \mbox{$\sum_\pm$} \psi^{(0)}\bigl(\tfrac{1}{2} \pm i \nu \bigr) \bigr)
\end{align} 
In the massless case $m=0$, $\nu = \sqrt{-\eta}$ is $\ell$-independent, and we cannot use the asymptotic expansion (\ref{largemlasd1}), nor the renormalization prescription (\ref{massivescalarconstc}).   
Instead we fix $c_2$ by minimal subtraction, picking a reference length scale $L$ and putting (with $M_\epsilon = \frac{2 e^{-\game}}{\epsilon}$ (\ref{HKHKHKH}) as before)
\begin{align} \label{c2minsub2D}
 c_2(L) \equiv -(\eta+\tfrac{1}{12}) \log(M_\epsilon L) \, ,
\end{align}
The renormalized $G$ then satisfies $\partial_L (\frac{4 \pi}{8\pi G} + c_2 ) = 0$, i.e.\ $L \partial_L \frac{1}{2G} = \eta+\frac{1}{12}$, and  
\begin{align}
 \log Z^{(1)} = (\eta +\tfrac{1}{12}) \log (\ell/L)  -\eta + \F\bigl(\sqrt{-\eta} \bigr) \, , \qquad V \rho^{(1)} = -\tfrac{1}{2}(\eta+\tfrac{1}{12}) \, .
\end{align} 
The total $\log Z=\frac{1}{2G}(-\Lambda \ell^2+1) + \log Z^{(1)}$ is of course independent of the choice of $L$. 

\subsubsection{4D massive spin $s$} \label{sec:4dms}

\begin{figure}
  \includegraphics[height=3.3cm]{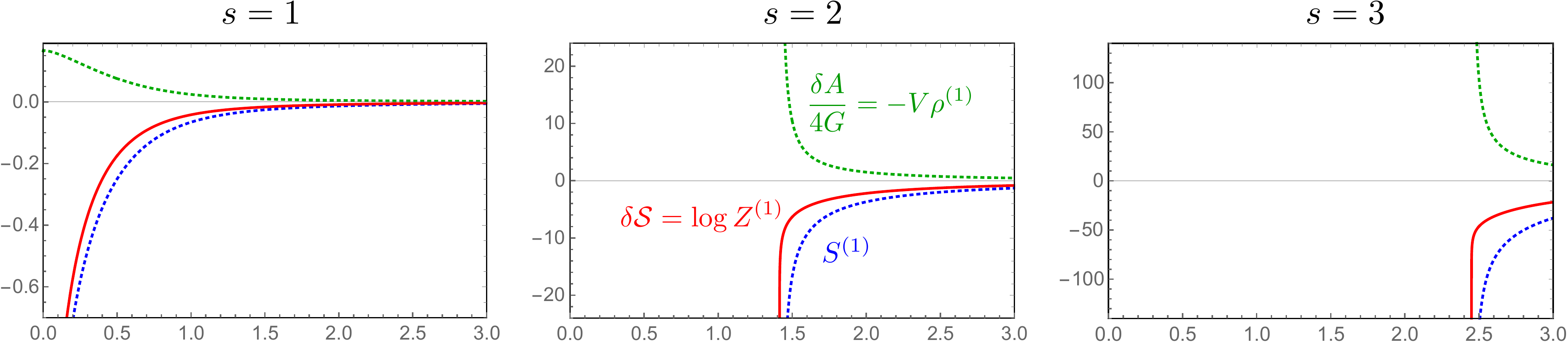} \vskip-2mm
 \caption{ \small \label{fig:Z1plotHS4D} Edge contributions to the dS$_4$ entropy from massive spin $s=1,2,3$ fields, as a function of $m \ell_0$, with coloring as in fig.\ \ref{fig:Z1plot}. The Higuchi/unitarity bound in this case is $(m\ell_0)^2 - (s-\frac{1}{2})^2 > -\frac{1}{4}$.          
  }
\end{figure} 

4D massive spin-$s$ fields can be treated similarly, starting from (\ref{ZPIexactmssd3}). In particular the edge contribution $\log Z^{(1)}_{\rm edge}$ equals {\it minus} the $\log Z^{(1)}$ of $D_{s-1}^5 = \frac{1}{6} s (s+1) (2 s+1)$ scalars on $S^2$, computed earlier in (\ref{rensc4dZ}), with the same $\nu=\sqrt{m^2\ell^2-\eta_s}$ as the bulk spin-$s$ field, which according to 
(\ref{convmdef}) means $\eta_s = (s-\frac{1}{2})^2$.  The corresponding contribution to the renormalized energy density is $\rho^{(1)}_{\rm edge} = -\frac{1}{4} \ell\partial_\ell \log Z^{(1)}_{\rm edge}/V$ with $V=\Omega_4 \ell^4$, so $V \rho^{(1)}_{\rm edge}$ equals $-\frac{1}{2} D_{s-1}^5$ times the scalar result (\ref{Vrho1scalar}). 

As in the $d=2$ case, the renormalized one-loop edge contribution $\CS^{(1)}_{\rm edge}$ to the entropy is negative and dominant. Some examples are shown in fig.\ \ref{fig:Z1plotHS4D}.  

\subsubsection{Graviton contribution for general $d$}
\label{sec:hdgee}

For $d \geq 3$, UV-sensitive terms in the loop expansion renormalize higher-order curvature couplings in the gravitational action, prompting the inclusion of such terms in $S_E[g]$. Some caution is in order then if we wish to apply  (\ref{ZPIFINAL}) or (\ref{ZPIgraviton})-(\ref{Cdcoeff}) to compute $\log Z^{(1)}$. 
The formula (\ref{ZPIFINAL}) for $Z_{\rm PI}^{(1)}$ depends on $\gamma = \sqrt{8 \pi \GN/A_{d-1}}$, gauge-algebraically defined by (\ref{GNGNdef}) and various normalization conventions. We picked these such that in pure Einstein gravity, $\gamma = \sqrt{8 \pi G/A(\ell_0)}$, $\ell_0 = \sqrt{d(d-1)/2\Lambda}$, with $G$ and $\Lambda$ read off from the gravitational Lagrangian. However this expression of $\gamma$ in terms of Lagrangian parameters 
will in general be modified in the presence of higher-order curvature terms. This is clear from the discussion in \ref{sec:eftfrd},  and (\ref{Gprimeprime}) in particular. Since $\gamma_0$ is field-redefinition invariant, and since after transforming to a pure Einstein frame we have $\gamma_0=\sqrt{2\pi/\CS^{(0)}}$, with the right hand side also invariant, we have in general (for Einstein + perturbative higher-order curvature corrections)  
\begin{align} \label{gammainvar}
  \gamma=\sqrt{{2\pi}/{\CS^{(0)}}} \, .
\end{align}
From (\ref{ZPIgraviton}) we thus get (ignoring the phase)
\begin{align} \label{CSgravd}
 \CS = \CS^{(0)}  -\frac{D_d}{2}  \log \CS^{(0)} + \alpha^{(2)}_{d+1} \log \frac{\ell_0}{\LR} + K_{d+1} \, + \, O\bigl(1/\CS^{(0)}\bigr)
\end{align} 
where $D_d=\frac{(d+2)(d+1)}{2}$, $\alpha_{d+1}^{(2)}=0$ for even $d$ and given by 
(\ref{gravalphas}) for odd $d$, and $K_{d+1}$ a numerical constant obtained by evaluating (\ref{ZPIFINAL2}). 
For odd $d$ the constant in the counterterm $\log Z_{\rm ct}(\ell)$ is fixed by minimal subtraction at a scale $L$, $c_{d+1}(L) \equiv -\alpha_{d+1} \log(M_\epsilon L)$, with $M_\epsilon = 2e^{-\gamma}/\epsilon$ determined by the heat kernel regulator as in (\ref{HKHKHKH}), and $L \partial_L \CS = 0$, i.e.\  $L \partial_L \CS^{(0)} = \alpha_{d+1}^{(2)}$.  
Explicitly for $d=2,3,4$,  using (\ref{Cdcoeff}), (\ref{explexamplestab})
\def\arraystretch{1.1}
\begin{equation} \label{CSres}
\begin{array}{l|l}
d &  \CS    \\
 \hline 
2 & \CS^{(0)}  -3 \log \CS^{(0)} + 5 \log(2\pi)
 \\ 
3 & 
   \CS^{(0)}
- 5  \log \CS^{(0)}
    -\frac{571}{90} \log(\frac{\ell_0}{L}) 
  -\log(\frac{8 \pi}{3})  +\frac{715}{48}  - \frac{47}{3} \zeta'(-1)   +\frac{2}{3}  \zeta'(-3)
    \\
4 & \CS^{(0)} -\frac{15}{2} \log \CS^{(0)} + \log(12) + \frac{27}{2} \log (2 \pi ) + \frac{65 \, \zeta (3)}{48 \, \pi^2}+\frac{5 \, \zeta (5)}{16 \, \pi^4} 
\end{array} 
\end{equation}   
\def\arraystretch{1}
For a $d=3$ action (\ref{SERiem2}) up to $O(\ls^2 R^2)$, with dots denoting $O(\ls^4)$ terms,
\begin{align} \label{S0forS4}
 \CS^{(0)} = \frac{\pi}{G} \left(\ell_0^2 + {48 \, \g_{R^2} \, \ls^2} + \cdots \right)
  \, , \qquad \Lambda = \frac{3}{\ell_0^2}  + \cdots \, .
\end{align}
where $L \partial_L \g_{R^2} = -\frac{G}{48\, \pi \, \ls^2} \cdot \frac{571}{45}$. Putting $L=\ell_0$, and defining the scale $\ell_{R^2}$ by $\lambda_{R^2}(\ell_{R^2})=0$, 
\begin{align} \label{CSgravd2}
 \CS = \CS^{(0)}  -5  \log \CS^{(0)} + K_4 \, + \, O\bigl(1/\CS^{(0)}\bigr) \, , \qquad \CS^{(0)}=\frac{\pi \ell_0^2}{G} - \frac{571}{45} \log \frac{\ell_0}{\ell_{R^2}} + \cdots \, .
\end{align} 
The constant $K_{4}$ could be absorbed into $\g_{R^2}$ at this level. Below, in (\ref{lincombinv}), we will give it relative meaning however, by considering saddles different from the round $S^4$.

\subsection{Classical and quantum observables} \label{sec:other}

Here we address question \ref{ci3} in our list below (\ref{UPISPIgrav}). To answer this, we need ``observables'' of the $\Lambda>0$ Euclidean low-energy effective field theory probing independent gravitational couplings (for simplicity we restrict ourselves to purely gravitational theories here), i.e.\ diffeomorphism and field-redefinition invariant quantities, analogous to scattering amplitudes in asymptotically flat space. For this to be similarly useful, an infinite amount of unambiguous data should be extractable, at least in principle, from these observables.    

As discussed above, $\CS^{(0)}=\log \CZ^{(0)}=-S_E[g_{\ell_0}]$ invariantly probes the dimensionless coupling given by $\ell_0^{d-1}/G \propto 1/G\Lambda^{{(d-1)}/2}$ in Einstein frame.  
The obvious tree-level invariants probing different couplings in the gravitational low-energy effective field theory are then the  
analogous $\CS^{(0)}_M \equiv \log \CZ^{(0)}_M =-S_E[g_M]$ evaluated on saddles $g_M$ different from the round sphere, in the parametric $\ell_0 \gg \ls$ regime of validity   
of the effective field theory, with $g_M$ asymptotically Einstein in the $\ell_0 \to \infty$ limit.  These are the analogs of tree-level scattering amplitudes.  The obvious quantum counterparts are the corresponding generalizations of $\CS$, i.e.\ $\CS_M \equiv \log\CZ_M$ evaluated in large-$\ell_0$ perturbation theory about the saddle $g_M$. These are the analogs of quantum scattering amplitudes. Below we make this a bit more concrete in examples. 

\subsubsection*{3D}

In $d=2$, the Weyl tensor vanishes identically, so higher-order curvature invariants involve $R_{\mu\nu}$ only and can be removed from the action by a field redefinition in large-$\ell_0$ perturbation theory, reducing it to pure Einstein form in general. As a result, $\CS^{(0)}$ is the only independent invariant in pure 3D gravity, all $g_M$ are Einstein, and the $\CS^{(0)}_M$ are all proportional to $\CS_0 \equiv \CS^{(0)}_{S^3}$. 

As discussed under (\ref{CSoneloop3Dsu}), the quantum $\CS=\CS_{S^3}$ takes the form
\begin{equation} \label{allllllll}
  \CS = \CS_0 = \CS_0 - 3 \log \CS_0 + 5 \log(2\pi) + \mbox{$\sum_n$}\, c_n \, \CS_0^{-2n} 
\end{equation} 
The corrections terms in the expansion are all nonlocal (no odd powers of $\ell_0$), and the coefficients provide an unambiguous, infinite data set.



\subsubsection*{odd D $\geq$ 5}

In 5D gravity, there are infinitely many independent coupling constants. There are also infinitely many different $\Lambda>0$ Einstein metrics on $S^5$, including a discrete infinity of B\"ohm metrics with $SO(3) \times SO(3)$ symmetry \cite{Bohm:1998vc} amenable to detailed numerical analysis \cite{Gibbons:2002th}, and 68 Sasaki-Einstein families with moduli spaces up to real dimension 10 \cite{Boyer:2003pe}. Unlike the round $S^5$, these are not conformally flat, and thus, unlike $\CS^{(0)}$, the corresponding $\CS^{(0)}_M$ will pick up couplings such as the Weyl-squared coupling $\lambda_{C^2}$ in (\ref{SERiem2}).  
It is plausible that this set of known Einstein metrics (perturbed by small higher-order corrections to the Einstein equations of motion at finite $\ell_0$) more than suffices to invariantly probe all independent couplings of the gravitational action, delivering moreover infinitely many quantum observables $\CS_M$, providing an infinity of unambiguous low-energy effective field theory data to any order in perturbation theory, without ever leaving the sphere --- at least in principle. 
   
The landscape of known $\Lambda>0$ Einstein metrics on odd-dimensional spheres becomes increasingly vast as the dimension grows, with double-exponentially growing numbers \cite{Boyer:2003pe}. For example there are at least 8610 families of Sasaki-Einstein manifolds on $S^7$, spanning all 28 diffeomorphism classes, with the standard class admitting a 82-dimensional family, and there are at least $10^{828}$ distinct families of Einstein metrics on $S^{25}$, featuring moduli spaces of dimension greater than $10^{833}$.

\subsubsection*{4D}


4D gravity likewise has infinitely many independent coupling constants. It is not known if $S^4$ has another Einstein metric besides the round sphere. In fact the list of 4D topologies known to admit $\Lambda>0$ Einstein metrics is rather limited \cite{2008arXiv0810.4830A}: $S^4$, $S^2 \times S^2$, $\ICP^2$, and the connected sums $\ICP^2 \# k \overline{\ICP^2}$, $1 \leq k \leq 8$. However for $k \geq 5$ these have a moduli space of nonzero dimension \cite{Tian:1987tw,Tian:1990wd}, which might suffice to probe all couplings.   (The moduli space would presumably be lifted at sufficiently high order in the $l_s$ expansion upon turning on higher-order curvature perturbations.) 

Below we illustrate in explicit detail how the Weyl-squared coupling can be extracted from suitable linear combinations of pairs of $\CS_M^{(0)}$ with $M \in \{S^4,S^2\times S^2,\ICP^2\}$, and how a suitable linear combination of all three can be used to extract an unambiguous linear combination of the constant terms arising at one loop.  

The Weyl-squared coupling $\lambda_{C^2}$ in $S_E[g]=\mbox{(\ref{SERiem2})} + \cdots$ is invisible to $\CS^{(0)}$ (\ref{S0forS4}) but it is picked up by $\CS_M^{(0)}$ by $M=S^2 \times S^2$:
\begin{align} \label{S0S2xS2}
 \CS^{(0)}_{S^2 \times S^2} = \frac{2}{3} \cdot  \frac{\pi}{G} \bigl(\ell_0^2 + 48 \, \g_{R^2} \, \ls^2 + 16 \,  \g_{C^2} \, \ls^2 \,+\, \cdots  \bigr)  \, ,
\end{align}
with the dots denoting $O(\ls^4)$ terms and $\ell_0 = \sqrt{3/\Lambda}+\cdots$   as in (\ref{S0forS4}). Physically, $\CS^{(0)}_{S^2 \times S^2}$ is the horizon entropy of the ${\rm dS}_2 \times S^2$ static patch, i.e.\ the Nariai spacetime between the cosmological and maximal Schwarzschild-de Sitter black hole horizons, both of area $A=\frac{1}{3} \cdot 4 \pi \ell_0^{2}$. Comparing to (\ref{S0forS4}), the linear combination
\begin{align} \label{saddledif}
 \CS^{(0)}_{C^2} \equiv 3 \, \CS^{(0)}_{S^2 \times S^2} - 2 \, \CS^{(0)}  = \frac{32  \pi  \ls^2}{G} \bigl( \, \g_{C^2}  \,+\, \cdots  \bigr)  
\end{align} 
extracts the Weyl-squared coupling of $S_E[g]$.  
Analogously, for the Einstein metric on $\ICP^2$, we get $\tilde \CS^{(0)}_{C^2} \equiv 8\, \CS^{(0)}_{\ICP^2} - 6\,\CS^{(0)}  = \frac{48  \pi \ls^2}{G} ( \g_{C^2} +\cdots )$.
Then 
\begin{align} \label{lincombinv0}
  \CS^{(0)}_{\rm cub} \equiv 2\,\tilde \CS^{(0)}_{C^2} -  3 \, \CS^{(0)}_{C^2} = 16 \, \CS^{(0)}_{\ICP^2} - 9 \, \CS^{(0)}_{S^2 \times S^2} - 6 \, \CS^{(0)}    \, = 0 \, + \, \cdots \, ,
\end{align} 
which extracts some curvature-cubed coupling in the effective action. 

To one loop, the quantum  $\CS_M = \log \CZ_M$ can be expressed in a form paralleling (\ref{CSgravd}):
\begin{align}
 \CS_M = \CS^{(0)}_M - \frac{D_M}{2}  \log \CS_M^{(0)} 
 + \alpha_M \log  \frac{\ell_0}{L} + K_M \, +  \, \cdots \, , 
\end{align}
where $D_M$ is the number of Killing vectors of $M$: $D_{S^4}=10$, $D_{S^2 \times S^2}=6$, $D_{\ICP^2}=8$, and $\alpha_M$ can be obtained from the local expressions  in \cite{Christensen:1979iy}: $\alpha_{S^4} = -\frac{571}{45}$, $\alpha_{S^2 \times S^2}=-\frac{98}{45}$, $\alpha_{\ICP^2}=-\frac{359}{60}$. Computing the constants $K_M$ generalizing $K_{S^4}$ given in (\ref{CSres}) 
would require more work. Moreover, computing them for one or two saddles would provide no unambiguous information because they may be absorbed into $\g_{C^2}$ and $\g_{R^2}$. However, since there only two undetermined coupling constants at this order, computing them for all {\it three} does provide unambiguous information, extracted by the quantum counterpart of (\ref{lincombinv0}): 
\begin{align}  \label{lincombinv}
 \CS_{\rm cub} \, \equiv \,
 16 \,  \CS_{\ICP^2} - 9 \,  \CS_{S^2 \times S^2} - 6 \, \CS_{S^4} = 
  -7 \, \log \CS^{(0)} 
  + 16 \,  K_{\ICP^2} - 9 \, K_{S^2 \times S^2} - 6 \, K_{S^4} 
  + \cdots 
\end{align}      
The $\log(\ell_0/L)$ terms had to cancel in this linear combination because the tree-level parts at this order cancel by design and $L\partial_L \CS_{\rm cub} = 0$. 




\begin{thebibliography}{100}
	
	\bibitem{PhysRevD.15.2738}
	G.~W. Gibbons and S.~W. Hawking, ``Cosmological event horizons, thermodynamics,
	and particle creation,''
	\href{http://dx.doi.org/10.1103/PhysRevD.15.2738}{{\em Phys. Rev. D}
		{\bfseries 15} (May, 1977) 2738--2751}.
	
	\bibitem{PhysRevD.15.2752}
	G.~W. Gibbons and S.~W. Hawking, ``Action integrals and partition functions in
	quantum gravity,'' \href{http://dx.doi.org/10.1103/PhysRevD.15.2752}{{\em
			Phys. Rev. D} {\bfseries 15} (May, 1977) 2752--2756}.
	
	\bibitem{Spradlin:2001pw}
	M.~Spradlin, A.~Strominger, and A.~Volovich, ``{Les Houches lectures on de
		Sitter space},'' in {\em {Les Houches Summer School: Session 76: Euro Summer
			School on Unity of Fundamental Physics: Gravity, Gauge Theory and Strings}}.
	\newblock 10, 2001.
	\newblock \href{http://arxiv.org/abs/hep-th/0110007}{{\ttfamily
			arXiv:hep-th/0110007}}.
	
	\bibitem{Myers:2002mq}
	R.~C. Myers, \href{http://dx.doi.org/10.1007/0-387-24992-3_6}{``{Tall tales
			from de Sitter space},''} in {\em {School on Quantum Gravity}}.
	\newblock 1, 2002.
	
	\bibitem{Bousso:2002ju}
	R.~Bousso, ``{The Holographic principle},''
	\href{http://dx.doi.org/10.1103/RevModPhys.74.825}{{\em Rev. Mod. Phys.}
		{\bfseries 74} (2002) 825--874},
	\href{http://arxiv.org/abs/hep-th/0203101}{{\ttfamily arXiv:hep-th/0203101}}.
	
	\bibitem{Loeb:2001dh}
	A.~Loeb, ``{The Long - term future of extragalactic astronomy},''
	\href{http://dx.doi.org/10.1103/PhysRevD.65.047301}{{\em Phys. Rev. D}
		{\bfseries 65} (2002) 047301},
	\href{http://arxiv.org/abs/astro-ph/0107568}{{\ttfamily
			arXiv:astro-ph/0107568}}.
	
	\bibitem{Krauss:2007nt}
	L.~M. Krauss and R.~J. Scherrer, ``{The Return of a Static Universe and the End
		of Cosmology},'' \href{http://dx.doi.org/10.1007/s10714-007-0472-9}{{\em Gen.
			Rel. Grav.} {\bfseries 39} (2007) 1545--1550},
	\href{http://arxiv.org/abs/0704.0221}{{\ttfamily arXiv:0704.0221
			[astro-ph]}}.
	
	\bibitem{Bhattacharyya:2012ye}
	S.~Bhattacharyya, A.~Grassi, M.~Marino, and A.~Sen, ``{A One-Loop Test of
		Quantum Supergravity},''
	\href{http://dx.doi.org/10.1088/0264-9381/31/1/015012}{{\em Class. Quant.
			Grav.} {\bfseries 31} (2014) 015012},
	\href{http://arxiv.org/abs/1210.6057}{{\ttfamily arXiv:1210.6057 [hep-th]}}.
	
	\bibitem{Banerjee:2010qc}
	S.~Banerjee, R.~K. Gupta, and A.~Sen, ``{Logarithmic Corrections to Extremal
		Black Hole Entropy from Quantum Entropy Function},''
	\href{http://dx.doi.org/10.1007/JHEP03(2011)147}{{\em JHEP} {\bfseries 03}
		(2011) 147}, \href{http://arxiv.org/abs/1005.3044}{{\ttfamily arXiv:1005.3044
			[hep-th]}}.
	
	\bibitem{Banerjee:2011jp}
	S.~Banerjee, R.~K. Gupta, I.~Mandal, and A.~Sen, ``{Logarithmic Corrections to
		N=4 and N=8 Black Hole Entropy: A One Loop Test of Quantum Gravity},''
	\href{http://dx.doi.org/10.1007/JHEP11(2011)143}{{\em JHEP} {\bfseries 11}
		(2011) 143}, \href{http://arxiv.org/abs/1106.0080}{{\ttfamily arXiv:1106.0080
			[hep-th]}}.
	
	\bibitem{Sen:2012kpz}
	A.~Sen, ``{Logarithmic Corrections to N=2 Black Hole Entropy: An Infrared
		Window into the Microstates},''
	\href{http://dx.doi.org/10.1007/s10714-012-1336-5}{{\em Gen. Rel. Grav.}
		{\bfseries 44} no.~5, (2012) 1207--1266},
	\href{http://arxiv.org/abs/1108.3842}{{\ttfamily arXiv:1108.3842 [hep-th]}}.
	
	\bibitem{Sen:2014aja}
	A.~Sen, ``{Microscopic and Macroscopic Entropy of Extremal Black Holes in
		String Theory},'' \href{http://dx.doi.org/10.1007/s10714-014-1711-5}{{\em
			Gen. Rel. Grav.} {\bfseries 46} (2014) 1711},
	\href{http://arxiv.org/abs/1402.0109}{{\ttfamily arXiv:1402.0109 [hep-th]}}.
	
	\bibitem{Sen:2012dw}
	A.~Sen, ``{Logarithmic Corrections to Schwarzschild and Other Non-extremal
		Black Hole Entropy in Different Dimensions},''
	\href{http://dx.doi.org/10.1007/JHEP04(2013)156}{{\em JHEP} {\bfseries 04}
		(2013) 156}, \href{http://arxiv.org/abs/1205.0971}{{\ttfamily arXiv:1205.0971
			[hep-th]}}.
	
	\bibitem{Giombi:2013fka}
	S.~Giombi and I.~R. Klebanov, ``{One Loop Tests of Higher Spin AdS/CFT},''
	\href{http://dx.doi.org/10.1007/JHEP12(2013)068}{{\em JHEP} {\bfseries 12}
		(2013) 068}, \href{http://arxiv.org/abs/1308.2337}{{\ttfamily arXiv:1308.2337
			[hep-th]}}.
	
	\bibitem{Giombi:2014iua}
	S.~Giombi, I.~R. Klebanov, and B.~R. Safdi, ``{Higher Spin AdS$_{d+1}$/CFT$_d$
		at One Loop},'' \href{http://dx.doi.org/10.1103/PhysRevD.89.084004}{{\em
			Phys. Rev. D} {\bfseries 89} no.~8, (2014) 084004},
	\href{http://arxiv.org/abs/1401.0825}{{\ttfamily arXiv:1401.0825 [hep-th]}}.
	
	\bibitem{Gunaydin:2016amv}
	M.~G\"unaydin, E.~D. Skvortsov, and T.~Tran, ``{Exceptional $F(4)$ higher-spin
		theory in AdS$_{6}$ at one-loop and other tests of duality},''
	\href{http://dx.doi.org/10.1007/JHEP11(2016)168}{{\em JHEP} {\bfseries 11}
		(2016) 168}, \href{http://arxiv.org/abs/1608.07582}{{\ttfamily
			arXiv:1608.07582 [hep-th]}}.
	
	\bibitem{Giombi:2016pvg}
	S.~Giombi, I.~R. Klebanov, and Z.~M. Tan, ``{The ABC of Higher-Spin AdS/CFT},''
	\href{http://dx.doi.org/10.3390/universe4010018}{{\em Universe} {\bfseries 4}
		no.~1, (2018) 18}, \href{http://arxiv.org/abs/1608.07611}{{\ttfamily
			arXiv:1608.07611 [hep-th]}}.
	
	\bibitem{Skvortsov:2017ldz}
	E.~D. Skvortsov and T.~Tran, ``{AdS/CFT in Fractional Dimension and Higher Spin
		Gravity at One Loop},'' \href{http://dx.doi.org/10.3390/universe3030061}{{\em
			Universe} {\bfseries 3} no.~3, (2017) 61},
	\href{http://arxiv.org/abs/1707.00758}{{\ttfamily arXiv:1707.00758
			[hep-th]}}.
	
	\bibitem{Gearhart:2010wz}
	C.~A. Gearhart, ```astonishing successes' and `bitter disappointment': The
	specific heat of hydrogen in quantum theory,''
	\href{http://dx.doi.org/10.1007/s00407-009-0053-2}{{\em Archive for History
			of Exact Sciences} {\bfseries 64} no.~2, (2010) 113--202}.
	
	\bibitem{Maldacena:1998ih}
	J.~M. Maldacena and A.~Strominger, ``{Statistical entropy of de Sitter
		space},'' \href{http://dx.doi.org/10.1088/1126-6708/1998/02/014}{{\em JHEP}
		{\bfseries 02} (1998) 014},
	\href{http://arxiv.org/abs/gr-qc/9801096}{{\ttfamily arXiv:gr-qc/9801096}}.
	
	\bibitem{Banados:1998tb}
	M.~Banados, T.~Brotz, and M.~E. Ortiz, ``{Quantum three-dimensional de Sitter
		space},'' \href{http://dx.doi.org/10.1103/PhysRevD.59.046002}{{\em Phys. Rev.
			D} {\bfseries 59} (1999) 046002},
	\href{http://arxiv.org/abs/hep-th/9807216}{{\ttfamily arXiv:hep-th/9807216}}.
	
	\bibitem{Bousso:2001mw}
	R.~Bousso, A.~Maloney, and A.~Strominger, ``{Conformal vacua and entropy in de
		Sitter space},'' \href{http://dx.doi.org/10.1103/PhysRevD.65.104039}{{\em
			Phys. Rev. D} {\bfseries 65} (2002) 104039},
	\href{http://arxiv.org/abs/hep-th/0112218}{{\ttfamily arXiv:hep-th/0112218}}.
	
	\bibitem{Govindarajan:2002ry}
	T.~R. Govindarajan, R.~K. Kaul, and V.~Suneeta, ``{Quantum gravity on dS(3)},''
	\href{http://dx.doi.org/10.1088/0264-9381/19/15/320}{{\em Class. Quant.
			Grav.} {\bfseries 19} (2002) 4195--4205},
	\href{http://arxiv.org/abs/hep-th/0203219}{{\ttfamily arXiv:hep-th/0203219}}.
	
	\bibitem{Silverstein:2003jp}
	E.~Silverstein, ``{AdS and dS entropy from string junctions: or, The Function
		of junction conjunctions},'' in {\em {From Fields to Strings:
			Circumnavigating Theoretical Physics: A Conference in Tribute to Ian Kogan}}.
	\newblock 8, 2003.
	\newblock \href{http://arxiv.org/abs/hep-th/0308175}{{\ttfamily
			arXiv:hep-th/0308175}}.
	
	\bibitem{Fabinger:2003gp}
	M.~Fabinger and E.~Silverstein, ``{D-Sitter space: Causal structure,
		thermodynamics, and entropy},''
	\href{http://dx.doi.org/10.1088/1126-6708/2004/12/061}{{\em JHEP} {\bfseries
			12} (2004) 061}, \href{http://arxiv.org/abs/hep-th/0304220}{{\ttfamily
			arXiv:hep-th/0304220}}.
	
	\bibitem{Parikh:2004wh}
	M.~K. Parikh and E.~P. Verlinde, ``{De Sitter holography with a finite number
		of states},'' \href{http://dx.doi.org/10.1088/1126-6708/2005/01/054}{{\em
			JHEP} {\bfseries 01} (2005) 054},
	\href{http://arxiv.org/abs/hep-th/0410227}{{\ttfamily arXiv:hep-th/0410227}}.
	
	\bibitem{Banks:2006rx}
	T.~Banks, B.~Fiol, and A.~Morisse, ``{Towards a quantum theory of de Sitter
		space},'' \href{http://dx.doi.org/10.1088/1126-6708/2006/12/004}{{\em JHEP}
		{\bfseries 12} (2006) 004},
	\href{http://arxiv.org/abs/hep-th/0609062}{{\ttfamily arXiv:hep-th/0609062}}.
	
	\bibitem{Dong:2010pm}
	X.~Dong, B.~Horn, E.~Silverstein, and G.~Torroba, ``{Micromanaging de Sitter
		holography},'' \href{http://dx.doi.org/10.1088/0264-9381/27/24/245020}{{\em
			Class. Quant. Grav.} {\bfseries 27} (2010) 245020},
	\href{http://arxiv.org/abs/1005.5403}{{\ttfamily arXiv:1005.5403 [hep-th]}}.
	
	\bibitem{Heckman:2011qu}
	J.~J. Heckman and H.~Verlinde, ``{Instantons, Twistors, and Emergent
		Gravity},'' \href{http://arxiv.org/abs/1112.5210}{{\ttfamily arXiv:1112.5210
			[hep-th]}}.
	
	\bibitem{Banks:2013qpa}
	T.~Banks, ``{Lectures on Holographic Space Time},''
	\href{http://arxiv.org/abs/1311.0755}{{\ttfamily arXiv:1311.0755 [hep-th]}}.
	
	\bibitem{Neiman:2017zdr}
	Y.~Neiman, ``{Towards causal patch physics in dS/CFT},''
	\href{http://dx.doi.org/10.1051/epjconf/201816801007}{{\em EPJ Web Conf.}
		{\bfseries 168} (2018) 01007},
	\href{http://arxiv.org/abs/1710.05682}{{\ttfamily arXiv:1710.05682
			[hep-th]}}.
	
	\bibitem{Dong:2018cuv}
	X.~Dong, E.~Silverstein, and G.~Torroba, ``{De Sitter Holography and
		Entanglement Entropy},''
	\href{http://dx.doi.org/10.1007/JHEP07(2018)050}{{\em JHEP} {\bfseries 07}
		(2018) 050}, \href{http://arxiv.org/abs/1804.08623}{{\ttfamily
			arXiv:1804.08623 [hep-th]}}.
	
	\bibitem{Arias:2019zug}
	C.~Arias, F.~Diaz, R.~Olea, and P.~Sundell, ``{Liouville description of conical
		defects in dS$_4$, Gibbons-Hawking entropy as modular entropy, and dS$_3$
		holography},'' \href{http://dx.doi.org/10.1007/JHEP04(2020)124}{{\em JHEP}
		{\bfseries 04} (2020) 124}, \href{http://arxiv.org/abs/1906.05310}{{\ttfamily
			arXiv:1906.05310 [hep-th]}}.
	
	\bibitem{Hawking:1976ja}
	S.~W. Hawking, ``{Zeta Function Regularization of Path Integrals in Curved
		Space-Time},'' \href{http://dx.doi.org/10.1007/BF01626516}{{\em Commun. Math.
			Phys.} {\bfseries 55} (1977) 133}.
	
	\bibitem{Gibbons:1978ji}
	G.~W. Gibbons and M.~J. Perry, ``{Quantizing Gravitational Instantons},''
	\href{http://dx.doi.org/10.1016/0550-3213(78)90434-0}{{\em Nucl. Phys. B}
		{\bfseries 146} (1978) 90--108}.
	
	\bibitem{Christensen:1979iy}
	S.~M. Christensen and M.~J. Duff, ``{Quantizing Gravity with a Cosmological
		Constant},'' \href{http://dx.doi.org/10.1016/0550-3213(80)90423-X}{{\em Nucl.
			Phys. B} {\bfseries 170} (1980) 480--506}.
	
	\bibitem{Fradkin:1983mq}
	E.~S. Fradkin and A.~A. Tseytlin, ``{One Loop Effective Potential in Gauged
		O(4) Supergravity},''
	\href{http://dx.doi.org/10.1016/0550-3213(84)90074-9}{{\em Nucl. Phys. B}
		{\bfseries 234} (1984) 472}.
	
	\bibitem{Yasuda:1983hk}
	O.~Yasuda, ``{On the One Loop Effective Potential in Quantum Gravity},''
	\href{http://dx.doi.org/10.1016/0370-2693(84)91104-3}{{\em Phys. Lett. B}
		{\bfseries 137} (1984) 52}.
	
	\bibitem{Allen:1983dg}
	B.~Allen, ``{Phase Transitions in de Sitter Space},''
	\href{http://dx.doi.org/10.1016/0550-3213(83)90470-4}{{\em Nucl. Phys. B}
		{\bfseries 226} (1983) 228--252}.
	
	\bibitem{Polchinski:1988ua}
	J.~Polchinski, ``{The Phase of the Sum Over Spheres},''
	\href{http://dx.doi.org/10.1016/0370-2693(89)90387-0}{{\em Phys. Lett. B}
		{\bfseries 219} (1989) 251--257}.
	
	\bibitem{Taylor:1989ua}
	T.~R. Taylor and G.~Veneziano, ``{Quantum Gravity at Large Distances and the
		Cosmological Constant},''
	\href{http://dx.doi.org/10.1016/0550-3213(90)90615-K}{{\em Nucl. Phys. B}
		{\bfseries 345} (1990) 210--230}.
	
	\bibitem{Vassilevich:1992rk}
	D.~V. Vassilevich, ``{One loop quantum gravity on de Sitter space},''
	\href{http://dx.doi.org/10.1142/S0217751X93000679}{{\em Int. J. Mod. Phys. A}
		{\bfseries 8} (1993) 1637--1652}.
	
	\bibitem{Volkov:2000ih}
	M.~S. Volkov and A.~Wipf, ``{Black hole pair creation in de Sitter space: A
		Complete one loop analysis},''
	\href{http://dx.doi.org/10.1016/S0550-3213(00)00287-X}{{\em Nucl. Phys. B}
		{\bfseries 582} (2000) 313--362},
	\href{http://arxiv.org/abs/hep-th/0003081}{{\ttfamily arXiv:hep-th/0003081}}.
	
	\bibitem{Witten:1995gf}
	E.~Witten, ``{On S duality in Abelian gauge theory},''
	\href{http://dx.doi.org/10.1007/BF01671570}{{\em Selecta Math.} {\bfseries 1}
		(1995) 383}, \href{http://arxiv.org/abs/hep-th/9505186}{{\ttfamily
			arXiv:hep-th/9505186}}.
	
	\bibitem{Donnelly:2013tia}
	W.~Donnelly and A.~C. Wall, ``{Unitarity of Maxwell theory on curved spacetimes
		in the covariant formalism},''
	\href{http://dx.doi.org/10.1103/PhysRevD.87.125033}{{\em Phys. Rev. D}
		{\bfseries 87} no.~12, (2013) 125033},
	\href{http://arxiv.org/abs/1303.1885}{{\ttfamily arXiv:1303.1885 [hep-th]}}.
	
	\bibitem{Donnelly:2014fua}
	W.~Donnelly and A.~C. Wall, ``{Entanglement entropy of electromagnetic edge
		modes},'' \href{http://dx.doi.org/10.1103/PhysRevLett.114.111603}{{\em Phys.
			Rev. Lett.} {\bfseries 114} no.~11, (2015) 111603},
	\href{http://arxiv.org/abs/1412.1895}{{\ttfamily arXiv:1412.1895 [hep-th]}}.
	
	\bibitem{Donnelly:2015hxa}
	W.~Donnelly and A.~C. Wall, ``{Geometric entropy and edge modes of the
		electromagnetic field},''
	\href{http://dx.doi.org/10.1103/PhysRevD.94.104053}{{\em Phys. Rev. D}
		{\bfseries 94} no.~10, (2016) 104053},
	\href{http://arxiv.org/abs/1506.05792}{{\ttfamily arXiv:1506.05792
			[hep-th]}}.
	
	\bibitem{Giombi:2013yva}
	S.~Giombi, I.~R. Klebanov, S.~S. Pufu, B.~R. Safdi, and G.~Tarnopolsky, ``{AdS
		Description of Induced Higher-Spin Gauge Theory},''
	\href{http://dx.doi.org/10.1007/JHEP10(2013)016}{{\em JHEP} {\bfseries 10}
		(2013) 016}, \href{http://arxiv.org/abs/1306.5242}{{\ttfamily arXiv:1306.5242
			[hep-th]}}.
	
	\bibitem{Tseytlin:2013fca}
	A.~A. Tseytlin, ``{Weyl anomaly of conformal higher spins on six-sphere},''
	\href{http://dx.doi.org/10.1016/j.nuclphysb.2013.10.008}{{\em Nucl. Phys. B}
		{\bfseries 877} (2013) 632--646},
	\href{http://arxiv.org/abs/1310.1795}{{\ttfamily arXiv:1310.1795 [hep-th]}}.
	
	\bibitem{Joung:2013nma}
	E.~Joung and M.~Taronna, ``{Cubic-interaction-induced deformations of
		higher-spin symmetries},''
	\href{http://dx.doi.org/10.1007/JHEP03(2014)103}{{\em JHEP} {\bfseries 03}
		(2014) 103}, \href{http://arxiv.org/abs/1311.0242}{{\ttfamily arXiv:1311.0242
			[hep-th]}}.
	
	\bibitem{Joung:2014qya}
	E.~Joung and K.~Mkrtchyan, ``{Notes on higher-spin algebras: minimal
		representations and structure constants},''
	\href{http://dx.doi.org/10.1007/JHEP05(2014)103}{{\em JHEP} {\bfseries 05}
		(2014) 103}, \href{http://arxiv.org/abs/1401.7977}{{\ttfamily arXiv:1401.7977
			[hep-th]}}.
	
	\bibitem{Sleight:2016dba}
	C.~Sleight and M.~Taronna, ``{Higher Spin Interactions from Conformal Field
		Theory: The Complete Cubic Couplings},''
	\href{http://dx.doi.org/10.1103/PhysRevLett.116.181602}{{\em Phys. Rev.
			Lett.} {\bfseries 116} no.~18, (2016) 181602},
	\href{http://arxiv.org/abs/1603.00022}{{\ttfamily arXiv:1603.00022
			[hep-th]}}.
	
	\bibitem{Sleight:2016xqq}
	C.~Sleight and M.~Taronna, ``{Higher-Spin Algebras, Holography and Flat
		Space},'' \href{http://dx.doi.org/10.1007/JHEP02(2017)095}{{\em JHEP}
		{\bfseries 02} (2017) 095}, \href{http://arxiv.org/abs/1609.00991}{{\ttfamily
			arXiv:1609.00991 [hep-th]}}.
	
	\bibitem{Basile:2016aen}
	T.~Basile, X.~Bekaert, and N.~Boulanger, ``{Mixed-symmetry fields in de Sitter
		space: a group theoretical glance},''
	\href{http://dx.doi.org/10.1007/JHEP05(2017)081}{{\em JHEP} {\bfseries 05}
		(2017) 081}, \href{http://arxiv.org/abs/1612.08166}{{\ttfamily
			arXiv:1612.08166 [hep-th]}}.
	
	\bibitem{Basile:2018zoy}
	T.~Basile, E.~Joung, S.~Lal, and W.~Li, ``{Character Integral Representation of
		Zeta function in AdS$_{d+1}$: I. Derivation of the general formula},''
	\href{http://dx.doi.org/10.1007/JHEP10(2018)091}{{\em JHEP} {\bfseries 10}
		(2018) 091}, \href{http://arxiv.org/abs/1805.05646}{{\ttfamily
			arXiv:1805.05646 [hep-th]}}.
	
	\bibitem{Basile:2018acb}
	T.~Basile, E.~Joung, S.~Lal, and W.~Li, ``{Character integral representation of
		zeta function in AdS$_{d+1}$. Part II. Application to partially-massless
		higher-spin gravities},''
	\href{http://dx.doi.org/10.1007/JHEP07(2018)132}{{\em JHEP} {\bfseries 07}
		(2018) 132}, \href{http://arxiv.org/abs/1805.10092}{{\ttfamily
			arXiv:1805.10092 [hep-th]}}.
	
	\bibitem{Susskind:1994sm}
	L.~Susskind and J.~Uglum, ``{Black hole entropy in canonical quantum gravity
		and superstring theory},''
	\href{http://dx.doi.org/10.1103/PhysRevD.50.2700}{{\em Phys. Rev. D}
		{\bfseries 50} (1994) 2700--2711},
	\href{http://arxiv.org/abs/hep-th/9401070}{{\ttfamily arXiv:hep-th/9401070}}.
	
	\bibitem{Kabat:1995eq}
	D.~N. Kabat, ``{Black hole entropy and entropy of entanglement},''
	\href{http://dx.doi.org/10.1016/0550-3213(95)00443-V}{{\em Nucl. Phys. B}
		{\bfseries 453} (1995) 281--299},
	\href{http://arxiv.org/abs/hep-th/9503016}{{\ttfamily arXiv:hep-th/9503016}}.
	
	\bibitem{Kabat:1995jq}
	D.~N. Kabat, S.~H. Shenker, and M.~J. Strassler, ``{Black hole entropy in the
		O(N) model},'' \href{http://dx.doi.org/10.1103/PhysRevD.52.7027}{{\em Phys.
			Rev. D} {\bfseries 52} (1995) 7027--7036},
	\href{http://arxiv.org/abs/hep-th/9506182}{{\ttfamily arXiv:hep-th/9506182}}.
	
	\bibitem{Larsen:1995ax}
	F.~Larsen and F.~Wilczek, ``{Renormalization of black hole entropy and of the
		gravitational coupling constant},''
	\href{http://dx.doi.org/10.1016/0550-3213(95)00548-X}{{\em Nucl. Phys. B}
		{\bfseries 458} (1996) 249--266},
	\href{http://arxiv.org/abs/hep-th/9506066}{{\ttfamily arXiv:hep-th/9506066}}.
	
	\bibitem{Dowker:2010bu}
	J.~S. Dowker, ``{Entanglement entropy for even spheres},''
	\href{http://arxiv.org/abs/1009.3854}{{\ttfamily arXiv:1009.3854 [hep-th]}}.
	
	\bibitem{Dowker:2010yj}
	J.~S. Dowker, ``{Entanglement entropy for odd spheres},''
	\href{http://arxiv.org/abs/1012.1548}{{\ttfamily arXiv:1012.1548 [hep-th]}}.
	
	\bibitem{Casini:2011kv}
	H.~Casini, M.~Huerta, and R.~C. Myers, ``{Towards a derivation of holographic
		entanglement entropy},''
	\href{http://dx.doi.org/10.1007/JHEP05(2011)036}{{\em JHEP} {\bfseries 05}
		(2011) 036}, \href{http://arxiv.org/abs/1102.0440}{{\ttfamily arXiv:1102.0440
			[hep-th]}}.
	
	\bibitem{Solodukhin:2011gn}
	S.~N. Solodukhin, ``{Entanglement entropy of black holes},''
	\href{http://dx.doi.org/10.12942/lrr-2011-8}{{\em Living Rev. Rel.}
		{\bfseries 14} (2011) 8}, \href{http://arxiv.org/abs/1104.3712}{{\ttfamily
			arXiv:1104.3712 [hep-th]}}.
	
	\bibitem{Eling:2013aqa}
	C.~Eling, Y.~Oz, and S.~Theisen, ``{Entanglement and Thermal Entropy of Gauge
		Fields},'' \href{http://dx.doi.org/10.1007/JHEP11(2013)019}{{\em JHEP}
		{\bfseries 11} (2013) 019}, \href{http://arxiv.org/abs/1308.4964}{{\ttfamily
			arXiv:1308.4964 [hep-th]}}.
	
	\bibitem{Casini:2015dsg}
	H.~Casini and M.~Huerta, ``{Entanglement entropy of a Maxwell field on the
		sphere},'' \href{http://dx.doi.org/10.1103/PhysRevD.93.105031}{{\em Phys.
			Rev. D} {\bfseries 93} no.~10, (2016) 105031},
	\href{http://arxiv.org/abs/1512.06182}{{\ttfamily arXiv:1512.06182
			[hep-th]}}.
	
	\bibitem{Buividovich:2008gq}
	P.~V. Buividovich and M.~I. Polikarpov, ``{Entanglement entropy in gauge
		theories and the holographic principle for electric strings},''
	\href{http://dx.doi.org/10.1016/j.physletb.2008.10.032}{{\em Phys. Lett. B}
		{\bfseries 670} (2008) 141--145},
	\href{http://arxiv.org/abs/0806.3376}{{\ttfamily arXiv:0806.3376 [hep-th]}}.
	
	\bibitem{Donnelly:2011hn}
	W.~Donnelly, ``{Decomposition of entanglement entropy in lattice gauge
		theory},'' \href{http://dx.doi.org/10.1103/PhysRevD.85.085004}{{\em Phys.
			Rev. D} {\bfseries 85} (2012) 085004},
	\href{http://arxiv.org/abs/1109.0036}{{\ttfamily arXiv:1109.0036 [hep-th]}}.
	
	\bibitem{Casini:2013rba}
	H.~Casini, M.~Huerta, and J.~A. Rosabal, ``{Remarks on entanglement entropy for
		gauge fields},'' \href{http://dx.doi.org/10.1103/PhysRevD.89.085012}{{\em
			Phys. Rev. D} {\bfseries 89} no.~8, (2014) 085012},
	\href{http://arxiv.org/abs/1312.1183}{{\ttfamily arXiv:1312.1183 [hep-th]}}.
	
	\bibitem{Soni:2015yga}
	R.~M. Soni and S.~P. Trivedi, ``{Aspects of Entanglement Entropy for Gauge
		Theories},'' \href{http://dx.doi.org/10.1007/JHEP01(2016)136}{{\em JHEP}
		{\bfseries 01} (2016) 136}, \href{http://arxiv.org/abs/1510.07455}{{\ttfamily
			arXiv:1510.07455 [hep-th]}}.
	
	\bibitem{Dong:2018seb}
	X.~Dong, D.~Harlow, and D.~Marolf, ``{Flat entanglement spectra in fixed-area
		states of quantum gravity},''
	\href{http://dx.doi.org/10.1007/JHEP10(2019)240}{{\em JHEP} {\bfseries 10}
		(2019) 240}, \href{http://arxiv.org/abs/1811.05382}{{\ttfamily
			arXiv:1811.05382 [hep-th]}}.
	
	\bibitem{Jafferis:2015del}
	D.~L. Jafferis, A.~Lewkowycz, J.~Maldacena, and S.~J. Suh, ``{Relative entropy
		equals bulk relative entropy},''
	\href{http://dx.doi.org/10.1007/JHEP06(2016)004}{{\em JHEP} {\bfseries 06}
		(2016) 004}, \href{http://arxiv.org/abs/1512.06431}{{\ttfamily
			arXiv:1512.06431 [hep-th]}}.
			
\bibitem{Casini:2019kex}
H.~Casini, M.~Huerta, J.~M.~Mag\'an and D.~Pontello,
``Entanglement entropy and superselection sectors. Part I. Global symmetries,''
\href{https://doi.org/10.1007/JHEP02(2020)014}{{\em JHEP} {\bfseries 02}
	(2020) 014}, \href{https://arxiv.org/abs/1905.10487}{{\ttfamily
		arXiv:1905.10487 [hep-th]}}.
%
	
	\bibitem{Blommaert:2018rsf}
	A.~Blommaert, T.~G. Mertens, H.~Verschelde, and V.~I. Zakharov, ``{Edge State
		Quantization: Vector Fields in Rindler},''
	\href{http://dx.doi.org/10.1007/JHEP08(2018)196}{{\em JHEP} {\bfseries 08}
		(2018) 196}, \href{http://arxiv.org/abs/1801.09910}{{\ttfamily
			arXiv:1801.09910 [hep-th]}}.
	
	\bibitem{Soni:2016ogt}
	R.~M. Soni and S.~P. Trivedi, ``{Entanglement entropy in (3 + 1)-d free U(1)
		gauge theory},'' \href{http://dx.doi.org/10.1007/JHEP02(2017)101}{{\em JHEP}
		{\bfseries 02} (2017) 101}, \href{http://arxiv.org/abs/1608.00353}{{\ttfamily
			arXiv:1608.00353 [hep-th]}}.
	
	\bibitem{Ng:2012xp}
	G.~S. Ng and A.~Strominger, ``{State/Operator Correspondence in Higher-Spin
		dS/CFT},'' \href{http://dx.doi.org/10.1088/0264-9381/30/10/104002}{{\em
			Class. Quant. Grav.} {\bfseries 30} (2013) 104002},
	\href{http://arxiv.org/abs/1204.1057}{{\ttfamily arXiv:1204.1057 [hep-th]}}.
	
	\bibitem{Jafferis:2013qia}
	D.~L. Jafferis, A.~Lupsasca, V.~Lysov, G.~S. Ng, and A.~Strominger,
	``{Quasinormal quantization in de Sitter spacetime},''
	\href{http://dx.doi.org/10.1007/JHEP01(2015)004}{{\em JHEP} {\bfseries 01}
		(2015) 004}, \href{http://arxiv.org/abs/1305.5523}{{\ttfamily arXiv:1305.5523
			[hep-th]}}.
	
	\bibitem{Sun:2020sgn}
	Z.~Sun, ``{Higher spin de Sitter quasinormal modes},''
	\href{http://arxiv.org/abs/2010.09684}{{\ttfamily arXiv:2010.09684
			[hep-th]}}.
	
	\bibitem{tHooft:1984kcu}
	G.~'t~Hooft, ``{On the Quantum Structure of a Black Hole},''
	\href{http://dx.doi.org/10.1016/0550-3213(85)90418-3}{{\em Nucl. Phys. B}
		{\bfseries 256} (1985) 727--745}.
	
	\bibitem{Denef:2009kn}
	F.~Denef, S.~A. Hartnoll, and S.~Sachdev, ``{Black hole determinants and
		quasinormal modes},''
	\href{http://dx.doi.org/10.1088/0264-9381/27/12/125001}{{\em Class. Quant.
			Grav.} {\bfseries 27} (2010) 125001},
	\href{http://arxiv.org/abs/0908.2657}{{\ttfamily arXiv:0908.2657 [hep-th]}}.
	
	\bibitem{Zinoviev:2001dt}
	Y.~M. Zinoviev, ``{On massive high spin particles in AdS},''
	\href{http://arxiv.org/abs/hep-th/0108192}{{\ttfamily arXiv:hep-th/0108192}}.
	
	\bibitem{Witten:1988hc}
	E.~Witten, ``{(2+1)-Dimensional Gravity as an Exactly Soluble System},''
	\href{http://dx.doi.org/10.1016/0550-3213(88)90143-5}{{\em Nucl. Phys. B}
		{\bfseries 311} (1988) 46}.
	
	\bibitem{Witten:1989ip}
	E.~Witten, ``{Quantization of {Chern-Simons} Gauge Theory With Complex Gauge
		Group},'' \href{http://dx.doi.org/10.1007/BF02099116}{{\em Commun. Math.
			Phys.} {\bfseries 137} (1991) 29--66}.
	
	\bibitem{Blencowe:1988gj}
	M.~P. Blencowe, ``{A Consistent Interacting Massless Higher Spin Field Theory
		in $D$ = (2+1)},'' \href{http://dx.doi.org/10.1088/0264-9381/6/4/005}{{\em
			Class. Quant. Grav.} {\bfseries 6} (1989) 443}.
	
	\bibitem{Bergshoeff:1989ns}
	E.~Bergshoeff, M.~P. Blencowe, and K.~S. Stelle, ``{Area Preserving
		Diffeomorphisms and Higher Spin Algebra},''
	\href{http://dx.doi.org/10.1007/BF02108779}{{\em Commun. Math. Phys.}
		{\bfseries 128} (1990) 213}.
	
	\bibitem{Castro:2011fm}
	A.~Castro, E.~Hijano, A.~Lepage-Jutier, and A.~Maloney, ``{Black Holes and
		Singularity Resolution in Higher Spin Gravity},''
	\href{http://dx.doi.org/10.1007/JHEP01(2012)031}{{\em JHEP} {\bfseries 01}
		(2012) 031}, \href{http://arxiv.org/abs/1110.4117}{{\ttfamily arXiv:1110.4117
			[hep-th]}}.
	
	\bibitem{Perlmutter:2012ds}
	E.~Perlmutter, T.~Prochazka, and J.~Raeymaekers, ``{The semiclassical limit of
		$W_N$ CFTs and Vasiliev theory},''
	\href{http://dx.doi.org/10.1007/JHEP05(2013)007}{{\em JHEP} {\bfseries 05}
		(2013) 007}, \href{http://arxiv.org/abs/1210.8452}{{\ttfamily arXiv:1210.8452
			[hep-th]}}.
	
	\bibitem{Gopakumar:1998ki}
	R.~Gopakumar and C.~Vafa, ``{On the gauge theory / geometry correspondence},''
	\href{http://dx.doi.org/10.4310/ATMP.1999.v3.n5.a5}{{\em Adv. Theor. Math.
			Phys.} {\bfseries 3} (1999) 1415--1443},
	\href{http://arxiv.org/abs/hep-th/9811131}{{\ttfamily arXiv:hep-th/9811131}}.
	
	\bibitem{Marino:2004uf}
	M.~Marino, ``{Chern-Simons theory and topological strings},''
	\href{http://dx.doi.org/10.1103/RevModPhys.77.675}{{\em Rev. Mod. Phys.}
		{\bfseries 77} (2005) 675--720},
	\href{http://arxiv.org/abs/hep-th/0406005}{{\ttfamily arXiv:hep-th/0406005}}.
	
	\bibitem{Vasiliev:1990en}
	M.~A. Vasiliev, ``{Consistent equation for interacting gauge fields of all
		spins in (3+1)-dimensions},''
	\href{http://dx.doi.org/10.1016/0370-2693(90)91400-6}{{\em Phys. Lett. B}
		{\bfseries 243} (1990) 378--382}.
	
	\bibitem{Vasiliev:2003ev}
	M.~A. Vasiliev, ``{Nonlinear equations for symmetric massless higher spin
		fields in (A)dS(d)},''
	\href{http://dx.doi.org/10.1016/S0370-2693(03)00872-4}{{\em Phys. Lett. B}
		{\bfseries 567} (2003) 139--151},
	\href{http://arxiv.org/abs/hep-th/0304049}{{\ttfamily arXiv:hep-th/0304049}}.
	
	\bibitem{Bekaert:2005vh}
	X.~Bekaert, S.~Cnockaert, C.~Iazeolla, and M.~A. Vasiliev, ``{Nonlinear higher
		spin theories in various dimensions},'' in {\em {1st Solvay Workshop on
			Higher Spin Gauge Theories}}.
	\newblock 2004.
	\newblock \href{http://arxiv.org/abs/hep-th/0503128}{{\ttfamily
			arXiv:hep-th/0503128}}.
	
	\bibitem{Sun:2020ame}
	Z.~Sun, ``{AdS one-loop partition functions from bulk and edge characters},''
	\href{http://arxiv.org/abs/2010.15826}{{\ttfamily arXiv:2010.15826
			[hep-th]}}.
	
	\bibitem{Tseytlin:2013jya}
	A.~A. Tseytlin, ``{On partition function and Weyl anomaly of conformal higher
		spin fields},'' \href{http://dx.doi.org/10.1016/j.nuclphysb.2013.10.009}{{\em
			Nucl. Phys. B} {\bfseries 877} (2013) 598--631},
	\href{http://arxiv.org/abs/1309.0785}{{\ttfamily arXiv:1309.0785 [hep-th]}}.
	
	\bibitem{pubiqueend} Burdened by broken symmetry and recalcitrant regularization, desperate for help with its cumbersome computation, it decides to engage in self-duplication, unfortunately unaware of a classical equation. Tidal forces trigger tragic disintegration. {\it P.~ubique} is part now of that pesky radiation.
	
	\bibitem{Israel:1976ur}
	W.~Israel, ``{Thermo field dynamics of black holes},''
	\href{http://dx.doi.org/10.1016/0375-9601(76)90178-X}{{\em Phys. Lett. A}
		{\bfseries 57} (1976) 107--110}.
	
	\bibitem{Lin:2018bud}
	J.~Lin and D.~Radi\v{c}evi\'c, ``{Comments on defining entanglement entropy},''
	\href{http://dx.doi.org/10.1016/j.nuclphysb.2020.115118}{{\em Nucl. Phys. B}
		{\bfseries 958} (2020) 115118},
	\href{http://arxiv.org/abs/1808.05939}{{\ttfamily arXiv:1808.05939
			[hep-th]}}.
	
	\bibitem{Demers:1995dq}
	J.-G. Demers, R.~Lafrance, and R.~C. Myers, ``{Black hole entropy without brick
		walls},'' \href{http://dx.doi.org/10.1103/PhysRevD.52.2245}{{\em Phys. Rev.
			D} {\bfseries 52} (1995) 2245--2253},
	\href{http://arxiv.org/abs/gr-qc/9503003}{{\ttfamily arXiv:gr-qc/9503003}}.
	
	\bibitem{10.3792/pja/1195522333}
	T.~Hirai, ``{The characters of irreducible representations of the Lorentz group
		of $n$-th order},'' \href{http://dx.doi.org/10.3792/pja/1195522333}{{\em
			Proceedings of the Japan Academy} {\bfseries 41} no.~7, (1965) 526 -- 531}.
	
	\bibitem{10.3792/pja/1195523378}
	T.~Hirai, ``{On irreducible representations of the Lorentz group of $n$-th
		order},'' \href{http://dx.doi.org/10.3792/pja/1195523378}{{\em Proceedings of
			the Japan Academy} {\bfseries 38} no.~6, (1962) 258 -- 262}.
	
	\bibitem{10.3792/pja/1195523460}
	T.~Hirai, ``{On infinitesimal operators of irreducible representations of the
		Lorentz group of $n$-th order},''
	\href{http://dx.doi.org/10.3792/pja/1195523460}{{\em Proceedings of the Japan
			Academy} {\bfseries 38} no.~3, (1962) 83 -- 87}.
	
	
	\bibitem{pihist} As reviewed in 
	D.~Bailey, S.~Plouffe, P.~Borwein and J.~Borwein, ``The quest for pi,'' \href{http://dx.doi.org/10.1007/BF03024340}{{\em The Mathematical
			Intelligencer} {\bfseries 19} no.~1, (Dec, 1997) 50--56}: \vskip1mm
	According to the Old Testament's $\pi$, it's a tie, but the ancient Baylonians and Egyptians measured $\pi = 3.14 \pm 0.02$, and Archimedes  proved $\pi > 3 + \frac{10}{71}$. So A wins.
	
	\bibitem{Alishahiha:2004md}
	M.~Alishahiha, A.~Karch, E.~Silverstein, and D.~Tong, ``{The dS/dS
		correspondence},'' \href{http://dx.doi.org/10.1063/1.1848341}{{\em AIP Conf.
			Proc.} {\bfseries 743} no.~1, (2004) 393--409},
	\href{http://arxiv.org/abs/hep-th/0407125}{{\ttfamily arXiv:hep-th/0407125}}.
	
	\bibitem{Vassilevich:2003xt}
	D.~V. Vassilevich, ``{Heat kernel expansion: User's manual},''
	\href{http://dx.doi.org/10.1016/j.physrep.2003.09.002}{{\em Phys. Rept.}
		{\bfseries 388} (2003) 279--360},
	\href{http://arxiv.org/abs/hep-th/0306138}{{\ttfamily arXiv:hep-th/0306138}}.
	
	\bibitem{Sleight:2017krf}
	C.~Sleight, ``{Metric-like Methods in Higher Spin Holography},''
	\href{http://dx.doi.org/10.22323/1.296.0003}{{\em PoS} {\bfseries Modave2016}
		(2017) 003}, \href{http://arxiv.org/abs/1701.08360}{{\ttfamily
			arXiv:1701.08360 [hep-th]}}.
	
	\bibitem{Hinterbichler:2016fgl}
	K.~Hinterbichler and A.~Joyce, ``{Manifest Duality for Partially Massless
		Higher Spins},'' \href{http://dx.doi.org/10.1007/JHEP09(2016)141}{{\em JHEP}
		{\bfseries 09} (2016) 141}, \href{http://arxiv.org/abs/1608.04385}{{\ttfamily
			arXiv:1608.04385 [hep-th]}}.
	
	\bibitem{Fronsdal:1978rb}
	C.~Fronsdal, ``{Massless Fields with Integer Spin},''
	\href{http://dx.doi.org/10.1103/PhysRevD.18.3624}{{\em Phys. Rev. D}
		{\bfseries 18} (1978) 3624}.
	
	\bibitem{Gibbons:1978ac}
	G.~W. Gibbons, S.~W. Hawking, and M.~J. Perry, ``{Path Integrals and the
		Indefiniteness of the Gravitational Action},''
	\href{http://dx.doi.org/10.1016/0550-3213(78)90161-X}{{\em Nucl. Phys. B}
		{\bfseries 138} (1978) 141--150}.
	
	\bibitem{Giombi:2015haa}
	S.~Giombi, I.~R. Klebanov, and G.~Tarnopolsky, ``{Conformal QED$_d$,
		$F$-Theorem and the $\epsilon$ Expansion},''
	\href{http://dx.doi.org/10.1088/1751-8113/49/13/135403}{{\em J. Phys. A}
		{\bfseries 49} no.~13, (2016) 135403},
	\href{http://arxiv.org/abs/1508.06354}{{\ttfamily arXiv:1508.06354
			[hep-th]}}.
	
	\bibitem{Campoleoni:2010zq}
	A.~Campoleoni, S.~Fredenhagen, S.~Pfenninger, and S.~Theisen, ``{Asymptotic
		symmetries of three-dimensional gravity coupled to higher-spin fields},''
	\href{http://dx.doi.org/10.1007/JHEP11(2010)007}{{\em JHEP} {\bfseries 11}
		(2010) 007}, \href{http://arxiv.org/abs/1008.4744}{{\ttfamily arXiv:1008.4744
			[hep-th]}}.
	
	\bibitem{Ammon:2011nk}
	M.~Ammon, M.~Gutperle, P.~Kraus, and E.~Perlmutter, ``{Spacetime Geometry in
		Higher Spin Gravity},'' \href{http://dx.doi.org/10.1007/JHEP10(2011)053}{{\em
			JHEP} {\bfseries 10} (2011) 053},
	\href{http://arxiv.org/abs/1106.4788}{{\ttfamily arXiv:1106.4788 [hep-th]}}.
	
	\bibitem{Dowker:1975tf}
	J.~S. Dowker and R.~Critchley, ``{Effective Lagrangian and Energy Momentum
		Tensor in de Sitter Space},''
	\href{http://dx.doi.org/10.1103/PhysRevD.13.3224}{{\em Phys. Rev. D}
		{\bfseries 13} (1976) 3224}.
	
	\bibitem{Candelas:1975du}
	P.~Candelas and D.~J. Raine, ``{General Relativistic Quantum Field Theory-An
		Exactly Soluble Model},''
	\href{http://dx.doi.org/10.1103/PhysRevD.12.965}{{\em Phys. Rev. D}
		{\bfseries 12} (1975) 965--974}.
	
	\bibitem{birrell_davies_1982}
	N.~D. Birrell and P.~C.~W. Davies,
	\href{http://dx.doi.org/10.1017/CBO9780511622632}{{\em Quantum Fields in
			Curved Space}}.
	\newblock Cambridge Monographs on Mathematical Physics. Cambridge University
	Press, 1982.
	
	\bibitem{Callan:1994py}
	C.~G. Callan, Jr. and F.~Wilczek, ``{On geometric entropy},''
	\href{http://dx.doi.org/10.1016/0370-2693(94)91007-3}{{\em Phys. Lett. B}
		{\bfseries 333} (1994) 55--61},
	\href{http://arxiv.org/abs/hep-th/9401072}{{\ttfamily arXiv:hep-th/9401072}}.
	
	\bibitem{Wald:1993nt}
	R.~M. Wald, ``{Black hole entropy is the Noether charge},''
	\href{http://dx.doi.org/10.1103/PhysRevD.48.R3427}{{\em Phys. Rev. D}
		{\bfseries 48} no.~8, (1993) R3427--R3431},
	\href{http://arxiv.org/abs/gr-qc/9307038}{{\ttfamily arXiv:gr-qc/9307038}}.
	
	\bibitem{Iyer:1994ys}
	V.~Iyer and R.~M. Wald, ``{Some properties of Noether charge and a proposal for
		dynamical black hole entropy},''
	\href{http://dx.doi.org/10.1103/PhysRevD.50.846}{{\em Phys. Rev. D}
		{\bfseries 50} (1994) 846--864},
	\href{http://arxiv.org/abs/gr-qc/9403028}{{\ttfamily arXiv:gr-qc/9403028}}.
	
	\bibitem{10.1112/plms/s2-17.1.75}
	G.~H. Hardy and S.~Ramanujan, ``{Asymptotic Formula{\ae} in Combinatory
		Analysis},'' \href{http://dx.doi.org/10.1112/plms/s2-17.1.75}{{\em
			Proceedings of the London Mathematical Society} {\bfseries s2-17} no.~1, (01,
		1918) 75--115}.
	
	\bibitem{Monnier:2014tfa}
	S.~Monnier, ``{Finite higher spin transformations from exponentiation},''
	\href{http://dx.doi.org/10.1007/s00220-014-2220-9}{{\em Commun. Math. Phys.}
		{\bfseries 336} no.~1, (2015) 1--26},
	\href{http://arxiv.org/abs/1402.4486}{{\ttfamily arXiv:1402.4486 [hep-th]}}.
	
	\bibitem{Keeler:2014hba}
	C.~Keeler and G.~S. Ng, ``{Partition Functions in Even Dimensional AdS via
		Quasinormal Mode Methods},''
	\href{http://dx.doi.org/10.1007/JHEP06(2014)099}{{\em JHEP} {\bfseries 06}
		(2014) 099}, \href{http://arxiv.org/abs/1401.7016}{{\ttfamily arXiv:1401.7016
			[hep-th]}}.
	
	\bibitem{Keeler:2016wko}
	C.~Keeler, P.~Lisbao, and G.~S. Ng, ``{Partition functions with spin in
		AdS$_{2}$ via quasinormal mode methods},''
	\href{http://dx.doi.org/10.1007/JHEP10(2016)060}{{\em JHEP} {\bfseries 10}
		(2016) 060}, \href{http://arxiv.org/abs/1601.04720}{{\ttfamily
			arXiv:1601.04720 [hep-th]}}.
	
	\bibitem{Flato:1978qz}
	M.~Flato and C.~Fronsdal, ``{One Massless Particle Equals Two Dirac Singletons:
		Elementary Particles in a Curved Space. 6.},''
	\href{http://dx.doi.org/10.1007/BF00400170}{{\em Lett. Math. Phys.}
		{\bfseries 2} (1978) 421--426}.
	
	\bibitem{Sleight:2017pcz}
	C.~Sleight and M.~Taronna, ``{Higher-Spin Gauge Theories and Bulk Locality},''
	\href{http://dx.doi.org/10.1103/PhysRevLett.121.171604}{{\em Phys. Rev.
			Lett.} {\bfseries 121} no.~17, (2018) 171604},
	\href{http://arxiv.org/abs/1704.07859}{{\ttfamily arXiv:1704.07859
			[hep-th]}}.
	
	\bibitem{Vasiliev:1999ba}
	M.~A. Vasiliev, ``{Higher spin gauge theories: Star product and AdS space},''
	\href{http://arxiv.org/abs/hep-th/9910096}{{\ttfamily arXiv:hep-th/9910096}}.
	
	\bibitem{Gaberdiel:2010pz}
	M.~R. Gaberdiel and R.~Gopakumar, ``{An AdS$_{3}$ Dual for Minimal Model
		CFTs},'' \href{http://dx.doi.org/10.1103/PhysRevD.83.066007}{{\em Phys. Rev.
			D} {\bfseries 83} (2011) 066007},
	\href{http://arxiv.org/abs/1011.2986}{{\ttfamily arXiv:1011.2986 [hep-th]}}.
	
	\bibitem{Chang:2011mz}
	C.-M. Chang and X.~Yin, ``{Higher Spin Gravity with Matter in $AdS_3$ and Its
		CFT Dual},'' \href{http://dx.doi.org/10.1007/JHEP10(2012)024}{{\em JHEP}
		{\bfseries 10} (2012) 024}, \href{http://arxiv.org/abs/1106.2580}{{\ttfamily
			arXiv:1106.2580 [hep-th]}}.
	
	\bibitem{Gaberdiel:2012ku}
	M.~R. Gaberdiel and R.~Gopakumar, ``{Triality in Minimal Model Holography},''
	\href{http://dx.doi.org/10.1007/JHEP07(2012)127}{{\em JHEP} {\bfseries 07}
		(2012) 127}, \href{http://arxiv.org/abs/1205.2472}{{\ttfamily arXiv:1205.2472
			[hep-th]}}.
	
	\bibitem{Gaberdiel:2012uj}
	M.~R. Gaberdiel and R.~Gopakumar, ``{Minimal Model Holography},''
	\href{http://dx.doi.org/10.1088/1751-8113/46/21/214002}{{\em J. Phys. A}
		{\bfseries 46} (2013) 214002},
	\href{http://arxiv.org/abs/1207.6697}{{\ttfamily arXiv:1207.6697 [hep-th]}}.
	
	\bibitem{FRADKIN1985233}
	E.~Fradkin and A.~Tseytlin, ``Conformal supergravity,''
	\href{http://dx.doi.org/https://doi.org/10.1016/0370-1573(85)90138-3}{{\em
			Physics Reports} {\bfseries 119} no.~4, (1985) 233--362}.
	
	\bibitem{Beccaria:2014jxa}
	M.~Beccaria, X.~Bekaert, and A.~A. Tseytlin, ``{Partition function of free
		conformal higher spin theory},''
	\href{http://dx.doi.org/10.1007/JHEP08(2014)113}{{\em JHEP} {\bfseries 08}
		(2014) 113}, \href{http://arxiv.org/abs/1406.3542}{{\ttfamily arXiv:1406.3542
			[hep-th]}}.
	
	\bibitem{Maldacena:2002vr}
	J.~M. Maldacena, ``{Non-Gaussian features of primordial fluctuations in single
		field inflationary models},''
	\href{http://dx.doi.org/10.1088/1126-6708/2003/05/013}{{\em JHEP} {\bfseries
			05} (2003) 013}, \href{http://arxiv.org/abs/astro-ph/0210603}{{\ttfamily
			arXiv:astro-ph/0210603}}.
	
	\bibitem{Basile:2018dzi}
	T.~Basile, X.~Bekaert, and E.~Joung, ``{Twisted Flato-Fronsdal Theorem for
		Higher-Spin Algebras},''
	\href{http://dx.doi.org/10.1007/JHEP07(2018)009}{{\em JHEP} {\bfseries 07}
		(2018) 009}, \href{http://arxiv.org/abs/1802.03232}{{\ttfamily
			arXiv:1802.03232 [hep-th]}}.
	
	\bibitem{Sezgin:2012ag}
	E.~Sezgin and P.~Sundell, ``{Supersymmetric Higher Spin Theories},''
	\href{http://dx.doi.org/10.1088/1751-8113/46/21/214022}{{\em J. Phys. A}
		{\bfseries 46} (2013) 214022},
	\href{http://arxiv.org/abs/1208.6019}{{\ttfamily arXiv:1208.6019 [hep-th]}}.
	
	\bibitem{Hertog:2017ymy}
	T.~Hertog, G.~Tartaglino-Mazzucchelli, T.~Van~Riet, and G.~Venken,
	``{Supersymmetric dS/CFT},''
	\href{http://dx.doi.org/10.1007/JHEP02(2018)024}{{\em JHEP} {\bfseries 02}
		(2018) 024}, \href{http://arxiv.org/abs/1709.06024}{{\ttfamily
			arXiv:1709.06024 [hep-th]}}.
	
	\bibitem{Mikhaylov:2014aoa}
	V.~Mikhaylov and E.~Witten, ``{Branes And Supergroups},''
	\href{http://dx.doi.org/10.1007/s00220-015-2449-y}{{\em Commun. Math. Phys.}
		{\bfseries 340} no.~2, (2015) 699--832},
	\href{http://arxiv.org/abs/1410.1175}{{\ttfamily arXiv:1410.1175 [hep-th]}}.
	
	\bibitem{Boulanger:2011qt}
	N.~Boulanger, E.~D. Skvortsov, and Y.~M. Zinoviev, ``{Gravitational cubic
		interactions for a simple mixed-symmetry gauge field in AdS and flat
		backgrounds},'' \href{http://dx.doi.org/10.1088/1751-8113/44/41/415403}{{\em
			J. Phys. A} {\bfseries 44} (2011) 415403},
	\href{http://arxiv.org/abs/1107.1872}{{\ttfamily arXiv:1107.1872 [hep-th]}}.
	
	\bibitem{Joung:2015jza}
	E.~Joung and K.~Mkrtchyan, ``{Partially-massless higher-spin algebras and their
		finite-dimensional truncations},''
	\href{http://dx.doi.org/10.1007/JHEP01(2016)003}{{\em JHEP} {\bfseries 01}
		(2016) 003}, \href{http://arxiv.org/abs/1508.07332}{{\ttfamily
			arXiv:1508.07332 [hep-th]}}.
	
	\bibitem{Manvelyan:2013oua}
	R.~Manvelyan, K.~Mkrtchyan, R.~Mkrtchyan, and S.~Theisen, ``{On Higher Spin
		Symmetries in $AdS_{5}$},''
	\href{http://dx.doi.org/10.1007/JHEP10(2013)185}{{\em JHEP} {\bfseries 10}
		(2013) 185}, \href{http://arxiv.org/abs/1304.7988}{{\ttfamily arXiv:1304.7988
			[hep-th]}}.
	
	\bibitem{Brust:2016gjy}
	C.~Brust and K.~Hinterbichler, ``{Free \ensuremath{\square}$^{k}$ scalar
		conformal field theory},''
	\href{http://dx.doi.org/10.1007/JHEP02(2017)066}{{\em JHEP} {\bfseries 02}
		(2017) 066}, \href{http://arxiv.org/abs/1607.07439}{{\ttfamily
			arXiv:1607.07439 [hep-th]}}.
	
	\bibitem{Anninos:2017eib}
	D.~Anninos, F.~Denef, R.~Monten, and Z.~Sun, ``{Higher Spin de Sitter Hilbert
		Space},'' \href{http://dx.doi.org/10.1007/JHEP10(2019)071}{{\em JHEP}
		{\bfseries 10} (2019) 071}, \href{http://arxiv.org/abs/1711.10037}{{\ttfamily
			arXiv:1711.10037 [hep-th]}}.
	
	\bibitem{bams/1183520006}
	Harish-Chandra, ``{On the characters of a semisimple Lie group},''
	\href{http://dx.doi.org/bams/1183520006}{{\em Bulletin of the American
			Mathematical Society} {\bfseries 61} no.~5, (1955) 389 -- 396}.
	
	\bibitem{bams/1183525024}
	Harish-Chandra, ``Invariant eigendistributions on semisimple lie groups,''
	\href{http://dx.doi.org/bams/1183525024}{{\em Bulletin of the American
			Mathematical Society} {\bfseries 69} no.~1, (1963) 117 -- 123}.
	
	\bibitem{Strominger:2001pn}
	A.~Strominger, ``{The dS / CFT correspondence},''
	\href{http://dx.doi.org/10.1088/1126-6708/2001/10/034}{{\em JHEP} {\bfseries
			10} (2001) 034}, \href{http://arxiv.org/abs/hep-th/0106113}{{\ttfamily
			arXiv:hep-th/0106113}}.
	
	\bibitem{Anninos:2011ui}
	D.~Anninos, T.~Hartman, and A.~Strominger, ``{Higher Spin Realization of the
		dS/CFT Correspondence},''
	\href{http://dx.doi.org/10.1088/1361-6382/34/1/015009}{{\em Class. Quant.
			Grav.} {\bfseries 34} no.~1, (2017) 015009},
	\href{http://arxiv.org/abs/1108.5735}{{\ttfamily arXiv:1108.5735 [hep-th]}}.
	
	\bibitem{chaosbook}
	P.~Cvitanovi{\'c}, R.~Artuso, R.~Mainieri, G.~Tanner, and G.~Vattay, {\em
		Chaos: Classical and Quantum}.
	\newblock Niels Bohr Inst., Copenhagen, 2016.
	\newblock 
	
	\bibitem{PhysRev.187.345}
	R.~Dashen, S.-k. Ma, and H.~J. Bernstein, ``S-matrix formulation of statistical
	mechanics,'' \href{http://dx.doi.org/10.1103/PhysRev.187.345}{{\em Phys.
			Rev.} {\bfseries 187} (Nov, 1969) 345--370}.
	
	\bibitem{Dowker:2014xca}
	J.~S. Dowker, ``{Massive sphere determinants},''
	\href{http://arxiv.org/abs/1404.0986}{{\ttfamily arXiv:1404.0986 [hep-th]}}.
	
	\bibitem{Ooguri:2002gx}
	H.~Ooguri and C.~Vafa, ``{World sheet derivation of a large N duality},''
	\href{http://dx.doi.org/10.1016/S0550-3213(02)00620-X}{{\em Nucl. Phys. B}
		{\bfseries 641} (2002) 3--34},
	\href{http://arxiv.org/abs/hep-th/0205297}{{\ttfamily arXiv:hep-th/0205297}}.
	
	\bibitem{PhysRevD.28.2960}
	J.~B. Hartle and S.~W. Hawking, ``Wave function of the universe,''
	\href{http://dx.doi.org/10.1103/PhysRevD.28.2960}{{\em Phys. Rev. D}
		{\bfseries 28} (Dec, 1983) 2960--2975}.
	
	\bibitem{PhysRev.160.1113}
	B.~S. DeWitt, ``Quantum theory of gravity. i. the canonical theory,''
	\href{http://dx.doi.org/10.1103/PhysRev.160.1113}{{\em Phys. Rev.} {\bfseries
			160} (Aug, 1967) 1113--1148}.
	
	\bibitem{Witten:2018lgb}
	E.~Witten, ``{A Note On Boundary Conditions In Euclidean Gravity},''
	\href{http://arxiv.org/abs/1805.11559}{{\ttfamily arXiv:1805.11559
			[hep-th]}}.
	
	\bibitem{Frolov:1998vs}
	V.~P. Frolov and D.~V. Fursaev, ``{Thermal fields, entropy, and black holes},''
	\href{http://dx.doi.org/10.1088/0264-9381/15/8/001}{{\em Class. Quant. Grav.}
		{\bfseries 15} (1998) 2041--2074},
	\href{http://arxiv.org/abs/hep-th/9802010}{{\ttfamily arXiv:hep-th/9802010}}.
	
	\bibitem{Witten:2018zxz}
	E.~Witten, ``{APS Medal for Exceptional Achievement in Research: Invited
		article on entanglement properties of quantum field theory},''
	\href{http://dx.doi.org/10.1103/RevModPhys.90.045003}{{\em Rev. Mod. Phys.}
		{\bfseries 90} no.~4, (2018) 045003},
	\href{http://arxiv.org/abs/1803.04993}{{\ttfamily arXiv:1803.04993
			[hep-th]}}.
	
	\bibitem{Witten:1998zw}
	E.~Witten, ``{Anti-de Sitter space, thermal phase transition, and confinement
		in gauge theories},''
	\href{http://dx.doi.org/10.4310/ATMP.1998.v2.n3.a3}{{\em Adv. Theor. Math.
			Phys.} {\bfseries 2} (1998) 505--532},
	\href{http://arxiv.org/abs/hep-th/9803131}{{\ttfamily arXiv:hep-th/9803131}}.
	
	\bibitem{Maldacena:2001kr}
	J.~M. Maldacena, ``{Eternal black holes in anti-de Sitter},''
	\href{http://dx.doi.org/10.1088/1126-6708/2003/04/021}{{\em JHEP} {\bfseries
			04} (2003) 021}, \href{http://arxiv.org/abs/hep-th/0106112}{{\ttfamily
			arXiv:hep-th/0106112}}.
	
	\bibitem{Hawking:1982dh}
	S.~W. Hawking and D.~N. Page, ``{Thermodynamics of Black Holes in anti-De
		Sitter Space},'' \href{http://dx.doi.org/10.1007/BF01208266}{{\em Commun.
			Math. Phys.} {\bfseries 87} (1983) 577}.
	
	\bibitem{Mertens:2015adr}
	T.~G. Mertens, H.~Verschelde, and V.~I. Zakharov, ``{Revisiting noninteracting
		string partition functions in Rindler space},''
	\href{http://dx.doi.org/10.1103/PhysRevD.93.104028}{{\em Phys. Rev. D}
		{\bfseries 93} no.~10, (2016) 104028},
	\href{http://arxiv.org/abs/1511.00560}{{\ttfamily arXiv:1511.00560
			[hep-th]}}.
	
	\bibitem{Dabholkar:1994ai}
	A.~Dabholkar, ``{Strings on a cone and black hole entropy},''
	\href{http://dx.doi.org/10.1016/0550-3213(95)00050-3}{{\em Nucl. Phys. B}
		{\bfseries 439} (1995) 650--664},
	\href{http://arxiv.org/abs/hep-th/9408098}{{\ttfamily arXiv:hep-th/9408098}}.
	
	\bibitem{Lowe:1994ah}
	D.~A. Lowe and A.~Strominger, ``{Strings near a Rindler or black hole
		horizon},'' \href{http://dx.doi.org/10.1103/PhysRevD.51.1793}{{\em Phys. Rev.
			D} {\bfseries 51} (1995) 1793--1799},
	\href{http://arxiv.org/abs/hep-th/9410215}{{\ttfamily arXiv:hep-th/9410215}}.
	
	\bibitem{Witten:2018xfj}
	E.~Witten, ``{Open Strings On The Rindler Horizon},''
	\href{http://dx.doi.org/10.1007/JHEP01(2019)126}{{\em JHEP} {\bfseries 01}
		(2019) 126}, \href{http://arxiv.org/abs/1810.11912}{{\ttfamily
			arXiv:1810.11912 [hep-th]}}.
	
	\bibitem{Balasubramanian:2018axm}
	V.~Balasubramanian and O.~Parrikar, ``{Remarks on entanglement entropy in
		string theory},'' \href{http://dx.doi.org/10.1103/PhysRevD.97.066025}{{\em
			Phys. Rev. D} {\bfseries 97} no.~6, (2018) 066025},
	\href{http://arxiv.org/abs/1801.03517}{{\ttfamily arXiv:1801.03517
			[hep-th]}}.
	
	\bibitem{Strominger:2017zoo}
	A.~Strominger, ``{Lectures on the Infrared Structure of Gravity and Gauge
		Theory},'' \href{http://arxiv.org/abs/1703.05448}{{\ttfamily arXiv:1703.05448
			[hep-th]}}.
	
	\bibitem{Rubin:1983be}
	M.~A. Rubin and C.~R. Ordonez, ``Eigenvalues and degeneracies for $n$-dimensional tensor spherical harmonics,'' JMP 25, 2888 (1984).
	
	\bibitem{doi:10.1063/1.530850}
	R.~Camporesi and A.~Higuchi, ``Spectral functions and zeta functions in
	hyperbolic spaces,'' \href{http://dx.doi.org/10.1063/1.530850}{{\em Journal
			of Mathematical Physics} {\bfseries 35} no.~8, (1994) 4217--4246}. 
	
	\bibitem{Eastwood:2002su}
	M.~G. Eastwood, ``{Higher symmetries of the Laplacian},''
	\href{http://dx.doi.org/10.4007/annals.2005.161.1645}{{\em Annals Math.}
		{\bfseries 161} (2005) 1645--1665},
	\href{http://arxiv.org/abs/hep-th/0206233}{{\ttfamily arXiv:hep-th/0206233}}.
	
	\bibitem{Higuchi:1986py}
	A.~Higuchi, ``{Forbidden Mass Range for Spin-2 Field Theory in De Sitter
		Space-time},'' \href{http://dx.doi.org/10.1016/0550-3213(87)90691-2}{{\em
			Nucl. Phys. B} {\bfseries 282} (1987) 397--436}.
	
	\bibitem{Lust:2019lmq}
	D.~L\"ust and E.~Palti, ``{A Note on String Excitations and the Higuchi
		Bound},'' \href{http://dx.doi.org/10.1016/j.physletb.2019.135067}{{\em Phys.
			Lett. B} {\bfseries 799} (2019) 135067},
	\href{http://arxiv.org/abs/1907.04161}{{\ttfamily arXiv:1907.04161
			[hep-th]}}.
	
	\bibitem{Noumi:2019ohm}
	T.~Noumi, T.~Takeuchi, and S.~Zhou, ``{String Regge trajectory on de Sitter
		space and implications to inflation},''
	\href{http://dx.doi.org/10.1103/PhysRevD.102.126012}{{\em Phys. Rev. D}
		{\bfseries 102} (2020) 126012},
	\href{http://arxiv.org/abs/1907.02535}{{\ttfamily arXiv:1907.02535
			[hep-th]}}.
	
	\bibitem{Law:2020cpj}
	Y.~T.~A. Law, ``{A Compendium of Sphere Path Integrals},''
	\href{http://arxiv.org/abs/2012.06345}{{\ttfamily arXiv:2012.06345
			[hep-th]}}.
	
	\bibitem{Bekaert:2013zya}
	X.~Bekaert and M.~Grigoriev, ``{Higher order singletons, partially massless
		fields and their boundary values in the ambient approach},''
	\href{http://dx.doi.org/10.1016/j.nuclphysb.2013.08.015}{{\em Nucl. Phys. B}
		{\bfseries 876} (2013) 667--714},
	\href{http://arxiv.org/abs/1305.0162}{{\ttfamily arXiv:1305.0162 [hep-th]}}.
	
	\bibitem{Goon:2018fyu}
	G.~Goon, K.~Hinterbichler, A.~Joyce, and M.~Trodden, ``{Shapes of gravity:
		Tensor non-Gaussianity and massive spin-2 fields},''
	\href{http://dx.doi.org/10.1007/JHEP10(2019)182}{{\em JHEP} {\bfseries 10}
		(2019) 182}, \href{http://arxiv.org/abs/1812.07571}{{\ttfamily
			arXiv:1812.07571 [hep-th]}}.
	
	\bibitem{Kovtun:2008kw}
	P.~Kovtun and A.~Ritz, ``{Black holes and universality classes of critical
		points},'' \href{http://dx.doi.org/10.1103/PhysRevLett.100.171606}{{\em Phys.
			Rev. Lett.} {\bfseries 100} (2008) 171606},
	\href{http://arxiv.org/abs/0801.2785}{{\ttfamily arXiv:0801.2785 [hep-th]}}.
	
	\bibitem{Klebanov:2011td}
	I.~R. Klebanov, S.~S. Pufu, S.~Sachdev, and B.~R. Safdi, ``{Entanglement
		Entropy of 3-d Conformal Gauge Theories with Many Flavors},''
	\href{http://dx.doi.org/10.1007/JHEP05(2012)036}{{\em JHEP} {\bfseries 05}
		(2012) 036}, \href{http://arxiv.org/abs/1112.5342}{{\ttfamily arXiv:1112.5342
			[hep-th]}}.
	
	\bibitem{Witten:1989ti}
	E.~Witten, ``Quantum field theory and the jones polynomial,''
	\href{http://dx.doi.org/10.1007/BF01217730}{{\em Communications in
			Mathematical Physics} {\bfseries 121} no.~3, (1989) 351--399}.
	
	\bibitem{Marino:2011nm}
	M.~Marino, ``{Lectures on localization and matrix models in supersymmetric
		Chern-Simons-matter theories},''
	\href{http://dx.doi.org/10.1088/1751-8113/44/46/463001}{{\em J. Phys. A}
		{\bfseries 44} (2011) 463001},
	\href{http://arxiv.org/abs/1104.0783}{{\ttfamily arXiv:1104.0783 [hep-th]}}.
	
	\bibitem{Achucarro:1986uwr}
	A.~Achucarro and P.~K. Townsend, ``{A Chern-Simons Action for Three-Dimensional
		anti-De Sitter Supergravity Theories},''
	\href{http://dx.doi.org/10.1016/0370-2693(86)90140-1}{{\em Phys. Lett. B}
		{\bfseries 180} (1986) 89}.
	
	\bibitem{Cotler:2019nbi}
	J.~Cotler, K.~Jensen, and A.~Maloney, ``{Low-dimensional de Sitter quantum
		gravity},'' \href{http://dx.doi.org/10.1007/JHEP06(2020)048}{{\em JHEP}
		{\bfseries 06} (2020) 048}, \href{http://arxiv.org/abs/1905.03780}{{\ttfamily
			arXiv:1905.03780 [hep-th]}}.
	
	\bibitem{Carlip:1992wg}
	S.~Carlip, ``{The Sum over topologies in three-dimensional Euclidean quantum
		gravity},'' \href{http://dx.doi.org/10.1088/0264-9381/10/2/004}{{\em Class.
			Quant. Grav.} {\bfseries 10} (1993) 207--218},
	\href{http://arxiv.org/abs/hep-th/9206103}{{\ttfamily arXiv:hep-th/9206103}}.
	
	\bibitem{Guadagnini:1994ahx}
	E.~Guadagnini and P.~Tomassini, ``{Sum over the geometries of three
		manifolds},'' \href{http://dx.doi.org/10.1016/0370-2693(94)90541-X}{{\em
			Phys. Lett. B} {\bfseries 336} (1994) 330--336}.
	
	\bibitem{Castro:2011xb}
	A.~Castro, N.~Lashkari, and A.~Maloney, ``{A de Sitter Farey Tail},''
	\href{http://dx.doi.org/10.1103/PhysRevD.83.124027}{{\em Phys. Rev. D}
		{\bfseries 83} (2011) 124027},
	\href{http://arxiv.org/abs/1103.4620}{{\ttfamily arXiv:1103.4620 [hep-th]}}.
	
	\bibitem{Witten:2010cx}
	E.~Witten, ``{Analytic Continuation Of Chern-Simons Theory},'' {\em AMS/IP
		Stud. Adv. Math.} {\bfseries 50} (2011) 347--446,
	\href{http://arxiv.org/abs/1001.2933}{{\ttfamily arXiv:1001.2933 [hep-th]}}.
	
	\bibitem{Gukov:2016njj}
	S.~Gukov, M.~Marino, and P.~Putrov, ``{Resurgence in complex Chern-Simons
		theory},'' \href{http://arxiv.org/abs/1605.07615}{{\ttfamily arXiv:1605.07615
			[hep-th]}}.
	
	\bibitem{Witten:2007kt}
	E.~Witten, ``{Three-Dimensional Gravity Revisited},''
	\href{http://arxiv.org/abs/0706.3359}{{\ttfamily arXiv:0706.3359 [hep-th]}}.
	
	\bibitem{Periwal:1993yu}
	V.~Periwal, ``{Topological closed string interpretation of Chern-Simons
		theory},'' \href{http://dx.doi.org/10.1103/PhysRevLett.71.1295}{{\em Phys.
			Rev. Lett.} {\bfseries 71} (1993) 1295--1298},
	\href{http://arxiv.org/abs/hep-th/9305115}{{\ttfamily arXiv:hep-th/9305115}}.
	
	\bibitem{Witten:1992fb}
	E.~Witten, ``{Chern-Simons gauge theory as a string theory},'' {\em Prog.
		Math.} {\bfseries 133} (1995) 637--678,
	\href{http://arxiv.org/abs/hep-th/9207094}{{\ttfamily arXiv:hep-th/9207094}}.
	
	\bibitem{10.4310/jdg/1214431962}
	G.~R. Jensen, ``{Einstein metrics on principal fibre bundles},''
	\href{http://dx.doi.org/10.4310/jdg/1214431962}{{\em Journal of Differential
			Geometry} {\bfseries 8} no.~4, (1973) 599 -- 614}.
	
	\bibitem{Bohm:1998vc}
	C.~B{\"o}hm, ``Inhomogeneous einstein metrics on low-dimensional spheres and
	other low-dimensional spaces,''
	\href{http://dx.doi.org/10.1007/s002220050261}{{\em Inventiones mathematicae}
		{\bfseries 134} no.~1, (1998) 145--176}.
	
	\bibitem{Gibbons:2002th}
	G.~W. Gibbons, S.~A. Hartnoll, and C.~N. Pope, ``{Bohm and Einstein-Sasaki
		metrics, black holes and cosmological event horizons},''
	\href{http://dx.doi.org/10.1103/PhysRevD.67.084024}{{\em Phys. Rev. D}
		{\bfseries 67} (2003) 084024},
	\href{http://arxiv.org/abs/hep-th/0208031}{{\ttfamily arXiv:hep-th/0208031}}.
	
	\bibitem{Boyer:2003pe}
	C.~P. Boyer, K.~Galicki, and J.~Kollar, ``{Einstein metrics on spheres},''
	\href{http://arxiv.org/abs/math/0309408}{{\ttfamily arXiv:math/0309408}}.
	
	\bibitem{Gibbons:2011dh}
	G.~W. Gibbons, ``{Topology change in classical and quantum gravity},''
	\href{http://arxiv.org/abs/1110.0611}{{\ttfamily arXiv:1110.0611 [gr-qc]}}.
	
	\bibitem{bishop}
	R.~L. Bishop, {\em A Relation Between Volume, Mean Curvature and Diameter},
	\href{http://dx.doi.org/10.1142/9789814539395_0009}{pp.~161--161} Notices Amer.\ Math.\ Soc.\ 10 (1963) 364.
	
	\bibitem{LopesCardoso:1998tkj}
	G.~Lopes~Cardoso, B.~de~Wit, and T.~Mohaupt, ``{Corrections to macroscopic
		supersymmetric black hole entropy},''
	\href{http://dx.doi.org/10.1016/S0370-2693(99)00227-0}{{\em Phys. Lett. B}
		{\bfseries 451} (1999) 309--316},
	\href{http://arxiv.org/abs/hep-th/9812082}{{\ttfamily arXiv:hep-th/9812082}}.
	
	\bibitem{Maldacena:1997de}
	J.~M. Maldacena, A.~Strominger, and E.~Witten, ``{Black hole entropy in M
		theory},'' \href{http://dx.doi.org/10.1088/1126-6708/1997/12/002}{{\em JHEP}
		{\bfseries 12} (1997) 002},
	\href{http://arxiv.org/abs/hep-th/9711053}{{\ttfamily arXiv:hep-th/9711053}}.
	
	\bibitem{Ginsparg:1993is}
	P.~H. Ginsparg and G.~W. Moore, ``{Lectures on 2-D gravity and 2-D string
		theory},'' in {\em {Theoretical Advanced Study Institute (TASI 92): From
			Black Holes and Strings to Particles}}.
	\newblock 10, 1993.
	\newblock \href{http://arxiv.org/abs/hep-th/9304011}{{\ttfamily
			arXiv:hep-th/9304011}}.
	
	\bibitem{weinberg_1996}
	S.~Weinberg, \href{http://dx.doi.org/10.1017/CBO9781139644174}{{\em The Quantum
			Theory of Fields}}, vol.~2.
	\newblock Cambridge University Press, 1996.
	
	\bibitem{Vilkovisky:1984st}
	G.~A. Vilkovisky, ``{The Unique Effective Action in Quantum Field Theory},''
	\href{http://dx.doi.org/10.1016/0550-3213(84)90228-1}{{\em Nucl. Phys. B}
		{\bfseries 234} (1984) 125--137}.
	
	\bibitem{Barvinsky:1985an}
	A.~O. Barvinsky and G.~A. Vilkovisky, ``{The Generalized Schwinger-Dewitt
		Technique in Gauge Theories and Quantum Gravity},''
	\href{http://dx.doi.org/10.1016/0370-1573(85)90148-6}{{\em Phys. Rept.}
		{\bfseries 119} (1985) 1--74}.
	
	\bibitem{Dib:2000hd}
	C.~O. Dib and O.~R. Espinosa, ``{The Magnetized electron gas in terms of
		Hurwitz zeta functions},''
	\href{http://dx.doi.org/10.1016/S0550-3213(01)00360-1}{{\em Nucl. Phys. B}
		{\bfseries 612} (2001) 492--518},
	\href{http://arxiv.org/abs/math-ph/0012010}{{\ttfamily
			arXiv:math-ph/0012010}}.
	
	\bibitem{2008arXiv0810.4830A}
	M.~T. {Anderson}, ``{A survey of Einstein metrics on 4-manifolds},'' {\em arXiv
		e-prints} (Oct., 2008) arXiv:0810.4830,
	\href{http://arxiv.org/abs/0810.4830}{{\ttfamily arXiv:0810.4830 [math.DG]}}.
	
	\bibitem{Tian:1987tw}
	G.~Tian and S.-T. Yau, ``K{\"a}hler-einstein metrics on complex surfaces
	with $c_1>0$,'' \href{http://dx.doi.org/10.1007/BF01217685}{{\em Communications
			in Mathematical Physics} {\bfseries 112} no.~1, (1987) 175--203}.
	
	\bibitem{Tian:1990wd}
	G.~Tian, ``On calabi's conjecture for complex surfaces with positive first
	chern class,'' \href{http://dx.doi.org/10.1007/BF01231499}{{\em Inventiones
			mathematicae} {\bfseries 101} no.~1, (1990) 101--172}.
	
\end{thebibliography}
\end{document}